\documentclass[twocolumn]{aastex63}

\usepackage{lineno}
\revised{\today}
\shorttitle{GECKO: GECKO Follow-up Observation of GW190425}
\shortauthors{Paek et al.}
\graphicspath{{./}{figures/}}

\begin{document}


\title{Gravitational-wave Electromagnetic Counterpart Korean Observatory (GECKO): GECKO Follow-up Observation of GW190425}

\correspondingauthor{Myungshin Im}
\email{gregorypaek94@gmail.com, myungshin.im@gmail.com}
\author[0000-0002-6639-6533]{Gregory S.H. Paek}
\affiliation{SNU Astronomy Research Center, Astronomy Program, Department of Physics \& Astronomy, Seoul National University, 1 Gwanak-ro, Gwanak-gu, Seoul,  08826, Republic of Korea} 

\author[0000-0002-8537-6714]{Myungshin Im}
\affiliation{SNU Astronomy Research Center, Astronomy Program, Department of Physics \& Astronomy, Seoul National University, 1 Gwanak-ro, Gwanak-gu, Seoul,  08826, Republic of Korea} 
\author[0000-0002-1294-168X]{Joonho Kim}
\affiliation{SNU Astronomy Research Center, Astronomy Program, Department of Physics \& Astronomy, Seoul National University, 1 Gwanak-ro, Gwanak-gu, Seoul,  08826, Republic of Korea} 
\affiliation{Daegu National Science Museum, 20, Techno-daero 6-gil, Yuga-myeon, Dalseong-gun, Daegu 43023, Republic of Korea}

\author[0000-0002-5760-8186]{Gu Lim}
\affiliation{SNU Astronomy Research Center, Astronomy Program, Department of Physics \& Astronomy, Seoul National University, 1 Gwanak-ro, Gwanak-gu, Seoul,  08826, Republic of Korea} 
\affiliation{Department of Earth Sciences, Pusan National University, Busan 46241, Republic of Korea}

\author[0009-0008-3499-3043]{Bomi Park}
\affiliation{SNU Astronomy Research Center, Astronomy Program, Department of Physics \& Astronomy, Seoul National University, 1 Gwanak-ro, Gwanak-gu, Seoul,  08826, Republic of Korea} 
\author[0000-0001-8642-1241]{Changsu Choi}
\affiliation{SNU Astronomy Research Center, Astronomy Program, Department of Physics \& Astronomy, Seoul National University, 1 Gwanak-ro, Gwanak-gu, Seoul,  08826, Republic of Korea} 
\affiliation{Korea Astronomy and Space Science Institute, 776 Daedeokdae-ro, Yuseong-gu, Daejeon 34055, Korea}


\author[0000-0002-0070-1582]{Sophia Kim}
\affiliation{SNU Astronomy Research Center, Astronomy Program, Department of Physics \& Astronomy, Seoul National University, 1 Gwanak-ro, Gwanak-gu, Seoul,  08826, Republic of Korea} 


\author{Claudio Barbieri}
\affiliation{Universit`a degli Studi di Milano-Bicocca, I-20126 Milano, Italy}
\affiliation{INFN, Sezione di Milano-Bicocca, I-20126 Milano, Italy}

\author{Om Sharan Salafia}
\affiliation{Universit`a degli Studi di Milano-Bicocca, I-20126 Milano, Italy}
\affiliation{INFN, Sezione di Milano-Bicocca, I-20126 Milano, Italy}
\affiliation{INAF, Osservatorio Astronomico di Brera sede di Merate, I-23807 Merate, Lecco, Italy}





\author[0000-0002-4627-4002]{Insu Paek}
\affiliation{SNU Astronomy Research Center, Astronomy Program, Department of Physics \& Astronomy, Seoul National University, 1 Gwanak-ro, Gwanak-gu, Seoul,  08826, Republic of Korea} 

\author[0000-0002-2188-4832]{Suhyun Shin}
\affiliation{SNU Astronomy Research Center, Astronomy Program, Department of Physics \& Astronomy, Seoul National University, 1 Gwanak-ro, Gwanak-gu, Seoul,  08826, Republic of Korea} 

\author{Jinguk Seo}
\affiliation{SNU Astronomy Research Center, Astronomy Program, Department of Physics \& Astronomy, Seoul National University, 1 Gwanak-ro, Gwanak-gu, Seoul,  08826, Republic of Korea} 
\author[0000-0003-4412-7161]{Hyung Mok Lee}
\affiliation{Astronomy Program, FPRD, Department of Physics \& Astronomy, Seoul National University, 1 Gwanak-ro, Gwanak-gu, Seoul 08826, Republic of Korea}

\author{Chung-Uk Lee}
\affiliation{Korea Astronomy and Space Science Institute, 776 Daedeokdae-ro, Yuseong-gu, Daejeon 34055, Korea}

\author[0000-0003-0562-5643]{Seung-Lee Kim}
\affiliation{Korea Astronomy and Space Science Institute, 776 Daedeokdae-ro, Yuseong-gu, Daejeon 34055, Korea}

\author[0000-0001-9515-3584]{Hyun-Il Sung}
\affiliation{Korea Astronomy and Space Science Institute, 776 Daedeokdae-ro, Yuseong-gu, Daejeon 34055, Korea}



\begin{abstract}

One of the keys to the success of multi-messenger astronomy is the rapid identification of the electromagnetic wave counterpart, kilonova, of the gravitational wave event. Despite its importance, it is hard to find a KN associated with a GW event, due to a poorly constrained GW localization map and numerous signals that could be confused as a KN. Here, we present the Gravitational-wave Electromagnetic wave Counterpart Korean Observatory (GECKO) project, the GECKO observation of GW190425, and prospects of GECKO in the fourth observing run (O4) of the GW detectors. We outline our follow-up observation strategies during O3. In particular, we describe our galaxy-targeted observation criteria that prioritize based on galaxy properties. Armed with this strategy, we performed optical/near-infrared follow-up observation of GW190425, the first binary neutron star merger event during the O3 run. Despite a vast localization area of $\rm 7,460\:deg^2$, we observed 621 host galaxy candidates, corresponding to 29.5\% of the scores we assigned, with most of them observed within the first 3 days of the GW event. Ten transients were discovered during this search, including a new transient with a host galaxy. No plausible KN was found, but we were still able to constrain the properties of potential KNe using upper limits. The GECKO observation demonstrates that GECKO can possibly uncover a GW170817-like KN at a distance $\rm <200$ Mpc if the localization area is of the order of hundreds $\rm deg^2$, providing a bright prospect for the identification of GW EM counterparts during the O4 run.

\end{abstract}


\section{Introduction} \label{sec:intro}

Compact binary coalescence (CBC) events such as a binary neutron star (BNS) merger, binary black hole (BBH) merger, and neutron star-black hole (NSBH) merger, are the promising gravitational-wave (GW) source that the network of gravitational-wave detectors with the Advanced Laser Interferometer Gravitational-wave Observatory \citep[aLIGO;][]{2015CQGra..32g4001L}, the Advanced Virgo \citep[aVirgo;][]{2015CQGra..32b4001A}, and the Kamioka Gravitational Wave Detector \citep[KAGRA;][]{2021PTEP.2021eA101A} can detect \citep{2020LRR....23....3A}. If CBC contains more than one neutron star (NS), the thermal emission by radioactive decay of heavy elements, called the kilonova (KN), is expected \citep{2010MNRAS.406.2650M,2012ApJ...746...48M}.


On 2017 August 17, the first BNS merger event was detected by the network of GW detectors, Advanced LIGO Livingston Observatory (aLLO), Advanced LIGO Handford Observatory (aLHO), and Advanced Virgo Observatory in observing run 2 (O2) \citep{2017PhRvL.119p1101A}. About 2 seconds after the GW signal detection, a short gamma-ray burst (SGRB) was detected, and after 11 hours, the GW optical counterpart was found in the massive galaxy, NGC 4993 \citep{2017ApJ...848L..19C,2017Sci...358.1556C,2017ApJ...848L..17C,2017Sci...358.1570D,2017Sci...358.1565E,2017Sci...358.1559K,2017Sci...358.1583K,2017ApJ...850L...1L,2017ApJ...848L..32M,2017Natur.551...67P,2017Sci...358.1574S,2017Natur.551...75S,2017PASJ...69..101U}. The identified electromagnetic wave (EM) counterpart was followed up by various observatories at different wavelengths \citep{2017ApJ...848L..12A}. 

This historical GW event opened a new way of exploring the universe, namely, GW multi-messenger astronomy (MMA). Above all, the optical/near-infrared (NIR) counterpart pinpointed the exact position of the GW source, providing the link between the BNS merger, SGRB, KN, and, GW. Not only that, the environment of the GW source was studied for the first time \citep{2017ApJ...849L..16I,2017ApJ...848L..28L}, also prompting the investigations of the host galaxies through simulations \citep{2018MNRAS.481.5324M,2019MNRAS.487.1675A} and SGRB host galaxies \citep{2022ApJ...940...57N,2022ApJ...940...56F}. The formation of lanthanide elements has been witnessed \citep{2017ApJ...848L..19C}. Furthermore, attempts have been made to measure the Hubble constant with GW standard siren technique \citep{2017Natur.551...85A}, through the combined information of the luminosity distance from the result of GW analysis and redshift measurement from optical spectroscopy. Such measurements from many GW events could potentially contribute to resolving the Hubble tension \citep{1986Natur.323..310S,2005ApJ...629...15H, 2019NatAs...3..891V}.

Despite the great opportunities of MMA, GW 170817 has remained the only GW event for which MMA studies were possible. This is because there are several main obstacles to identifying the GW EM counterpart. First, KNe are $\rm \sim 3\:mag$ fainter than Type Ia SN, and it decays fast ($\rm \sim 1\:mag/day$) \citep{2017Natur.551...64A,2017Natur.551...80K,2019MNRAS.489.5037B}. Second, Due to the current precision of GW detectors to date, the GW localization areas typically range from hundreds to thousands $\rm deg^{2}$ depending on the distance of the GW source and duty cycles of GW detectors \citep{2020LRR....23....3A}. In the O4 run, it is expected that the GW localization will improve to tens to hundreds of $\rm deg^{2}$ \citep{2020LRR....23....3A}. However, such localization areas are much larger than the typical field of view (FOV) of optical telescopes ($\rm FOV < 1\:deg^{2}$). 

In order to catch KNe associated with GW sources, we have established a global network of telescopes, GW EM-Counterpart Korean Observatories \citep[GECKO;][]{2020grbg.conf...25I}. GECKO is made of nearly twenty telescopes around the world. This allows us to perform follow-up observations both at the northern and southern hemispheres 24 hrs a day. Several GECKO facilities are wide-field telescopes (FOV $\rm > 1\:deg^{2}$), capable of covering wide localization areas of GW events. The rest are narrow-field telescopes of 0.35 - 1.5 $\rm m$ aperture sizes. We have used the GECKO telescopes during the O2 and O3 runs for follow-up observations of GW sources, including studies of GW 170817 \citep{2017ApJ...849L..16I,2017Natur.551...71T,2021ApJ...916...47K}.

GW190425 is a CBC GW event that was detected on 2019-04-25 at 08:18:05 UT \citep{2020ApJ...892L...3A}. This event was interpreted as involving either two neutron stars or one neutron star and a black hole, yet was detected with a poorly defined preliminary localization region of $\rm 10,182 \, deg^{2}$ with a 90\% confidence level and a distance of $\rm 156 \pm 45 \, Mpc$. At that time, aLHO was offline, so only aLLO detected the GW signal with the sub-threshold detection with the Virgo detector. Therefore, the GW skymap localization was dominated by the antenna pattern of LLO and, as a result, poorly constrained with a localization area of thousands of square degrees. Nonetheless, the False alarm rate (FAR) is only about 1 per $\rm 7 \times 10^{4}\:yr$, showing that the event was due to an astrophysical source. With GECKO, we performed intensive follow-up observations of this event \citep{2019GCN.24183....1I,2019GCN.24188....1P,2019GCN.24216....1K,2019GCN.24318....1I}. Combined with other global efforts to find the KN, GW190425 provided an excellent opportunity to employ GECKO to find the KN associated with the event and place useful constraints on its physical properties. 
Unfortunately, there has not been a reported significant counterpart of GW190425 so far \citep{2019ApJ...885L..19C,2019ApJ...880L...4H}.

In this paper, we describe the observational strategy employed during the O3 run by GECKO and report the results of the GW190425 follow-up observation. We will also show prospects of future GECKO observations, including what kind of constraints we could have imposed on the KN properties (e.g., ejecta mass and velocities) associated with GW190425 with improved observing strategies. Our GECKO O3 run observation strategy is now modified for the O4 run, but the strategy described here serves as the basis for the O4 and future run GECKO strategies.

In Section \ref{sec:GECKO}, we introduce the GECKO project. In Section \ref{sec:observation}, we present the strategy for rapid identification of GW EM counterpart. In Section \ref{sec:GW190425} and \ref{sec:transient}, we show the result of the follow-up observation of GW190425, including the transients identified in the area covered by us. In Section \ref{sec:nature}, we compare the upper limits of our follow-up observation with GW170817-like KN models and discuss the nature of the transients observed by the GECKO facilities. Section \ref{sec:prospect} discusses the prospects of the GECKO observations for future GW events, such as the expected area coverage by GECKO and the constraints we could put on KN properties even if there is no detection. In Section \ref{sec:summary}, we provide a summary and discussion of this work. In this paper, we adopt the AB magnitude system unless described otherwise.

\section{GECKO} \label{sec:GECKO}
\begin{figure}[ht!]
\plotone{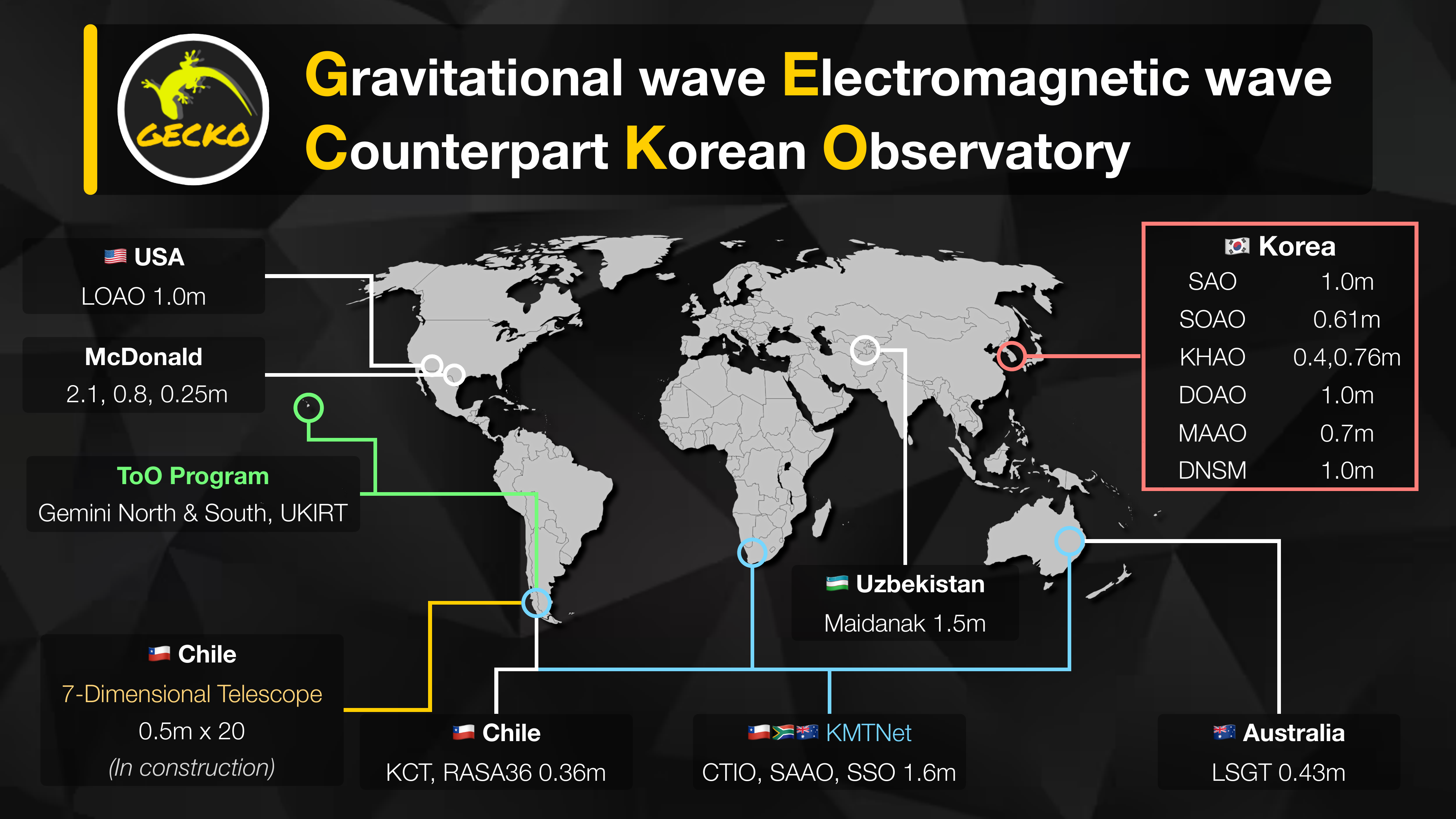}
\caption{The locations of the telescopes in the GECKO project.
\label{fig:gecko}}

\end{figure}

The GECKO facilities are shown in Figure \ref{fig:gecko} and summarized in Table \ref{tab:facility}. Many of the GECKO telescopes are a part of the Intensive Monitoring Survey of Nearby Galaxies \citep[IMSNG;][]{2015JKAS...48..207I} program that monitors nearby galaxies with a high cadence ($\rm <1\:day$) to catch early light curve (LC) of supernovae (SNe) and they are described in detail in \citet{2015JKAS...48..207I} and \citet{2019JKAS...52...11I}.


One noteworthy facility is the Korea Microlensing Telescope Network \citep[KMTNet;][]{2016JKAS...49...37K}. KMTNet consists of three 1.6-m telescopes having a wide FOV ($\rm \sim4.0\:deg^{2}$) located at three different longitudes in the southern hemisphere --- Cerro Tololo Inter-American Observatory (Chile, KMTNet-CTIO), South Africa Astronomical Observatory (South Africa, KMTNet-SAAO), and Siding Spring Observatory (Australia, KMTNet-SSO). Its primary goal is to find exoplanets in the Galactic bulge by adopting the gravitational microlensing technique \citep{2016JKAS...49...37K}. The large aperture, wide FOV, and continuous coverage of the southern sky are also suitable for GW optical follow-up and hence, we have been using KMTNet for GW follow-up observations \citep{2017Natur.551...71T,2021ApJ...916...47K} in the O2 through O4 runs. 

One of the most powerful advantages of our facilities in the GECKO project is the network of various telescopes located in wide ranges of latitudes and longitudes. This network allows us to promptly respond to the GW follow-up observations, irrespective of the location and trigger time denoted on the GW skymap.

GECKO facilities include spectroscopic instruments. We have facilities with medium-band observation capabilities that can provide low-resolution spectra of the target galaxy. LSGT \citep{2015JKAS...48..207I,2017JKAS...50...71C} and SQUEAN on the 2.1m telescope \citep{2016PASP..128k5004K} and WIT \citep{2018AAS...23144203J} at the McDonald Observatory have such capabilities. We are also constructing a new multiple telescope system in Chile, the 7-Dimensional Telescope (7DT) in Chile (M. Im et al., in preparation). 7DT will have wide field IFU-type low-resolution spectral mapping capabilities and is capable of classifying KNe from other transients using a single epoch data (G. S. H. Paek et al., in preparation). We have been regularly accessing the Gemini telescopes for deep spectroscopy and used the United Kingdom Infrared Telescope in the past for NIR follow-up of GW sources during the O3 run.

\section{Observing Strategy} \label{sec:observation}

We adopt two follow-up strategies for searching for the GW EM counterpart. First, galaxy-targeted observation targets host galaxy candidates where the GW source may exist \citep[e.g.,][]{2012ApJ...759...22K,2016ApJ...820..136G,2017PASJ...69....9Y,2017ApJ...848L..33A,2019ApJ...875...59Y,2021PTEP.2021eA104S}. We adopt this strategy mainly for narrow FOV telescopes. Second, tiling observation covers the GW localization area with wide FOV telescopes as much as possible \citep[e.g.,][]{2019ApJS..245...15K,2019ApJ...881L..26L,2020ApJ...894..127W,2021PASJ...73..350O,2023ApJ...947....9O}. For poorly constrained GW localization, tiling observation is more efficient than galaxy-targeted observation. More details about the tiling observation strategy with KMTNet are in \citep{2021ApJ...916...47K}. In this paper, we focus on describing the galaxy-targeted observation strategy that was used in following up the GW190425. 

\subsection{GW host galaxy candidate selection} \label{subsec:selection}
We developed an automatic pipeline to extract the information about the GW source for arranging the optical follow-up observation. A preliminary analysis of GW signal is distributed through $\texttt{PyGCN}$\footnote{\url{https://github.com/nasa-gcn/pygcn}} as a HEALPix format \citep{2005ApJ...622..759G}. First, the pipeline determines whether to follow up based on several factors: the inclusion of a neutron star in the GW progenitor system, GW distance ranges ($\rm d < 500\:Mpc$), false alarm rate ($\rm FAR < 1/yr$), and the size of GW localization ($\rm S_{gw} < 10,000\:deg^{2}$). For instance, a GW event that involves a progenitor system with at least one NS to potentially produce a bright EM counterpart would be considered a high-priority candidate for follow-up observation.

Next, we limit the host galaxy candidates to be in the volume within the 90\% confidence region of the GW localization area and $\rm \pm{3}\: \sigma$ of the estimated GW luminosity distance. Specifically, we select these candidates from the GLADE catalog \citep{2018MNRAS.479.2374D}, which was implemented during O3 and is useful for EM follow-up observations, containing information for 3 million galaxies. Depending on the number of host galaxy candidates, the spatial constraint was adjusted. For example, if the number of host galaxy candidates is more than thousands, we loosen the limit of the confidence area from 90\% to 50\%.

To prioritize which galaxies to observe first, understanding galaxy properties that correlate with BNS merger rates is important. However, the properties of the galaxy that control the BNS merger rates are yet uncertain. Previous studies used the B-band luminosity of the galaxy as the proxy for both the star formation rate (SFR) \citep{2017Sci...358.1556C}, based on theoretical approaches that higher SFR has a higher BNS merger rate \citep{1991ApJ...380L..17P,2002ApJ...572..407B}. From NGC 4993, the host galaxy of GW170817, one may guess that massive, and quiescent galaxies are host galaxies of BNS merger. Indeed, recent simulation works, which incorporate stellar evolutionary models into galaxy simulation, along with observational studies of SGRBs, suggest that the stellar mass of the host galaxy seems to be an important parameter controlling the BNS merger rate \citep{2018ApJ...856..101M,2019MNRAS.487.1675A,2020MNRAS.495.1841A,2020MNRAS.491.3419A}. However, we caution that the observational studies of SGRB host galaxies show that many of them ($\sim 84 \%$) are star-forming galaxies and other parameters such as SFR may be as important as the stellar mass in controlling the BNS merger rate (\citealt{2022ApJ...940...57N}; M. Jeong et al., in preparation). Given these considerations and no readily available SFRs in the GLADE catalog, we consider the stellar mass to be proportional to the BNS merger rate for the target prioritization. Since the stellar mass was not available in the GLADE catalog at the time of the observations, we use K-band luminosity which is known as a proxy for the stellar mass \citep[e.g.][]{2012AAS...21944127K}.

\begin{figure}
\plotone{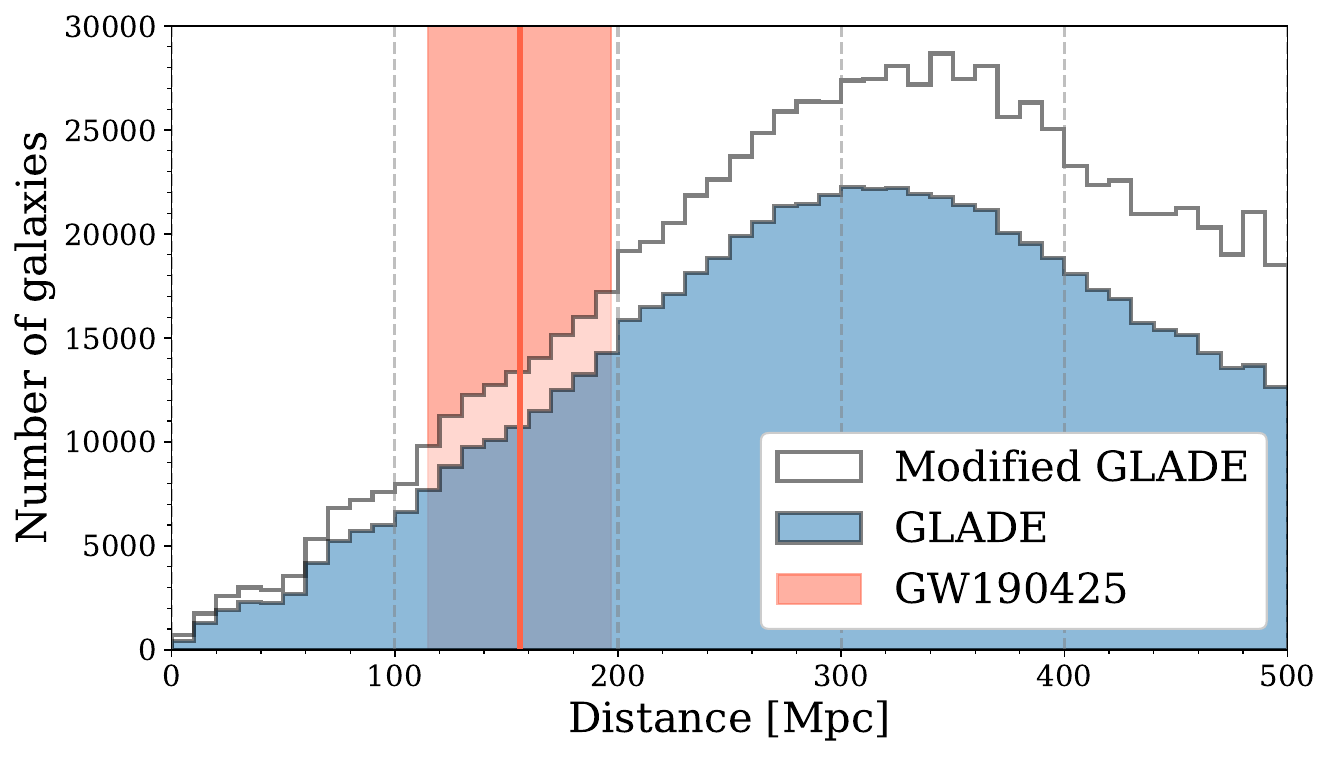}
\caption{Distance histogram of galaxies selected for the galaxy-targeted observation. The blue-shaded region shows galaxies with the $K$-band data originally in the GLADE catalog and the hollow region shows galaxies in the modified GLADE catalog with newly added K-band information from 2MASS PSC detailed in Section \ref{subsec:selection}. The red line and red shaded region represent the mean distance and distance range of GW190425, $\rm 156\pm{41} Mpc$. Galaxies in the histogram in the red-shaded area were prioritized for the galaxy-targeted observation.
\label{fig:glade}}
\end{figure}

The $K$-band luminosity is from the 2 Micron All-Sky Survey Extended Source Catalog (2MASS XSC; \citealt{2000AJ....119.2498J}) and hence many sources miss $K$-band luminosities. Therefore, we additionally match the GLADE sources without $K$-band luminosity entries  with the 2MASS Point Source Catalog (2MASS PSC; \citealt{2006AJ....131.1163S}) to implement the missing information. $K$-band magnitudes are cross-matched utilizing $\texttt{astroquery}$\footnote{\url{https://github.com/astropy/astroquery}} \citep{2019AJ....157...98G}. As a result, 798,796 additional $K$-band magnitudes are added to the GLADE catalog, among which 21,242 galaxies are at $\rm 156 \pm{41\:Mpc}$, the distance range of GW190425 (Figure \ref{fig:glade}). 

\subsection{Score} \label{subsec:score}
We assign a score (S) to each galaxy in the search volume by giving weights based on $K$-band luminosity and the localization probability at the position of the galaxy. Besides $K$-band luminosity being a good proxy of stellar mass, the $K$-band has the advantage of being less affected by dust extinction.

We modify the equation form for the ranking system of DLT40 \citep{2017Sci...358.1556C} and replace the $B$-band luminosity term with the $K$-band luminosity term. We calculated the $K$-band proxy, $\rm \tilde{L}_{K}$, for each host galaxy candidate.
\begin{equation}
\rm \tilde{L}_{K} = D_{gal}^{2}\:10^{-0.4\:m_{K}} ,
\end{equation}
where $\rm D_{gal}$ represents the luminosity distance to the galaxy and $\rm m_{K}$ stands for the $K$-band magnitude uncorrected for Milky Way (MW) extinction.
\begin{equation}
\rm S=k^{-1} \times P_{2D} \times \left( 1-erf \left( \frac{ \left\vert D_{gal} - D_{GW} \right\vert }{ \sigma_{D,gal}^2 + \sigma_{D,GW}^2 } \right) \right) \times \tilde{L}_{K} ,
\end{equation}

where $\rm k^{-1}$ is the normalization factor, $\rm P_{2D}$ represents the localization map probability at the position of the galaxy, $\rm D_{GW}$ is the mean luminosity distance of the GW, $\rm \sigma_{D,gal}$ is the uncertainty of the galaxy's distance, $\rm \sigma_{D,GW}$ is the standard deviation of the GW luminosity distance.

Since not all objects in the catalog have $K$-band magnitudes even after supplementing the catalog with the additional $K$-band data, we divide the objects into three priority groups. First, if both $K$-band and distance information are available, we consider them as the priority 1 group and calculate their score using equation 2. Second, if galaxies do not have the $K$-band magnitude but have the $B$-band and distance information, we consider them as the priority 2 group and convert the $K$-band luminosity term to $B$-band term luminosity on equation (1) and then, calculate their scores using equation (2). This process of the normalization of the score is added since the $B$-band luminosity is not as good as the $K$-band luminosity as a mass proxy and the luminosity scale of $B$-band and $K$-band data are vastly different. Furthermore, galaxies are likely to be low-mass objects if the $K$-band is not available, and such objects have a lower probability of hosting KN events than high-mass objects. We normalize the score of the priority 2 group objects with the lowest score of the priority 1 objects. Third, if objects have only distance information available, we consider them as the priority 3 group objects. The luminosity term on equation (2) is set by 1, and similar to the priority 2 group, we normalize the score with the lowest score of the priority 2 objects. As a result, we can assign scores to all galaxies with the distance information in the galaxy catalog. 

Note that we did not apply the MW dust extinction correction to the $K$-band luminosity. Our decision was based on the low impact of dust extinction in the NIR range, as our analysis showed negligible differences in the scores with or without the correction. Additionally, uncertainties introduced by using the $K$-band luminosity as a stellar mass proxy, and the employment of the 2MASS PSC catalog, are significant enough to make the correction for the MW extinction less critical. Nevertheless, we plan to incorporate $K$-band luminosity for MW extinction in future work.

\subsection{Application to GW170817} \label{subsec:gw170817}
\begin{figure}
\plotone{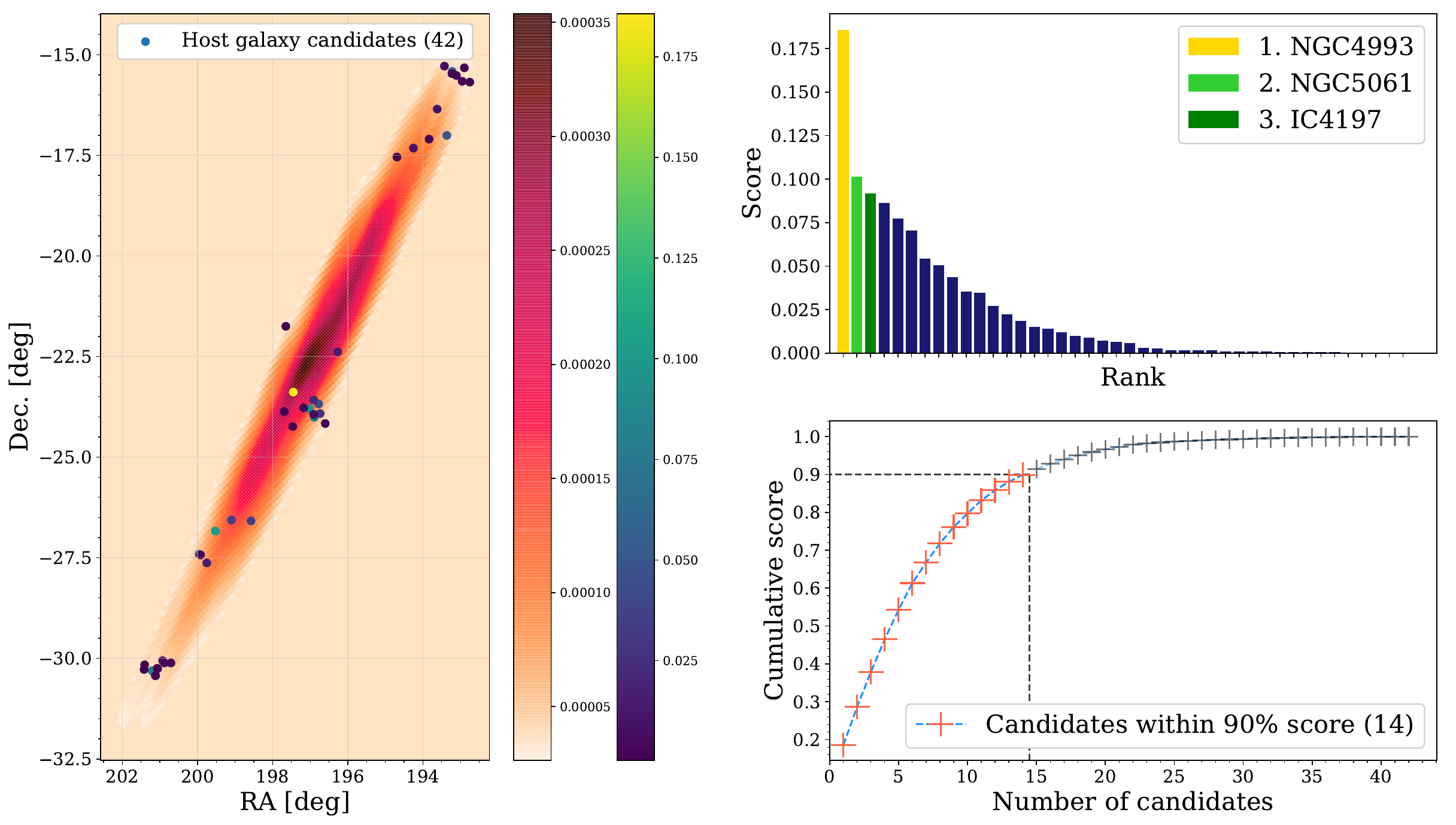}
\caption{Host galaxy selection and prioritization for GW170817 using the method described in Section \ref{subsec:gw170817}. (Left) 90\% confidence region of skymap localization (shaded region) and 42 host galaxy candidates (solid circles) are shown. Host galaxy candidates are selected within this constrained region and $\rm 3\sigma$ of distance range. (Right) The top bar graph shows the rank of host galaxy candidates. Including second and third-ranked host candidates, the first-rank NGC4993 is presented with different colors. (Left) The bottom shows the cumulative score distribution. The 14 highest scores galaxies cover more than 90\% of the cumulative score. To discriminate them from others, They are colored with a blue dotted line and red cross point.
\label{fig:gw170817}}
\end{figure}

As a validation, we applied our prioritization method to the case of GW170817 (Figure \ref{fig:gw170817}). The number of host galaxy candidates falling within 3 standard deviations from the GW-based mean distance and a 90\% confidence level of GW skymap localization amounts to 42 (Table \ref{tab:gw170817}). We assign scores to them, and the galaxy with the highest score is NGC4993 where the EM counterpart of GW170817 was found. Cumulative score distribution shows the top one-third of GW host galaxy candidates (14) cover more than 90\% of the total score. Therefore, we expect that our scoring system will be efficient for finding the EM counterpart of the GW170817-like events.
\section{Follow-up observation of GW190425} \label{sec:GW190425}
\begin{figure}
\plotone{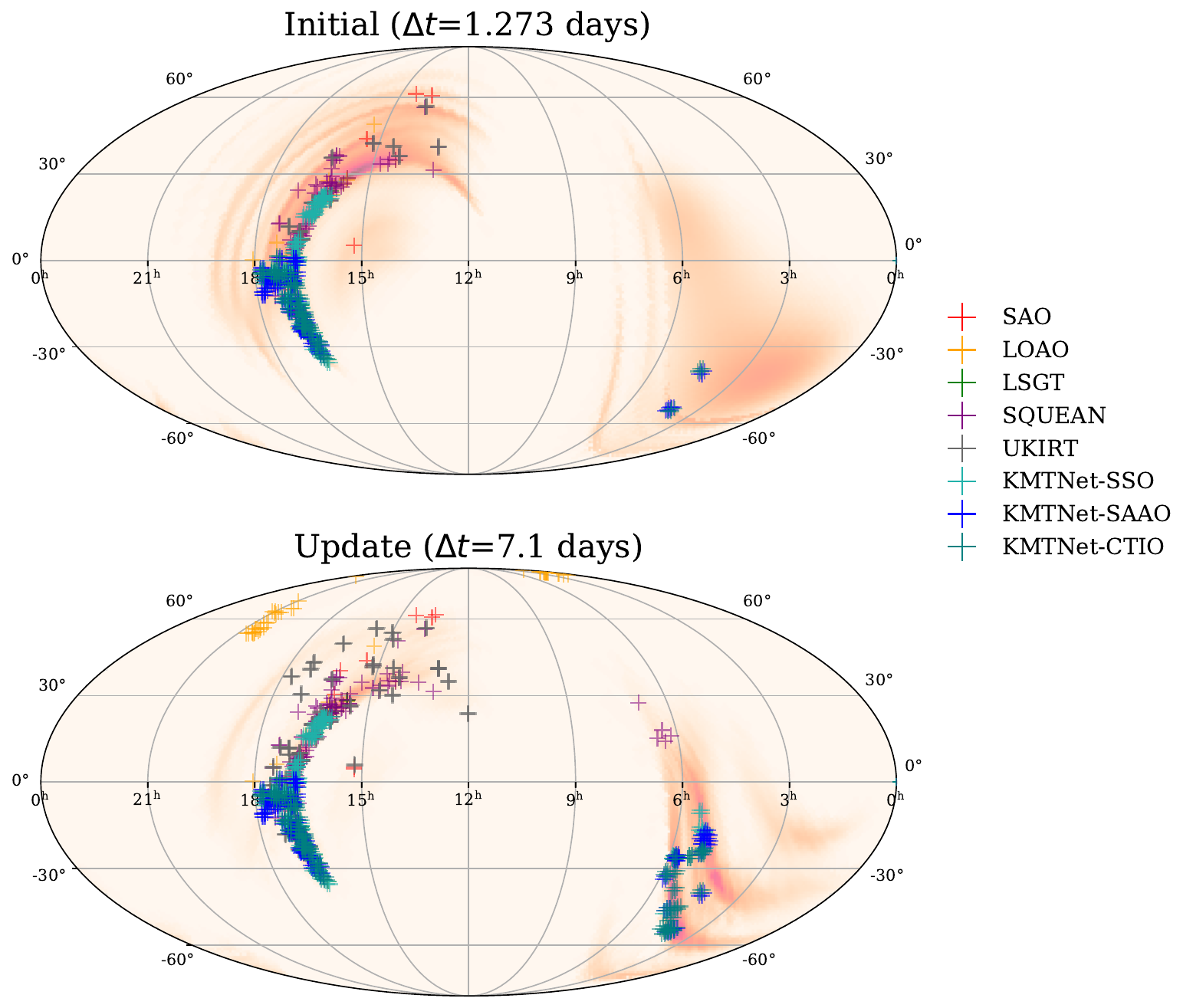}
\caption{Skymap localization of GW190425 and follow-up coverage from GECKO project. The colored cross shows the pointing coordinate on the skymap localization for each observatory. (Top) Initial skymap localization and coverage until 1.273 days after the GW event emerged are shown. (Bottom) Updated skymap localization area and coverage until 7.1 days after GW event emerged when follow-up observation was ended are shown. The dominant region is changed to the Southern Hemisphere.
\label{fig:gw190425}}
\end{figure}

In this section, we present the results of the follow-up observation of GW190425 during the O3 run. For this event, there were several versions of the localization maps, with two noteworthy maps - one is the initial localization map, and another is the final updated localization map (Figure \ref{fig:gw190425}). The updated localization map was distributed 30 hours after the initial signal distribution. There are two main differences between them. First, the localization region is constrained more tightly from $\rm 10,182\ deg^2$ to $\rm 7,460\ deg^2$. Second, the most probable region is shifted from the Northern hemisphere to the Southern hemisphere. So, after the updated signal, we focused our follow-up observation on the Southern hemisphere. Automatic pipeline selected and prioritized 5,419 (7,639) GW host galaxy candidates for the updated (initial) as described in Section \ref{subsec:selection} and \ref{subsec:score}.
\subsection{Observation} \label{subsec:observation}

We initiated the follow-up observation 90 minutes after the GW trigger, using the SQUEAN instrument on the McDonald Observatory's 2.1-m telescope in $i$-band. The starting time of GECKO observation is similar to those of other facilities \citep[e.g.][]{2019ApJ...880L...4H,2019ApJ...885L..19C}. This was followed by observations with LOAO, KMTNet-SSO, LSGT \citep{2015JKAS...48..207I}, UKIRT, KMTNet-SAAO, KMTNet-CTIO, and SAO \citep{2021JKAS...54...89I} following the GECKO observation strategy. All these facilities adopted the galaxy-targeted observation strategy. Given the reddening of the KN \citep{2017Natur.551...80K} and the sensitivity of our instruments, we chose to use redder optical and NIR wavelength filters available on each telescope ($R$-band; SAO, LOAO, KMTNet, $r$-band; LSGT, $i$-band; SQUEAN, $J$, and $K$-bands). The details of the observation are summarized in Table \ref{tab:mediandepth}.

We managed to cover 621 (649) out of the 5,419 (7,639) host galaxy candidates for the updated (initial) GW signal. While we only covered 11.5\% in terms of the number of objects, the portion of the total score covered was higher, at 29.5\%. Also, we calculated the area coverage of our observations by matching image WCS information with the HEALPix map of the GW event localization. We adopted this approach for two main reasons: First, some FOVs lie outside the localization area (partially or fully); Second, some FOVs overlap with each other. One pixel of HEALPix corresponds to 0.052$\rm deg^{2}$. Most telescopes have FOVs comparable with or larger than the HEALPix map resolution, except for SQUEAN. Observation pointings of SQUEAN are within the 90\% confidence area of GW190425, so we count the number of pointings and multiply it by the FOV of SQUEAN. According to this calculation, we covered about 410 $\rm deg^2$ (393 $\rm deg^2$). KMTNet observation makes up most of the total coverage ($\rm <$97\%). 

The overlap between host galaxy candidates identified in the initial and updated GW signals is noteworthy. Specifically, there are 4,567 overlapping candidates, which constitute 84.3\% of the total in the updated GW signal. Out of these, 562 were actually observed, accounting for 90.5\% of the observed candidates in the updated signal. In terms of the northern hemisphere, the most probable area showed no significant difference between the initial and updated localizations. Considering that we observed 649 galaxies in the GW localization area, we wasted about 15\% of the time observing galaxies in the area of the initial alert.

\begin{figure}
\plotone{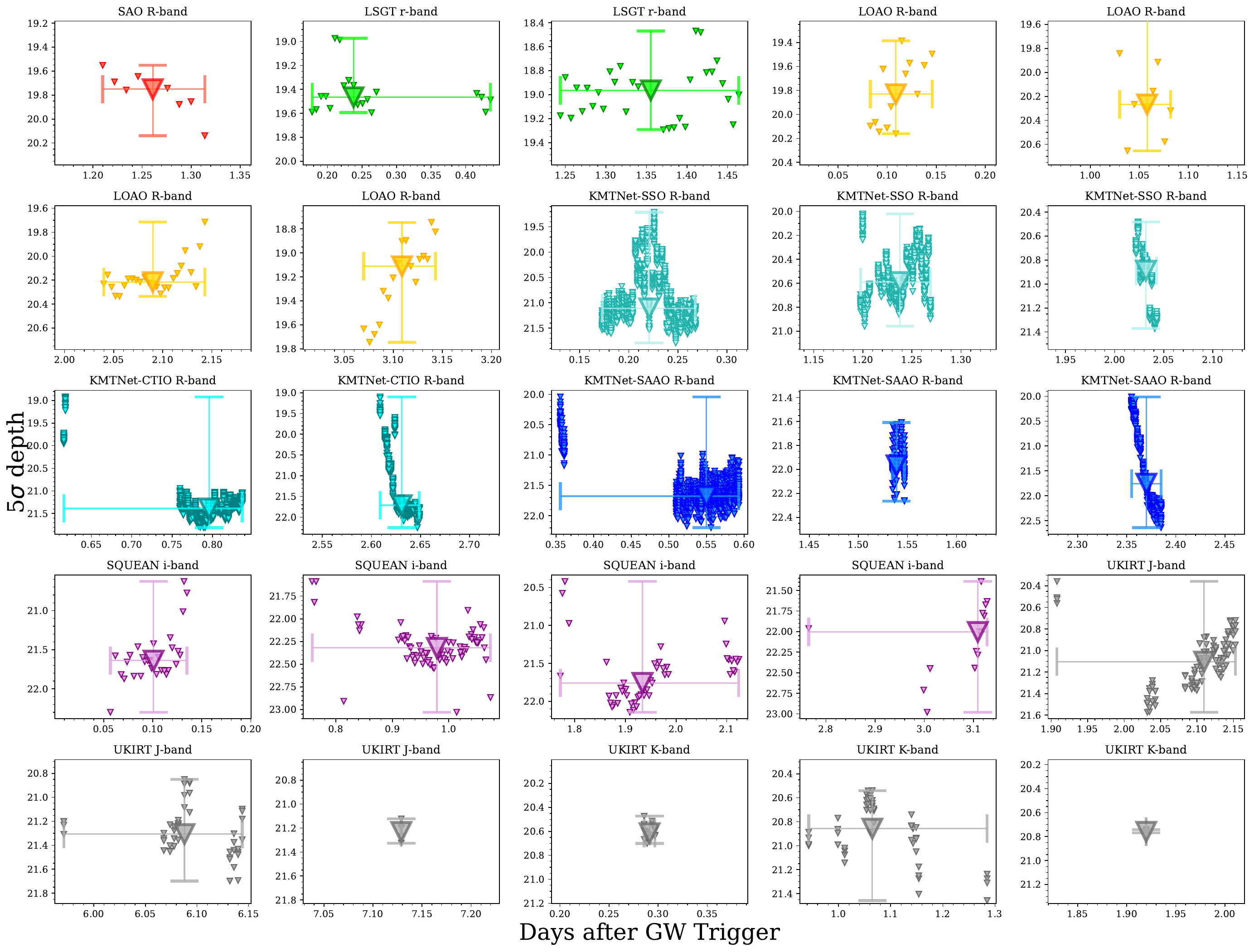}
\caption{5$\sigma$ depth variation over a single night for follow-up observation for GW190425 for each observatory in the GECKO project. Small points show a single observation pointing. Large points show the median value of all observation time at night and 5 $\sigma$ depth and the errorbar represents the range from minimum to maximum for time and depth. Note that the KMTNet image consists of 4 chips and they consist of 8 channels, so there are 32 similar depths at the same epoch.
\label{fig:obs}}
\end{figure}

Each observation pointing is shown in Figure \ref{fig:gw190425}. Figure \ref{fig:obs} summarizes the point-source $5\sigma$ detection limits of different facilities at various epochs. The varying depths for the same facilities are due to varying weather conditions. In particular, we started the second KMTNet-CTIO and the third KMTNet-SAAO observations during twilight, which shows up as gradual increases in the image depths. For some facilities, primarily KMTNet, there exist gaps in the observation times due to the scheduling of other observation programs with higher priorities. 

We also performed follow-up observations on three potential KN candidates reported by other groups. ZTF19aarykkb and ZTF19aarzaod reported by Zwicky Transient Facility \citep{2019GCN.24191....1K} were observed on 2019-04-26 08:46:29 UTC and 2019-04-26-08:38:01 UTC respectively by SQUEAN, and PS19qp reported by Pan-STARRs \citep{2019GCN.24210....1S} was observed on 2019-04-26 15:09:15 UTC by UKIRT. However, they all turned out to be SNe (ZTF19aarykkb; \citealt{2019GCN.24204....1P,2019GCN.24220....1D, 2019GCN.24260....1C}, ZTF19aarzaod; \citealt{2019GCN.24205....1B,2019GCN.24209....1W}, PS19qp; \citealt{2019GCN.24358....1D,2019GCN.24295....1M,2019GCN.24241....1L,2019GCN.24230....1M}).

\subsection{Data Reduction and Photometry} \label{subsec:photometry}
We utilized the Python library \texttt{ccdproc}\footnote{\url{https://ccdproc.readthedocs.io/en/latest}} to manage the basic calibrations of our optical data. This library facilitated effective and efficient handling of basic preprocessing tasks such as bias and dark subtraction, as well as flat field correction. We conducted astrometry using \texttt{astrometry.net}\footnote{\url{https://astrometry.net}}. The whole procedure was executed within the automatic image process pipeline \citep[\texttt{gpPy};][]{gppy}.

The object detection and photometry were done using SExtractor \citep[SE;][]{1996A&AS..117..393B}. Here, we performed photometric calibration using stars cataloged in Pan-STARRS1 (PS1) $\rm 3\pi$ survey \citep{2016arXiv161205560C} and the AAVSO Photometric All-Sky Survey \citep{2009AAS...21440702H}. For the UKIRT NIR image, we use point sources in the 2MASS PSC. 

In order to calculate the zero-points (ZP) using reference stars, we applied several selection criteria to choose the most suitable stars from our image. We select the stars with the following criteria:

\begin{itemize}
\item The stars in the reference catalog should have magnitude ranges not to be saturated.

\item The stars should have magnitude errors not exceeding 0.05 in both the reference catalog and the science image.

\item From the SE output, we selected stars that had a \texttt{FLAGS} value of 0, implying they are not saturated or blended by another source.

\item The chosen stars should be located within a radius equivalent to 75\% of the image size from the image center.
\end{itemize}

In cases where the image lacked an adequate number of stars that meet the specified conditions, we employed a more relaxed criterion to include additional stars for the calibration process. 

For the KMTNet images, we have taken an additional step to determine the image ZPs. KMTNet images are composed of four CCD chips and eight readout channels. Each readout channel has similar background levels and $\rm sigma$ \citep{2016JKAS...49...37K}, but ZPs are slightly different from each other. Therefore, we split the 8 channels and calculated ZPs for each of them.

Since we are interested in point source detection and photometry, we adopt the aperture magnitude for photometry. The aperture diameter is defined twice seeing measured by SE which gives a nearly optimal signal-to-noise ratio (SNR).

To calculate the 5$\sigma$ depth for the point source detection, we use the following expression:

$$\rm (5 \sigma \: depth) \:=\: ZP\: - 2.5\: \log{ \:5 \sigma \sqrt{{ \pi r^{2}} } }, $$

where $\rm ZP$ is the zero point, $\rm \sigma$ is the standard deviation of the sky background, and $\rm r$ means the aperture radius, which corresponds to the seeing FWHM value.

\section{Transients from GECKO Observation} \label{sec:transient}
We conducted image subtraction with reference images taken in $2009-2012$ from the PS1 survey using $\texttt{HOTPANTS}$ software \citep{2015ascl.soft04004B} for all the images taken except for UKIRT and KMTNet images at $Dec < -30$ degree where the coverage of PS1 survey. For the KMTNet images taken at $Dec < -30$ degree, images taken at later epochs (a few days) were used as reference images. Then, we searched each subtracted image for transient candidates by detecting sources in the subtraction images and visually inspecting them. During this process, many non-astronomical detections (bogus) were made due to imperfect subtraction and cosmic rays. To remove them, we adopted the following procedures:

\begin{enumerate}
\item To remove cosmic rays and signals arising from diffraction spikes, objects were excluded if their elongations or ellipticities were larger than the corresponding mean values plus the standard deviation of point sources in the image catalog.

\item Then, we selected sources in the subtracted image using the SE \texttt{FLAGS} parameter (\texttt{FLAGS}=0) to avoid the objects that are on bad or saturated pixels.

\item Imperfect subtraction often produces clumps of negative pixels. To remove false detection caused by such features, we multiplied the subtraction image by -1 and detected positive signals. Then, we excluded detections whose locations match with detections in the negative images.

\item The objects that survived the above filtering procedures were visually inspected to remove spurious detections.
\end{enumerate}


As most of the UKIRT images had no deeper reference images, we monitored the variability for each source across multiple epochs. Transients were searched as objects that are flux-varying point sources (\texttt{CLASS\_STAR} $> 0.5$) and away from the edges of the images. We used the VISTA Hemisphere Survey \citep[VHS;][]{2013Msngr.154...35M} $K$-band images taken in 2009 as reference images for some of the UKIRT images for which a proper image subtraction was needed to measure fluxes of transients accurately.

Selected candidates were matched with known transient catalogs. Moving objects in the solar system were identified using $\texttt{astroquery}$ from the \texttt{Skybot} archive (\texttt{Skybot}; \citealt{2006ASPC..351..367B}), and previously reported transients like SNe were found from the transient name server \citep[TNS;][]{2021AAS...23742305G}.

As a result, 11,581 filtered transient candidates were filtered out, among which 3,637 were images taken by small telescopes and 7944 were images taken with KMTNet. The final list of transients is 1,336 solar system objects matched with \texttt{Skybot}, eight previously reported transients with unknown nature, three reported SNe (Section \ref{subsec:observation}), and new one transient named GECKO190427a from the KMTNet-SAAO images. 

\section{Nature of Transients and Constraints on KN Property} \label{sec:nature}


\subsection{GECKO Depths vs KN Models} \label{subsec:gw170817like}
\begin{figure}
\plotone{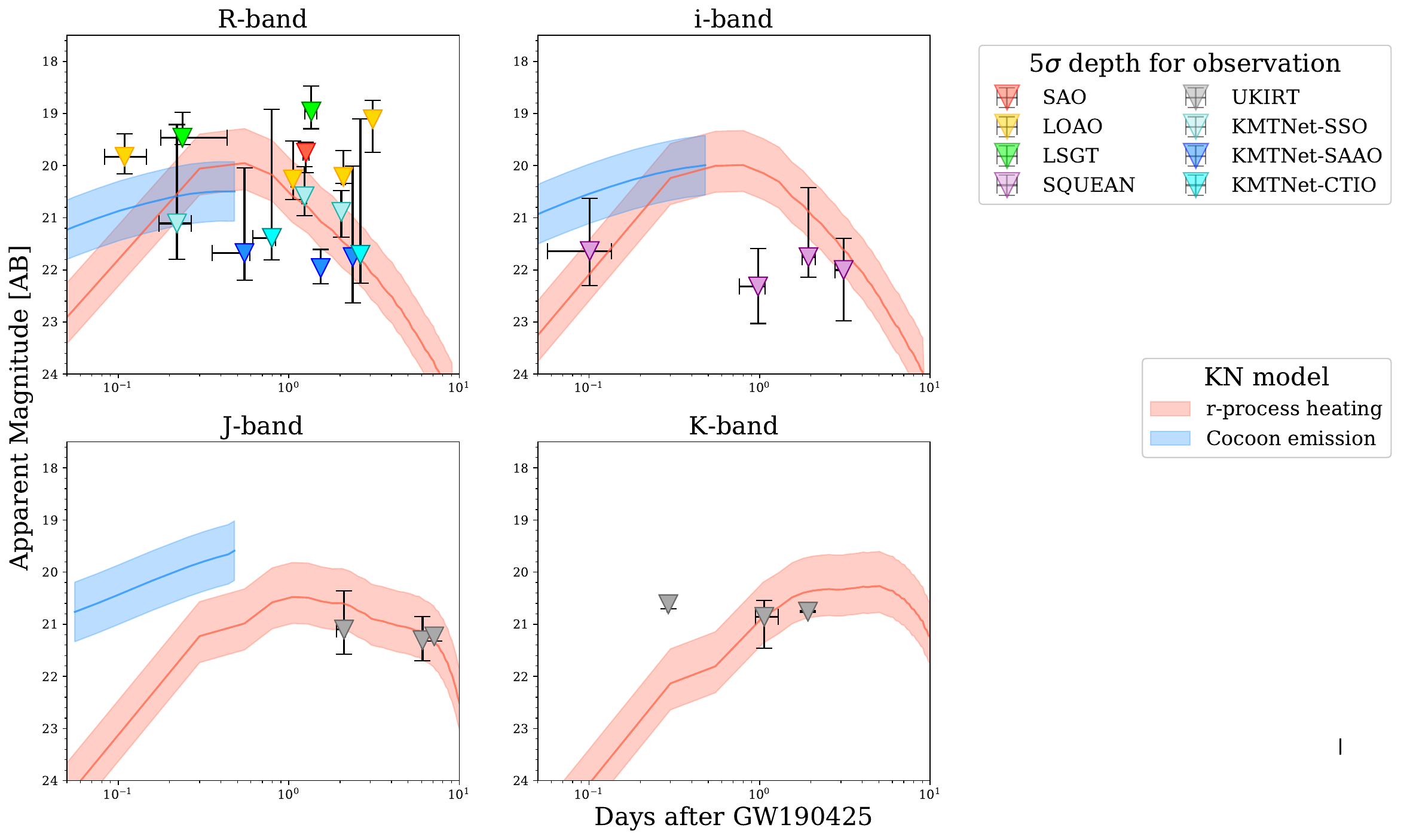}
\caption{$\rm 5\sigma$ depth as a function of time after emergence of GW190425 and model KN light curves in $R, i, J$, and $K$-bands.
The inverted triangles are the median $5\sigma$ depths at the median times of the observation sequences. The error bar represents the range of the depth and the time of the images in the corresponding observation sequence. Different colors represent different observatories. Note that the LSGT observation depths are indicated for $r$-band. We plot two KN models scaled to the distance of GW190425, and the shaded region corresponds to the distance range of GW190425 ($\rm 156\pm41Mpc$). The red curve is GW170817-like KN from K17. The blue one shows a shock cooling early emission in the $r$, $i$, and $J$-band from PK18.
\label{fig:depth}}
\end{figure}

To illustrate the sensitivity of the GECKO observations, we compared the detection upper limits with GW170817-like KN models, scaled at the GW luminosity distance of GW190425 of $\rm 156 \pm 41\:Mpc$ \citep{2020ApJ...892L...3A}.
First, according to the \citet[][hereafter K17]{2017Natur.551...80K}, KN emission from GW170817 consists of two components. Blue emission is from the ejecta composed of mainly light r-process elements having low opacity and the other is a red component from the ejecta composed of relatively heavy r-process elements having high opacity. We combine these two components of the GW170817-like spectrum provided by K17 and convert the combined spectrum into AB magnitude in observed bands.
Second is the shock cooling or cocoon emission model \citep[hereafter PK18]{2018ApJ...855..103P}. According to PK18, a distinctive characteristic of shock cooling emission is blue excess emission in the early phase. We take GW170817-like shock cooling emission. Assuming $\rm 37.7\pm8.7\:Mpc$ as the distance of GW170817 \citep{2017ApJ...849L..16I}, we scaled both models to the updated distance of GW190425 (Figure \ref{fig:depth}).

Overall, the follow-up observation depths sufficiently cover GW170817-like KN model LCs. In the $R$-band, the 1-m class telescopes provide data that are slightly deeper than the model LC. On the other hand, KMTNet gives image depths that are deeper than both model LCs for several days. The SQUEAN observation could cover not only the initial cooling emissions but also the emissions by r-process heating.

\subsection{Comparison of LCs of transients with KN models} \label{subsec:tc}
\begin{figure}
\plotone{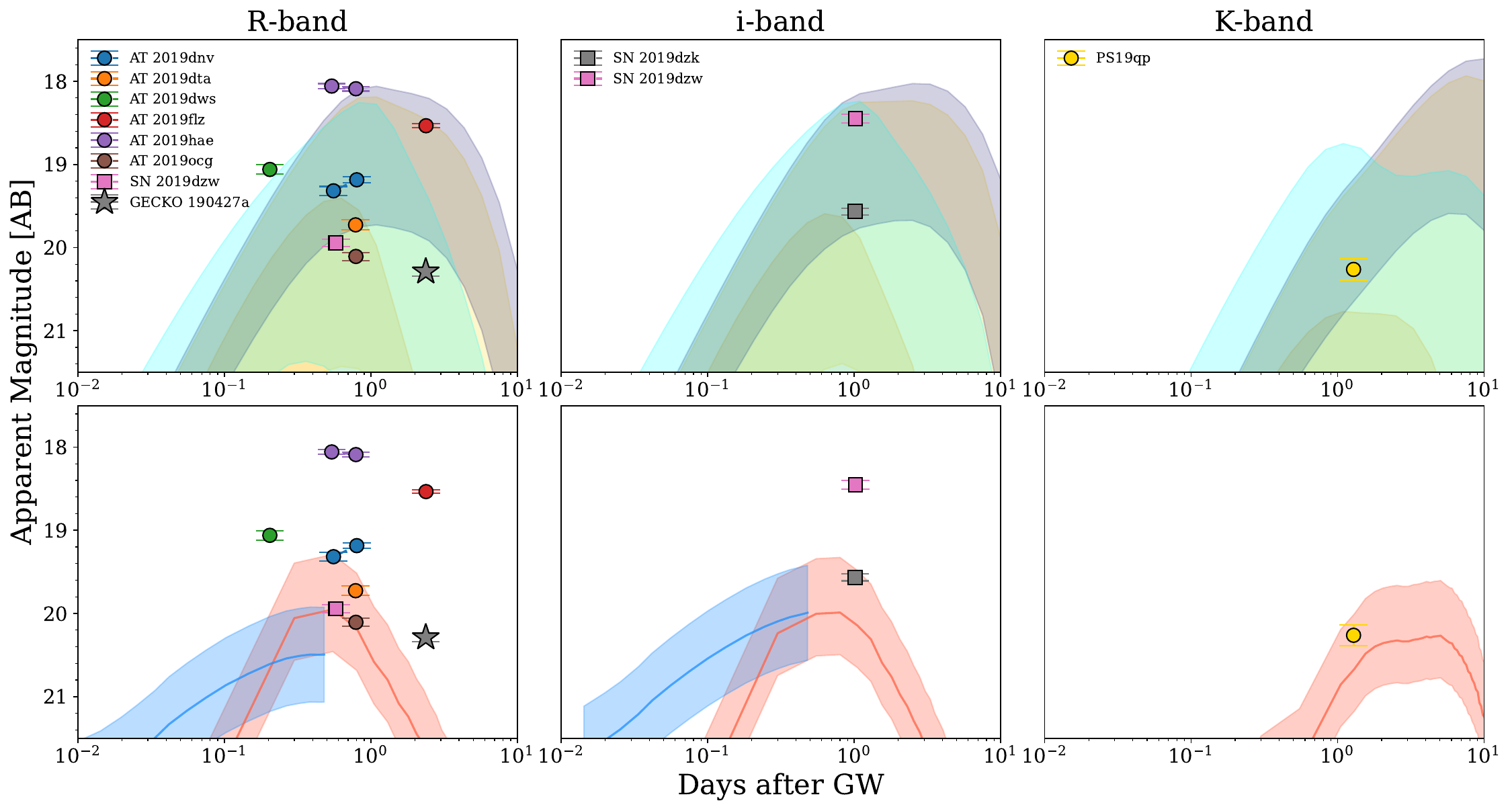}
\caption{Comparison of LCs of the transients found in our observation and the KN model LCs in $R$ and $i$-bands. (Top) KN models from \citep{2020arXiv200209395B}. SN 2019dzw and SN 2019dzk spectroscopically classified as SN are expressed in square points and others are expressed in circles. GECKO190427a which is found in this follow-up observation is shown as a star-shaped point. Yellow and orange shaded regions show the KN model LCs assuming the softest equation of state emerged by NSBH merger and NSNS merger respectively. Purple and cyan shaded regions show the KN model LCs assuming the stiffest equation of state emerged by the NSBH merger and NSNS merger respectively. AT 2019dws and AT 2019hae are out of the range of these KN models. (Bottom) GW170817-like KN models from K17 (red line) and PK18 (blue line). The LCs of SN 2019dzw, SN 2019dzk, AT 2019dta, AT 2019ocg, and marginally AT 2019dnv are within the KN models.
\label{fig:transient}}
\end{figure}

In Figure \ref{fig:transient}, we compare the LCs of eight transients with unknown nature and three SNe, and GECKO190427a including the KN models in $R$ and $i$-bands to find the GW EM counterpart. KN model LCs from \citet[][hereafter B20]{2020arXiv200209395B} and the GW170817-like KN model of K17 and PK 18 are shown in the top and bottom panels respectively. Note that B20 provides KN model LCs depending on the equation of state (EoS) and BH spin originating from both progenitor systems. We take component mass configurations of BNS and NSBH consistent with the chirp mass of GW190425 (Table 2 in B20) with the stiffest and softest EoS (DD2 and APR4 for each).

Figure \ref{fig:transient} illustrates the challenge of determining the nature of transients based on a limited number of data points. Among the detected transients, most of them appear to be brighter than the GW170817-like kilonova (K17). Particularly, the brightness of AT2019hae in the $R$-band exceeds the expected brightness from the KN LCs with extreme EoS, suggesting that it is not a KN. On the other hand, PS19qp, which has been classified as a SN based on spectroscopy \citep{2019GCN.24358....1D,2019GCN.24295....1M,2019GCN.24230....1M}, shows consistent brightness in the $K$-band with both K17 and B20 model LCs. Therefore, it remains challenging to distinguish KNe from unknown transients based on a single photometric data point. It is crucial to utilize multi-epoch data and consider the decay rate of brightness or the color evolution of the transients to improve the identification of KNe.


\subsection{Nature of GECKO190427a} \label{subsec:gck}
\begin{figure}
\plotone{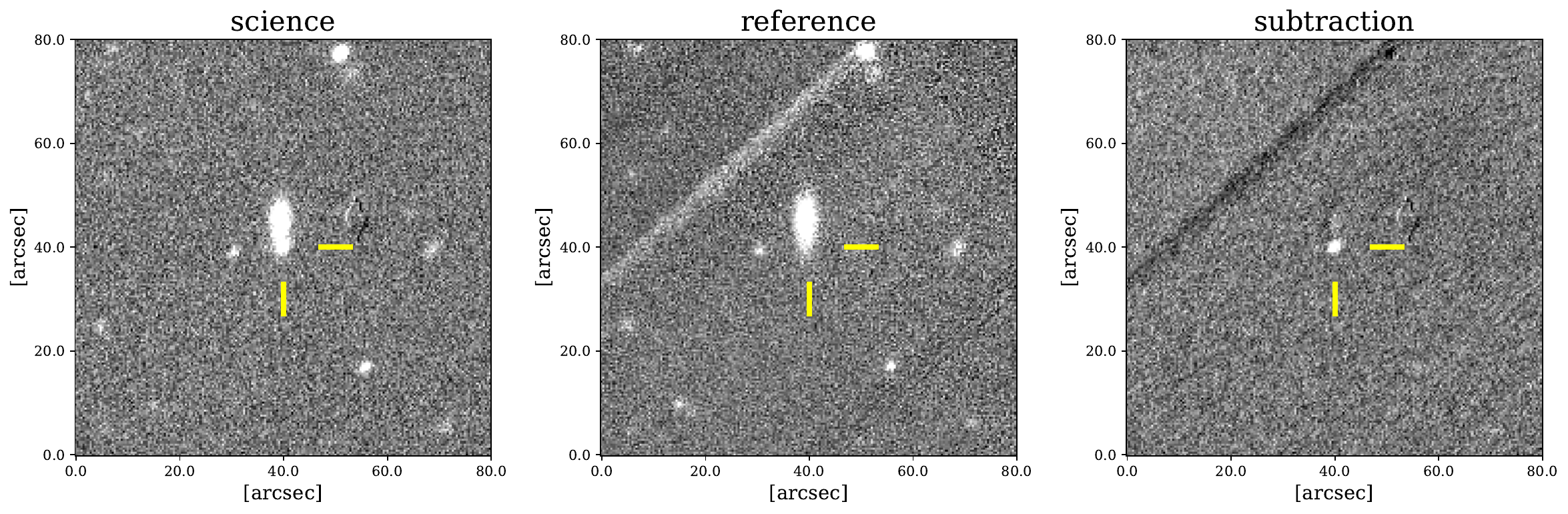}
\caption{Images GECKO190427a taken by KMTNet-SAAO in $R$-band at 2019-04-27 17:12:57 UTC. The left panel is the observed image, the middle panel shows the PS1 reference image in $r$-band \citep{2016arXiv161205560C}, and the right panel shows the subtraction image (the observed image minus the reference image). The yellow tick mark indicates the location of GECKO190427a.
\label{fig:gck_snap}}
\end{figure}

In KMTNet-SAAO $R$-band observation taken on 2019-04-27 17:12:57 UTC ($\rm \sim 2.4$ days after the GW event), we found GECKO190427a in the Southern hemisphere of the GW skymap localization, with an apparent magnitude of $R$=$\rm{20.29 \pm 0.05}$ (Figure \ref{fig:gck_snap}). No match was found with reported transients on GCN or TNS. As seen in Figure \ref{fig:gck_snap}, it is located on a host galaxy, WISEA J050339.09-234733.9 \citep{2013wise.rept....1C} with a 4.7 $\arcsec$ offset in the North direction. Its host galaxy is not contained in the modified GLADE catalog. Details are listed in Table \ref{tab:transient}.

We found the archival image with the closest epoch taken by Dark Energy Camera (DECam) in $r$-band at 2019-01-09 04:20:23 UTC, which shows no significant signal at the location of GECKO190427a with $(\rm 5 \sigma \: depth)\:=\: 23.9$ mag, confirming that this is a transient. Since GECKO190427a was discovered during the post-event processing of the data, no intensive follow-up observation was carried out.

First, we compare its apparent magnitude with different KN models. As shown in Figure \ref{fig:transient}, while GECKO190427a is consistent with the KN models of a BNS merger with DD2 EoS and an NSBH merger with APR4 from B20 (upper left panel), it is too bright to be a GW170817-like KN as modeled by K17.

Second, since there was no redshift information for its host galaxy in the modified GLADE catalog, we measured its photometric redshift with the \texttt{EAZY}\footnote{\url{https://github.com/gbrammer/eazy-photoz}} code \citep{2008ApJ...686.1503B}. We collected optical and NIR images in $g$, $r$, $i$, $z$, $y$, $J$, and $H$ ($g$, $r$, $i$, $z$, $y$-band from PS1; \citealt{2020ApJS..251....7F}, and $J$, $H$-band from the 2MASS All-Sky Atlas Image Service; \citealt{ipac_2mass_2020}). We used SE to measure their magnitudes (\texttt{MAG\_AUTO}) directly from the retrieved images, and made corrections for the Galactic extinction \citep{2019ApJ...877..116W} while the photometric results in $W1$, and $W2$ are compiled from \citet{wright_allwise_2019}. Photometries are listed in Table \ref{tab:GECKO190427a}.

\begin{figure}
\plotone{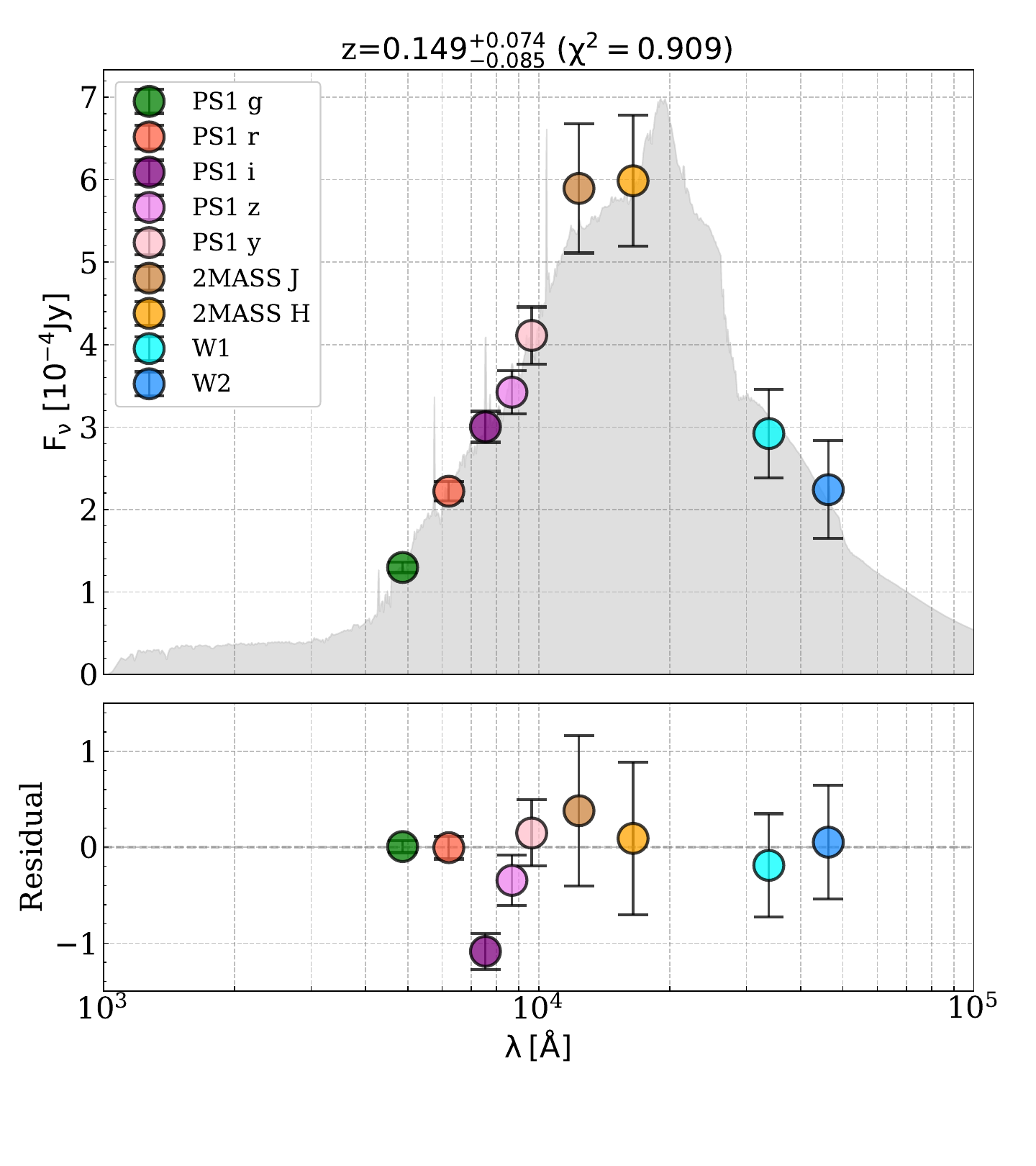}
\caption{Result for photometric redshift estimation with the \texttt{EAZY} code for host galaxy of GECKO190427a. Each colored points represent the magnitude in each filter from PS1, 2MASS, and WISE. The WISE data points are queried from NED, and others are measured by SE. All magnitudes are MW-corrected with \citep{2019ApJ...877..116W}. (Top) The gray-shaded region shows photometric points (Bottom) Difference between the SED fitting result and photometric points is shown.
\label{fig:sed}}
\end{figure}

\begin{figure}
\plotone{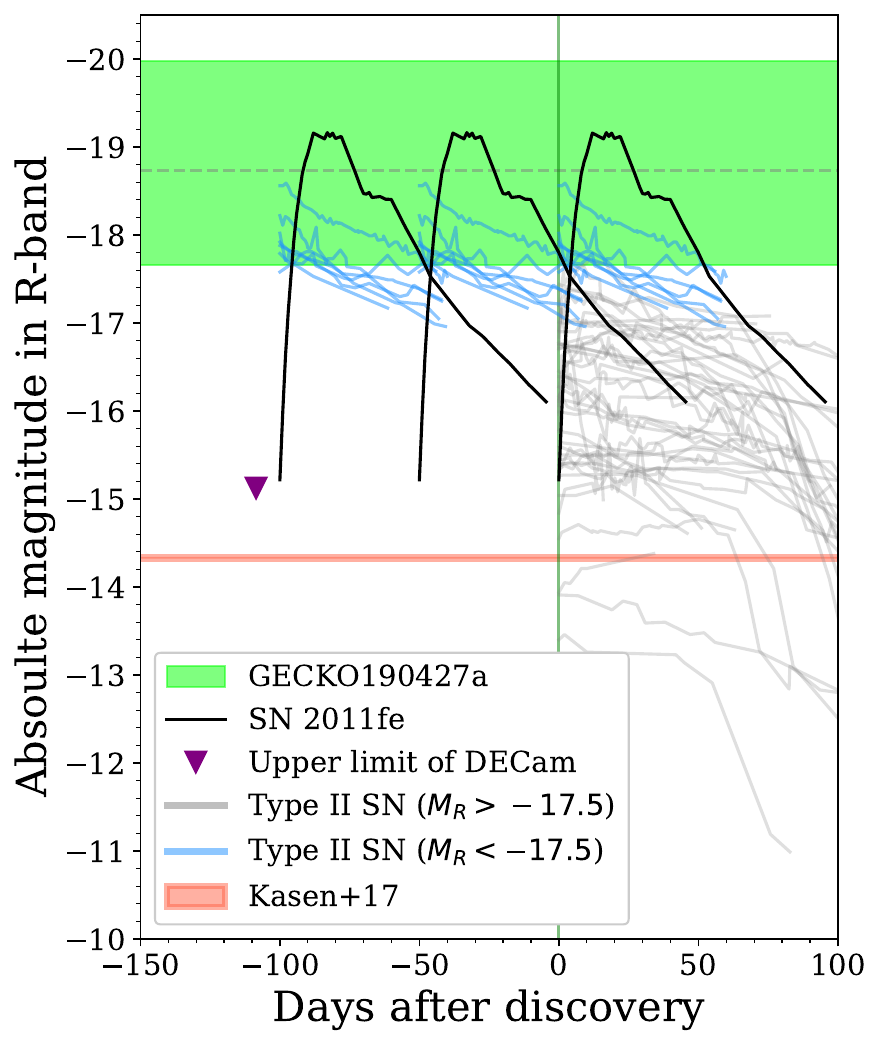}
\caption{Comparison of the range of absolute magnitude of GECKO190427a with different transient LCs in R-band. The grey-dotted horizontal line and green-shaded region show the absolute magnitude and its range of GECKO190427a scaled by the photometric redshift range of its host galaxy (Figure \ref{fig:sed}). The red line shows the expected brightness of a KN based on the KN model from K17 at the time when GECKO190427a was found. The black line shows the LC of a Type Ia SN, SN 2011fe. Blue lines show LCs of bright Type II SNe, with $M_{R}<-17.5$ at the first epoch, and grey lines show LCs of Type II SNe with $M_{R}>-17.5$ at the first epoch, as listed in \citet{2019MNRAS.490.2799D}. LCs for both types of SNe have been shifted by $\rm t_{0}-50$ days and $\rm t_{0}-100$ days for comparison. The inverted purple triangle indicates the upper detection limit from DECam at $\rm t_{0}-108.5$ days.
\label{fig:gck_comp}}
\end{figure}

Figure \ref{fig:sed} shows the best-fit spectral energy distribution (SED) and the measured redshift is $0.149_{-0.073}^{+0.064}$, corresponding to $\rm 638_{-313}^{+274}\: Mpc$ in luminosity distance\footnote{Peculiar velocity of galaxy is negligible at the given redshift.}. We found that the host galaxy is contained in the GLADE+ catalog \citep{2022MNRAS.514.1403D}, and it has a consistent redshift ($\rm z \sim 0.166$), corresponding to $\rm 711\:Mpc$ in luminosity distance. Both values are too far from GW's luminosity distance of $\rm 156 \pm 41 Mpc$. In Figure \ref{fig:gck_comp}, to illustrate this better, we convert the apparent magnitude of GECKO190427a to the absolute magnitude with the measured distance and compare it with the expected brightness of a KN in the model from K17, a typical and well-studied Type Ia SN, SN 2011fe, and Type II SNe (Figure \ref{fig:gck_comp}). Our analysis indicates that the brightness of GECKO190427a, which exceeds typical KN brightness, is similar to that of SNe near their peak brightness, suggesting that GECKO190427a is likely an SN. A direct constraint from the DECam upper limit suggests that the outbreak should have occurred at $\rm t_{0}>-108.5$ days. Furthermore, the LCs of Type Ia SN, and bright Type II SNe ($\rm M_{R}<-17.5$) suggest the onset of the transient happened at $\rm -50$ days $\rm< t_{0} <$ $\rm 0$ days if GECKO190427a is a Type Ia SN, or a bright Type II SN. In conclusion, GECKO190427a is unlikely to be related to GW190425. The distance of the host galaxy played a critical role in this assessment, demonstrating the importance of having this information in advance. GECKO190427a is probably an SN, but we do not exclude the possibility of being a dim solar system object accidentally passing through the galaxy.

\section{Prospects of GECKO Observations}\label{sec:prospect}

\subsection{Outlook for GECKO Observations in O4}\label{subsec:o4}

The wide-field telescopes such as KMTNet of GECKO can be used more effectively than we have done for GW 190425. During the GW190425 campaign, the lack of sufficient preparatory works forced us to do galaxy-targeted observation even with KMTNet, making an unfortunate overlap of field coverage.
Here, we will estimate what portion of the localization area could have been observed with KMTNet if we performed wide-field tiling observations. As discussed in the previous section, the depths we achieved with KMTNet are sufficient to detect many types of KNe. Our GW190425 observation with KMTNet took an average of about 6 minutes per field (including two dither points), with the actual exposure time being 4 minutes.

Under the same depth conditions as the GW190425 observation, observations would commence immediately after an alert from the available site - in this case, the KMTNet at CTIO. All KMTNet sites are subject to a minimum altitude limit of 30 degrees. We assume that tiling observations are to be carried out sequentially, starting from the tiles with the highest $\rm P_{2D}$ sum. Observations could take place during astronomical twilight when the Sun's altitude is less than -18 degrees, and they could continue for up to 2.5 days after the GW trigger. The declination limit is roughly $\rm Dec = 30$ degrees for the northern hemisphere targets.

\begin{figure}
\plotone{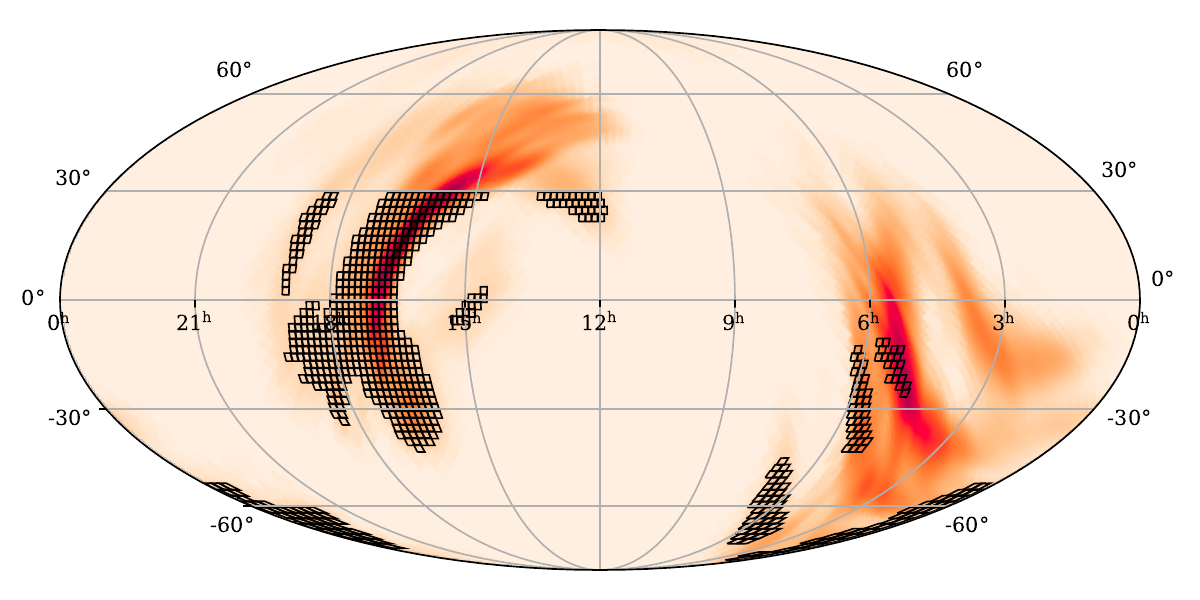}
\caption{Simulated coverage map of KMTNet tiling observation for GW190425. The color map shows the final localization probability map from GWTC-2 \citep{2020ApJ...892L...3A}. There are 735 black boxes each of which corresponds to the FOV of KMTNet ($\rm \sim 4\:deg^{2}$). 
\label{fig:tile}}
\end{figure}

The localization area of GW190425 is dispersed with a considerate offset in the RA direction and divided into two main regions at RA = 17 hr and 6 hr and an additional weaker peak at 3 hr. KMTNet could have covered most of the area around RA = 17 hr, but it could have only observed a partial area for the southern hemisphere around RA = 6 hr due to the visibility issue. Despite the limitation, with observations from three sites over three consecutive nights, we anticipate $\rm 2,940\:deg^{2}$ of area coverage with 735 tiles (Figure \ref{fig:tile}), equating to 32.3\% of $\rm P_{2D}$ coverage.

\begin{figure}
\plotone{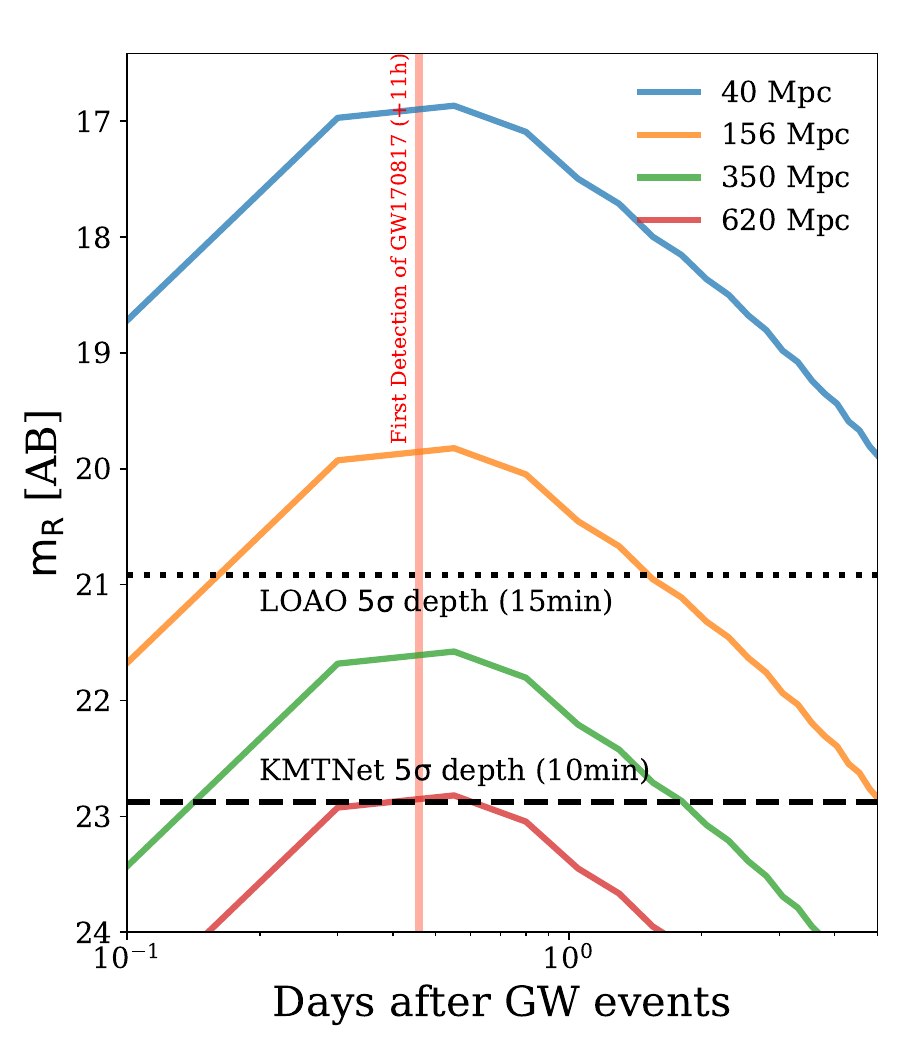}
\caption{Expected kilonova light curve (KN LC), scaled to four different distances in R-band from K17, over the days following the emergence of GW events, and the $\rm 5\sigma$ detection depth of LOAO, one of the 1-m class telescope and KMTNet, a 1.6-m telescope in the GECKO project. The x-axis is expressed on a log scale to illustrate how GECKO facilities can capture the early LC of a KN. Red vertical lines indicate the time when the EM counterpart of GW170817 was discovered after the GW event. The dotted line and dashed line represent the $\rm 5\sigma$ depths of an LOAO with a 15-minute exposure time, and KMTNet with a 10-minute exposure time, respectively. The blue line shows the KN models at a distance of 40 Mpc, where GW170817 is located. The dark green line indicates the model at a distance of 156 Mpc, the median luminosity distance of GW190425. The light green line represents the model at 350 Mpc, the distance within which GW detectors would detect 90\% of BNS merger events in the O4 run. Finally, the red line shows the model at 620 Mpc, the distance within which GW detectors would detect 90\% of BNS merger events in the O5 run.
\label{fig:o4}}
\end{figure}

The expected median 90 \% credible region (CR) in O4 varies with studies, from $\rm 33\:deg^2$ in \citet{2020LRR....23....3A} to $\rm 1,820\:deg^2$ in \citet{2022ApJ...924...54P}. This is mainly due to the difference in the SNR threshold. Even adopting the pessimistic but perhaps more realistic estimate of \citet{2022ApJ...924...54P} of thousands $\rm deg^2$, the expected median CR is comparable to the area covered by KMTNet's tiling observation for up to 2.5 days after the GW trigger. This suggests that KMTNet can cover most of the observable area within 2.5 days during O4, reaching depths sufficient to detect a GW170817-like KN at $\rm 350\:Mpc$ (Figure \ref{fig:o4}), which is the expected median luminosity distance for O4 \citep{2022ApJ...924...54P}. Coordinating with other facilities is expected to provide more robust coverage for poorly localized areas (i.e., areas larger than \(3,000\: \text{deg}^{2}\)) within a few days.

Our GECKO observations started as fast as about 90 minutes after the GW trigger, similar to the follow-up observations carried out by other groups. Since then, we have upgraded our follow-up alert software \citep[\texttt{GeckoDigestor};][]{geckodigestor} and can possibly produce observation commands as soon as 10 minutes after the GW trigger. This does not mean that we would start observations at 10 minutes because the updated localization area of GW events could be significantly less than the initial estimates. Hence, in O4, we expect to start follow-up observations about an hour after the GW trigger or could be much sooner if the GW event localization area is well constrained. 

\begin{figure}
\plotone{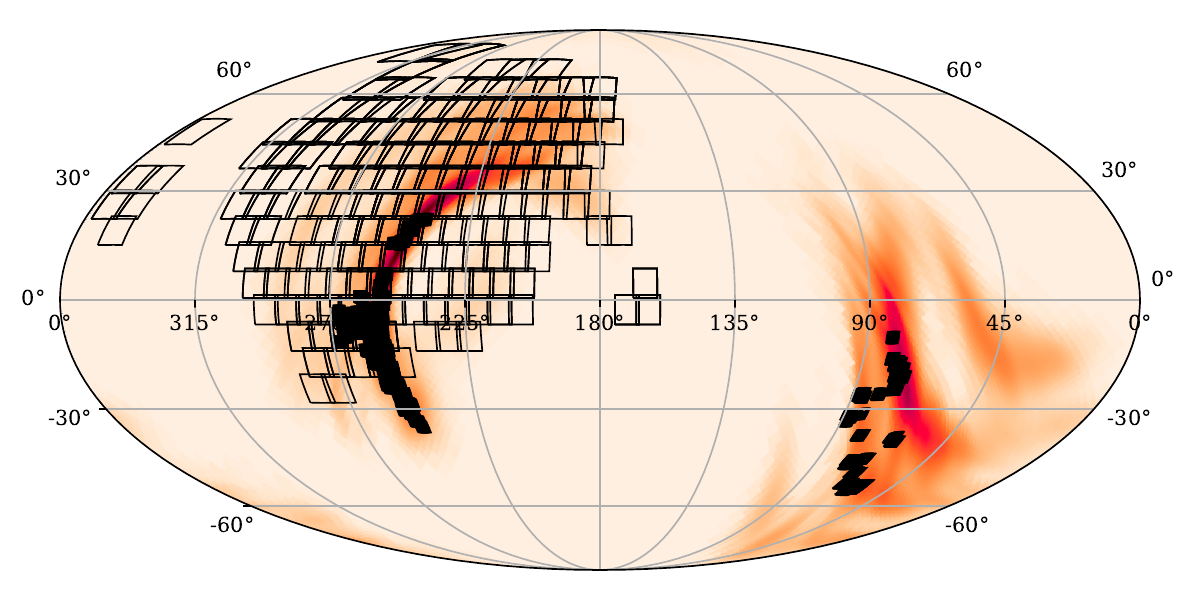}
\caption{Coverage of GW190425 by ZTF in GROWTH and KMTNet in GECKO. The larger FOV ($\rm 47\:deg^{2}$) shows the coverage of ZTF and the smaller FOV ($\rm \sim 4\:deg^{2}$) shows the coverage of KMTNet. KMTNet observation covered both the Northern and Southern hemispheres using the galaxy-targeted observation. Therefore, a large overlapped area exists because the host galaxy candidates with high probability are located in the nearby field. On the other hand, the ZTF observations covered most of the Northern hemisphere with the tiling strategy.
\label{fig:allcoverage}}
\end{figure}

Figure \ref{fig:allcoverage} shows the combined coverage of KMTNet and Zwicky Transient Facility (ZTF) of 3,300 $\rm deg^2$ in the Global Relay of Observatories Watching Transients Happen (GROWTH; \citealt{2019ApJ...885L..19C}). However, in combination with the northern hemisphere facilities using an improved GECKO strategy, covering the localization areas beyond $\rm 5,000\:deg^2$ is not out of scope. For example, the ZTF covered about 3,100 $\rm deg^2$ area in the northern hemisphere to a depth of 21 AB mag in the $g$- and $r$-bands. Combining our anticipated GECKO coverage of 2,940 $\rm deg^2$, we can cover most of the GW190425 90\% CR.

On the other hand, GW190425 tells us that the GW localization is often bipolar when the signal is weak (GW200105, and GW200115; \citealt{2021ApJ...915L...5A}). If these bipolar regions are aligned in the direction of a similar RA, the optical follow-up can be carried out effectively. On the other hand, if the bipolar regions are aligned toward a similar Dec direction like GW190425, the RA separation of the two significant localizations could be large and only one region could be observed with optical telescopes due to the visibility constraints. In conclusion, KMTNet, when used for tiling observations, can effectively cover large, poorly constrained localization areas.

\subsection{GECKO Prospects for Constraints on KN Models} \label{subsec:constraint}
The expected properties of KN in GW190425-like events can be potentially inferred by comparing the upper limits from our follow-up observations with the KN LCs modeled by a wide range of parameter grids, assuming that the GW source is located in our coverage. We used KN LCs predicted by axisymmetric 2D KN model, varying mass, velocity, morphology, viewing angle, and composition \citep{2021ApJ...918...10W}. In this model, the ejected materials consist of the dynamical ejecta which is made of the heavy r-process elements, and the wind ejecta which is composed of the light r-process elements, these two kinds of ejecta are distributed in different longitudes along with the axis of the system. Therefore, the predicted KN SED shape for an observer varies depending on the viewing angle from on-axis (0 deg) to edge-on (180 deg), the dynamical ejecta mass ($\rm m_{dyn}=0.001, 0.003, 0.01, 0.03, 0.1\:M_{\odot}$), the wind ejecta mass ($\rm m_{wind}=0.001, 0.003, 0.01, 0.03, 0.1\:M_{\odot}$), dynamical ejecta velocity ($\rm m_{wind}=0.05, 0.15, 0.3\:c$), and the wind ejecta velocity ($\rm v_{wind}=0.05, 0.15, 0.3\:c$).

In this analysis, we consider constraints on KN properties from two cases; KMTNet+ZTF and SQUEAN+ZTF observations. In the former case, we assume that the CR of GW190425 was covered by both facilities as that could have happened (see Section \ref{subsec:gw170817like}). In the latter case, we assume that instruments with filter sets like SQUEAN and ZTF covered the entire CR or galaxy targets. To consider realistic cases, we took the median value of the upper limits taken on the same night as the data point at a certain epoch. The LCs of KMTNet in the $R$-band, ZTF in $g$- and $r$-band are sensitive to the wind ejecta which produce the bluer light in the early phase. In contrast, the upper limits in the $i$-band can constrain the dynamical ejecta mass effectively \citep{2022ApJ...927..163C}.

\begin{figure}
\plotone{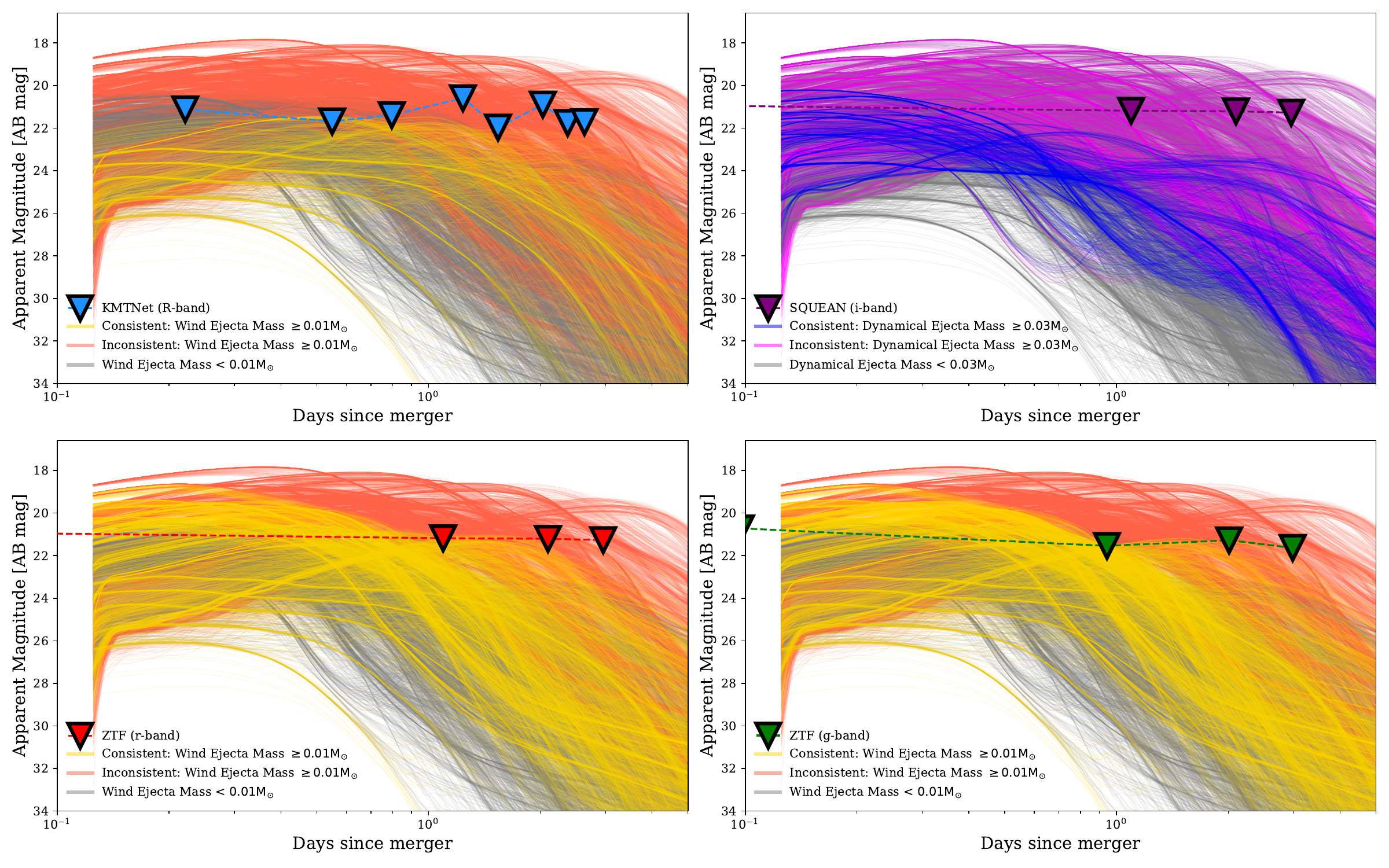}
\caption{Comparison of 2D kilonova models with the upper limit of KMTNet, SQUEAN, and ZTF follow-up observation of GW190425. (Top) The triangle points on the left show the upper limit of KMTNet observation in the $R$-band. The right shows the comparison with the SQUEAN observation in the $i$-band. (Bottom) The left and right show the comparison with the ZTF observations in the $g$- and $r$-band, respectively. Red-colored lines are consistent model LCs with the upper limit and have $\rm m_{wind} \geq 0.01 M_{\odot}$, yellow-colored lines are inconsistent model LCs with the upper limit and have $\rm m_{wind} \geq 0.01 M_{\odot}$, and grey colored lines are consistent model LCs having $\rm m_{wind} < 0.01 M_{\odot}$.
\label{fig:model_lc}}
\end{figure}

In Figure \ref{fig:model_lc}, we compare the upper limits with model LCs derived from various combinations of parameters, which include 54 viewing angles ranging from 0 to 180 degrees. If the predicted brightness from a model is brighter than the upper limit at an epoch, we reject the model and vice versa. Note that we did not consider the upper limits taken earlier than $\rm \Delta t < 0.2\:days$ because the KN model does not provide LCs at such epochs. For that reason, the earliest upper limits of ZTF and SQUEAN observations (ZTF $g$-band; 2.34 hours, $r$-band; 0.95 hours, and SQUEAN $i$-band 2.43 hours from BNS merger event) were not used in this analysis.

\begin{figure}
\plotone{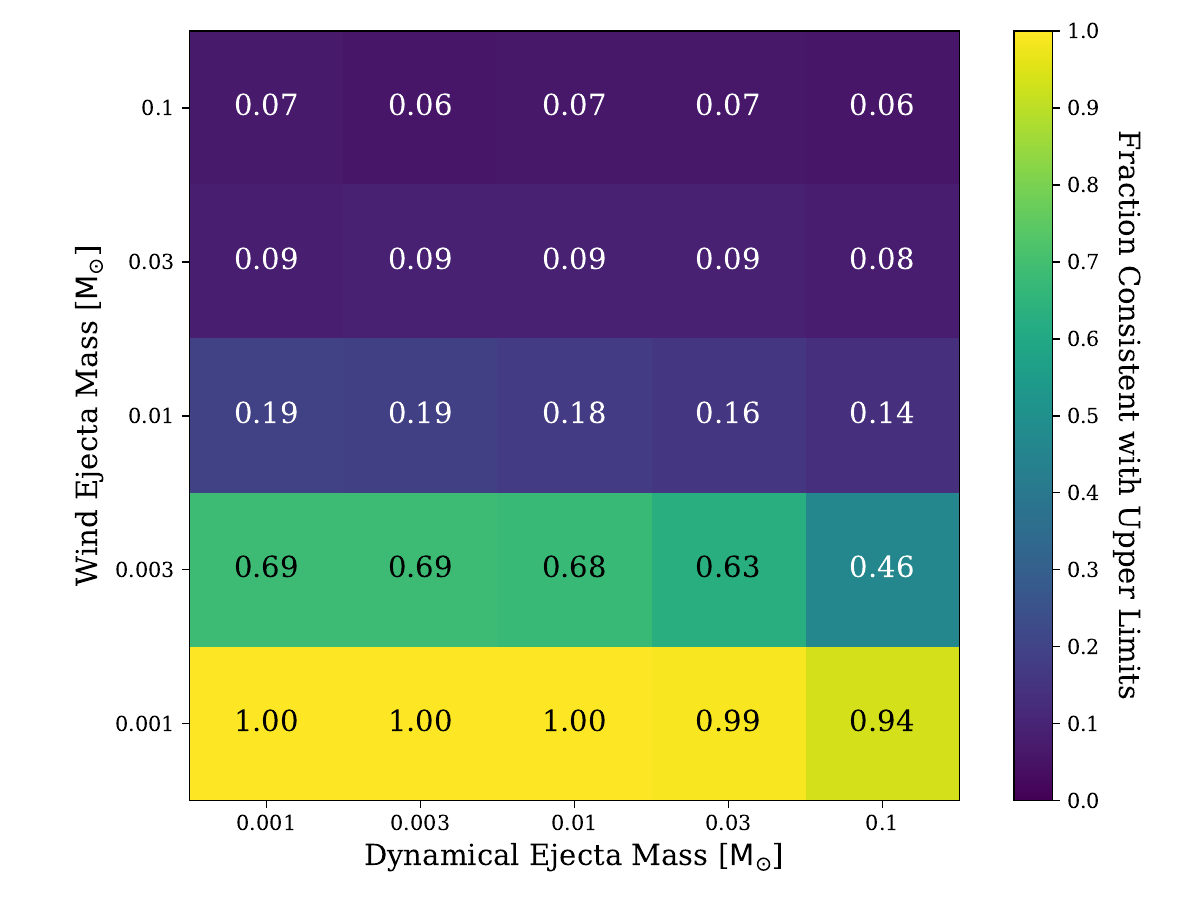}
\caption{Matrix showing the fraction of the consistent models with the KMTNet and ZTF upper limits. Each bin has 396 models for the dynamical ejecta mass (x-axis) and the wind ejecta mass (y-axis). The color shows the fraction. The yellower color shows most of the models are consistent with observations, and the bluer color shows few or none of the models are consistent with observations. If the fraction is less than 0.3, we regard it as a less probable model, and those bins are written with white text.
\label{fig:kn_mass}}
\end{figure}

\begin{figure}
\plotone{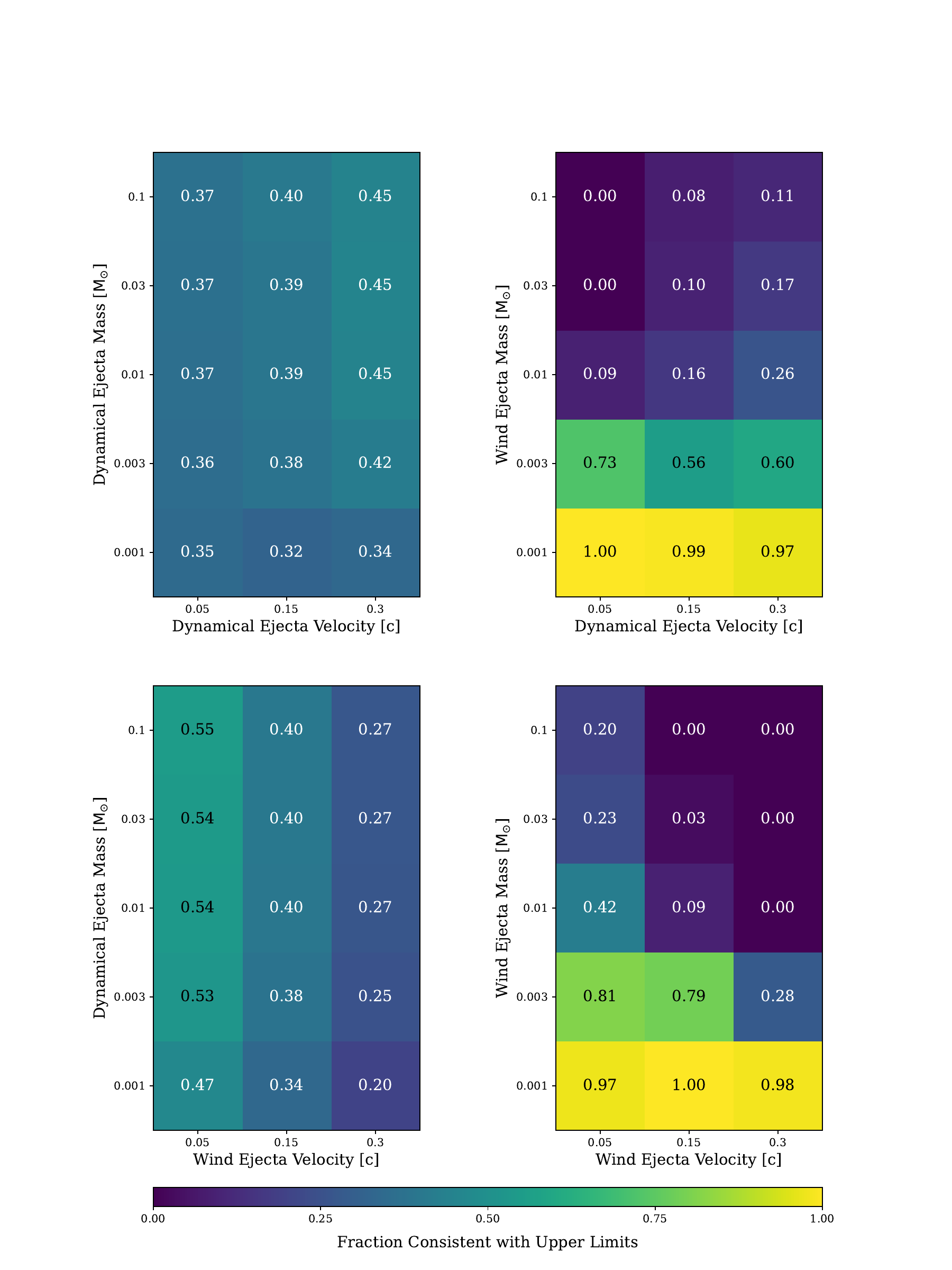}
\caption{Matrix showing the fraction of the consistent models with the KMTNet and ZTF upper limits. Combinations of ejecta velocity and mass are shown in each panel. Each bin has 660 models. The color shows the fraction and the yellower color show most of the models are consistent with observations, and the bluer color shows few or none of the models are consistent with observations. If the fraction is less than 0.3, we regard it as a less probable model, and those bins are written with white text. (Top left) the dynamical ejecta velocity (x-axis) and the dynamical ejecta mass (y-axis), (Top right) the dynamical ejecta velocity (x-axis) and the wind ejecta mass (y-axis), (Bottom left) the wind ejecta velocity (x-axis) and the dynamical ejecta mass (y-axis), and (Bottom right) the wind ejecta velocity (x-axis) and the wind ejecta mass (y-axis).
\label{fig:kn_v_vs_m}}
\end{figure}

Figures \ref{fig:kn_mass} and \ref{fig:kn_v_vs_m} show how the combination of KMTNet and ZTF observations can constrain various combinations of model parameters. Figure \ref{fig:kn_mass} shows the fraction of consistent KN models with the upper limits of KMTNet and ZTF for the combinations of $\rm m_{dyn}$ and $\rm m_{wind}$. Models with high wind ejecta masses ($\rm m_{wind} > 0.01 M_{\odot}$) have less than 20\% for most grids of $\rm m_{dyn}$ due to the non-detection in $g$ or $R$-bands in early epochs, while the upper limits allow low $\rm m_{wind}$ is of $\rm m_{wind} < 0.003 \, M_{\odot}$. The value of $\rm m_{dyn}$ is less tightly constrained than $\rm m_{wind}$, with most models for the entire range of $\rm m_{\rm dyn}$ allowed as long as $\rm m_{\rm wind}$ values are low ($\rm < 0.003\, M_{\odot}$). Figure \ref{fig:kn_v_vs_m} shows the fraction of allowed models in ejecta mass versus ejecta velocity space. Overall, it is difficult to constrain the dynamical or wind ejecta velocities, but there is a tendency for lower wind ejecta velocity to be favored ($< 0.015\, c$). Combining Figures \ref{fig:kn_mass} and \ref{fig:kn_v_vs_m}, we draw a conclusion that KN models with high wind ejecta mass with high wind velocity ($\rm >0.01c$) could be disfavored if the GW source exists in the area covered by KMTNet and ZTF. As explained in \citet{2022ApJ...927..163C}, the $R$- and $i$-bands can constrain the properties of wind ejecta, but dynamical ejecta can not be ruled out with upper limits in those bands.

\begin{figure}
\plotone{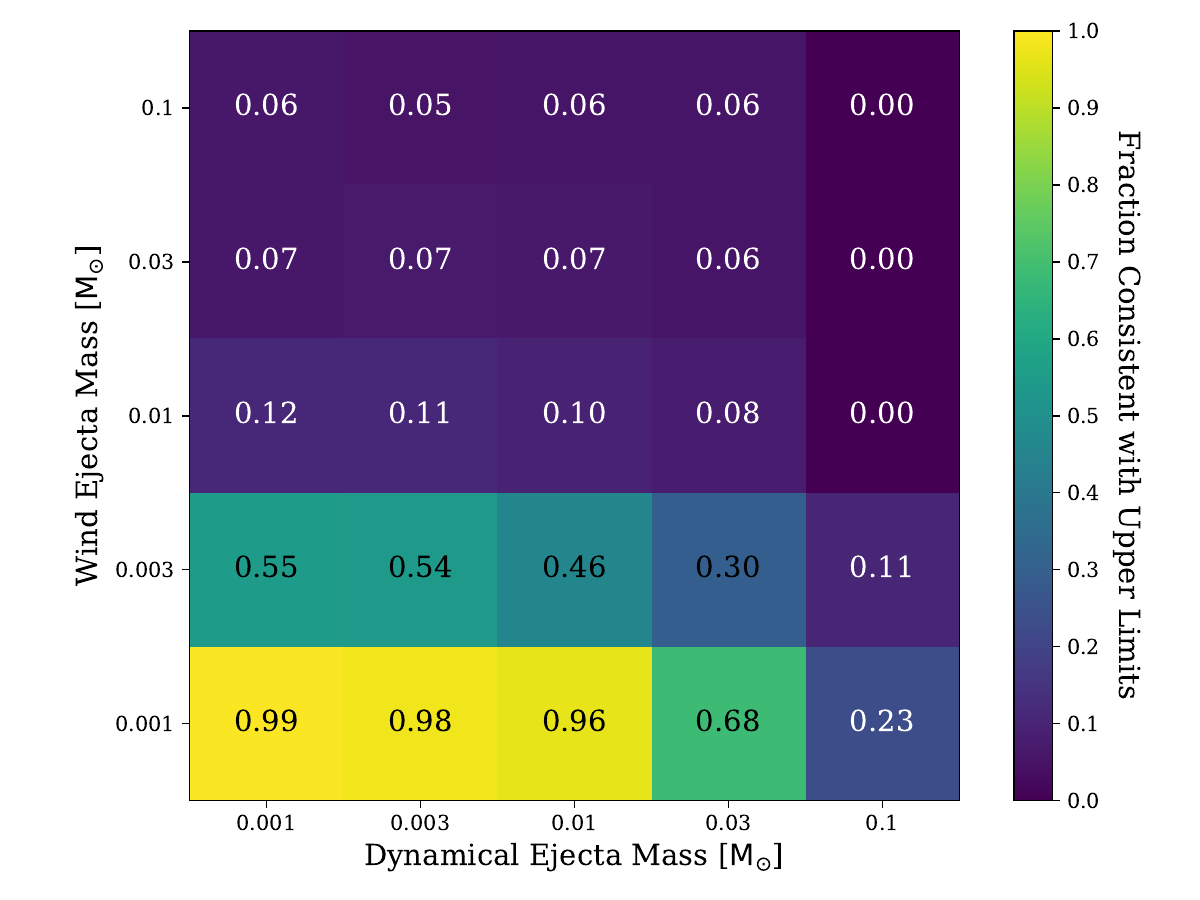}
\caption{Matrix showing the fraction of the consistent models with the SQUEAN and ZTF upper limits. Each bin has 396 models for the dynamical ejecta mass (x-axis) and the wind ejecta mass (y-axis). The meaning of color is the same as Figure \ref{fig:kn_mass}.
\label{fig:kn_mass_sqztf}}
\end{figure}

\begin{figure}
\plotone{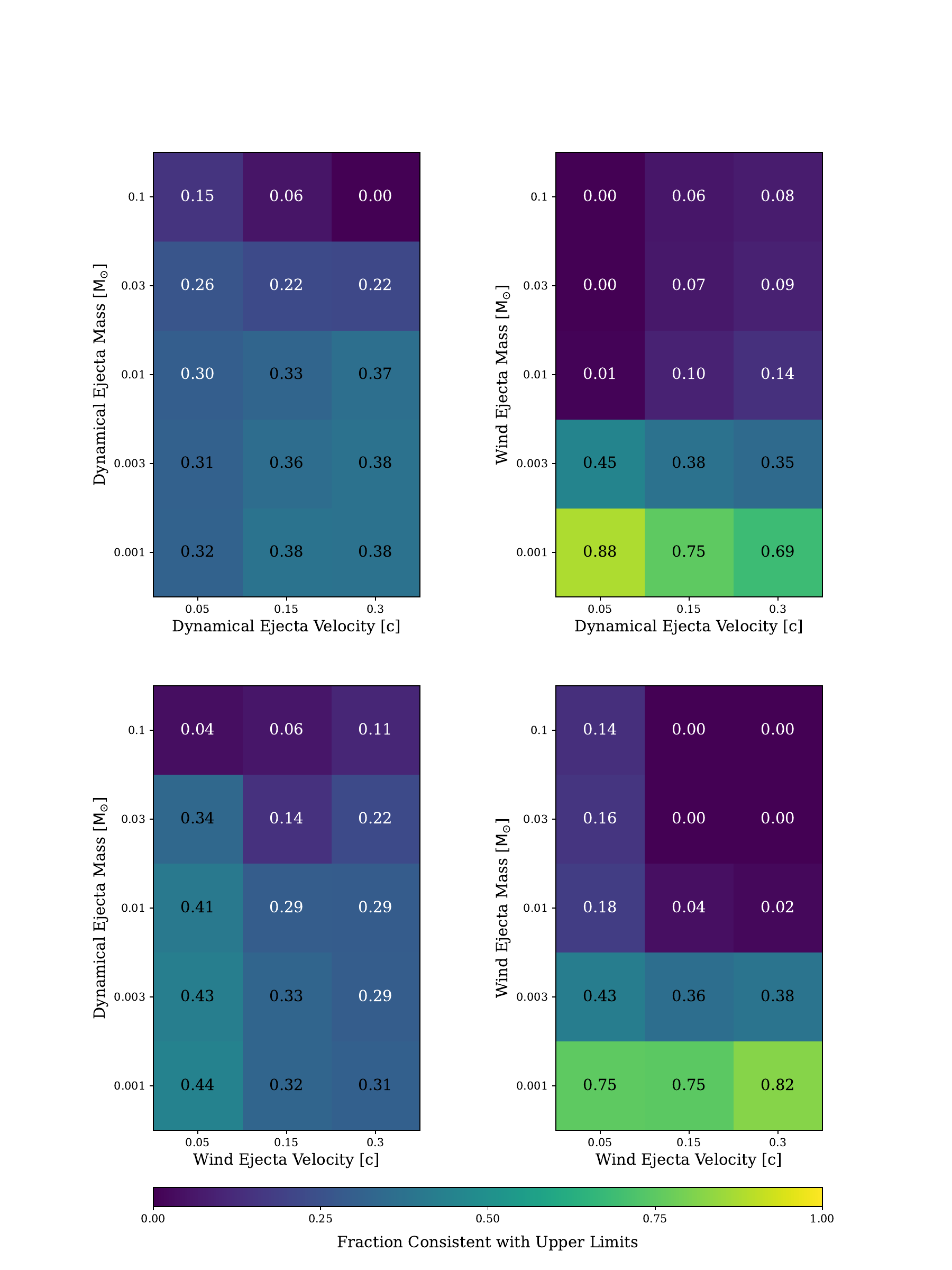}
\caption{Matrix showing the fraction of the consistent models with the SQUEAN and ZTF upper limits. Combinations of ejecta velocity and mass are shown in each panel. Each bin has 660 models. The meaning of color is the same as Figure \ref{fig:kn_v_vs_m}.
\label{fig:kn_v_vs_m_sqztf}}
\end{figure}

With the combination of SQUEAN and ZTF photometry information, the constraint on dynamical ejecta mass improves (Figures \ref{fig:kn_mass_sqztf} \& \ref{fig:kn_v_vs_m_sqztf}). This is because the SQUEAN data point represents a NIR photometry ($i$), and $m_{\rm dyn}$ controls the strength of the red component of the KN LC -- the more $m_{\rm dyn}$, the stronger the red KN LC. With the $i$-band data, the non-detection of KN can constrain the dynamical ejecta mass to be $m_{\rm dyn} \lesssim 0.03 \, M_{\odot}$ if the GW event occurred in the galaxies that both SQUEAN and ZTF covered. For the ejecta velocities, no useful constraints can be obtained, similar to the case of the ZTF+KMTNet observation combination.

These comparisons show that useful constraints can be obtained on some of the KN properties even when there is no detection if the entire GW localization area is covered by GECKO and/or ZTF. In particular, a NIR band such as $i$-band can help better constrain the dynamical ejecta mass, while the blue optical bands can be useful for constraining the wind component. Multiple epoch observations can help narrow down the plausible KN parameter space compared to a single epoch observation.
\section{Summary and Conclusion}\label{sec:summary}
In this paper, we described GECKO which aims to catch the GW EM counterpart, and presented the follow-up observation of the binary neutron star merger event, GW190425. Based on observational results, we also presented the prospects of GECKO observations of GW190425-like events.

We described the GECKO strategy for the galaxy-targeted observation for which galaxies are prioritized in the following way. First, we used the GW luminosity distance and GW skymap localization to select the host galaxy candidates where the GW source might exist from the galaxy catalog. Each host galaxy candidate was assigned a score considering the location, distance, and $K$-band luminosity. Here, we regard the $K$-band luminosity as a proxy of stellar mass which is known to be approximately proportional to the BNS merger rate based on the latest simulation results. As a result of applying our prioritization to GW170817, NGC4993, which is the host galaxy, was ranked first with an unrivaled score. When the GW alert is distributed through \texttt{PyGCN}\footnote{\url{https://github.com/nasa-gcn/pygcn}}, an automated pipeline selects and prioritizes GW host galaxy candidates, generates a target list, and then delivers it to the staff at each observation site within ten minutes.


GECKO observation of GW190425 shows that the number of covered host galaxy candidates is about 10\% of the total number. However, the score coverage stands at about 30\%. We show that the imaging depths of GECKO, especially KMTNet, are deep enough for detecting a GW170817-like KN at $\rm 150\:Mpc$, demonstrating the effectiveness of KMTNet for GW optical counterpart search.

In our search area, we found a number of solar system moving objects, eight previously reported transients, and one new transient, GECKO190427a. We estimated the distance to the host galaxy of GECKO190427a, which turned out to be too far from the estimated distance from GW. We suggest that GECKO190427a is likely to be a SN.

Based on our observations of GW190425, we discussed prospects of GECKO observation of GW190425-like events. First, if KMTNet were used to do tiling observations, it could have covered about 3,000 $\rm deg^2$. Combining with other telescopes (e.g., ZTF) observations, it would be possible to cover 90\% CR as large as $>$ 5,000 $\rm deg^2$ to detect EM counterparts of events out to 350 Mpc in several days. Adopting two or three filters, wide-field follow-up observations can exclude KN models with a significant amount of wind ejecta ($\rm m_{wind} > 0.01\:M_{\odot}$).

In the ongoing O4 run, we anticipate annually receiving around a hundred GW alerts, including perhaps a dozen cases of BNS and NSBH mergers. With the lessons learned from O3 such as the observation of GW190425, we are expanding our facilities (e.g., 7DT) and have upgraded software and strategies that implement the up-to-date host galaxy studies and machine-learning techniques. This should allow us to discover and study optical counterparts of GW events during O4 with improved efficiency.

\acknowledgments
This work was supported by the National Research Foundation of Korea (NRF) grants, No. 2020R1A2C3011091, and No. 2021M3F7A1084525, funded by the Korea government (MSIT).
G.L. acknowledges support from the Basic Science Research Program through NRF funded by MSIT (No. 2022R1A6A3A01085930).
We thank the operators at LOAO, DOAO, SAO, and KMTNet for performing the requested observations.
This research has made use of the KMTNet system operated by the Korea Astronomy and Space Science Institute (KASI) at three host sites CTIO in Chile, SAAO in South Africa, and SSO in Australia. Data transfer from the host site to KASI was supported by the Korea Research Environment Open NETwork (KREONET).
Based on observations obtained with The United Kingdom Infrared Telescope (UKIRT), which is supported by NASA and operated under an agreement among the University of Hawaii, the University of Arizona, and the Lockheed Martin Advanced Technology Center; operations are enabled through the cooperation of the Joint Astronomy Centre of the Science and Technology Facilities Council of the U.K. This paper includes data taken at the McDonald Observatory of the University of Texas at Austin. This research
made use of the data taken with LOAO which is operated
by the Korea Astronomy and Space Science Institute (KASI),
and the data from the DOAO of the National Youth Space Center (NYSC).
We also thank iTelecope.Net and its staff Brad Moore, Pete Poulos, and Ian Leeder for their management and support for LSGT observation.
We would like to thank Dr. Albino Perego for his help during the development of the kilonova model, building upon his previous works.

\vspace{5mm}
\facilities{SAO(1-m), LOAO(1-m), LSGT(0.43-m), Struve(2.1-m, SQUEAN), UKIRT(3.8-m), KMTNet(1.6-m; SAAO, SSO, CTIO)}


\software{
        astropy \citep{2013A&A...558A..33A,2018AJ....156..123A,2022ApJ...935..167A}
        astroquery \citep{2019AJ....157...98G},
        Astrometry.net \citep{2010AJ....139.1782L},
        ccdproc \citep{2015ascl.soft10007C},
        EAZY \citep{2008ApJ...686.1503B}, 
        HOTPANTS \citep{2015ascl.soft04004B}, 
        SExtractor \citep{1996A&AS..117..393B},
        SCAMP \citep{2010ascl.soft10063B},
        }






\newpage
\bibliography{sample63}{}

\begin{thebibliography}{}
\expandafter\ifx\csname natexlab\endcsname\relax\def\natexlab#1{#1}\fi
\providecommand{\url}[1]{\href{#1}{#1}}
\providecommand{\dodoi}[1]{doi:~\href{http://doi.org/#1}{\nolinkurl{#1}}}
\providecommand{\doeprint}[1]{\href{http://ascl.net/#1}{\nolinkurl{http://ascl.net/#1}}}
\providecommand{\doarXiv}[1]{\href{https://arxiv.org/abs/#1}{\nolinkurl{https://arxiv.org/abs/#1}}}

\bibitem[{{Abbott} {et~al.}(2017{\natexlab{a}}){Abbott}, {Abbott}, {Abbott}, {Acernese}, {Ackley}, {Adams}, {Adams}, {Addesso}, {Adhikari}, {Adya}, {Affeldt}, {Afrough}, {Agarwal}, {Agathos}, {Agatsuma}, {Aggarwal}, {Aguiar}, {Aiello}, {Ain}, {Ajith}, {Allen}, {Allen}, {Allocca}, {Altin}, {Amato}, {Ananyeva}, {Anderson}, {Anderson}, {Angelova}, {Antier}, {Appert}, {Arai}, {Araya}, {Areeda}, {Arnaud}, {Arun}, {Ascenzi}, {Ashton}, {Ast}, {Aston}, {Astone}, {Atallah}, {Aufmuth}, {Aulbert}, {AultONeal}, {Austin}, {Avila-Alvarez}, {Babak}, {Bacon}, {Bader}, {Bae}, {Bailes}, {Baker}, {Baldaccini}, {Ballardin}, {Ballmer}, {Banagiri}, {Barayoga}, {Barclay}, {Barish}, {Barker}, {Barkett}, {Barone}, {Barr}, {Barsotti}, {Barsuglia}, {Barta}, {Barthelmy}, {Bartlett}, {Bartos}, {Bassiri}, {Basti}, {Batch}, {Bawaj}, {Bayley}, {Bazzan}, {B{\'e}csy}, {Beer}, {Bejger}, {Belahcene}, {Bell}, {Berger}, {Bergmann}, {Bernuzzi}, {Bero}, {Berry}, {Bersanetti}, {Bertolini}, {Betzwieser}, {Bhagwat}, {Bhandare}, {Bilenko},
  {Billingsley}, {Billman}, {Birch}, {Birney}, {Birnholtz}, {Biscans}, {Biscoveanu}, {Bisht}, {Bitossi}, {Biwer}, {Bizouard}, {Blackburn}, {Blackman}, {Blair}, {Blair}, {Blair}, {Bloemen}, {Bock}, {Bode}, {Boer}, {Bogaert}, {Bohe}, {Bondu}, {Bonilla}, {Bonnand}, {Boom}, {Bork}, {Boschi}, {Bose}, {Bossie}, {Bouffanais}, {Bozzi}, {Bradaschia}, {Brady}, {Branchesi}, {Brau}, {Briant}, {Brillet}, {Brinkmann}, {Brisson}, {Brockill}, {Broida}, {Brooks}, {Brown}, {Brown}, {Brunett}, {Buchanan}, {Buikema}, {Bulik}, {Bulten}, {Buonanno}, {Buskulic}, {Buy}, {Byer}, {Cabero}, {Cadonati}, {Cagnoli}, {Cahillane}, {Calder{\'o}n Bustillo}, {Callister}, {Calloni}, {Camp}, {Canepa}, {Canizares}, {Cannon}, {Cao}, {Cao}, {Capano}, {Capocasa}, {Carbognani}, {Caride}, {Carney}, {Carullo}, {Casanueva Diaz}, {Casentini}, {Caudill}, {Cavagli{\`a}}, {Cavalier}, {Cavalieri}, {Cella}, {Cepeda}, {Cerd{\'a}-Dur{\'a}n}, {Cerretani}, {Cesarini}, {Chamberlin}, {Chan}, {Chao}, {Charlton}, {Chase}, {Chassande-Mottin}, {Chatterjee},
  {Chatziioannou}, {Cheeseboro}, {Chen}, {Chen}, {Chen}, {Cheng}, {Chia}, {Chincarini}, {Chiummo}, {Chmiel}, {Cho}, {Cho}, {Chow}, {Christensen}, {Chu}, {Chua}, {Chua}, {Chung}, {Chung}, {Ciani}, {Ciolfi}, {Cirelli}, {Cirone}, {Clara}, {Clark}, {Clearwater}, {Cleva}, {Cocchieri}, {Coccia}, {Cohadon}, {Cohen}, {Colla}, {Collette}, {Cominsky}, {Constancio}, {Conti}, {Cooper}, {Corban}, {Corbitt}, {Cordero-Carri{\'o}n}, {Corley}, {Cornish}, {Corsi}, {Cortese}, {Costa}, {Coughlin}, {Coughlin}, {Coulon}, {Countryman}, {Couvares}, {Covas}, {Cowan}, {Coward}, {Cowart}, {Coyne}, {Coyne}, {Creighton}, {Creighton}, {Cripe}, {Crowder}, {Cullen}, {Cumming}, {Cunningham}, {Cuoco}, {Dal Canton}, {D{\'a}lya}, {Danilishin}, {D'Antonio}, {Danzmann}, {Dasgupta}, {Da Silva Costa}, {Dattilo}, {Dave}, {Davier}, {Davis}, {Daw}, {Day}, {De}, {DeBra}, {Degallaix}, {De Laurentis}, {Del{\'e}glise}, {Del Pozzo}, {Demos}, {Denker}, {Dent}, {De Pietri}, {Dergachev}, {De Rosa}, {DeRosa}, {De Rossi}, {DeSalvo}, {de Varona}, {Devenson},
  {Dhurandhar}, {D{\'\i}az}, {Dietrich}, {Di Fiore}, {Di Giovanni}, {Di Girolamo}, {Di Lieto}, {Di Pace}, {Di Palma}, {Di Renzo}, {Doctor}, {Dolique}, {Donovan}, {Dooley}, {Doravari}, {Dorrington}, {Douglas}, {Dovale {\'A}lvarez}, {Downes}, {Drago}, {Dreissigacker}, {Driggers}, {Du}, {Ducrot}, {Dudi}, {Dupej}, {Dwyer}, {Edo}, {Edwards}, {Effler}, {Eggenstein}, {Ehrens}, {Eichholz}, {Eikenberry}, {Eisenstein}, {Essick}, {Estevez}, {Etienne}, {Etzel}, {Evans}, {Evans}, {Factourovich}, {Fafone}, {Fair}, {Fairhurst}, {Fan}, {Farinon}, {Farr}, {Farr}, {Fauchon-Jones}, {Favata}, {Fays}, {Fee}, {Fehrmann}, {Feicht}, {Fejer}, {Fernandez-Galiana}, {Ferrante}, {Ferreira}, {Ferrini}, {Fidecaro}, {Finstad}, {Fiori}, {Fiorucci}, {Fishbach}, {Fisher}, {Fitz-Axen}, {Flaminio}, {Fletcher}, {Fong}, {Font}, {Forsyth}, {Forsyth}, {Fournier}, {Frasca}, {Frasconi}, {Frei}, {Freise}, {Frey}, {Frey}, {Fries}, {Fritschel}, {Frolov}, {Fulda}, {Fyffe}, {Gabbard}, {Gadre}, {Gaebel}, {Gair}, {Gammaitoni}, {Ganija}, {Gaonkar},
  {Garcia-Quiros}, {Garufi}, {Gateley}, {Gaudio}, {Gaur}, {Gayathri}, {Gehrels}, {Gemme}, {Genin}, {Gennai}, {George}, {George}, {Gergely}, {Germain}, {Ghonge}, {Ghosh}, {Ghosh}, {Ghosh}, {Giaime}, {Giardina}, {Giazotto}, {Gill}, {Glover}, {Goetz}, {Goetz}, {Gomes}, {Goncharov}, {Gonz{\'a}lez}, {Gonzalez Castro}, {Gopakumar}, {Gorodetsky}, {Gossan}, {Gosselin}, {Gouaty}, {Grado}, {Graef}, {Granata}, {Grant}, {Gras}, {Gray}, {Greco}, {Green}, {Gretarsson}, {Groot}, {Grote}, {Grunewald}, {Gruning}, {Guidi}, {Guo}, {Gupta}, {Gupta}, {Gushwa}, {Gustafson}, {Gustafson}, {Halim}, {Hall}, {Hall}, {Hamilton}, {Hammond}, {Haney}, {Hanke}, {Hanks}, {Hanna}, {Hannam}, {Hannuksela}, {Hanson}, {Hardwick}, {Harms}, {Harry}, {Harry}, {Hart}, {Haster}, {Haughian}, {Healy}, {Heidmann}, {Heintze}, {Heitmann}, {Hello}, {Hemming}, {Hendry}, {Heng}, {Hennig}, {Heptonstall}, {Heurs}, {Hild}, {Hinderer}, {Ho}, {Hoak}, {Hofman}, {Holt}, {Holz}, {Hopkins}, {Horst}, {Hough}, {Houston}, {Howell}, {Hreibi}, {Hu}, {Huerta}, {Huet},
  {Hughey}, {Husa}, {Huttner}, {Huynh-Dinh}, {Indik}, {Inta}, {Intini}, {Isa}, {Isac}, {Isi}, {Iyer}, {Izumi}, {Jacqmin}, {Jani}, {Jaranowski}, {Jawahar}, {Jim{\'e}nez-Forteza}, {Johnson}, {Johnson-McDaniel}, {Jones}, {Jones}, {Jonker}, {Ju}, {Junker}, {Kalaghatgi}, {Kalogera}, {Kamai}, {Kandhasamy}, {Kang}, {Kanner}, {Kapadia}, {Karki}, {Karvinen}, {Kasprzack}, {Kastaun}, {Katolik}, {Katsavounidis}, {Katzman}, {Kaufer}, {Kawabe}, {K{\'e}f{\'e}lian}, {Keitel}, {Kemball}, {Kennedy}, {Kent}, {Key}, {Khalili}, {Khan}, {Khan}, {Khan}, {Khazanov}, {Kijbunchoo}, {Kim}, {Kim}, {Kim}, {Kim}, {Kim}, {Kim}, {Kimbrell}, {King}, {King}, {Kinley-Hanlon}, {Kirchhoff}, {Kissel}, {Kleybolte}, {Klimenko}, {Knowles}, {Koch}, {Koehlenbeck}, {Koley}, {Kondrashov}, {Kontos}, {Korobko}, {Korth}, {Kowalska}, {Kozak}, {Kr{\"a}mer}, {Kringel}, {Krishnan}, {Kr{\'o}lak}, {Kuehn}, {Kumar}, {Kumar}, {Kumar}, {Kuo}, {Kutynia}, {Kwang}, {Lackey}, {Lai}, {Landry}, {Lang}, {Lange}, {Lantz}, {Lanza}, {Larson}, {Lartaux-Vollard}, {Lasky},
  {Laxen}, {Lazzarini}, {Lazzaro}, {Leaci}, {Leavey}, {Lee}, {Lee}, {Lee}, {Lee}, {Lee}, {Lehmann}, {Lenon}, {Leon}, {Leonardi}, {Leroy}, {Letendre}, {Levin}, {Li}, {Linker}, {Littenberg}, {Liu}, {Liu}, {Lo}, {Lockerbie}, {London}, {Lord}, {Lorenzini}, {Loriette}, {Lormand}, {Losurdo}, {Lough}, {Lousto}, {Lovelace}, {L{\"u}ck}, {Lumaca}, {Lundgren}, {Lynch}, {Ma}, {Macas}, {Macfoy}, {Machenschalk}, {MacInnis}, {Macleod}, {Maga{\~n}a Hernandez}, {Maga{\~n}a-Sandoval}, {Maga{\~n}a Zertuche}, {Magee}, {Majorana}, {Maksimovic}, {Man}, {Mandic}, {Mangano}, {Mansell}, {Manske}, {Mantovani}, {Marchesoni}, {Marion}, {M{\'a}rka}, {M{\'a}rka}, {Markakis}, {Markosyan}, {Markowitz}, {Maros}, {Marquina}, {Marsh}, {Martelli}, {Martellini}, {Martin}, {Martin}, {Martynov}, {Marx}, {Mason}, {Massera}, {Masserot}, {Massinger}, {Masso-Reid}, {Mastrogiovanni}, {Matas}, {Matichard}, {Matone}, {Mavalvala}, {Mazumder}, {McCarthy}, {McClelland}, {McCormick}, {McCuller}, {McGuire}, {McIntyre}, {McIver}, {McManus}, {McNeill}, {McRae},
  {McWilliams}, {Meacher}, {Meadors}, {Mehmet}, {Meidam}, {Mejuto-Villa}, {Melatos}, {Mendell}, {Mercer}, {Merilh}, {Merzougui}, {Meshkov}, {Messenger}, {Messick}, {Metzdorff}, {Meyers}, {Miao}, {Michel}, {Middleton}, {Mikhailov}, {Milano}, {Miller}, {Miller}, {Miller}, {Millhouse}, {Milovich-Goff}, {Minazzoli}, {Minenkov}, {Ming}, {Mishra}, {Mitra}, {Mitrofanov}, {Mitselmakher}, {Mittleman}, {Moffa}, {Moggi}, {Mogushi}, {Mohan}, {Mohapatra}, {Molina}, {Montani}, {Moore}, {Moraru}, {Moreno}, {Morisaki}, {Morriss}, {Mours}, {Mow-Lowry}, {Mueller}, {Muir}, {Mukherjee}, {Mukherjee}, {Mukherjee}, {Mukund}, {Mullavey}, {Munch}, {Mu{\~n}iz}, {Muratore}, {Murray}, {Nagar}, {Napier}, {Nardecchia}, {Naticchioni}, {Nayak}, {Neilson}, {Nelemans}, {Nelson}, {Nery}, {Neunzert}, {Nevin}, {Newport}, {Newton}, {Ng}, {Nguyen}, {Nguyen}, {Nichols}, {Nielsen}, {Nissanke}, {Nitz}, {Noack}, {Nocera}, {Nolting}, {North}, {Nuttall}, {Oberling}, {O'Dea}, {Ogin}, {Oh}, {Oh}, {Ohme}, {Okada}, {Oliver}, {Oppermann}, {Oram}, {O'Reilly},
  {Ormiston}, {Ortega}, {O'Shaughnessy}, {Ossokine}, {Ottaway}, {Overmier}, {Owen}, {Pace}, {Page}, {Page}, {Pai}, {Pai}, {Palamos}, {Palashov}, {Palomba}, {Pal-Singh}, {Pan}, {Pan}, {Pang}, {Pang}, {Pankow}, {Pannarale}, {Pant}, {Paoletti}, {Paoli}, {Papa}, {Parida}, {Parker}, {Pascucci}, {Pasqualetti}, {Passaquieti}, {Passuello}, {Patil}, {Patricelli}, {Pearlstone}, {Pedraza}, {Pedurand}, {Pekowsky}, {Pele}, {Penn}, {Perez}, {Perreca}, {Perri}, {Pfeiffer}, {Phelps}, {Piccinni}, {Pichot}, {Piergiovanni}, {Pierro}, {Pillant}, {Pinard}, {Pinto}, {Pirello}, {Pitkin}, {Poe}, {Poggiani}, {Popolizio}, {Porter}, {Post}, {Powell}, {Prasad}, {Pratt}, {Pratten}, {Predoi}, {Prestegard}, {Prijatelj}, {Principe}, {Privitera}, {Prix}, {Prodi}, {Prokhorov}, {Puncken}, {Punturo}, {Puppo}, {P{\"u}rrer}, {Qi}, {Quetschke}, {Quintero}, {Quitzow-James}, {Raab}, {Rabeling}, {Radkins}, {Raffai}, {Raja}, {Rajan}, {Rajbhandari}, {Rakhmanov}, {Ramirez}, {Ramos-Buades}, {Rapagnani}, {Raymond}, {Razzano}, {Read}, {Regimbau}, {Rei},
  {Reid}, {Reitze}, {Ren}, {Reyes}, {Ricci}, {Ricker}, {Rieger}, {Riles}, {Rizzo}, {Robertson}, {Robie}, {Robinet}, {Rocchi}, {Rolland}, {Rollins}, {Roma}, {Romano}, {Romano}, {Romel}, {Romie}, {Rosi{\'n}ska}, {Ross}, {Rowan}, {R{\"u}diger}, {Ruggi}, {Rutins}, {Ryan}, {Sachdev}, {Sadecki}, {Sadeghian}, {Sakellariadou}, {Salconi}, {Saleem}, {Salemi}, {Samajdar}, {Sammut}, {Sampson}, {Sanchez}, {Sanchez}, {Sanchis-Gual}, {Sandberg}, {Sanders}, {Sassolas}, {Sathyaprakash}, {Saulson}, {Sauter}, {Savage}, {Sawadsky}, {Schale}, {Scheel}, {Scheuer}, {Schmidt}, {Schmidt}, {Schnabel}, {Schofield}, {Sch{\"o}nbeck}, {Schreiber}, {Schuette}, {Schulte}, {Schutz}, {Schwalbe}, {Scott}, {Scott}, {Seidel}, {Sellers}, {Sengupta}, {Sentenac}, {Sequino}, {Sergeev}, {Shaddock}, {Shaffer}, {Shah}, {Shahriar}, {Shaner}, {Shao}, {Shapiro}, {Shawhan}, {Sheperd}, {Shoemaker}, {Shoemaker}, {Siellez}, {Siemens}, {Sieniawska}, {Sigg}, {Silva}, {Singer}, {Singh}, {Singhal}, {Sintes}, {Slagmolen}, {Smith}, {Smith}, {Smith}, {Somala},
  {Son}, {Sonnenberg}, {Sorazu}, {Sorrentino}, {Souradeep}, {Spencer}, {Srivastava}, {Staats}, {Staley}, {Steinke}, {Steinlechner}, {Steinlechner}, {Steinmeyer}, {Stevenson}, {Stone}, {Stops}, {Strain}, {Stratta}, {Strigin}, {Strunk}, {Sturani}, {Stuver}, {Summerscales}, {Sun}, {Sunil}, {Suresh}, {Sutton}, {Swinkels}, {Szczepa{\'n}czyk}, {Tacca}, {Tait}, {Talbot}, {Talukder}, {Tanner}, {T{\'a}pai}, {Taracchini}, {Tasson}, {Taylor}, {Taylor}, {Tewari}, {Theeg}, {Thies}, {Thomas}, {Thomas}, {Thomas}, {Thorne}, {Thorne}, {Thrane}, {Tiwari}, {Tiwari}, {Tokmakov}, {Toland}, {Tonelli}, {Tornasi}, {Torres-Forn{\'e}}, {Torrie}, {T{\"o}yr{\"a}}, {Travasso}, {Traylor}, {Trinastic}, {Tringali}, {Trozzo}, {Tsang}, {Tse}, {Tso}, {Tsukada}, {Tsuna}, {Tuyenbayev}, {Ueno}, {Ugolini}, {Unnikrishnan}, {Urban}, {Usman}, {Vahlbruch}, {Vajente}, {Valdes}, {Vallisneri}, {van Bakel}, {van Beuzekom}, {van den Brand}, {Van Den Broeck}, {Vander-Hyde}, {van der Schaaf}, {van Heijningen}, {van Veggel}, {Vardaro}, {Varma}, {Vass},
  {Vas{\'u}th}, {Vecchio}, {Vedovato}, {Veitch}, {Veitch}, {Venkateswara}, {Venugopalan}, {Verkindt}, {Vetrano}, {Vicer{\'e}}, {Viets}, {Vinciguerra}, {Vine}, {Vinet}, {Vitale}, {Vo}, {Vocca}, {Vorvick}, {Vyatchanin}, {Wade}, {Wade}, {Wade}, {Walet}, {Walker}, {Wallace}, {Walsh}, {Wang}, {Wang}, {Wang}, {Wang}, {Wang}, {Ward}, {Warner}, {Was}, {Watchi}, {Weaver}, {Wei}, {Weinert}, {Weinstein}, {Weiss}, {Wen}, {Wessel}, {We{\ss}els}, {Westerweck}, {Westphal}, {Wette}, {Whelan}, {Whitcomb}, {Whiting}, {Whittle}, {Wilken}, {Williams}, {Williams}, {Williamson}, {Willis}, {Willke}, {Wimmer}, {Winkler}, {Wipf}, {Wittel}, {Woan}, {Woehler}, {Wofford}, {Wong}, {Worden}, {Wright}, {Wu}, {Wysocki}, {Xiao}, {Yamamoto}, {Yancey}, {Yang}, {Yap}, {Yazback}, {Yu}, {Yu}, {Yvert}, {Zadro{\.Z}ny}, {Zanolin}, {Zelenova}, {Zendri}, {Zevin}, {Zhang}, {Zhang}, {Zhang}, {Zhang}, {Zhao}, {Zhou}, {Zhou}, {Zhu}, {Zhu}, {Zimmerman}, {Zucker}, {Zweizig}, {LIGO Scientific Collaboration}, \& {Virgo Collaboration}}]{2017PhRvL.119p1101A}
{Abbott}, B.~P., {Abbott}, R., {Abbott}, T.~D., {et~al.} 2017{\natexlab{a}}, \prl, 119, 161101, \dodoi{10.1103/PhysRevLett.119.161101}

\bibitem[{{Abbott} {et~al.}(2017{\natexlab{b}}){Abbott}, {Abbott}, {Abbott}, {Acernese}, {Ackley}, {Adams}, {Adams}, {Addesso}, {Adhikari}, {Adya}, {Affeldt}, {Afrough}, {Agarwal}, {Agathos}, {Agatsuma}, {Aggarwal}, {Aguiar}, {Aiello}, {Ain}, {Ajith}, {Allen}, {Allen}, {Allocca}, {Altin}, {Amato}, {Ananyeva}, {Anderson}, {Anderson}, {Angelova}, {Antier}, {Appert}, {Arai}, {Araya}, {Areeda}, {Arnaud}, {Arun}, {Ascenzi}, {Ashton}, {Ast}, {Aston}, {Astone}, {Atallah}, {Aufmuth}, {Aulbert}, {AultONeal}, {Austin}, {Avila-Alvarez}, {Babak}, {Bacon}, {Bader}, {Bae}, {Baker}, {Baldaccini}, {Ballardin}, {Ballmer}, {Banagiri}, {Barayoga}, {Barclay}, {Barish}, {Barker}, {Barkett}, {Barone}, {Barr}, {Barsotti}, {Barsuglia}, {Barta}, {Barthelmy}, {Bartlett}, {Bartos}, {Bassiri}, {Basti}, {Batch}, {Bawaj}, {Bayley}, {Bazzan}, {B{\'e}csy}, {Beer}, {Bejger}, {Belahcene}, {Bell}, {Berger}, {Bergmann}, {Bero}, {Berry}, {Bersanetti}, {Bertolini}, {Betzwieser}, {Bhagwat}, {Bhandare}, {Bilenko}, {Billingsley}, {Billman}, {Birch},
  {Birney}, {Birnholtz}, {Biscans}, {Biscoveanu}, {Bisht}, {Bitossi}, {Biwer}, {Bizouard}, {Blackburn}, {Blackman}, {Blair}, {Blair}, {Blair}, {Bloemen}, {Bock}, {Bode}, {Boer}, {Bogaert}, {Bohe}, {Bondu}, {Bonilla}, {Bonnand}, {Boom}, {Bork}, {Boschi}, {Bose}, {Bossie}, {Bouffanais}, {Bozzi}, {Bradaschia}, {Brady}, {Branchesi}, {Brau}, {Briant}, {Brillet}, {Brinkmann}, {Brisson}, {Brockill}, {Broida}, {Brooks}, {Brown}, {Brown}, {Brunett}, {Buchanan}, {Buikema}, {Bulik}, {Bulten}, {Buonanno}, {Buskulic}, {Buy}, {Byer}, {Cabero}, {Cadonati}, {Cagnoli}, {Cahillane}, {Calder{\'o}n Bustillo}, {Callister}, {Calloni}, {Camp}, {Canepa}, {Canizares}, {Cannon}, {Cao}, {Cao}, {Capano}, {Capocasa}, {Carbognani}, {Caride}, {Carney}, {Casanueva Diaz}, {Casentini}, {Caudill}, {Cavagli{\`a}}, {Cavalier}, {Cavalieri}, {Cella}, {Cepeda}, {Cerd{\'a}-Dur{\'a}n}, {Cerretani}, {Cesarini}, {Chamberlin}, {Chan}, {Chao}, {Charlton}, {Chase}, {Chassande-Mottin}, {Chatterjee}, {Chatziioannou}, {Cheeseboro}, {Chen}, {Chen}, {Chen},
  {Cheng}, {Chia}, {Chincarini}, {Chiummo}, {Chmiel}, {Cho}, {Cho}, {Chow}, {Christensen}, {Chu}, {Chua}, {Chua}, {Chung}, {Chung}, {Ciani}, {Ciolfi}, {Cirelli}, {Cirone}, {Clara}, {Clark}, {Clearwater}, {Cleva}, {Cocchieri}, {Coccia}, {Cohadon}, {Cohen}, {Colla}, {Collette}, {Cominsky}, {Constancio}, {Conti}, {Cooper}, {Corban}, {Corbitt}, {Cordero-Carri{\'o}n}, {Corley}, {Cornish}, {Corsi}, {Cortese}, {Costa}, {Coughlin}, {Coughlin}, {Coulon}, {Countryman}, {Couvares}, {Covas}, {Cowan}, {Coward}, {Cowart}, {Coyne}, {Coyne}, {Creighton}, {Creighton}, {Cripe}, {Crowder}, {Cullen}, {Cumming}, {Cunningham}, {Cuoco}, {Dal Canton}, {D{\'a}lya}, {Danilishin}, {D'Antonio}, {Danzmann}, {Dasgupta}, {Da Silva Costa}, {Dattilo}, {Dave}, {Davier}, {Davis}, {Daw}, {Day}, {De}, {DeBra}, {Degallaix}, {De Laurentis}, {Del{\'e}glise}, {Del Pozzo}, {Demos}, {Denker}, {Dent}, {De Pietri}, {Dergachev}, {De Rosa}, {DeRosa}, {De Rossi}, {DeSalvo}, {de Varona}, {Devenson}, {Dhurandhar}, {D{\'\i}az}, {Di Fiore}, {Di Giovanni}, {Di
  Girolamo}, {Di Lieto}, {Di Pace}, {Di Palma}, {Di Renzo}, {Doctor}, {Dolique}, {Donovan}, {Dooley}, {Doravari}, {Dorrington}, {Douglas}, {Dovale {\'A}lvarez}, {Downes}, {Drago}, {Dreissigacker}, {Driggers}, {Du}, {Ducrot}, {Dupej}, {Dwyer}, {Edo}, {Edwards}, {Effler}, {Ehrens}, {Eichholz}, {Eikenberry}, {Eisenstein}, {Essick}, {Estevez}, {Etienne}, {Etzel}, {Evans}, {Evans}, {Factourovich}, {Fafone}, {Fair}, {Fairhurst}, {Fan}, {Farinon}, {Farr}, {Farr}, {Fauchon-Jones}, {Favata}, {Fays}, {Fee}, {Fehrmann}, {Feicht}, {Fejer}, {Fernandez-Galiana}, {Ferrante}, {Ferreira}, {Ferrini}, {Fidecaro}, {Finstad}, {Fiori}, {Fiorucci}, {Fishbach}, {Fisher}, {Fitz-Axen}, {Flaminio}, {Fletcher}, {Fong}, {Font}, {Forsyth}, {Forsyth}, {Fournier}, {Frasca}, {Frasconi}, {Frei}, {Freise}, {Frey}, {Frey}, {Fries}, {Fritschel}, {Frolov}, {Fulda}, {Fyffe}, {Gabbard}, {Gadre}, {Gaebel}, {Gair}, {Gammaitoni}, {Ganija}, {Gaonkar}, {Garcia-Quiros}, {Garufi}, {Gateley}, {Gaudio}, {Gaur}, {Gayathri}, {Gehrels}, {Gemme}, {Genin},
  {Gennai}, {George}, {George}, {Gergely}, {Germain}, {Ghonge}, {Ghosh}, {Ghosh}, {Ghosh}, {Giaime}, {Giardina}, {Giazotto}, {Gill}, {Glover}, {Goetz}, {Goetz}, {Gomes}, {Goncharov}, {Gonz{\'a}lez}, {Gonzalez Castro}, {Gopakumar}, {Gorodetsky}, {Gossan}, {Gosselin}, {Gouaty}, {Grado}, {Graef}, {Granata}, {Grant}, {Gras}, {Gray}, {Greco}, {Green}, {Gretarsson}, {Griswold}, {Groot}, {Grote}, {Grunewald}, {Gruning}, {Guidi}, {Guo}, {Gupta}, {Gupta}, {Gushwa}, {Gustafson}, {Gustafson}, {Halim}, {Hall}, {Hall}, {Hamilton}, {Hammond}, {Haney}, {Hanke}, {Hanks}, {Hanna}, {Hannam}, {Hannuksela}, {Hanson}, {Hardwick}, {Harms}, {Harry}, {Harry}, {Hart}, {Haster}, {Haughian}, {Healy}, {Heidmann}, {Heintze}, {Heitmann}, {Hello}, {Hemming}, {Hendry}, {Heng}, {Hennig}, {Heptonstall}, {Heurs}, {Hild}, {Hinderer}, {Hoak}, {Hofman}, {Holt}, {Holz}, {Hopkins}, {Horst}, {Hough}, {Houston}, {Howell}, {Hreibi}, {Hu}, {Huerta}, {Huet}, {Hughey}, {Husa}, {Huttner}, {Huynh-Dinh}, {Indik}, {Inta}, {Intini}, {Isa}, {Isac}, {Isi},
  {Iyer}, {Izumi}, {Jacqmin}, {Jani}, {Jaranowski}, {Jawahar}, {Jim{\'e}nez-Forteza}, {Johnson}, {Jones}, {Jones}, {Jonker}, {Ju}, {Junker}, {Kalaghatgi}, {Kalogera}, {Kamai}, {Kandhasamy}, {Kang}, {Kanner}, {Kapadia}, {Karki}, {Karvinen}, {Kasprzack}, {Katolik}, {Katsavounidis}, {Katzman}, {Kaufer}, {Kawabe}, {K{\'e}f{\'e}lian}, {Keitel}, {Kemball}, {Kennedy}, {Kent}, {Key}, {Khalili}, {Khan}, {Khan}, {Khan}, {Khazanov}, {Kijbunchoo}, {Kim}, {Kim}, {Kim}, {Kim}, {Kim}, {Kim}, {Kimbrell}, {King}, {King}, {Kinley-Hanlon}, {Kirchhoff}, {Kissel}, {Kleybolte}, {Klimenko}, {Knowles}, {Koch}, {Koehlenbeck}, {Koley}, {Kondrashov}, {Kontos}, {Korobko}, {Korth}, {Kowalska}, {Kozak}, {Kr{\"a}mer}, {Kringel}, {Krishnan}, {Kr{\'o}lak}, {Kuehn}, {Kumar}, {Kumar}, {Kumar}, {Kuo}, {Kutynia}, {Kwang}, {Lackey}, {Lai}, {Landry}, {Lang}, {Lange}, {Lantz}, {Lanza}, {Larson}, {Lartaux-Vollard}, {Lasky}, {Laxen}, {Lazzarini}, {Lazzaro}, {Leaci}, {Leavey}, {Lee}, {Lee}, {Lee}, {Lee}, {Lee}, {Lehmann}, {Lenon}, {Leonardi}, {Leroy},
  {Letendre}, {Levin}, {Li}, {Linker}, {Littenberg}, {Liu}, {Lo}, {Lockerbie}, {London}, {Lord}, {Lorenzini}, {Loriette}, {Lormand}, {Losurdo}, {Lough}, {Lousto}, {Lovelace}, {L{\"u}ck}, {Lumaca}, {Lundgren}, {Lynch}, {Ma}, {Macas}, {Macfoy}, {Machenschalk}, {MacInnis}, {Macleod}, {Maga{\~n}a Hernandez}, {Maga{\~n}a-Sandoval}, {Maga{\~n}a Zertuche}, {Magee}, {Majorana}, {Maksimovic}, {Man}, {Mandic}, {Mangano}, {Mansell}, {Manske}, {Mantovani}, {Marchesoni}, {Marion}, {M{\'a}rka}, {M{\'a}rka}, {Markakis}, {Markosyan}, {Markowitz}, {Maros}, {Marquina}, {Marsh}, {Martelli}, {Martellini}, {Martin}, {Martin}, {Martynov}, {Mason}, {Massera}, {Masserot}, {Massinger}, {Masso-Reid}, {Mastrogiovanni}, {Matas}, {Matichard}, {Matone}, {Mavalvala}, {Mazumder}, {McCarthy}, {McClelland}, {McCormick}, {McCuller}, {McGuire}, {McIntyre}, {McIver}, {McManus}, {McNeill}, {McRae}, {McWilliams}, {Meacher}, {Meadors}, {Mehmet}, {Meidam}, {Mejuto-Villa}, {Melatos}, {Mendell}, {Mercer}, {Merilh}, {Merzougui}, {Meshkov}, {Messenger},
  {Messick}, {Metzdorff}, {Meyers}, {Miao}, {Michel}, {Middleton}, {Mikhailov}, {Milano}, {Miller}, {Miller}, {Miller}, {Millhouse}, {Milovich-Goff}, {Minazzoli}, {Minenkov}, {Ming}, {Mishra}, {Mitra}, {Mitrofanov}, {Mitselmakher}, {Mittleman}, {Moffa}, {Moggi}, {Mogushi}, {Mohan}, {Mohapatra}, {Montani}, {Moore}, {Moraru}, {Moreno}, {Morriss}, {Mours}, {Mow-Lowry}, {Mueller}, {Muir}, {Mukherjee}, {Mukherjee}, {Mukherjee}, {Mukund}, {Mullavey}, {Munch}, {Mu{\~n}iz}, {Muratore}, {Murray}, {Napier}, {Nardecchia}, {Naticchioni}, {Nayak}, {Neilson}, {Nelemans}, {Nelson}, {Nery}, {Neunzert}, {Nevin}, {Newport}, {Newton}, {Ng}, {Nguyen}, {Nguyen}, {Nichols}, {Nielsen}, {Nissanke}, {Nitz}, {Noack}, {Nocera}, {Nolting}, {North}, {Nuttall}, {Oberling}, {O'Dea}, {Ogin}, {Oh}, {Oh}, {Ohme}, {Okada}, {Oliver}, {Oppermann}, {Oram}, {O'Reilly}, {Ormiston}, {Ortega}, {O'Shaughnessy}, {Ossokine}, {Ottaway}, {Overmier}, {Owen}, {Pace}, {Page}, {Page}, {Pai}, {Pai}, {Palamos}, {Palashov}, {Palomba}, {Pal-Singh}, {Pan}, {Pan},
  {Pang}, {Pang}, {Pankow}, {Pannarale}, {Pant}, {Paoletti}, {Paoli}, {Papa}, {Parida}, {Parker}, {Pascucci}, {Pasqualetti}, {Passaquieti}, {Passuello}, {Patil}, {Patricelli}, {Pearlstone}, {Pedraza}, {Pedurand}, {Pekowsky}, {Pele}, {Penn}, {Perez}, {Perreca}, {Perri}, {Pfeiffer}, {Phelps}, {Piccinni}, {Pichot}, {Piergiovanni}, {Pierro}, {Pillant}, {Pinard}, {Pinto}, {Pirello}, {Pitkin}, {Poe}, {Poggiani}, {Popolizio}, {Porter}, {Post}, {Powell}, {Prasad}, {Pratt}, {Pratten}, {Predoi}, {Prestegard}, {Price}, {Prijatelj}, {Principe}, {Privitera}, {Prodi}, {Prokhorov}, {Puncken}, {Punturo}, {Puppo}, {P{\"u}rrer}, {Qi}, {Quetschke}, {Quintero}, {Quitzow-James}, {Raab}, {Rabeling}, {Radkins}, {Raffai}, {Raja}, {Rajan}, {Rajbhandari}, {Rakhmanov}, {Ramirez}, {Ramos-Buades}, {Rapagnani}, {Raymond}, {Razzano}, {Read}, {Regimbau}, {Rei}, {Reid}, {Reitze}, {Ren}, {Reyes}, {Ricci}, {Ricker}, {Rieger}, {Riles}, {Rizzo}, {Robertson}, {Robie}, {Robinet}, {Rocchi}, {Rolland}, {Rollins}, {Roma}, {Romano}, {Romel}, {Romie},
  {Rosi{\'n}ska}, {Ross}, {Rowan}, {R{\"u}diger}, {Ruggi}, {Rutins}, {Ryan}, {Sachdev}, {Sadecki}, {Sadeghian}, {Sakellariadou}, {Salconi}, {Saleem}, {Salemi}, {Samajdar}, {Sammut}, {Sampson}, {Sanchez}, {Sanchez}, {Sanchis-Gual}, {Sandberg}, {Sanders}, {Sassolas}, {Sathyaprakash}, {Saulson}, {Sauter}, {Savage}, {Sawadsky}, {Schale}, {Scheel}, {Scheuer}, {Schmidt}, {Schmidt}, {Schnabel}, {Schofield}, {Sch{\"o}nbeck}, {Schreiber}, {Schuette}, {Schulte}, {Schutz}, {Schwalbe}, {Scott}, {Scott}, {Seidel}, {Sellers}, {Sengupta}, {Sentenac}, {Sequino}, {Sergeev}, {Shaddock}, {Shaffer}, {Shah}, {Shahriar}, {Shaner}, {Shao}, {Shapiro}, {Shawhan}, {Sheperd}, {Shoemaker}, {Shoemaker}, {Siellez}, {Siemens}, {Sieniawska}, {Sigg}, {Silva}, {Singer}, {Singh}, {Singhal}, {Sintes}, {Slagmolen}, {Smith}, {Smith}, {Smith}, {Somala}, {Son}, {Sonnenberg}, {Sorazu}, {Sorrentino}, {Souradeep}, {Spencer}, {Srivastava}, {Staats}, {Staley}, {Steinke}, {Steinlechner}, {Steinlechner}, {Steinmeyer}, {Stevenson}, {Stone}, {Stops},
  {Strain}, {Stratta}, {Strigin}, {Strunk}, {Sturani}, {Stuver}, {Summerscales}, {Sun}, {Sunil}, {Suresh}, {Sutton}, {Swinkels}, {Szczepa{\'n}czyk}, {Tacca}, {Tait}, {Talbot}, {Talukder}, {Tanner}, {T{\'a}pai}, {Taracchini}, {Tasson}, {Taylor}, {Taylor}, {Tewari}, {Theeg}, {Thies}, {Thomas}, {Thomas}, {Thomas}, {Thorne}, {Thorne}, {Thrane}, {Tiwari}, {Tiwari}, {Tokmakov}, {Toland}, {Tonelli}, {Tornasi}, {Torres-Forn{\'e}}, {Torrie}, {T{\"o}yr{\"a}}, {Travasso}, {Traylor}, {Trinastic}, {Tringali}, {Trozzo}, {Tsang}, {Tse}, {Tso}, {Tsukada}, {Tsuna}, {Tuyenbayev}, {Ueno}, {Ugolini}, {Unnikrishnan}, {Urban}, {Usman}, {Vahlbruch}, {Vajente}, {Valdes}, {van Bakel}, {van Beuzekom}, {van den Brand}, {Van Den Broeck}, {Vander-Hyde}, {van der Schaaf}, {van Heijningen}, {van Veggel}, {Vardaro}, {Varma}, {Vass}, {Vas{\'u}th}, {Vecchio}, {Vedovato}, {Veitch}, {Veitch}, {Venkateswara}, {Venugopalan}, {Verkindt}, {Vetrano}, {Vicer{\'e}}, {Viets}, {Vinciguerra}, {Vine}, {Vinet}, {Vitale}, {Vo}, {Vocca}, {Vorvick},
  {Vyatchanin}, {Wade}, {Wade}, {Wade}, {Walet}, {Walker}, {Wallace}, {Walsh}, {Wang}, {Wang}, {Wang}, {Wang}, {Wang}, {Ward}, {Warner}, {Was}, {Watchi}, {Weaver}, {Wei}, {Weinert}, {Weinstein}, {Weiss}, {Wen}, {Wessel}, {Wessels}, {Westerweck}, {Westphal}, {Wette}, {Whelan}, {Whitcomb}, {Whiting}, {Whittle}, {Wilken}, {Williams}, {Williams}, {Williamson}, {Willis}, {Willke}, {Wimmer}, {Winkler}, {Wipf}, {Wittel}, {Woan}, {Woehler}, {Wofford}, {Wong}, {Worden}, {Wright}, {Wu}, {Wysocki}, {Xiao}, {Yamamoto}, {Yancey}, {Yang}, {Yap}, {Yazback}, {Yu}, {Yu}, {Yvert}, {Zadro{\.z}ny}, {Zanolin}, {Zelenova}, {Zendri}, {Zevin}, {Zhang}, {Zhang}, {Zhang}, {Zhang}, {Zhao}, {Zhou}, {Zhou}, {Zhu}, {Zhu}, {Zimmerman}, {Zucker}, {Zweizig}, {LIGO Scientific Collaboration}, {Virgo Collaboration}, {Wilson-Hodge}, {Bissaldi}, {Blackburn}, {Briggs}, {Burns}, {Cleveland}, {Connaughton}, {Gibby}, {Giles}, {Goldstein}, {Hamburg}, {Jenke}, {Hui}, {Kippen}, {Kocevski}, {McBreen}, {Meegan}, {Paciesas}, {Poolakkil}, {Preece},
  {Racusin}, {Roberts}, {Stanbro}, {Veres}, {von Kienlin}, {GBM}, {Savchenko}, {Ferrigno}, {Kuulkers}, {Bazzano}, {Bozzo}, {Brandt}, {Chenevez}, {Courvoisier}, {Diehl}, {Domingo}, {Hanlon}, {Jourdain}, {Laurent}, {Lebrun}, {Lutovinov}, {Martin-Carrillo}, {Mereghetti}, {Natalucci}, {Rodi}, {Roques}, {Sunyaev}, {Ubertini}, {INTEGRAL}, {Aartsen}, {Ackermann}, {Adams}, {Aguilar}, {Ahlers}, {Ahrens}, {Samarai}, {Altmann}, {Andeen}, {Anderson}, {Ansseau}, {Anton}, {Arg{\"u}elles}, {Auffenberg}, {Axani}, {Bagherpour}, {Bai}, {Barron}, {Barwick}, {Baum}, {Bay}, {Beatty}, {Becker Tjus}, {Bernardini}, {Besson}, {Binder}, {Bindig}, {Blaufuss}, {Blot}, {Bohm}, {B{\"o}rner}, {Bos}, {Bose}, {B{\"o}ser}, {Botner}, {Bourbeau}, {Bourbeau}, {Bradascio}, {Braun}, {Brayeur}, {Brenzke}, {Bretz}, {Bron}, {Brostean-Kaiser}, {Burgman}, {Carver}, {Casey}, {Casier}, {Cheung}, {Chirkin}, {Christov}, {Clark}, {Classen}, {Coenders}, {Collin}, {Conrad}, {Cowen}, {Cross}, {Day}, {de Andr{\'e}}, {De Clercq}, {DeLaunay}, {Dembinski}, {De
  Ridder}, {Desiati}, {de Vries}, {de Wasseige}, {de With}, {DeYoung}, {D{\'\i}az-V{\'e}lez}, {di Lorenzo}, {Dujmovic}, {Dumm}, {Dunkman}, {Dvorak}, {Eberhardt}, {Ehrhardt}, {Eichmann}, {Eller}, {Evenson}, {Fahey}, {Fazely}, {Felde}, {Filimonov}, {Finley}, {Flis}, {Franckowiak}, {Friedman}, {Fuchs}, {Gaisser}, {Gallagher}, {Gerhardt}, {Ghorbani}, {Giang}, {Glauch}, {Gl{\"u}senkamp}, {Goldschmidt}, {Gonzalez}, {Grant}, {Griffith}, {Haack}, {Hallgren}, {Halzen}, {Hanson}, {Hebecker}, {Heereman}, {Helbing}, {Hellauer}, {Hickford}, {Hignight}, {Hill}, {Hoffman}, {Hoffmann}, {Hokanson-Fasig}, {Hoshina}, {Huang}, {Huber}, {Hultqvist}, {H{\"u}nnefeld}, {In}, {Ishihara}, {Jacobi}, {Japaridze}, {Jeong}, {Jero}, {Jones}, {Kalaczynski}, {Kang}, {Kappes}, {Karg}, {Karle}, {Kauer}, {Keivani}, {Kelley}, {Kheirandish}, {Kim}, {Kim}, {Kintscher}, {Kiryluk}, {Kittler}, {Klein}, {Kohnen}, {Koirala}, {Kolanoski}, {K{\"o}pke}, {Kopper}, {Kopper}, {Koschinsky}, {Koskinen}, {Kowalski}, {Krings}, {Kroll}, {Kr{\"u}ckl}, {Kunnen},
  {Kunwar}, {Kurahashi}, {Kuwabara}, {Kyriacou}, {Labare}, {Lanfranchi}, {Larson}, {Lauber}, {Lesiak-Bzdak}, {Leuermann}, {Liu}, {Lu}, {L{\"u}nemann}, {Luszczak}, {Madsen}, {Maggi}, {Mahn}, {Mancina}, {Maruyama}, {Mase}, {Maunu}, {McNally}, {Meagher}, {Medici}, {Meier}, {Menne}, {Merino}, {Meures}, {Miarecki}, {Micallef}, {Moment{\'e}}, {Montaruli}, {Moore}, {Moulai}, {Nahnhauer}, {Nakarmi}, {Naumann}, {Neer}, {Niederhausen}, {Nowicki}, {Nygren}, {Obertacke Pollmann}, {Olivas}, {O'Murchadha}, {Palczewski}, {Pandya}, {Pankova}, {Peiffer}, {Pepper}, {P{\'e}rez de los Heros}, {Pieloth}, {Pinat}, {Price}, {Przybylski}, {Raab}, {R{\"a}del}, {Rameez}, {Rawlins}, {Rea}, {Reimann}, {Relethford}, {Relich}, {Resconi}, {Rhode}, {Richman}, {Robertson}, {Rongen}, {Rott}, {Ruhe}, {Ryckbosch}, {Rysewyk}, {S{\"a}lzer}, {Sanchez Herrera}, {Sandrock}, {Sandroos}, {Santander}, {Sarkar}, {Sarkar}, {Satalecka}, {Schlunder}, {Schmidt}, {Schneider}, {Schoenen}, {Sch{\"o}neberg}, {Schumacher}, {Seckel}, {Seunarine}, {Soedingrekso},
  {Soldin}, {Song}, {Spiczak}, {Spiering}, {Stachurska}, {Stamatikos}, {Stanev}, {Stasik}, {Stettner}, {Steuer}, {Stezelberger}, {Stokstad}, {St{\"o}ssl}, {Strotjohann}, {Stuttard}, {Sullivan}, {Sutherland}, {Taboada}, {Tatar}, {Tenholt}, {Ter-Antonyan}, {Terliuk}, {Te{\v{s}}i{\'c}}, {Tilav}, {Toale}, {Tobin}, {Toscano}, {Tosi}, {Tselengidou}, {Tung}, {Turcati}, {Turley}, {Ty}, {Unger}, {Usner}, {Vandenbroucke}, {Van Driessche}, {van Eijndhoven}, {Vanheule}, {van Santen}, {Vehring}, {Vogel}, {Vraeghe}, {Walck}, {Wallace}, {Wallraff}, {Wandler}, {Wandkowsky}, {Waza}, {Weaver}, {Weiss}, {Wendt}, {Werthebach}, {Whelan}, {Wiebe}, {Wiebusch}, {Wille}, {Williams}, {Wills}, {Wolf}, {Wood}, {Woolsey}, {Woschnagg}, {Xu}, {Xu}, {Xu}, {Yanez}, {Yodh}, {Yoshida}, {Yuan}, {Zoll}, {IceCube Collaboration}, {Balasubramanian}, {Mate}, {Bhalerao}, {Bhattacharya}, {Vibhute}, {Dewangan}, {Rao}, {Vadawale}, {AstroSat Cadmium Zinc Telluride Imager Team}, {Svinkin}, {Hurley}, {Aptekar}, {Frederiks}, {Golenetskii}, {Kozlova},
  {Lysenko}, {Oleynik}, {Tsvetkova}, {Ulanov}, {Cline}, {IPN Collaboration}, {Li}, {Xiong}, {Zhang}, {Lu}, {Song}, {Cao}, {Chang}, {Chen}, {Chen}, {Chen}, {Chen}, {Chen}, {Chen}, {Cui}, {Cui}, {Deng}, {Dong}, {Du}, {Fu}, {Gao}, {Gao}, {Gao}, {Ge}, {Gu}, {Guan}, {Guo}, {Han}, {Hu}, {Huang}, {Huo}, {Jia}, {Jiang}, {Jiang}, {Jin}, {Jin}, {Li}, {Li}, {Li}, {Li}, {Li}, {Li}, {Li}, {Li}, {Li}, {Li}, {Li}, {Liang}, {Liao}, {Liu}, {Liu}, {Liu}, {Liu}, {Liu}, {Liu}, {Liu}, {Lu}, {Lu}, {Luo}, {Ma}, {Meng}, {Nang}, {Nie}, {Ou}, {Qu}, {Sai}, {Sun}, {Tan}, {Tao}, {Tao}, {Tuo}, {Wang}, {Wang}, {Wang}, {Wang}, {Wang}, {Wen}, {Wu}, {Wu}, {Xiao}, {Xu}, {Xu}, {Yan}, {Yang}, {Yang}, {Yang}, {Zhang}, {Zhang}, {Zhang}, {Zhang}, {Zhang}, {Zhang}, {Zhang}, {Zhang}, {Zhang}, {Zhang}, {Zhang}, {Zhang}, {Zhang}, {Zhang}, {Zhang}, {Zhang}, {Zhang}, {Zhang}, {Zhao}, {Zhao}, {Zhao}, {Zheng}, {Zhu}, {Zhu}, {Zou}, {Insight-HXMT Collaboration}, {Albert}, {Andr{\'e}}, {Anghinolfi}, {Ardid}, {Aubert}, {Aublin}, {Avgitas}, {Baret},
  {Barrios-Mart{\'\i}}, {Basa}, {Belhorma}, {Bertin}, {Biagi}, {Bormuth}, {Bourret}, {Bouwhuis}, {Br{\^a}nza{\c{s}}}, {Bruijn}, {Brunner}, {Busto}, {Capone}, {Caramete}, {Carr}, {Celli}, {Cherkaoui El Moursli}, {Chiarusi}, {Circella}, {Coelho}, {Coleiro}, {Coniglione}, {Costantini}, {Coyle}, {Creusot}, {D{\'\i}az}, {Deschamps}, {De Bonis}, {Distefano}, {Di Palma}, {Domi}, {Donzaud}, {Dornic}, {Drouhin}, {Eberl}, {El Bojaddaini}, {El Khayati}, {Els{\"a}sser}, {Enzenh{\"o}fer}, {Ettahiri}, {Fassi}, {Felis}, {Fusco}, {Gay}, {Giordano}, {Glotin}, {Gr{\'e}goire}, {Ruiz}, {Graf}, {Hallmann}, {van Haren}, {Heijboer}, {Hello}, {Hern{\'a}ndez-Rey}, {H{\"o}ssl}, {Hofest{\"a}dt}, {Hugon}, {Illuminati}, {James}, {de Jong}, {Jongen}, {Kadler}, {Kalekin}, {Katz}, {Kiessling}, {Kouchner}, {Kreter}, {Kreykenbohm}, {Kulikovskiy}, {Lachaud}, {Lahmann}, {Lef{\`e}vre}, {Leonora}, {Lotze}, {Loucatos}, {Marcelin}, {Margiotta}, {Marinelli}, {Mart{\'\i}nez-Mora}, {Mele}, {Melis}, {Michael}, {Migliozzi}, {Moussa}, {Navas}, {Nezri},
  {Organokov}, {P{\u{a}}v{\u{a}}la{\c{s}}}, {Pellegrino}, {Perrina}, {Piattelli}, {Popa}, {Pradier}, {Quinn}, {Racca}, {Riccobene}, {S{\'a}nchez-Losa}, {Salda{\~n}a}, {Salvadori}, {Samtleben}, {Sanguineti}, {Sapienza}, {Sieger}, {Spurio}, {Stolarczyk}, {Taiuti}, {Tayalati}, {Trovato}, {Turpin}, {T{\"o}nnis}, {Vallage}, {Van Elewyck}, {Versari}, {Vivolo}, {Vizzoca}, {Wilms}, {Zornoza}, {Z{\'u}{\~n}iga}, {ANTARES Collaboration}, {Beardmore}, {Breeveld}, {Burrows}, {Cenko}, {Cusumano}, {D'A{\`\i}}, {de Pasquale}, {Emery}, {Evans}, {Giommi}, {Gronwall}, {Kennea}, {Krimm}, {Kuin}, {Lien}, {Marshall}, {Melandri}, {Nousek}, {Oates}, {Osborne}, {Pagani}, {Page}, {Palmer}, {Perri}, {Siegel}, {Sbarufatti}, {Tagliaferri}, {Tohuvavohu}, {Swift Collaboration}, {Tavani}, {Verrecchia}, {Bulgarelli}, {Evangelista}, {Pacciani}, {Feroci}, {Pittori}, {Giuliani}, {Del Monte}, {Donnarumma}, {Argan}, {Trois}, {Ursi}, {Cardillo}, {Piano}, {Longo}, {Lucarelli}, {Munar-Adrover}, {Fuschino}, {Labanti}, {Marisaldi}, {Minervini},
  {Fioretti}, {Parmiggiani}, {Gianotti}, {Trifoglio}, {Di Persio}, {Antonelli}, {Barbiellini}, {Caraveo}, {Cattaneo}, {Costa}, {Colafrancesco}, {D'Amico}, {Ferrari}, {Morselli}, {Paoletti}, {Picozza}, {Pilia}, {Rappoldi}, {Soffitta}, {Vercellone}, {AGILE Team}, {Foley}, {Coulter}, {Kilpatrick}, {Drout}, {Piro}, {Shappee}, {Siebert}, {Simon}, {Ulloa}, {Kasen}, {Madore}, {Murguia-Berthier}, {Pan}, {Prochaska}, {Ramirez-Ruiz}, {Rest}, {Rojas-Bravo}, {1M2H Team}, {Berger}, {Soares-Santos}, {Annis}, {Alexander}, {Allam}, {Balbinot}, {Blanchard}, {Brout}, {Butler}, {Chornock}, {Cook}, {Cowperthwaite}, {Diehl}, {Drlica-Wagner}, {Drout}, {Durret}, {Eftekhari}, {Finley}, {Fong}, {Frieman}, {Fryer}, {Garc{\'\i}a-Bellido}, {Gruendl}, {Hartley}, {Herner}, {Kessler}, {Lin}, {Lopes}, {Louren{\c{c}}o}, {Margutti}, {Marshall}, {Matheson}, {Medina}, {Metzger}, {Mu{\~n}oz}, {Muir}, {Nicholl}, {Nugent}, {Palmese}, {Paz-Chinch{\'o}n}, {Quataert}, {Sako}, {Sauseda}, {Schlegel}, {Scolnic}, {Secco}, {Smith}, {Sobreira}, {Villar},
  {Vivas}, {Wester}, {Williams}, {Yanny}, {Zenteno}, {Zhang}, {Abbott}, {Banerji}, {Bechtol}, {Benoit-L{\'e}vy}, {Bertin}, {Brooks}, {Buckley-Geer}, {Burke}, {Capozzi}, {Carnero Rosell}, {Carrasco Kind}, {Castander}, {Crocce}, {Cunha}, {D'Andrea}, {da Costa}, {Davis}, {DePoy}, {Desai}, {Dietrich}, {Eifler}, {Fernandez}, {Flaugher}, {Fosalba}, {Gaztanaga}, {Gerdes}, {Giannantonio}, {Goldstein}, {Gruen}, {Gschwend}, {Gutierrez}, {Honscheid}, {James}, {Jeltema}, {Johnson}, {Johnson}, {Kent}, {Krause}, {Kron}, {Kuehn}, {Lahav}, {Lima}, {Maia}, {March}, {Martini}, {McMahon}, {Menanteau}, {Miller}, {Miquel}, {Mohr}, {Nichol}, {Ogando}, {Plazas}, {Romer}, {Roodman}, {Rykoff}, {Sanchez}, {Scarpine}, {Schindler}, {Schubnell}, {Sevilla-Noarbe}, {Sheldon}, {Smith}, {Smith}, {Stebbins}, {Suchyta}, {Swanson}, {Tarle}, {Thomas}, {Troxel}, {Tucker}, {Vikram}, {Walker}, {Wechsler}, {Weller}, {Carlin}, {Gill}, {Li}, {Marriner}, {Neilsen}, {Dark Energy Camera GW-EM Collaboration}, {DES Collaboration}, {Haislip}, {Kouprianov},
  {Reichart}, {Sand}, {Tartaglia}, {Valenti}, {Yang}, {DLT40 Collaboration}, {Benetti}, {Brocato}, {Campana}, {Cappellaro}, {Covino}, {D'Avanzo}, {D'Elia}, {Getman}, {Ghirlanda}, {Ghisellini}, {Limatola}, {Nicastro}, {Palazzi}, {Pian}, {Piranomonte}, {Possenti}, {Rossi}, {Salafia}, {Tomasella}, {Amati}, {Antonelli}, {Bernardini}, {Bufano}, {Capaccioli}, {Casella}, {Dadina}, {De Cesare}, {Di Paola}, {Giuffrida}, {Giunta}, {Israel}, {Lisi}, {Maiorano}, {Mapelli}, {Masetti}, {Pescalli}, {Pulone}, {Salvaterra}, {Schipani}, {Spera}, {Stamerra}, {Stella}, {Testa}, {Turatto}, {Vergani}, {Aresu}, {Bachetti}, {Buffa}, {Burgay}, {Buttu}, {Caria}, {Carretti}, {Casasola}, {Castangia}, {Carboni}, {Casu}, {Concu}, {Corongiu}, {Deiana}, {Egron}, {Fara}, {Gaudiomonte}, {Gusai}, {Ladu}, {Loru}, {Leurini}, {Marongiu}, {Melis}, {Melis}, {Migoni}, {Milia}, {Navarrini}, {Orlati}, {Ortu}, {Palmas}, {Pellizzoni}, {Perrodin}, {Pisanu}, {Poppi}, {Righini}, {Saba}, {Serra}, {Serrau}, {Stagni}, {Surcis}, {Vacca}, {Vargiu}, {Hunt},
  {Jin}, {Klose}, {Kouveliotou}, {Mazzali}, {M{\o}ller}, {Nava}, {Piran}, {Selsing}, {Vergani}, {Wiersema}, {Toma}, {Higgins}, {Mundell}, {di Serego Alighieri}, {G{\'o}tz}, {Gao}, {Gomboc}, {Kaper}, {Kobayashi}, {Kopac}, {Mao}, {Starling}, {Steele}, {van der Horst}, {GRAWITA: GRAvitational Wave Inaf TeAm}, {Acero}, {Atwood}, {Baldini}, {Barbiellini}, {Bastieri}, {Berenji}, {Bellazzini}, {Bissaldi}, {Blandford}, {Bloom}, {Bonino}, {Bottacini}, {Bregeon}, {Buehler}, {Buson}, {Cameron}, {Caputo}, {Caraveo}, {Cavazzuti}, {Chekhtman}, {Cheung}, {Chiang}, {Ciprini}, {Cohen-Tanugi}, {Cominsky}, {Costantin}, {Cuoco}, {D'Ammando}, {de Palma}, {Digel}, {Di Lalla}, {Di Mauro}, {Di Venere}, {Dubois}, {Fegan}, {Focke}, {Franckowiak}, {Fukazawa}, {Funk}, {Fusco}, {Gargano}, {Gasparrini}, {Giglietto}, {Giordano}, {Giroletti}, {Glanzman}, {Green}, {Grondin}, {Guillemot}, {Guiriec}, {Harding}, {Horan}, {J{\'o}hannesson}, {Kamae}, {Kensei}, {Kuss}, {La Mura}, {Latronico}, {Lemoine-Goumard}, {Longo}, {Loparco}, {Lovellette},
  {Lubrano}, {Magill}, {Maldera}, {Manfreda}, {Mazziotta}, {McEnery}, {Meyer}, {Michelson}, {Mirabal}, {Monzani}, {Moretti}, {Morselli}, {Moskalenko}, {Negro}, {Nuss}, {Ojha}, {Omodei}, {Orienti}, {Orlando}, {Palatiello}, {Paliya}, {Paneque}, {Pesce-Rollins}, {Piron}, {Porter}, {Principe}, {Rain{\`o}}, {Rando}, {Razzano}, {Razzaque}, {Reimer}, {Reimer}, {Reposeur}, {Rochester}, {Saz Parkinson}, {Sgr{\`o}}, {Siskind}, {Spada}, {Spandre}, {Suson}, {Takahashi}, {Tanaka}, {Thayer}, {Thayer}, {Thompson}, {Tibaldo}, {Torres}, {Torresi}, {Troja}, {Venters}, {Vianello}, {Zaharijas}, {Fermi Large Area Telescope Collaboration}, {Allison}, {Bannister}, {Dobie}, {Kaplan}, {Lenc}, {Lynch}, {Murphy}, {Sadler}, {Australia Telescope Compact Array}, {Hotan}, {James}, {Oslowski}, {Raja}, {Shannon}, {Whiting}, {Australian SKA Pathfinder}, {Arcavi}, {Howell}, {McCully}, {Hosseinzadeh}, {Hiramatsu}, {Poznanski}, {Barnes}, {Zaltzman}, {Vasylyev}, {Maoz}, {Las Cumbres Observatory Group}, {Cooke}, {Bailes}, {Wolf}, {Deller},
  {Lidman}, {Wang}, {Gendre}, {Andreoni}, {Ackley}, {Pritchard}, {Bessell}, {Chang}, {M{\"o}ller}, {Onken}, {Scalzo}, {Ridden-Harper}, {Sharp}, {Tucker}, {Farrell}, {Elmer}, {Johnston}, {Venkatraman Krishnan}, {Keane}, {Green}, {Jameson}, {Hu}, {Ma}, {Sun}, {Wu}, {Wang}, {Shang}, {Hu}, {Ashley}, {Yuan}, {Li}, {Tao}, {Zhu}, {Zhang}, {Suntzeff}, {Zhou}, {Yang}, {Orange}, {Morris}, {Cucchiara}, {Giblin}, {Klotz}, {Staff}, {Thierry}, {Schmidt}, {OzGrav}, {(Deeper}, {Wider}, {program}, {AST3}, {CAASTRO Collaborations}, {Tanvir}, {Levan}, {Cano}, {de Ugarte-Postigo}, {Gonz{\'a}lez-Fern{\'a}ndez}, {Greiner}, {Hjorth}, {Irwin}, {Kr{\"u}hler}, {Mandel}, {Milvang-Jensen}, {O'Brien}, {Rol}, {Rosetti}, {Rosswog}, {Rowlinson}, {Steeghs}, {Th{\"o}ne}, {Ulaczyk}, {Watson}, {Bruun}, {Cutter}, {Figuera Jaimes}, {Fujii}, {Fruchter}, {Gompertz}, {Jakobsson}, {Hodosan}, {J{\`e}rgensen}, {Kangas}, {Kann}, {Rabus}, {Schr{\o}der}, {Stanway}, {Wijers}, {VINROUGE Collaboration}, {Lipunov}, {Gorbovskoy}, {Kornilov}, {Tyurina},
  {Balanutsa}, {Kuznetsov}, {Vlasenko}, {Podesta}, {Lopez}, {Podesta}, {Levato}, {Saffe}, {Mallamaci}, {Budnev}, {Gress}, {Kuvshinov}, {Gorbunov}, {Vladimirov}, {Zimnukhov}, {Gabovich}, {Yurkov}, {Sergienko}, {Rebolo}, {Serra-Ricart}, {Tlatov}, {Ishmuhametova}, {MASTER Collaboration}, {Abe}, {Aoki}, {Aoki}, {Asakura}, {Baar}, {Barway}, {Bond}, {Doi}, {Finet}, {Fujiyoshi}, {Furusawa}, {Honda}, {Itoh}, {Kanda}, {Kawabata}, {Kawabata}, {Kim}, {Koshida}, {Kuroda}, {Lee}, {Liu}, {Matsubayashi}, {Miyazaki}, {Morihana}, {Morokuma}, {Motohara}, {Murata}, {Nagai}, {Nagashima}, {Nagayama}, {Nakaoka}, {Nakata}, {Ohsawa}, {Ohshima}, {Ohta}, {Okita}, {Saito}, {Saito}, {Sako}, {Sekiguchi}, {Sumi}, {Tajitsu}, {Takahashi}, {Takayama}, {Tamura}, {Tanaka}, {Tanaka}, {Terai}, {Tominaga}, {Tristram}, {Uemura}, {Utsumi}, {Yamaguchi}, {Yasuda}, {Yoshida}, {Zenko}, {J-GEM}, {Adams}, {Anupama}, {Bally}, {Barway}, {Bellm}, {Blagorodnova}, {Cannella}, {Chandra}, {Chatterjee}, {Clarke}, {Cobb}, {Cook}, {Copperwheat}, {De}, {Emery},
  {Feindt}, {Foster}, {Fox}, {Frail}, {Fremling}, {Frohmaier}, {Garcia}, {Ghosh}, {Giacintucci}, {Goobar}, {Gottlieb}, {Grefenstette}, {Hallinan}, {Harrison}, {Heida}, {Helou}, {Ho}, {Horesh}, {Hotokezaka}, {Ip}, {Itoh}, {Jacobs}, {Jencson}, {Kasen}, {Kasliwal}, {Kassim}, {Kim}, {Kiran}, {Kuin}, {Kulkarni}, {Kupfer}, {Lau}, {Madsen}, {Mazzali}, {Miller}, {Miyasaka}, {Mooley}, {Myers}, {Nakar}, {Ngeow}, {Nugent}, {Ofek}, {Palliyaguru}, {Pavana}, {Perley}, {Peters}, {Pike}, {Piran}, {Qi}, {Quimby}, {Rana}, {Rosswog}, {Rusu}, {Sadler}, {Van Sistine}, {Sollerman}, {Xu}, {Yan}, {Yatsu}, {Yu}, {Zhang}, {Zhao}, {GROWTH}, {JAGWAR}, {Caltech-NRAO}, {TTU-NRAO}, {NuSTAR Collaborations}, {Chambers}, {Huber}, {Schultz}, {Bulger}, {Flewelling}, {Magnier}, {Lowe}, {Wainscoat}, {Waters}, {Willman}, {Pan-STARRS}, {Ebisawa}, {Hanyu}, {Harita}, {Hashimoto}, {Hidaka}, {Hori}, {Ishikawa}, {Isobe}, {Iwakiri}, {Kawai}, {Kawai}, {Kawamuro}, {Kawase}, {Kitaoka}, {Makishima}, {Matsuoka}, {Mihara}, {Morita}, {Morita}, {Nakahira},
  {Nakajima}, {Nakamura}, {Negoro}, {Oda}, {Sakamaki}, {Sasaki}, {Serino}, {Shidatsu}, {Shimomukai}, {Sugawara}, {Sugita}, {Sugizaki}, {Tachibana}, {Takao}, {Tanimoto}, {Tomida}, {Tsuboi}, {Tsunemi}, {Ueda}, {Ueno}, {Yamada}, {Yamaoka}, {Yamauchi}, {Yatabe}, {Yoneyama}, {Yoshii}, {MAXI Team}, {Coward}, {Crisp}, {Macpherson}, {Andreoni}, {Laugier}, {Noysena}, {Klotz}, {Gendre}, {Thierry}, {Turpin}, {Consortium}, {Im}, {Choi}, {Kim}, {Yoon}, {Lim}, {Lee}, {Lee}, {Kim}, {Ko}, {Joe}, {Kwon}, {Kim}, {Lim}, {Choi}, {KU Collaboration}, {Fynbo}, {Malesani}, {Xu}, {Optical Telescope}, {Smartt}, {Jerkstrand}, {Kankare}, {Sim}, {Fraser}, {Inserra}, {Maguire}, {Leloudas}, {Magee}, {Shingles}, {Smith}, {Young}, {Kotak}, {Gal-Yam}, {Lyman}, {Homan}, {Agliozzo}, {Anderson}, {Angus}, {Ashall}, {Barbarino}, {Bauer}, {Berton}, {Botticella}, {Bulla}, {Cannizzaro}, {Cartier}, {Cikota}, {Clark}, {De Cia}, {Della Valle}, {Dennefeld}, {Dessart}, {Dimitriadis}, {Elias-Rosa}, {Firth}, {Fl{\"o}rs}, {Frohmaier}, {Galbany},
  {Gonz{\'a}lez-Gait{\'a}n}, {Gromadzki}, {Guti{\'e}rrez}, {Hamanowicz}, {Harmanen}, {Heintz}, {Hernandez}, {Hodgkin}, {Hook}, {Izzo}, {James}, {Jonker}, {Kerzendorf}, {Kostrzewa-Rutkowska}, {Kromer}, {Kuncarayakti}, {Lawrence}, {Manulis}, {Mattila}, {McBrien}, {M{\"u}ller}, {Nordin}, {O'Neill}, {Onori}, {Palmerio}, {Pastorello}, {Patat}, {Pignata}, {Podsiadlowski}, {Razza}, {Reynolds}, {Roy}, {Ruiter}, {Rybicki}, {Salmon}, {Pumo}, {Prentice}, {Seitenzahl}, {Smith}, {Sollerman}, {Sullivan}, {Szegedi}, {Taddia}, {Taubenberger}, {Terreran}, {Van Soelen}, {Vos}, {Walton}, {Wright}, {Wyrzykowski}, {Yaron}, {pre=''(''>ePESSTO}, {Chen}, {Kr{\"u}hler}, {Schady}, {Wiseman}, {Greiner}, {Rau}, {Schweyer}, {Klose}, {Nicuesa Guelbenzu}, {GROND}, {Palliyaguru}, {Tech University}, {Shara}, {Williams}, {Vaisanen}, {Potter}, {Romero Colmenero}, {Crawford}, {Buckley}, {Mao}, {SALT Group}, {D{\'\i}az}, {Macri}, {Garc{\'\i}a Lambas}, {Mendes de Oliveira}, {Nilo Castell{\'o}n}, {Ribeiro}, {S{\'a}nchez}, {Schoenell}, {Abramo},
  {Akras}, {Alcaniz}, {Artola}, {Beroiz}, {Bonoli}, {Cabral}, {Camuccio}, {Chavushyan}, {Coelho}, {Colazo}, {Costa-Duarte}, {Cuevas Larenas}, {Dom{\'\i}nguez Romero}, {Dultzin}, {Fern{\'a}ndez}, {Garc{\'\i}a}, {Girardini}, {Gon{\c{c}}alves}, {Gon{\c{c}}alves}, {Gurovich}, {Jim{\'e}nez-Teja}, {Kanaan}, {Lares}, {Lopes de Oliveira}, {L{\'o}pez-Cruz}, {Melia}, {Molino}, {Padilla}, {Pe{\~n}uela}, {Placco}, {Qui{\~n}ones}, {Ram{\'\i}rez Rivera}, {Renzi}, {Riguccini}, {R{\'\i}os-L{\'o}pez}, {Rodriguez}, {Sampedro}, {Schneiter}, {Sodr{\'e}}, {Starck}, {Torres-Flores}, {Tornatore}, {Zadro{\.z}ny}, {Castillo}, {TOROS: Transient Robotic Observatory of South Collaboration}, {Castro-Tirado}, {Tello}, {Hu}, {Zhang}, {Cunniffe}, {Castell{\'o}n}, {Hiriart}, {Caballero-Garc{\'\i}a}, {Jel{\'\i}nek}, {Kub{\'a}nek}, {P{\'e}rez del Pulgar}, {Park}, {Jeong}, {Castro Cer{\'o}n}, {Pandey}, {Yock}, {Querel}, {Fan}, {Wang}, {BOOTES Collaboration}, {Beardsley}, {Brown}, {Crosse}, {Emrich}, {Franzen}, {Gaensler}, {Horsley},
  {Johnston-Hollitt}, {Kenney}, {Morales}, {Pallot}, {Sokolowski}, {Steele}, {Tingay}, {Trott}, {Walker}, {Wayth}, {Williams}, {Wu}, {Murchison Widefield Array}, {Yoshida}, {Sakamoto}, {Kawakubo}, {Yamaoka}, {Takahashi}, {Asaoka}, {Ozawa}, {Torii}, {Shimizu}, {Tamura}, {Ishizaki}, {Cherry}, {Ricciarini}, {Penacchioni}, {Marrocchesi}, {CALET Collaboration}, {Pozanenko}, {Volnova}, {Mazaeva}, {Minaev}, {Krugov}, {Kusakin}, {Reva}, {Moskvitin}, {Rumyantsev}, {Inasaridze}, {Klunko}, {Tungalag}, {Schmalz}, {Burhonov}, {IKI-GW Follow-up Collaboration}, {Abdalla}, {Abramowski}, {Aharonian}, {Ait Benkhali}, {Ang{\"u}ner}, {Arakawa}, {Arrieta}, {Aubert}, {Backes}, {Balzer}, {Barnard}, {Becherini}, {Becker Tjus}, {Berge}, {Bernhard}, {Bernl{\"o}hr}, {Blackwell}, {B{\"o}ttcher}, {Boisson}, {Bolmont}, {Bonnefoy}, {Bordas}, {Bregeon}, {Brun}, {Brun}, {Bryan}, {B{\"u}chele}, {Bulik}, {Capasso}, {Caroff}, {Carosi}, {Casanova}, {Cerruti}, {Chakraborty}, {Chaves}, {Chen}, {Chevalier}, {Colafrancesco}, {Condon}, {Conrad},
  {Davids}, {Decock}, {Deil}, {Devin}, {deWilt}, {Dirson}, {Djannati-Ata{\"\i}}, {Donath}, {O'C. Drury}, {Dutson}, {Dyks}, {Edwards}, {Egberts}, {Emery}, {Ernenwein}, {Eschbach}, {Farnier}, {Fegan}, {Fernandes}, {Fiasson}, {Fontaine}, {Funk}, {F{\"u}ssling}, {Gabici}, {Gallant}, {Garrigoux}, {Gat{\'e}}, {Giavitto}, {Giebels}, {Glawion}, {Glicenstein}, {Gottschall}, {Grondin}, {Hahn}, {Haupt}, {Hawkes}, {Heinzelmann}, {Henri}, {Hermann}, {Hinton}, {Hofmann}, {Hoischen}, {Holch}, {Holler}, {Horns}, {Ivascenko}, {Iwasaki}, {Jacholkowska}, {Jamrozy}, {Jankowsky}, {Jankowsky}, {Jingo}, {Jouvin}, {Jung-Richardt}, {Kastendieck}, {Katarzy{\'n}ski}, {Katsuragawa}, {Kerszberg}, {Khangulyan}, {Kh{\'e}lifi}, {King}, {Klepser}, {Klochkov}, {Klu{\'z}niak}, {Komin}, {Kosack}, {Krakau}, {Kraus}, {Kr{\"u}ger}, {Laffon}, {Lamanna}, {Lau}, {Lees}, {Lefaucheur}, {Lemi{\`e}re}, {Lemoine-Goumard}, {Lenain}, {Leser}, {Lohse}, {Lorentz}, {Liu}, {Lypova}, {Malyshev}, {Marandon}, {Marcowith}, {Mariaud}, {Marx}, {Maurin}, {Maxted},
  {Mayer}, {Meintjes}, {Meyer}, {Mitchell}, {Moderski}, {Mohamed}, {Mohrmann}, {Mor{\r{a}}}, {Moulin}, {Murach}, {Nakashima}, {de Naurois}, {Ndiyavala}, {Niederwanger}, {Niemiec}, {Oakes}, {O'Brien}, {Odaka}, {Ohm}, {Ostrowski}, {Oya}, {Padovani}, {Panter}, {Parsons}, {Pekeur}, {Pelletier}, {Perennes}, {Petrucci}, {Peyaud}, {Piel}, {Pita}, {Poireau}, {Poon}, {Prokhorov}, {Prokoph}, {P{\"u}hlhofer}, {Punch}, {Quirrenbach}, {Raab}, {Rauth}, {Reimer}, {Reimer}, {Renaud}, {de los Reyes}, {Rieger}, {Rinchiuso}, {Romoli}, {Rowell}, {Rudak}, {Rulten}, {Sahakian}, {Saito}, {Sanchez}, {Santangelo}, {Sasaki}, {Schlickeiser}, {Sch{\"u}ssler}, {Schulz}, {Schwanke}, {Schwemmer}, {Seglar-Arroyo}, {Settimo}, {Seyffert}, {Shafi}, {Shilon}, {Shiningayamwe}, {Simoni}, {Sol}, {Spanier}, {Spir-Jacob}, {Stawarz}, {Steenkamp}, {Stegmann}, {Steppa}, {Sushch}, {Takahashi}, {Tavernet}, {Tavernier}, {Taylor}, {Terrier}, {Tibaldo}, {Tiziani}, {Tluczykont}, {Trichard}, {Tsirou}, {Tsuji}, {Tuffs}, {Uchiyama}, {van der Walt}, {van Eldik},
  {van Rensburg}, {van Soelen}, {Vasileiadis}, {Veh}, {Venter}, {Viana}, {Vincent}, {Vink}, {Voisin}, {V{\"o}lk}, {Vuillaume}, {Wadiasingh}, {Wagner}, {Wagner}, {Wagner}, {White}, {Wierzcholska}, {Willmann}, {W{\"o}rnlein}, {Wouters}, {Yang}, {Zaborov}, {Zacharias}, {Zanin}, {Zdziarski}, {Zech}, {Zefi}, {Ziegler}, {Zorn}, {{\.Z}ywucka}, {H.~E.~S.~S. Collaboration}, {Fender}, {Broderick}, {Rowlinson}, {Wijers}, {Stewart}, {ter Veen}, {Shulevski}, {LOFAR Collaboration}, {Kavic}, {Simonetti}, {League}, {Tsai}, {Obenberger}, {Nathaniel}, {Taylor}, {Dowell}, {Liebling}, {Estes}, {Lippert}, {Sharma}, {Vincent}, {Farella}, {Wavelength Array}, {Abeysekara}, {Albert}, {Alfaro}, {Alvarez}, {Arceo}, {Arteaga-Vel{\'a}zquez}, {Avila Rojas}, {Ayala Solares}, {Barber}, {Becerra Gonzalez}, {Becerril}, {Belmont-Moreno}, {BenZvi}, {Berley}, {Bernal}, {Braun}, {Brisbois}, {Caballero-Mora}, {Capistr{\'a}n}, {Carrami{\~n}ana}, {Casanova}, {Castillo}, {Cotti}, {Cotzomi}, {Couti{\~n}o de Le{\'o}n}, {De Le{\'o}n}, {De la Fuente},
  {Diaz Hernandez}, {Dichiara}, {Dingus}, {DuVernois}, {D{\'\i}az-V{\'e}lez}, {Ellsworth}, {Engel}, {Enr{\'\i}quez-Rivera}, {Fiorino}, {Fleischhack}, {Fraija}, {Garc{\'\i}a-Gonz{\'a}lez}, {Garfias}, {Gerhardt}, {Gonz{\~o}lez Mu{\~n}oz}, {Gonz{\'a}lez}, {Goodman}, {Hampel-Arias}, {Harding}, {Hernandez}, {Hernandez-Almada}, {Hona}, {H{\"u}ntemeyer}, {Iriarte}, {Jardin-Blicq}, {Joshi}, {Kaufmann}, {Kieda}, {Lara}, {Lauer}, {Lennarz}, {Le{\'o}n Vargas}, {Linnemann}, {Longinotti}, {Raya}, {Luna-Garc{\'\i}a}, {L{\'o}pez-Coto}, {Malone}, {Marinelli}, {Martinez}, {Martinez-Castellanos}, {Mart{\'\i}nez-Castro}, {Mart{\'\i}nez-Huerta}, {Matthews}, {Miranda-Romagnoli}, {Moreno}, {Mostaf{\'a}}, {Nellen}, {Newbold}, {Nisa}, {Noriega-Papaqui}, {Pelayo}, {Pretz}, {P{\'e}rez-P{\'e}rez}, {Ren}, {Rho}, {Rivi{\`e}re}, {Rosa-Gonz{\'a}lez}, {Rosenberg}, {Ruiz-Velasco}, {Salazar}, {Salesa Greus}, {Sandoval}, {Schneider}, {Schoorlemmer}, {Sinnis}, {Smith}, {Springer}, {Surajbali}, {Tibolla}, {Tollefson}, {Torres}, {Ukwatta},
  {Weisgarber}, {Westerhoff}, {Wisher}, {Wood}, {Yapici}, {Yodh}, {Younk}, {Zhou}, {{\'A}lvarez}, {HAWC Collaboration}, {Aab}, {Abreu}, {Aglietta}, {Albuquerque}, {Albury}, {Allekotte}, {Almela}, {Alvarez Castillo}, {Alvarez-Mu{\~n}iz}, {Anastasi}, {Anchordoqui}, {Andrada}, {Andringa}, {Aramo}, {Arsene}, {Asorey}, {Assis}, {Avila}, {Badescu}, {Balaceanu}, {Barbato}, {Barreira Luz}, {Becker}, {Bellido}, {Berat}, {Bertaina}, {Bertou}, {Biermann}, {Biteau}, {Blaess}, {Blanco}, {Blazek}, {Bleve}, {Boh{\'a}{\v{c}}ov{\'a}}, {Bonifazi}, {Borodai}, {Botti}, {Brack}, {Brancus}, {Bretz}, {Bridgeman}, {Briechle}, {Buchholz}, {Bueno}, {Buitink}, {Buscemi}, {Caballero-Mora}, {Caccianiga}, {Cancio}, {Canfora}, {Caruso}, {Castellina}, {Catalani}, {Cataldi}, {Cazon}, {Chavez}, {Chinellato}, {Chudoba}, {Clay}, {Cobos Cerutti}, {Colalillo}, {Coleman}, {Collica}, {Coluccia}, {Concei{\c{c}}{\~a}o}, {Consolati}, {Contreras}, {Cooper}, {Coutu}, {Covault}, {Cronin}, {D'Amico}, {Daniel}, {Dasso}, {Daumiller}, {Dawson}, {Day}, {de
  Almeida}, {de Jong}, {De Mauro}, {de Mello Neto}, {De Mitri}, {de Oliveira}, {de Souza}, {Debatin}, {Deligny}, {D{\'\i}az Castro}, {Diogo}, {Dobrigkeit}, {D'Olivo}, {Dorosti}, {Dos Anjos}, {Dova}, {Dundovic}, {Ebr}, {Engel}, {Erdmann}, {Erfani}, {Escobar}, {Espadanal}, {Etchegoyen}, {Falcke}, {Farmer}, {Farrar}, {Fauth}, {Fazzini}, {Feldbusch}, {Fenu}, {Fick}, {Figueira}, {Filip{\v{c}}i{\v{c}}}, {Freire}, {Fujii}, {Fuster}, {Ga{\"\i}or}, {Garc{\'\i}a}, {Gat{\'e}}, {Gemmeke}, {Gherghel-Lascu}, {Ghia}, {Giaccari}, {Giammarchi}, {Giller}, {G{\l}as}, {Glaser}, {Golup}, {G{\'o}mez Berisso}, {G{\'o}mez Vitale}, {Gonz{\'a}lez}, {Gorgi}, {Gottowik}, {Grillo}, {Grubb}, {Guarino}, {Guedes}, {Halliday}, {Hampel}, {Hansen}, {Harari}, {Harrison}, {Harvey}, {Haungs}, {Hebbeker}, {Heck}, {Heimann}, {Herve}, {Hill}, {Hojvat}, {Holt}, {Homola}, {H{\"o}randel}, {Horvath}, {Hrabovsk{\'y}}, {Huege}, {Hulsman}, {Insolia}, {Isar}, {Jandt}, {Johnsen}, {Josebachuili}, {Jurysek}, {K{\"a}{\"a}p{\"a}}, {Kampert}, {Keilhauer},
  {Kemmerich}, {Kemp}, {Kieckhafer}, {Klages}, {Kleifges}, {Kleinfeller}, {Krause}, {Krohm}, {Kuempel}, {Kukec Mezek}, {Kunka}, {Kuotb Awad}, {Lago}, {LaHurd}, {Lang}, {Lauscher}, {Legumina}, {Leigui de Oliveira}, {Letessier-Selvon}, {Lhenry-Yvon}, {Link}, {Lo Presti}, {Lopes}, {L{\'o}pez}, {L{\'o}pez Casado}, {Lorek}, {Luce}, {Lucero}, {Malacari}, {Mallamaci}, {Mandat}, {Mantsch}, {Mariazzi}, {Maris}, {Marsella}, {Martello}, {Martinez}, {Mart{\'\i}nez Bravo}, {Mas{\'\i}as Meza}, {Mathes}, {Mathys}, {Matthews}, {Matthiae}, {Mayotte}, {Mazur}, {Medina}, {Medina-Tanco}, {Melo}, {Menshikov}, {Merenda}, {Michal}, {Micheletti}, {Middendorf}, {Miramonti}, {Mitrica}, {Mockler}, {Mollerach}, {Montanet}, {Morello}, {Morlino}, {M{\"u}ller}, {M{\"u}ller}, {Muller}, {M{\"u}ller}, {Mussa}, {Naranjo}, {Nguyen}, {Niculescu-Oglinzanu}, {Niechciol}, {Niemietz}, {Niggemann}, {Nitz}, {Nosek}, {Novotny}, {No{\v{z}}ka}, {N{\'u}{\~n}ez}, {Oikonomou}, {Olinto}, {Palatka}, {Pallotta}, {Papenbreer}, {Parente}, {Parra}, {Paul},
  {Pech}, {Pedreira}, {P{\c{e}}kala}, {Pe{\~n}a-Rodriguez}, {Pereira}, {Perlin}, {Perrone}, {Peters}, {Petrera}, {Phuntsok}, {Pierog}, {Pimenta}, {Pirronello}, {Platino}, {Plum}, {Poh}, {Porowski}, {Prado}, {Privitera}, {Prouza}, {Quel}, {Querchfeld}, {Quinn}, {Ramos-Pollan}, {Rautenberg}, {Ravignani}, {Ridky}, {Riehn}, {Risse}, {Ristori}, {Rizi}, {Rodrigues de Carvalho}, {Rodriguez Fernandez}, {Rodriguez Rojo}, {Roncoroni}, {Roth}, {Roulet}, {Rovero}, {Ruehl}, {Saffi}, {Saftoiu}, {Salamida}, {Salazar}, {Saleh}, {Salina}, {S{\'a}nchez}, {Sanchez-Lucas}, {Santos}, {Santos}, {Sarazin}, {Sarmento}, {Sarmiento-Cano}, {Sato}, {Schauer}, {Scherini}, {Schieler}, {Schimp}, {Schmidt}, {Scholten}, {Schov{\'a}nek}, {Schr{\"o}der}, {Schr{\"o}der}, {Schulz}, {Schumacher}, {Sciutto}, {Segreto}, {Shadkam}, {Shellard}, {Sigl}, {Silli}, {{\v{S}}m{\'\i}da}, {Snow}, {Sommers}, {Sonntag}, {Soriano}, {Squartini}, {Stanca}, {Stani{\v{c}}}, {Stasielak}, {Stassi}, {Stolpovskiy}, {Strafella}, {Streich}, {Suarez}, {Suarez-Dur{\'a}n},
  {Sudholz}, {Suomij{\"a}rvi}, {Supanitsky}, {{\v{S}}up{\'\i}k}, {Swain}, {Szadkowski}, {Taboada}, {Taborda}, {Timmermans}, {Todero Peixoto}, {Tomankova}, {Tom{\'e}}, {Torralba Elipe}, {Travnicek}, {Trini}, {Tueros}, {Ulrich}, {Unger}, {Urban}, {Vald{\'e}s Galicia}, {Vali{\~n}o}, {Valore}, {van Aar}, {van Bodegom}, {van den Berg}, {van Vliet}, {Varela}, {Vargas C{\'a}rdenas}, {V{\'a}zquez}, {Veberi{\v{c}}}, {Ventura}, {Vergara Quispe}, {Verzi}, {Vicha}, {Villase{\~n}or}, {Vorobiov}, {Wahlberg}, {Wainberg}, {Walz}, {Watson}, {Weber}, {Weindl}, {Wiede{\'n}ski}, {Wiencke}, {Wilczy{\'n}ski}, {Wirtz}, {Wittkowski}, {Wundheiler}, {Yang}, {Yushkov}, {Zas}, {Zavrtanik}, {Zavrtanik}, {Zepeda}, {Zimmermann}, {Ziolkowski}, {Zong}, {Zuccarello}, {Pierre Auger Collaboration}, {Kim}, {Schulze}, {Bauer}, {Corral-Santana}, {de Gregorio-Monsalvo}, {Gonz{\'a}lez-L{\'o}pez}, {Hartmann}, {Ishwara-Chandra}, {Mart{\'\i}n}, {Mehner}, {Misra}, {Micha{\l}owski}, {Resmi}, {ALMA Collaboration}, {Paragi}, {Agudo}, {An}, {Beswick},
  {Casadio}, {Frey}, {Jonker}, {Kettenis}, {Marcote}, {Moldon}, {Szomoru}, {van Langevelde}, {Yang}, {Euro VLBI Team}, {Cwiek}, {Cwiok}, {Czyrkowski}, {Dabrowski}, {Kasprowicz}, {Mankiewicz}, {Nawrocki}, {Opiela}, {Piotrowski}, {Wrochna}, {Zaremba}, {{\.Z}arnecki}, {Pi of Sky Collaboration}, {Haggard}, {Nynka}, {Ruan}, {Chandra Team at McGill University}, {Bland}, {Booler}, {Devillepoix}, {de Gois}, {Hancock}, {Howie}, {Paxman}, {Sansom}, {Towner}, {Desert Fireball Network}, {Tonry}, {Coughlin}, {Stubbs}, {Denneau}, {Heinze}, {Stalder}, {Weiland}, {ATLAS}, {Eatough}, {Kramer}, {Kraus}, {Time Resolution Universe Survey}, {Troja}, {Piro}, {Becerra Gonz{\'a}lez}, {Butler}, {Fox}, {Khandrika}, {Kutyrev}, {Lee}, {Ricci}, {Ryan}, {S{\'a}nchez-Ram{\'\i}rez}, {Veilleux}, {Watson}, {Wieringa}, {Burgess}, {van Eerten}, {Fontes}, {Fryer}, {Korobkin}, {Wollaeger}, {RIMAS}, {RATIR}, {Camilo}, {Foley}, {Goedhart}, {Makhathini}, {Oozeer}, {Smirnov}, {Fender}, {Woudt}, \& {South Africa/MeerKAT}}]{2017ApJ...848L..12A}
---. 2017{\natexlab{b}}, \apjl, 848, L12, \dodoi{10.3847/2041-8213/aa91c9}

\bibitem[{{Abbott} {et~al.}(2017{\natexlab{c}}){Abbott}, {Abbott}, {Abbott}, {Acernese}, {Ackley}, {Adams}, {Adams}, {Addesso}, {Adhikari}, {Adya}, {Affeldt}, {Afrough}, {Agarwal}, {Agathos}, {Agatsuma}, {Aggarwal}, {Aguiar}, {Aiello}, {Ain}, {Ajith}, {Allen}, {Allen}, {Allocca}, {Altin}, {Amato}, {Ananyeva}, {Anderson}, {Anderson}, {Angelova}, {Antier}, {Appert}, {Arai}, {Araya}, {Areeda}, {Arnaud}, {Arun}, {Ascenzi}, {Ashton}, {Ast}, {Aston}, {Astone}, {Atallah}, {Aufmuth}, {Aulbert}, {Aultoneal}, {Austin}, {Avila-Alvarez}, {Babak}, {Bacon}, {Bader}, {Bae}, {Baker}, {Baldaccini}, {Ballardin}, {Ballmer}, {Banagiri}, {Barayoga}, {Barclay}, {Barish}, {Barker}, {Barkett}, {Barone}, {Barr}, {Barsotti}, {Barsuglia}, {Barta}, {Bartlett}, {Bartos}, {Bassiri}, {Basti}, {Batch}, {Bawaj}, {Bayley}, {Bazzan}, {B{\'e}csy}, {Beer}, {Bejger}, {Belahcene}, {Bell}, {Berger}, {Bergmann}, {Bero}, {Berry}, {Bersanetti}, {Bertolini}, {Betzwieser}, {Bhagwat}, {Bhandare}, {Bilenko}, {Billingsley}, {Billman}, {Birch}, {Birney},
  {Birnholtz}, {Biscans}, {Biscoveanu}, {Bisht}, {Bitossi}, {Biwer}, {Bizouard}, {Blackburn}, {Blackman}, {Blair}, {Blair}, {Blair}, {Bloemen}, {Bock}, {Bode}, {Boer}, {Bogaert}, {Bohe}, {Bondu}, {Bonilla}, {Bonnand}, {Boom}, {Bork}, {Boschi}, {Bose}, {Bossie}, {Bouffanais}, {Bozzi}, {Bradaschia}, {Brady}, {Branchesi}, {Brau}, {Briant}, {Brillet}, {Brinkmann}, {Brisson}, {Brockill}, {Broida}, {Brooks}, {Brown}, {Brown}, {Brunett}, {Buchanan}, {Buikema}, {Bulik}, {Bulten}, {Buonanno}, {Buskulic}, {Buy}, {Byer}, {Cabero}, {Cadonati}, {Cagnoli}, {Cahillane}, {Bustillo}, {Callister}, {Calloni}, {Camp}, {Canepa}, {Canizares}, {Cannon}, {Cao}, {Cao}, {Capano}, {Capocasa}, {Carbognani}, {Caride}, {Carney}, {Diaz}, {Casentini}, {Caudill}, {Cavagli{\`a}}, {Cavalier}, {Cavalieri}, {Cella}, {Cepeda}, {Cerd{\'a}-Dur{\'a}n}, {Cerretani}, {Cesarini}, {Chamberlin}, {Chan}, {Chao}, {Charlton}, {Chase}, {Chassande-Mottin}, {Chatterjee}, {Chatziioannou}, {Cheeseboro}, {Chen}, {Chen}, {Chen}, {Cheng}, {Chia}, {Chincarini},
  {Chiummo}, {Chmiel}, {Cho}, {Cho}, {Chow}, {Christensen}, {Chu}, {Chua}, {Chua}, {Chung}, {Chung}, {Ciani}, {Ciolfi}, {Cirelli}, {Cirone}, {Clara}, {Clark}, {Clearwater}, {Cleva}, {Cocchieri}, {Coccia}, {Cohadon}, {Cohen}, {Colla}, {Collette}, {Cominsky}, {Constancio}, {Conti}, {Cooper}, {Corban}, {Corbitt}, {Cordero-Carri{\'o}n}, {Corley}, {Cornish}, {Corsi}, {Cortese}, {Costa}, {Coughlin}, {Coughlin}, {Coulon}, {Countryman}, {Couvares}, {Covas}, {Cowan}, {Coward}, {Cowart}, {Coyne}, {Coyne}, {Creighton}, {Creighton}, {Cripe}, {Crowder}, {Cullen}, {Cumming}, {Cunningham}, {Cuoco}, {Dal Canton}, {D{\'a}lya}, {Danilishin}, {D'Antonio}, {Danzmann}, {Dasgupta}, {da Silva Costa}, {Datrier}, {Dattilo}, {Dave}, {Davier}, {Davis}, {Daw}, {Day}, {de}, {Debra}, {Degallaix}, {de Laurentis}, {Del{\'e}glise}, {Del Pozzo}, {Demos}, {Denker}, {Dent}, {de Pietri}, {Dergachev}, {De Rosa}, {Derosa}, {de Rossi}, {Desalvo}, {de Varona}, {Devenson}, {Dhurandhar}, {D{\'\i}az}, {di Fiore}, {di Giovanni}, {di Girolamo}, {di
  Lieto}, {di Pace}, {di Palma}, {di Renzo}, {Doctor}, {Dolique}, {Donovan}, {Dooley}, {Doravari}, {Dorrington}, {Douglas}, {Dovale {\'A}lvarez}, {Downes}, {Drago}, {Dreissigacker}, {Driggers}, {Du}, {Ducrot}, {Dupej}, {Dwyer}, {Edo}, {Edwards}, {Effler}, {Eggenstein}, {Ehrens}, {Eichholz}, {Eikenberry}, {Eisenstein}, {Essick}, {Estevez}, {Etienne}, {Etzel}, {Evans}, {Evans}, {Factourovich}, {Fafone}, {Fair}, {Fairhurst}, {Fan}, {Farinon}, {Farr}, {Farr}, {Fauchon-Jones}, {Favata}, {Fays}, {Fee}, {Fehrmann}, {Feicht}, {Fejer}, {Fernandez-Galiana}, {Ferrante}, {Ferreira}, {Ferrini}, {Fidecaro}, {Finstad}, {Fiori}, {Fiorucci}, {Fishbach}, {Fisher}, {Fitz-Axen}, {Flaminio}, {Fletcher}, {Fong}, {Font}, {Forsyth}, {Forsyth}, {Fournier}, {Frasca}, {Frasconi}, {Frei}, {Freise}, {Frey}, {Frey}, {Fries}, {Fritschel}, {Frolov}, {Fulda}, {Fyffe}, {Gabbard}, {Gadre}, {Gaebel}, {Gair}, {Gammaitoni}, {Ganija}, {Gaonkar}, {Garcia-Quiros}, {Garufi}, {Gateley}, {Gaudio}, {Gaur}, {Gayathri}, {Gehrels}, {Gemme}, {Genin},
  {Gennai}, {George}, {George}, {Gergely}, {Germain}, {Ghonge}, {Ghosh}, {Ghosh}, {Ghosh}, {Giaime}, {Giardina}, {Giazotto}, {Gill}, {Glover}, {Goetz}, {Goetz}, {Gomes}, {Goncharov}, {Gonz{\'a}lez}, {Castro}, {Gopakumar}, {Gorodetsky}, {Gossan}, {Gosselin}, {Gouaty}, {Grado}, {Graef}, {Granata}, {Grant}, {Gras}, {Gray}, {Greco}, {Green}, {Gretarsson}, {Groot}, {Grote}, {Grunewald}, {Gruning}, {Guidi}, {Guo}, {Gupta}, {Gupta}, {Gushwa}, {Gustafson}, {Gustafson}, {Halim}, {Hall}, {Hall}, {Hamilton}, {Hammond}, {Haney}, {Hanke}, {Hanks}, {Hanna}, {Hannam}, {Hannuksela}, {Hanson}, {Hardwick}, {Harms}, {Harry}, {Harry}, {Hart}, {Haster}, {Haughian}, {Healy}, {Heidmann}, {Heintze}, {Heitmann}, {Hello}, {Hemming}, {Hendry}, {Heng}, {Hennig}, {Heptonstall}, {Heurs}, {Hild}, {Hinderer}, {Hoak}, {Hofman}, {Holt}, {Holz}, {Hopkins}, {Horst}, {Hough}, {Houston}, {Howell}, {Hreibi}, {Hu}, {Huerta}, {Huet}, {Hughey}, {Husa}, {Huttner}, {Huynh-Dinh}, {Indik}, {Inta}, {Intini}, {Isa}, {Isac}, {Isi}, {Iyer}, {Izumi},
  {Jacqmin}, {Jani}, {Jaranowski}, {Jawahar}, {Jim{\'e}nez-Forteza}, {Johnson}, {Jones}, {Jones}, {Jonker}, {Ju}, {Junker}, {Kalaghatgi}, {Kalogera}, {Kamai}, {Kandhasamy}, {Kang}, {Kanner}, {Kapadia}, {Karki}, {Karvinen}, {Kasprzack}, {Katolik}, {Katsavounidis}, {Katzman}, {Kaufer}, {Kawabe}, {K{\'e}f{\'e}lian}, {Keitel}, {Kemball}, {Kennedy}, {Kent}, {Key}, {Khalili}, {Khan}, {Khan}, {Khan}, {Khazanov}, {Kijbunchoo}, {Kim}, {Kim}, {Kim}, {Kim}, {Kim}, {Kim}, {Kimbrell}, {King}, {King}, {Kinley-Hanlon}, {Kirchhoff}, {Kissel}, {Kleybolte}, {Klimenko}, {Knowles}, {Koch}, {Koehlenbeck}, {Koley}, {Kondrashov}, {Kontos}, {Korobko}, {Korth}, {Kowalska}, {Kozak}, {Kr{\"a}mer}, {Kringel}, {Krishnan}, {Kr{\'o}lak}, {Kuehn}, {Kumar}, {Kumar}, {Kumar}, {Kuo}, {Kutynia}, {Kwang}, {Lackey}, {Lai}, {Landry}, {Lang}, {Lange}, {Lantz}, {Lanza}, {Lartaux-Vollard}, {Lasky}, {Laxen}, {Lazzarini}, {Lazzaro}, {Leaci}, {Leavey}, {Lee}, {Lee}, {Lee}, {Lee}, {Lee}, {Lehmann}, {Lenon}, {Leonardi}, {Leroy}, {Letendre}, {Levin}, {Li},
  {Linker}, {Littenberg}, {Liu}, {Liu}, {Lo}, {Lockerbie}, {London}, {Lord}, {Lorenzini}, {Loriette}, {Lormand}, {Losurdo}, {Lough}, {Lousto}, {Lovelace}, {L{\"u}ck}, {Lumaca}, {Lundgren}, {Lynch}, {Ma}, {Macas}, {Macfoy}, {Machenschalk}, {Macinnis}, {MacLeod}, {Hernandez}, {Maga{\~n}a-Sandoval}, {Zertuche}, {Magee}, {Majorana}, {Maksimovic}, {Man}, {Mandic}, {Mangano}, {Mansell}, {Manske}, {Mantovani}, {Marchesoni}, {Marion}, {M{\'a}rka}, {M{\'a}rka}, {Markakis}, {Markosyan}, {Markowitz}, {Maros}, {Marquina}, {Martelli}, {Martellini}, {Martin}, {Martin}, {Martynov}, {Mason}, {Massera}, {Masserot}, {Massinger}, {Masso-Reid}, {Mastrogiovanni}, {Matas}, {Matichard}, {Matone}, {Mavalvala}, {Mazumder}, {McCarthy}, {McClelland}, {McCormick}, {McCuller}, {McGuire}, {McIntyre}, {McIver}, {McManus}, {McNeill}, {McRae}, {McWilliams}, {Meacher}, {Meadors}, {Mehmet}, {Meidam}, {Mejuto-Villa}, {Melatos}, {Mendell}, {Mercer}, {Merilh}, {Merzougui}, {Meshkov}, {Messenger}, {Messick}, {Metzdorff}, {Meyers}, {Miao},
  {Michel}, {Middleton}, {Mikhailov}, {Milano}, {Miller}, {Miller}, {Miller}, {Millhouse}, {Milovich-Goff}, {Minazzoli}, {Minenkov}, {Ming}, {Mishra}, {Mitra}, {Mitrofanov}, {Mitselmakher}, {Mittleman}, {Moffa}, {Moggi}, {Mogushi}, {Mohan}, {Mohapatra}, {Montani}, {Moore}, {Moraru}, {Moreno}, {Morriss}, {Mours}, {Mow-Lowry}, {Mueller}, {Muir}, {Mukherjee}, {Mukherjee}, {Mukherjee}, {Mukund}, {Mullavey}, {Munch}, {Mu{\~n}iz}, {Muratore}, {Murray}, {Napier}, {Nardecchia}, {Naticchioni}, {Nayak}, {Neilson}, {Nelemans}, {Nelson}, {Nery}, {Neunzert}, {Nevin}, {Newport}, {Newton}, {Ng}, {Nguyen}, {Nichols}, {Nielsen}, {Nissanke}, {Nitz}, {Noack}, {Nocera}, {Nolting}, {North}, {Nuttall}, {Oberling}, {O'Dea}, {Ogin}, {Oh}, {Oh}, {Ohme}, {Okada}, {Oliver}, {Oppermann}, {Oram}, {O'Reilly}, {Ormiston}, {Ortega}, {O'Shaughnessy}, {Ossokine}, {Ottaway}, {Overmier}, {Owen}, {Pace}, {Page}, {Page}, {Pai}, {Pai}, {Palamos}, {Palashov}, {Palomba}, {Pal-Singh}, {Pan}, {Pan}, {Pang}, {Pang}, {Pankow}, {Pannarale}, {Pant},
  {Paoletti}, {Paoli}, {Papa}, {Parida}, {Parker}, {Pascucci}, {Pasqualetti}, {Passaquieti}, {Passuello}, {Patil}, {Patricelli}, {Pearlstone}, {Pedraza}, {Pedurand}, {Pekowsky}, {Pele}, {Penn}, {Perez}, {Perreca}, {Perri}, {Pfeiffer}, {Phelps}, {Piccinni}, {Pichot}, {Piergiovanni}, {Pierro}, {Pillant}, {Pinard}, {Pinto}, {Pirello}, {Pitkin}, {Poe}, {Poggiani}, {Popolizio}, {Porter}, {Post}, {Powell}, {Prasad}, {Pratt}, {Pratten}, {Predoi}, {Prestegard}, {Prijatelj}, {Principe}, {Privitera}, {Prodi}, {Prokhorov}, {Puncken}, {Punturo}, {Puppo}, {P{\"u}rrer}, {Qi}, {Quetschke}, {Quintero}, {Quitzow-James}, {Raab}, {Rabeling}, {Radkins}, {Raffai}, {Raja}, {Rajan}, {Rajbhandari}, {Rakhmanov}, {Ramirez}, {Ramos-Buades}, {Rapagnani}, {Raymond}, {Razzano}, {Read}, {Regimbau}, {Rei}, {Reid}, {Reitze}, {Ren}, {Reyes}, {Ricci}, {Ricker}, {Rieger}, {Riles}, {Rizzo}, {Robertson}, {Robie}, {Robinet}, {Rocchi}, {Rolland}, {Rollins}, {Roma}, {Romano}, {Romano}, {Romel}, {Romie}, {Rosi{\'n}ska}, {Ross}, {Rowan},
  {R{\"u}diger}, {Ruggi}, {Rutins}, {Ryan}, {Sachdev}, {Sadecki}, {Sadeghian}, {Sakellariadou}, {Salconi}, {Saleem}, {Salemi}, {Samajdar}, {Sammut}, {Sampson}, {Sanchez}, {Sanchez}, {Sanchis-Gual}, {Sandberg}, {Sanders}, {Sassolas}, {Sathyaprakash}, {Saulson}, {Sauter}, {Savage}, {Sawadsky}, {Schale}, {Scheel}, {Scheuer}, {Schmidt}, {Schmidt}, {Schnabel}, {Schofield}, {Sch{\"o}nbeck}, {Schreiber}, {Schuette}, {Schulte}, {Schutz}, {Schwalbe}, {Scott}, {Scott}, {Seidel}, {Sellers}, {Sengupta}, {Sentenac}, {Sequino}, {Sergeev}, {Shaddock}, {Shaffer}, {Shah}, {Shahriar}, {Shaner}, {Shao}, {Shapiro}, {Shawhan}, {Sheperd}, {Shoemaker}, {Shoemaker}, {Siellez}, {Siemens}, {Sieniawska}, {Sigg}, {Silva}, {Singer}, {Singh}, {Singhal}, {Sintes}, {Slagmolen}, {Smith}, {Smith}, {Smith}, {Somala}, {Son}, {Sonnenberg}, {Sorazu}, {Sorrentino}, {Souradeep}, {Spencer}, {Srivastava}, {Staats}, {Staley}, {Steer}, {Steinke}, {Steinlechner}, {Steinlechner}, {Steinmeyer}, {Stevenson}, {Stone}, {Stops}, {Strain}, {Stratta},
  {Strigin}, {Strunk}, {Sturani}, {Stuver}, {Summerscales}, {Sun}, {Sunil}, {Suresh}, {Sutton}, {Swinkels}, {Szczepa{\'n}czyk}, {Tacca}, {Tait}, {Talbot}, {Talukder}, {Tanner}, {T{\'a}pai}, {Taracchini}, {Tasson}, {Taylor}, {Taylor}, {Tewari}, {Theeg}, {Thies}, {Thomas}, {Thomas}, {Thomas}, {Thorne}, {Thrane}, {Tiwari}, {Tiwari}, {Tokmakov}, {Toland}, {Tonelli}, {Tornasi}, {Torres-Forn{\'e}}, {Torrie}, {T{\"o}yr{\"a}}, {Travasso}, {Traylor}, {Trinastic}, {Tringali}, {Trozzo}, {Tsang}, {Tse}, {Tso}, {Tsukada}, {Tsuna}, {Tuyenbayev}, {Ueno}, {Ugolini}, {Unnikrishnan}, {Urban}, {Usman}, {Vahlbruch}, {Vajente}, {Valdes}, {van Bakel}, {van Beuzekom}, {van den Brand}, {van den Broeck}, {Vander-Hyde}, {van der Schaaf}, {van Heijningen}, {van Veggel}, {Vardaro}, {Varma}, {Vass}, {Vas{\'u}th}, {Vecchio}, {Vedovato}, {Veitch}, {Veitch}, {Venkateswara}, {Venugopalan}, {Verkindt}, {Vetrano}, {Vicer{\'e}}, {Viets}, {Vinciguerra}, {Vine}, {Vinet}, {Vitale}, {Vo}, {Vocca}, {Vorvick}, {Vyatchanin}, {Wade}, {Wade}, {Wade},
  {Walet}, {Walker}, {Wallace}, {Walsh}, {Wang}, {Wang}, {Wang}, {Wang}, {Wang}, {Ward}, {Warner}, {Was}, {Watchi}, {Weaver}, {Wei}, {Weinert}, {Weinstein}, {Weiss}, {Wen}, {Wessel}, {We{\ss}els}, {Westerweck}, {Westphal}, {Wette}, {Whelan}, {Whitcomb}, {Whiting}, {Whittle}, {Wilken}, {Williams}, {Williams}, {Williamson}, {Willis}, {Willke}, {Wimmer}, {Winkler}, {Wipf}, {Wittel}, {Woan}, {Woehler}, {Wofford}, {Wong}, {Worden}, {Wright}, {Wu}, {Wysocki}, {Xiao}, {Yamamoto}, {Yancey}, {Yang}, {Yap}, {Yazback}, {Yu}, {Yu}, {Yvert}, {Zadro{\.z}ny}, {Zanolin}, {Zelenova}, {Zendri}, {Zevin}, {Zhang}, {Zhang}, {Zhang}, {Zhang}, {Zhao}, {Zhou}, {Zhou}, {Zhu}, {Zhu}, {Zimmerman}, {Zucker}, {Zweizig}, {Foley}, {Coulter}, {Drout}, {Kasen}, {Kilpatrick}, {Madore}, {Murguia-Berthier}, {Pan}, {Piro}, {Prochaska}, {Ramirez-Ruiz}, {Rest}, {Rojas-Bravo}, {Shappee}, {Siebert}, {Simon}, {Ulloa}, {Annis}, {Soares-Santos}, {Brout}, {Scolnic}, {Diehl}, {Frieman}, {Berger}, {Alexander}, {Allam}, {Balbinot}, {Blanchard}, {Butler},
  {Chornock}, {Cook}, {Cowperthwaite}, {Drlica-Wagner}, {Drout}, {Durret}, {Eftekhari}, {Finley}, {Fong}, {Fryer}, {Garc{\'\i}a-Bellido}, {Gill}, {Gruendl}, {Hanna}, {Hartley}, {Herner}, {Huterer}, {Kasen}, {Kessler}, {Li}, {Lin}, {Lopes}, {Louren{\c{c}}o}, {Margutti}, {Marriner}, {Marshall}, {Matheson}, {Medina}, {Metzger}, {Mu{\~n}oz}, {Muir}, {Nicholl}, {Nugent}, {Palmese}, {Paz-Chinch{\'o}n}, {Quataert}, {Sako}, {Sauseda}, {Schlegel}, {Secco}, {Smith}, {Sobreira}, {Stebbins}, {Villar}, {Vivas}, {Wester}, {Williams}, {Yanny}, {Zenteno}, {Abbott}, {Abdalla}, {Bechtol}, {Benoit-L{\'e}vy}, {Bertin}, {Bridle}, {Brooks}, {Buckley-Geer}, {Burke}, {Rosell}, {Kind}, {Carretero}, {Castander}, {Cunha}, {D'Andrea}, {da Costa}, {Davis}, {Depoy}, {Desai}, {Dietrich}, {Estrada}, {Fernandez}, {Flaugher}, {Fosalba}, {Gaztanaga}, {Gerdes}, {Giannantonio}, {Goldstein}, {Gruen}, {Gutierrez}, {Hartley}, {Honscheid}, {Jain}, {James}, {Jeltema}, {Johnson}, {Kent}, {Krause}, {Kron}, {Kuehn}, {Kuhlmann}, {Kuropatkin}, {Lahav},
  {Lima}, {Maia}, {March}, {Miller}, {Miquel}, {Neilsen}, {Nord}, {Ogando}, {Plazas}, {Romer}, {Roodman}, {Rykoff}, {Sanchez}, {Scarpine}, {Schubnell}, {Sevilla-Noarbe}, {Smith}, {Smith}, {Suchyta}, {Tarle}, {Thomas}, {Thomas}, {Troxel}, {Tucker}, {Vikram}, {Walker}, {Weller}, {Zhang}, {Haislip}, {Kouprianov}, {Reichart}, {Tartaglia}, {Sand}, {Valenti}, {Yang}, {Arcavi}, {Hosseinzadeh}, {Howell}, {McCully}, {Poznanski}, {Vasylyev}, {Tanvir}, {Levan}, {Hjorth}, {Cano}, {Copperwheat}, {de Ugarte-Postigo}, {Evans}, {Fynbo}, {Gonz{\'a}lez-Fern{\'a}ndez}, {Greiner}, {Irwin}, {Lyman}, {Mandel}, {McMahon}, {Milvang-Jensen}, {O'Brien}, {Osborne}, {Perley}, {Pian}, {Palazzi}, {Rol}, {Rosetti}, {Rosswog}, {Rowlinson}, {Schulze}, {Steeghs}, {Th{\"o}ne}, {Ulaczyk}, {Watson}, {Wiersema}, {Lipunov}, {Gorbovskoy}, {Kornilov}, {Tyurina}, {Balanutsa}, {Vlasenko}, {Gorbunov}, {Podesta}, {Levato}, {Saffe}, {Buckley}, {Budnev}, {Gress}, {Yurkov}, {Rebolo}, \& {Serra-Ricart}}]{2017Natur.551...85A}
---. 2017{\natexlab{c}}, \nat, 551, 85, \dodoi{10.1038/nature24471}

\bibitem[{{Abbott} {et~al.}(2020{\natexlab{a}}){Abbott}, {Abbott}, {Abbott}, {Abraham}, {Acernese}, {Ackley}, {Adams}, {Adya}, {Affeldt}, {Agathos}, {Agatsuma}, {Aggarwal}, {Aguiar}, {Aiello}, {Ain}, {Ajith}, {Akutsu}, {Allen}, {Allocca}, {Aloy}, {Altin}, {Amato}, {Ananyeva}, {Anderson}, {Anderson}, {Ando}, {Angelova}, {Antier}, {Appert}, {Arai}, {Arai}, {Arai}, {Araki}, {Araya}, {Araya}, {Areeda}, {Ar{\`e}ne}, {Aritomi}, {Arnaud}, {Arun}, {Ascenzi}, {Ashton}, {Aso}, {Aston}, {Astone}, {Aubin}, {Aufmuth}, {Aultoneal}, {Austin}, {Avendano}, {Avila-Alvarez}, {Babak}, {Bacon}, {Badaracco}, {Bader}, {Bae}, {Bae}, {Baiotti}, {Bajpai}, {Baker}, {Baldaccini}, {Ballardin}, {Ballmer}, {Banagiri}, {Barayoga}, {Barclay}, {Barish}, {Barker}, {Barkett}, {Barnum}, {Barone}, {Barr}, {Barsotti}, {Barsuglia}, {Barta}, {Bartlett}, {Barton}, {Bartos}, {Bassiri}, {Basti}, {Bawaj}, {Bayley}, {Bazzan}, {B{\'e}csy}, {Bejger}, {Belahcene}, {Bell}, {Beniwal}, {Berger}, {Bergmann}, {Bernuzzi}, {Bero}, {Berry}, {Bersanetti},
  {Bertolini}, {Betzwieser}, {Bhandare}, {Bidler}, {Bilenko}, {Bilgili}, {Billingsley}, {Birch}, {Birney}, {Birnholtz}, {Biscans}, {Biscoveanu}, {Bisht}, {Bitossi}, {Bizouard}, {Blackburn}, {Blair}, {Blair}, {Blair}, {Bloemen}, {Bode}, {Boer}, {Boetzel}, {Bogaert}, {Bondu}, {Bonilla}, {Bonnand}, {Booker}, {Boom}, {Booth}, {Bork}, {Boschi}, {Bose}, {Bossie}, {Bossilkov}, {Bosveld}, {Bouffanais}, {Bozzi}, {Bradaschia}, {Brady}, {Bramley}, {Branchesi}, {Brau}, {Briant}, {Briggs}, {Brighenti}, {Brillet}, {Brinkmann}, {Brisson}, {Brockill}, {Brooks}, {Brown}, {Brown}, {Brunett}, {Buikema}, {Bulik}, {Bulten}, {Buonanno}, {Buskulic}, {Buy}, {Byer}, {Cabero}, {Cadonati}, {Cagnoli}, {Cahillane}, {Bustillo}, {Callister}, {Calloni}, {Camp}, {Campbell}, {Canepa}, {Cannon}, {Cannon}, {Cao}, {Cao}, {Capocasa}, {Carbognani}, {Caride}, {Carney}, {Carullo}, {Casanueva Diaz}, {Casentini}, {Caudill}, {Cavagli{\`a}}, {Cavalier}, {Cavalieri}, {Cella}, {Cerd{\'a}-Dur{\'a}n}, {Cerretani}, {Cesarini}, {Chaibi}, {Chakravarti},
  {Chamberlin}, {Chan}, {Chan}, {Chao}, {Charlton}, {Chase}, {Chassande-Mottin}, {Chatterjee}, {Chaturvedi}, {Chatziioannou}, {Cheeseboro}, {Chen}, {Chen}, {Chen}, {Chen}, {Chen}, {Chen}, {Cheng}, {Cheong}, {Chia}, {Chincarini}, {Chiummo}, {Cho}, {Cho}, {Cho}, {Christensen}, {Chu}, {Chu}, {Chu}, {Chua}, {Chung}, {Chung}, {Ciani}, {Ciobanu}, {Ciolfi}, {Cipriano}, {Cirone}, {Clara}, {Clark}, {Clearwater}, {Cleva}, {Cocchieri}, {Coccia}, {Cohadon}, {Cohen}, {Colgan}, {Colleoni}, {Collette}, {Collins}, {Cominsky}, {Constancio}, {Conti}, {Cooper}, {Corban}, {Corbitt}, {Cordero-Carri{\'o}n}, {Corley}, {Cornish}, {Corsi}, {Cortese}, {Costa}, {Cotesta}, {Coughlin}, {Coughlin}, {Coulon}, {Countryman}, {Couvares}, {Covas}, {Cowan}, {Coward}, {Cowart}, {Coyne}, {Coyne}, {Creighton}, {Creighton}, {Cripe}, {Croquette}, {Crowder}, {Cullen}, {Cumming}, {Cunningham}, {Cuoco}, {Dal Canton}, {D{\'a}lya}, {Danilishin}, {D'Antonio}, {Danzmann}, {Dasgupta}, {da Silva Costa}, {Datrier}, {Dattilo}, {Dave}, {Davier}, {Davis}, {Daw},
  {Debra}, {Deenadayalan}, {Degallaix}, {de Laurentis}, {Del{\'e}glise}, {Pozzo}, {Demarchi}, {Demos}, {Dent}, {de Pietri}, {Derby}, {De Rosa}, {de Rossi}, {Desalvo}, {de Varona}, {Dhurandhar}, {D{\'\i}az}, {Dietrich}, {di Fiore}, {di Giovanni}, {di Girolamo}, {di Lieto}, {Ding}, {di Pace}, {di Palma}, {di Renzo}, {Dmitriev}, {Doctor}, {Doi}, {Donovan}, {Dooley}, {Doravari}, {Dorrington}, {Downes}, {Drago}, {Driggers}, {Du}, {Ducoin}, {Dupej}, {Dwyer}, {Easter}, {Edo}, {Edwards}, {Effler}, {Eguchi}, {Ehrens}, {Eichholz}, {Eikenberry}, {Eisenmann}, {Eisenstein}, {Enomoto}, {Essick}, {Estelles}, {Estevez}, {Etienne}, {Etzel}, {Evans}, {Evans}, {Fafone}, {Fair}, {Fairhurst}, {Fan}, {Farinon}, {Farr}, {Farr}, {Fauchon-Jones}, {Favata}, {Fays}, {Fazio}, {Fee}, {Feicht}, {Fejer}, {Feng}, {Fernandez-Galiana}, {Ferrante}, {Ferreira}, {Ferreira}, {Ferrini}, {Fidecaro}, {Fiori}, {Fiorucci}, {Fishbach}, {Fisher}, {Fishner}, {Fitz-Axen}, {Flaminio}, {Fletcher}, {Flynn}, {Fong}, {Font}, {Forsyth}, {Fournier}, {Frasca},
  {Frasconi}, {Frei}, {Freise}, {Frey}, {Frey}, {Fritschel}, {Frolov}, {Fujii}, {Fukunaga}, {Fukushima}, {Fulda}, {Fyffe}, {Gabbard}, {Gadre}, {Gaebel}, {Gair}, {Gammaitoni}, {Ganija}, {Gaonkar}, {Garcia}, {Garc{\'\i}a-Quir{\'o}s}, {Garufi}, {Gateley}, {Gaudio}, {Gaur}, {Gayathri}, {Ge}, {Gemme}, {Genin}, {Gennai}, {George}, {George}, {Gergely}, {Germain}, {Ghonge}, {Ghosh}, {Ghosh}, {Ghosh}, {Giacomazzo}, {Giaime}, {Giardina}, {Giazotto}, {Gill}, {Giordano}, {Glover}, {Godwin}, {Goetz}, {Goetz}, {Goncharov}, {Gonz{\'a}lez}, {Gonzalez Castro}, {Gopakumar}, {Gorodetsky}, {Gossan}, {Gosselin}, {Gouaty}, {Grado}, {Graef}, {Granata}, {Grant}, {Gras}, {Grassia}, {Gray}, {Gray}, {Greco}, {Green}, {Green}, {Gretarsson}, {Groot}, {Grote}, {Grunewald}, {Gruning}, {Guidi}, {Gulati}, {Guo}, {Gupta}, {Gupta}, {Gustafson}, {Gustafson}, {Haegel}, {Hagiwara}, {Haino}, {Halim}, {Hall}, {Hall}, {Hamilton}, {Hammond}, {Haney}, {Hanke}, {Hanks}, {Hanna}, {Hannam}, {Hannuksela}, {Hanson}, {Hardwick}, {Haris}, {Harms}, {Harry},
  {Harry}, {Hasegawa}, {Haster}, {Haughian}, {Hayakawa}, {Hayama}, {Hayes}, {Healy}, {Heidmann}, {Heintze}, {Heitmann}, {Hello}, {Hemming}, {Hendry}, {Heng}, {Hennig}, {Heptonstall}, {Heurs}, {Hild}, {Himemoto}, {Hinderer}, {Hiranuma}, {Hirata}, {Hirose}, {Hoak}, {Hochheim}, {Hofman}, {Holgado}, {Holland}, {Holt}, {Holz}, {Hong}, {Hopkins}, {Horst}, {Hough}, {Howell}, {Hoy}, {Hreibi}, {Hsieh}, {Huang}, {Huang}, {Huang}, {Huerta}, {Huet}, {Hughey}, {Hulko}, {Husa}, {Huttner}, {Huynh-Dinh}, {Idzkowski}, {Iess}, {Ikenoue}, {Imam}, {Inayoshi}, {Ingram}, {Inoue}, {Inta}, {Intini}, {Ioka}, {Irwin}, {Isa}, {Isac}, {Isi}, {Itoh}, {Iyer}, {Izumi}, {Jacqmin}, {Jadhav}, {Jani}, {Janthalur}, {Jaranowski}, {Jenkins}, {Jiang}, {Johnson}, {Jones}, {Jones}, {Jones}, {Jonker}, {Ju}, {Jung}, {Jung}, {Junker}, {Kajita}, {Kalaghatgi}, {Kalogera}, {Kamai}, {Kamiizumi}, {Kanda}, {Kandhasamy}, {Kang}, {Kanner}, {Kapadia}, {Karki}, {Karvinen}, {Kashyap}, {Kasprzack}, {Katsanevas}, {Katsavounidis}, {Katzman}, {Kaufer}, {Kawabe},
  {Kawaguchi}, {Kawai}, {Kawasaki}, {Keerthana}, {K{\'e}f{\'e}lian}, {Keitel}, {Kennedy}, {Key}, {Khalili}, {Khan}, {Khan}, {Khan}, {Khan}, {Khazanov}, {Khursheed}, {Kijbunchoo}, {Kim}, {Kim}, {Kim}, {Kim}, {Kim}, {Kim}, {Kim}, {Kim}, {Kimball}, {Kimura}, {King}, {King}, {Kinley-Hanlon}, {Kirchhoff}, {Kissel}, {Kita}, {Kitazawa}, {Kleybolte}, {Klika}, {Klimenko}, {Knowles}, {Knyazev}, {Koch}, {Koehlenbeck}, {Koekoek}, {Kojima}, {Kokeyama}, {Koley}, {Komori}, {Kondrashov}, {Kong}, {Kontos}, {Koper}, {Korobko}, {Korth}, {Kotake}, {Kowalska}, {Kozak}, {Kozakai}, {Kozu}, {Kringel}, {Krishnendu}, {Kr{\'o}lak}, {Kuehn}, {Kumar}, {Kumar}, {Kumar}, {Kumar}, {Kumar}, {Kume}, {Kuo}, {Kuo}, {Kuo}, {Kuroyanagi}, {Kusayanagi}, {Kutynia}, {Kwak}, {Kwang}, {Lackey}, {Lai}, {Lam}, {Landry}, {Lane}, {Lang}, {Lange}, {Lantz}, {Lanza}, {Lartaux-Vollard}, {Lasky}, {Laxen}, {Lazzarini}, {Lazzaro}, {Leaci}, {Leavey}, {Lecoeuche}, {Lee}, {Lee}, {Lee}, {Lee}, {Lee}, {Lee}, {Lee}, {Lehmann}, {Lenon}, {Leonardi}, {Leroy}, {Letendre},
  {Levin}, {Li}, {Li}, {Li}, {Li}, {Lin}, {Lin}, {Lin}, {Lin}, {Linde}, {Linker}, {Littenberg}, {Liu}, {Liu}, {Liu}, {Lo}, {Lockerbie}, {London}, {Longo}, {Lorenzini}, {Loriette}, {Lormand}, {Losurdo}, {Lough}, {Lousto}, {Lovelace}, {Lower}, {L{\"u}ck}, {Lumaca}, {Lundgren}, {Luo}, {Lynch}, {Ma}, {Macas}, {Macfoy}, {Macinnis}, {MacLeod}, {Macquet}, {Maga{\~n}a-Sandoval}, {Zertuche}, {Magee}, {Majorana}, {Maksimovic}, {Malik}, {Man}, {Mandic}, {Mangano}, {Mansell}, {Manske}, {Mantovani}, {Marchesoni}, {Marchio}, {Marion}, {M{\'a}rka}, {M{\'a}rka}, {Markakis}, {Markosyan}, {Markowitz}, {Maros}, {Marquina}, {Marsat}, {Martelli}, {Martin}, {Martin}, {Martynov}, {Mason}, {Massera}, {Masserot}, {Massinger}, {Masso-Reid}, {Mastrogiovanni}, {Matas}, {Matichard}, {Matone}, {Mavalvala}, {Mazumder}, {McCann}, {McCarthy}, {McClelland}, {McCormick}, {McCuller}, {McGuire}, {McIver}, {McManus}, {McRae}, {McWilliams}, {Meacher}, {Meadors}, {Mehmet}, {Mehta}, {Meidam}, {Melatos}, {Mendell}, {Mercer}, {Mereni}, {Merilh},
  {Merzougui}, {Meshkov}, {Messenger}, {Messick}, {Metzdorff}, {Meyers}, {Miao}, {Michel}, {Michimura}, {Middleton}, {Mikhailov}, {Milano}, {Miller}, {Miller}, {Millhouse}, {Mills}, {Milovich-Goff}, {Minazzoli}, {Minenkov}, {Mio}, {Mishkin}, {Mishra}, {Mistry}, {Mitra}, {Mitrofanov}, {Mitselmakher}, {Mittleman}, {Miyakawa}, {Miyamoto}, {Miyazaki}, {Miyo}, {Miyoki}, {Mo}, {Moffa}, {Mogushi}, {Mohapatra}, {Montani}, {Moore}, {Moraru}, {Moreno}, {Morisaki}, {Moriwaki}, {Mours}, {Mow-Lowry}, {Mukherjee}, {Mukherjee}, {Mukherjee}, {Mukund}, {Mullavey}, {Munch}, {Mu{\~n}iz}, {Muratore}, {Murray}, {Nagano}, {Nagano}, {Nagar}, {Nakamura}, {Nakano}, {Nakano}, {Nakashima}, {Nardecchia}, {Narikawa}, {Naticchioni}, {Nayak}, {Negishi}, {Neilson}, {Nelemans}, {Nelson}, {Nery}, {Neunzert}, {Ng}, {Ng}, {Nguyen}, {Ni}, {Nichols}, {Nishizawa}, {Nissanke}, {Nocera}, {North}, {Nuttall}, {Obergaulinger}, {Oberling}, {O'Brien}, {Obuchi}, {O'Dea}, {Ogaki}, {Ogin}, {Oh}, {Oh}, {Ohashi}, {Ohishi}, {Ohkawa}, {Ohme}, {Ohta}, {Okada},
  {Okutomi}, {Oliver}, {Oohara}, {Ooi}, {Oppermann}, {Oram}, {O'Reilly}, {Ormiston}, {Ortega}, {O'Shaughnessy}, {Oshino}, {Ossokine}, {Ottaway}, {Overmier}, {Owen}, {Pace}, {Pagano}, {Page}, {Pai}, {Pai}, {Palamos}, {Palashov}, {Palomba}, {Pal-Singh}, {Pan}, {Pan}, {Pang}, {Pang}, {Pang}, {Pankow}, {Pannarale}, {Pant}, {Paoletti}, {Paoli}, {Papa}, {Parida}, {Park}, {Parker}, {Pascucci}, {Pasqualetti}, {Passaquieti}, {Passuello}, {Patil}, {Patricelli}, {Pearlstone}, {Pedersen}, {Pedraza}, {Pedurand}, {Pele}, {Arellano}, {Penn}, {Perez}, {Perreca}, {Pfeiffer}, {Phelps}, {Phukon}, {Piccinni}, {Pichot}, {Piergiovanni}, {Pillant}, {Pinard}, {Pinto}, {Pirello}, {Pitkin}, {Poggiani}, {Pong}, {Ponrathnam}, {Popolizio}, {Porter}, {Powell}, {Prajapati}, {Prasad}, {Prasai}, {Prasanna}, {Pratten}, {Prestegard}, {Privitera}, {Prodi}, {Prokhorov}, {Puncken}, {Punturo}, {Puppo}, {P{\"u}rrer}, {Qi}, {Quetschke}, {Quinonez}, {Quintero}, {Quitzow-James}, {Raab}, {Radkins}, {Radulescu}, {Raffai}, {Raja}, {Rajan}, {Rajbhandari},
  {Rakhmanov}, {Ramirez}, {Ramos-Buades}, {Rana}, {Rao}, {Rapagnani}, {Raymond}, {Razzano}, {Read}, {Regimbau}, {Rei}, {Reid}, {Reitze}, {Ren}, {Ricci}, {Richardson}, {Richardson}, {Ricker}, {Riles}, {Rizzo}, {Robertson}, {Robie}, {Robinet}, {Rocchi}, {Rolland}, {Rollins}, {Roma}, {Romanelli}, {Romano}, {Romel}, {Romie}, {Rose}, {Rosi{\'n}ska}, {Rosofsky}, {Ross}, {Rowan}, {R{\"u}diger}, {Ruggi}, {Rutins}, {Ryan}, {Sachdev}, {Sadecki}, {Sago}, {Saito}, {Saito}, {Sakai}, {Sakai}, {Sakamoto}, {Sakellariadou}, {Sakuno}, {Salconi}, {Saleem}, {Samajdar}, {Sammut}, {Sanchez}, {Sanchez}, {Sanchis-Gual}, {Sandberg}, {Sanders}, {Santiago}, {Sarin}, {Sassolas}, {Sathyaprakash}, {Sato}, {Sato}, {Sauter}, {Savage}, {Sawada}, {Schale}, {Scheel}, {Scheuer}, {Schmidt}, {Schnabel}, {Schofield}, {Sch{\"o}nbeck}, {Schreiber}, {Schulte}, {Schutz}, {Schwalbe}, {Scott}, {Scott}, {Seidel}, {Sekiguchi}, {Sekiguchi}, {Sellers}, {Sengupta}, {Sennett}, {Sentenac}, {Sequino}, {Sergeev}, {Setyawati}, {Shaddock}, {Shaffer}, {Shahriar},
  {Shaner}, {Shao}, {Sharma}, {Shawhan}, {Shen}, {Shibagaki}, {Shimizu}, {Shimoda}, {Shimode}, {Shink}, {Shinkai}, {Shishido}, {Shoda}, {Shoemaker}, {Shoemaker}, {Shyamsundar}, {Siellez}, {Sieniawska}, {Sigg}, {Silva}, {Singer}, {Singh}, {Singhal}, {Sintes}, {Sitmukhambetov}, {Skliris}, {Slagmolen}, {Slaven-Blair}, {Smith}, {Smith}, {Somala}, {Somiya}, {Son}, {Sorazu}, {Sorrentino}, {Sotani}, {Souradeep}, {Sowell}, {Spencer}, {Srivastava}, {Srivastava}, {Staats}, {Stachie}, {Standke}, {Steer}, {Steinke}, {Steinlechner}, {Steinlechner}, {Steinmeyer}, {Stevenson}, {Stocks}, {Stone}, {Stops}, {Strain}, {Stratta}, {Strigin}, {Strunk}, {Sturani}, {Stuver}, {Sudhir}, {Sugimoto}, {Summerscales}, {Sun}, {Sunil}, {Suresh}, {Sutton}, {Suzuki}, {Suzuki}, {Swinkels}, {Szczepa{\'n}czyk}, {Tacca}, {Tagoshi}, {Tait}, {Takahashi}, {Takahashi}, {Takamori}, {Takano}, {Takeda}, {Takeda}, {Talbot}, {Talukder}, {Tanaka}, {Tanaka}, {Tanaka}, {Tanaka}, {Tanaka}, {Tanioka}, {Tanner}, {T{\'a}pai}, {Tapia San Martin}, {Taracchini},
  {Tasson}, {Taylor}, {Telada}, {Thies}, {Thomas}, {Thomas}, {Thondapu}, {Thorne}, {Thrane}, {Tiwari}, {Tiwari}, {Tiwari}, {Toland}, {Tomaru}, {Tomigami}, {Tomura}, {Tonelli}, {Tornasi}, {Torres-Forn{\'e}}, {Torrie}, {T{\"o}yr{\"a}}, {Travasso}, {Traylor}, {Tringali}, {Trovato}, {Trozzo}, {Trudeau}, {Tsang}, {Tsang}, {Tse}, {Tso}, {Tsubono}, {Tsuchida}, {Tsukada}, {Tsuna}, {Tsuzuki}, {Tuyenbayev}, {Uchikata}, {Uchiyama}, {Ueda}, {Uehara}, {Ueno}, {Ueshima}, {Ugolini}, {Unnikrishnan}, {Uraguchi}, {Urban}, {Ushiba}, {Usman}, {Vahlbruch}, {Vajente}, {Valdes}, {van Bakel}, {van Beuzekom}, {van den Brand}, {van den Broeck}, {Vander-Hyde}, {van der Schaaf}, {van Heijningen}, {van Putten}, {van Veggel}, {Vardaro}, {Varma}, {Vass}, {Vas{\'u}th}, {Vecchio}, {Vedovato}, {Veitch}, {Veitch}, {Venkateswara}, {Venugopalan}, {Verkindt}, {Vetrano}, {Vicer{\'e}}, {Viets}, {Vine}, {Vinet}, {Vitale}, {Vivanco}, {Vo}, {Vocca}, {Vorvick}, {Vyatchanin}, {Wade}, {Wade}, {Wade}, {Walet}, {Walker}, {Wallace}, {Walsh}, {Wang}, {Wang},
  {Wang}, {Wang}, {Wang}, {Wang}, {Ward}, {Warden}, {Warner}, {Was}, {Watchi}, {Weaver}, {Wei}, {Weinert}, {Weinstein}, {Weiss}, {Wellmann}, {Wen}, {Wessel}, {We{\ss}els}, {Westhouse}, {Wette}, {Whelan}, {Whiting}, {Whittle}, {Wilken}, {Williams}, {Williamson}, {Willis}, {Willke}, {Wimmer}, {Winkler}, {Wipf}, {Wittel}, {Woan}, {Woehler}, {Wofford}, {Worden}, {Wright}, {Wu}, {Wu}, {Wu}, {Wu}, {Wysocki}, {Xiao}, {Xu}, {Yamada}, {Yamamoto}, {Yamamoto}, {Yamamoto}, {Yamamoto}, {Yancey}, {Yang}, {Yap}, {Yazback}, {Yeeles}, {Yokogawa}, {Yokoyama}, {Yokozawa}, {Yoshioka}, {Yu}, {Yu}, {Yuen}, {Yuzurihara}, {Yvert}, {Zadro{\.z}ny}, {Zanolin}, {Zeidler}, {Zelenova}, {Zendri}, {Zevin}, {Zhang}, {Zhang}, {Zhang}, {Zhao}, {Zhao}, {Zhou}, {Zhou}, {Zhu}, {Zhu}, {Zimmerman}, {Zucker}, {Zweizig}, {Kagra Collaboration}, \& {VIRGO Collaboration}}]{2020LRR....23....3A}
---. 2020{\natexlab{a}}, Living Reviews in Relativity, 23, 3, \dodoi{10.1007/s41114-020-00026-9}

\bibitem[{{Abbott} {et~al.}(2020{\natexlab{b}}){Abbott}, {Abbott}, {Abbott}, {Abraham}, {Acernese}, {Ackley}, {Adams}, {Adhikari}, {Adya}, {Affeldt}, {Agathos}, {Agatsuma}, {Aggarwal}, {Aguiar}, {Aiello}, {Ain}, {Ajith}, {Allen}, {Allocca}, {Aloy}, {Altin}, {Amato}, {Anand}, {Ananyeva}, {Anderson}, {Anderson}, {Angelova}, {Antier}, {Appert}, {Arai}, {Araya}, {Areeda}, {Ar{\`e}ne}, {Arnaud}, {Aronson}, {Arun}, {Ascenzi}, {Ashton}, {Aston}, {Astone}, {Aubin}, {Aufmuth}, {AultONeal}, {Austin}, {Avendano}, {Avila-Alvarez}, {Babak}, {Bacon}, {Badaracco}, {Bader}, {Bae}, {Baird}, {Baker}, {Baldaccini}, {Ballardin}, {Ballmer}, {Bals}, {Banagiri}, {Barayoga}, {Barbieri}, {Barclay}, {Barish}, {Barker}, {Barkett}, {Barnum}, {Barone}, {Barr}, {Barsotti}, {Barsuglia}, {Barta}, {Bartlett}, {Bartos}, {Bassiri}, {Basti}, {Bawaj}, {Bayley}, {Baylor}, {Bazzan}, {B{\'e}csy}, {Bejger}, {Belahcene}, {Bell}, {Beniwal}, {Benjamin}, {Berger}, {Bergmann}, {Bernuzzi}, {Berry}, {Bersanetti}, {Bertolini}, {Betzwieser}, {Bhandare},
  {Bidler}, {Biggs}, {Bilenko}, {Bilgili}, {Billingsley}, {Birney}, {Birnholtz}, {Biscans}, {Bischi}, {Biscoveanu}, {Bisht}, {Bitossi}, {Bizouard}, {Blackburn}, {Blackman}, {Blair}, {Blair}, {Blair}, {Bloemen}, {Bobba}, {Bode}, {Boer}, {Boetzel}, {Bogaert}, {Bondu}, {Bonnand}, {Booker}, {Boom}, {Bork}, {Boschi}, {Bose}, {Bossilkov}, {Bosveld}, {Bouffanais}, {Bozzi}, {Bradaschia}, {Brady}, {Bramley}, {Branchesi}, {Brau}, {Breschi}, {Briant}, {Briggs}, {Brighenti}, {Brillet}, {Brinkmann}, {Brockill}, {Brooks}, {Brooks}, {Brown}, {Brunett}, {Buikema}, {Bulik}, {Bulten}, {Buonanno}, {Buskulic}, {Buy}, {Byer}, {Cabero}, {Cadonati}, {Cagnoli}, {Cahillane}, {Calder{\'o}n Bustillo}, {Callister}, {Calloni}, {Camp}, {Campbell}, {Canepa}, {Cannon}, {Cao}, {Cao}, {Carapella}, {Carbognani}, {Caride}, {Carney}, {Carullo}, {Casanueva Diaz}, {Casentini}, {Caudill}, {Cavagli{\`a}}, {Cavalier}, {Cavalieri}, {Cella}, {Cerd{\'a}-Dur{\'a}n}, {Cesarini}, {Chaibi}, {Chakravarti}, {Chamberlin}, {Chan}, {Chao}, {Charlton}, {Chase},
  {Chassande-Mottin}, {Chatterjee}, {Chaturvedi}, {Chatziioannou}, {Cheeseboro}, {Chen}, {Chen}, {Chen}, {Cheng}, {Cheong}, {Chia}, {Chiadini}, {Chincarini}, {Chiummo}, {Cho}, {Cho}, {Cho}, {Christensen}, {Chu}, {Chua}, {Chung}, {Chung}, {Ciani}, {Cie{\'s}lar}, {Ciobanu}, {Ciolfi}, {Cipriano}, {Cirone}, {Clara}, {Clark}, {Clearwater}, {Cleva}, {Coccia}, {Cohadon}, {Cohen}, {Colleoni}, {Collette}, {Collins}, {Colpi}, {Cominsky}, {Constancio}, {Conti}, {Cooper}, {Corban}, {Corbitt}, {Cordero-Carri{\'o}n}, {Corezzi}, {Corley}, {Cornish}, {Corre}, {Corsi}, {Cortese}, {Costa}, {Cotesta}, {Coughlin}, {Coughlin}, {Coulon}, {Countryman}, {Couvares}, {Covas}, {Cowan}, {Coward}, {Cowart}, {Coyne}, {Coyne}, {Creighton}, {Creighton}, {Cripe}, {Croquette}, {Crowder}, {Cullen}, {Cumming}, {Cunningham}, {Cuoco}, {Dal Canton}, {D{\'a}lya}, {D'Angelo}, {Danilishin}, {D'Antonio}, {Danzmann}, {Dasgupta}, {Da Silva Costa}, {Datrier}, {Dattilo}, {Dave}, {Davier}, {Davis}, {Daw}, {DeBra}, {Deenadayalan}, {Degallaix}, {De
  Laurentis}, {Del{\'e}glise}, {De Lillo}, {Del Pozzo}, {DeMarchi}, {Demos}, {Dent}, {De Pietri}, {De Rosa}, {De Rossi}, {DeSalvo}, {de Varona}, {Dhurandhar}, {D{\'\i}az}, {Dietrich}, {Di Fiore}, {DiFronzo}, {Di Giorgio}, {Di Giovanni}, {Di Giovanni}, {Di Girolamo}, {Di Lieto}, {Ding}, {Di Pace}, {Di Palma}, {Di Renzo}, {Divakarla}, {Dmitriev}, {Doctor}, {Donovan}, {Dooley}, {Doravari}, {Dorrington}, {Downes}, {Drago}, {Driggers}, {Du}, {Ducoin}, {Dudi}, {Dupej}, {Durante}, {Dwyer}, {Easter}, {Eddolls}, {Edo}, {Effler}, {Ehrens}, {Eichholz}, {Eikenberry}, {Eisenmann}, {Eisenstein}, {Errico}, {Essick}, {Estelles}, {Estevez}, {Etienne}, {Etzel}, {Evans}, {Evans}, {Fafone}, {Fairhurst}, {Fan}, {Farinon}, {Farr}, {Farr}, {Fauchon-Jones}, {Favata}, {Fays}, {Fazio}, {Fee}, {Feicht}, {Fejer}, {Feng}, {Fernandez-Galiana}, {Ferrante}, {Ferreira}, {Ferreira}, {Fidecaro}, {Fiori}, {Fiorucci}, {Fishbach}, {Fisher}, {Fishner}, {Fittipaldi}, {Fitz-Axen}, {Fiumara}, {Flaminio}, {Fletcher}, {Floden}, {Flynn}, {Fong}, {Font},
  {Forsyth}, {Fournier}, {Vivanco}, {Frasca}, {Frasconi}, {Frei}, {Freise}, {Frey}, {Frey}, {Fritschel}, {Frolov}, {Fronz{\`e}}, {Fulda}, {Fyffe}, {Gabbard}, {Gadre}, {Gaebel}, {Gair}, {Gamba}, {Gammaitoni}, {Gaonkar}, {Garc{\'\i}a-Quir{\'o}s}, {Garufi}, {Gateley}, {Gaudio}, {Gaur}, {Gayathri}, {Gemme}, {Genin}, {Gennai}, {George}, {George}, {George}, {Gergely}, {Ghonge}, {Ghosh}, {Ghosh}, {Ghosh}, {Giacomazzo}, {Giaime}, {Giardina}, {Gibson}, {Gill}, {Glover}, {Gniesmer}, {Godwin}, {Goetz}, {Goetz}, {Goncharov}, {Gonz{\'a}lez}, {Castro}, {Gopakumar}, {Gossan}, {Gosselin}, {Gouaty}, {Grace}, {Grado}, {Granata}, {Grant}, {Gras}, {Grassia}, {Gray}, {Gray}, {Greco}, {Green}, {Green}, {Gretarsson}, {Grimaldi}, {Grimm}, {Groot}, {Grote}, {Grunewald}, {Gruning}, {Guidi}, {Gulati}, {Guo}, {Gupta}, {Gupta}, {Gupta}, {Gustafson}, {Gustafson}, {Haegel}, {Halim}, {Hall}, {Hall}, {Hamilton}, {Hammond}, {Haney}, {Hanke}, {Hanks}, {Hanna}, {Hannam}, {Hannuksela}, {Hansen}, {Hanson}, {Harder}, {Hardwick}, {Haris}, {Harms},
  {Harry}, {Harry}, {Hasskew}, {Haster}, {Haughian}, {Hayes}, {Healy}, {Heidmann}, {Heintze}, {Heitmann}, {Hellman}, {Hello}, {Hemming}, {Hendry}, {Heng}, {Hennig}, {Heurs}, {Hild}, {Hinderer}, {Ho}, {Hochheim}, {Hofman}, {Holgado}, {Holland}, {Holt}, {Holz}, {Hopkins}, {Horst}, {Hough}, {Howell}, {Hoy}, {Huang}, {H{\"u}bner}, {Huerta}, {Huet}, {Hughey}, {Hui}, {Husa}, {Huttner}, {Huynh-Dinh}, {Idzkowski}, {Iess}, {Inchauspe}, {Ingram}, {Inta}, {Intini}, {Irwin}, {Isa}, {Isac}, {Isi}, {Iyer}, {Jacqmin}, {Jadhav}, {Jani}, {Janthalur}, {Jaranowski}, {Jariwala}, {Jenkins}, {Jiang}, {Johnson}, {Johnson-McDaniel}, {Jones}, {Jones}, {Jones}, {Jones}, {Jonker}, {Ju}, {Junker}, {Kalaghatgi}, {Kalogera}, {Kamai}, {Kandhasamy}, {Kang}, {Kanner}, {Kapadia}, {Karki}, {Kashyap}, {Kasprzack}, {Kastaun}, {Katsanevas}, {Katsavounidis}, {Katzman}, {Kaufer}, {Kawabe}, {Keerthana}, {K{\'e}f{\'e}lian}, {Keitel}, {Kennedy}, {Key}, {Khalili}, {Khan}, {Khan}, {Khazanov}, {Khetan}, {Khursheed}, {Kijbunchoo}, {Kim}, {Kim}, {Kim},
  {Kim}, {Kim}, {Kim}, {Kimball}, {King}, {Kinley-Hanlon}, {Kirchhoff}, {Kissel}, {Kleybolte}, {Klika}, {Klimenko}, {Knowles}, {Koch}, {Koehlenbeck}, {Koekoek}, {Koley}, {Kondrashov}, {Kontos}, {Koper}, {Korobko}, {Korth}, {Kovalam}, {Kozak}, {Kr{\"a}mer}, {Kringel}, {Krishnendu}, {Kr{\'o}lak}, {Krupinski}, {Kuehn}, {Kumar}, {Kumar}, {Kumar}, {Kumar}, {Kuo}, {Kutynia}, {Kwang}, {Lackey}, {Laghi}, {Lai}, {Lam}, {Landry}, {Landry}, {Lane}, {Lang}, {Lange}, {Lantz}, {Lanza}, {Lartaux-Vollard}, {Lasky}, {Laxen}, {Lazzarini}, {Lazzaro}, {Leaci}, {Leavey}, {Lecoeuche}, {Lee}, {Lee}, {Lee}, {Lee}, {Lee}, {Lee}, {Lehmann}, {Lenon}, {Leroy}, {Letendre}, {Levin}, {Li}, {Li}, {Li}, {Li}, {Li}, {Lin}, {Linde}, {Linker}, {Littenberg}, {Liu}, {Liu}, {Llorens-Monteagudo}, {Lo}, {London}, {Longo}, {Lorenzini}, {Loriette}, {Lormand}, {Losurdo}, {Lough}, {Lousto}, {Lovelace}, {Lower}, {Lucaccioni}, {L{\"u}ck}, {Lumaca}, {Lundgren}, {Lynch}, {Ma}, {Macas}, {Macfoy}, {MacInnis}, {Macleod}, {Macquet}, {Maga{\~n}a Hernandez},
  {Maga{\~n}a-Sandoval}, {Magee}, {Majorana}, {Maksimovic}, {Malik}, {Man}, {Mandic}, {Mangano}, {Mansell}, {Manske}, {Mantovani}, {Mapelli}, {Marchesoni}, {Marion}, {M{\'a}rka}, {M{\'a}rka}, {Markakis}, {Markosyan}, {Markowitz}, {Maros}, {Marquina}, {Marsat}, {Martelli}, {Martin}, {Martin}, {Martinez}, {Martynov}, {Masalehdan}, {Mason}, {Massera}, {Masserot}, {Massinger}, {Masso-Reid}, {Mastrogiovanni}, {Matas}, {Matichard}, {Matone}, {Mavalvala}, {McCann}, {McCarthy}, {McClelland}, {McCormick}, {McCuller}, {McGuire}, {McIsaac}, {McIver}, {McManus}, {McRae}, {McWilliams}, {Meacher}, {Meadors}, {Mehmet}, {Mehta}, {Meidam}, {Mejuto Villa}, {Melatos}, {Mendell}, {Mercer}, {Mereni}, {Merfeld}, {Merilh}, {Merzougui}, {Meshkov}, {Messenger}, {Messick}, {Messina}, {Metzdorff}, {Meyers}, {Meylahn}, {Miani}, {Miao}, {Michel}, {Middleton}, {Milano}, {Miller}, {Millhouse}, {Mills}, {Milovich-Goff}, {Minazzoli}, {Minenkov}, {Mishkin}, {Mishra}, {Mistry}, {Mitra}, {Mitrofanov}, {Mitselmakher}, {Mittleman}, {Mo}, {Moffa},
  {Mogushi}, {Mohapatra}, {Molina-Ruiz}, {Mondin}, {Montani}, {Moore}, {Moraru}, {Morawski}, {Moreno}, {Morisaki}, {Mours}, {Mow-Lowry}, {Muciaccia}, {Mukherjee}, {Mukherjee}, {Mukherjee}, {Mukherjee}, {Mukund}, {Mullavey}, {Munch}, {Mu{\~n}iz}, {Muratore}, {Murray}, {Nagar}, {Nardecchia}, {Naticchioni}, {Nayak}, {Neil}, {Neilson}, {Nelemans}, {Nelson}, {Nery}, {Neunzert}, {Nevin}, {Ng}, {Ng}, {Nguyen}, {Nguyen}, {Nichols}, {Nichols}, {Nissanke}, {Nocera}, {North}, {Nuttall}, {Obergaulinger}, {Oberling}, {O'Brien}, {Oganesyan}, {Ogin}, {Oh}, {Oh}, {Ohme}, {Ohta}, {Okada}, {Oliver}, {Oppermann}, {Oram}, {O'Reilly}, {Ormiston}, {Ortega}, {O'Shaughnessy}, {Ossokine}, {Ottaway}, {Overmier}, {Owen}, {Pace}, {Pagano}, {Page}, {Pagliaroli}, {Pai}, {Pai}, {Palamos}, {Palashov}, {Palomba}, {Pan}, {Panda}, {Pang}, {Pankow}, {Pannarale}, {Pant}, {Paoletti}, {Paoli}, {Parida}, {Parker}, {Pascucci}, {Pasqualetti}, {Passaquieti}, {Passuello}, {Patil}, {Patricelli}, {Payne}, {Pearlstone}, {Pechsiri}, {Pedersen}, {Pedraza},
  {Pedurand}, {Pele}, {Penn}, {Perego}, {Perez}, {P{\'e}rigois}, {Perreca}, {Petermann}, {Pfeiffer}, {Phelps}, {Phukon}, {Piccinni}, {Pichot}, {Piergiovanni}, {Pierro}, {Pillant}, {Pinard}, {Pinto}, {Pirello}, {Pitkin}, {Plastino}, {Poggiani}, {Pong}, {Ponrathnam}, {Popolizio}, {Porter}, {Powell}, {Prajapati}, {Prasad}, {Prasai}, {Prasanna}, {Pratten}, {Prestegard}, {Principe}, {Prodi}, {Prokhorov}, {Punturo}, {Puppo}, {P{\"u}rrer}, {Qi}, {Quetschke}, {Quinonez}, {Raab}, {Raaijmakers}, {Radkins}, {Radulesco}, {Raffai}, {Raja}, {Rajan}, {Rajbhandari}, {Rakhmanov}, {Ramirez}, {Ramos-Buades}, {Rana}, {Rao}, {Rapagnani}, {Raymond}, {Razzano}, {Read}, {Regimbau}, {Rei}, {Reid}, {Reitze}, {Rettegno}, {Ricci}, {Richardson}, {Richardson}, {Ricker}, {Riemenschneider}, {Riles}, {Rizzo}, {Robertson}, {Robinet}, {Rocchi}, {Rolland}, {Rollins}, {Roma}, {Romanelli}, {Romano}, {Romel}, {Romie}, {Rose}, {Rose}, {Rose}, {Rosell}, {Rosi{\'n}ska}, {Rosofsky}, {Ross}, {Rowan}, {Roy}, {R{\"u}diger}, {Ruggi}, {Rutins}, {Ryan},
  {Sachdev}, {Sadecki}, {Sakellariadou}, {Salafia}, {Salconi}, {Saleem}, {Samajdar}, {Sammut}, {Sanchez}, {Sanchez}, {Sanchis-Gual}, {Sanders}, {Santiago}, {Santos}, {Sarin}, {Sassolas}, {Sathyaprakash}, {Sauter}, {Savage}, {Schale}, {Scheel}, {Scheuer}, {Schmidt}, {Schnabel}, {Schofield}, {Sch{\"o}nbeck}, {Schreiber}, {Schulte}, {Schutz}, {Scott}, {Scott}, {Seidel}, {Sellers}, {Sengupta}, {Sennett}, {Sentenac}, {Sequino}, {Sergeev}, {Setyawati}, {Shaddock}, {Shaffer}, {Shahriar}, {Shaner}, {Sharma}, {Sharma}, {Shawhan}, {Shen}, {Shink}, {Shoemaker}, {Shoemaker}, {Shukla}, {ShyamSundar}, {Siellez}, {Sieniawska}, {Sigg}, {Singer}, {Singh}, {Singh}, {Singhal}, {Sintes}, {Sitmukhambetov}, {Skliris}, {Slagmolen}, {Slaven-Blair}, {Smith}, {Smith}, {Somala}, {Son}, {Soni}, {Sorazu}, {Sorrentino}, {Souradeep}, {Sowell}, {Spencer}, {Spera}, {Srivastava}, {Srivastava}, {Staats}, {Stachie}, {Standke}, {Steer}, {Steinke}, {Steinlechner}, {Steinlechner}, {Steinmeyer}, {Stevenson}, {Stocks}, {Stone}, {Stops}, {Strain},
  {Stratta}, {Strigin}, {Strunk}, {Sturani}, {Stuver}, {Sudhir}, {Summerscales}, {Sun}, {Sunil}, {Sur}, {Suresh}, {Sutton}, {Swinkels}, {Szczepa{\'n}czyk}, {Tacca}, {Tait}, {Talbot}, {Tanner}, {Tao}, {T{\'a}pai}, {Tapia}, {Tasson}, {Taylor}, {Tenorio}, {Terkowski}, {Thomas}, {Thomas}, {Thondapu}, {Thorne}, {Thrane}, {Tiwari}, {Tiwari}, {Tiwari}, {Toland}, {Tonelli}, {Tornasi}, {Torres-Forn{\'e}}, {Torrie}, {T{\"o}yr{\"a}}, {Travasso}, {Traylor}, {Tringali}, {Tripathee}, {Trovato}, {Trozzo}, {Tsang}, {Tse}, {Tso}, {Tsukada}, {Tsuna}, {Tsutsui}, {Tuyenbayev}, {Ueno}, {Ugolini}, {Unnikrishnan}, {Urban}, {Usman}, {Vahlbruch}, {Vajente}, {Valdes}, {Valentini}, {van Bakel}, {van Beuzekom}, {van den Brand}, {Van Den Broeck}, {Vander-Hyde}, {van der Schaaf}, {VanHeijningen}, {van Veggel}, {Vardaro}, {Varma}, {Vass}, {Vas{\'u}th}, {Vecchio}, {Vedovato}, {Veitch}, {Veitch}, {Venkateswara}, {Venugopalan}, {Verkindt}, {Vetrano}, {Vicer{\'e}}, {Viets}, {Vinciguerra}, {Vine}, {Vinet}, {Vitale}, {Vo}, {Vocca}, {Vorvick},
  {Vyatchanin}, {Wade}, {Wade}, {Wade}, {Walet}, {Walker}, {Wallace}, {Walsh}, {Wang}, {Wang}, {Wang}, {Wang}, {Ward}, {Warden}, {Warner}, {Was}, {Watchi}, {Weaver}, {Wei}, {Weinert}, {Weinstein}, {Weiss}, {Wellmann}, {Wen}, {Wessel}, {We{\ss}els}, {Westhouse}, {Wette}, {Whelan}, {White}, {Whiting}, {Whittle}, {Wilken}, {Williams}, {Williamson}, {Willis}, {Willke}, {Winkler}, {Wipf}, {Wittel}, {Woan}, {Woehler}, {Wofford}, {Wright}, {Wu}, {Wysocki}, {Xiao}, {Xu}, {Yamamoto}, {Yancey}, {Yang}, {Yang}, {Yang}, {Yap}, {Yazback}, {Yeeles}, {Yu}, {Yu}, {Yuen}, {Zadro{\.z}ny}, {Zadro{\.z}ny}, {Zanolin}, {Zelenova}, {Zendri}, {Zevin}, {Zhang}, {Zhang}, {Zhang}, {Zhao}, {Zhao}, {Zhou}, {Zhou}, {Zhu}, {Zimmerman}, {Zucker}, \& {Zweizig}}]{2020ApJ...892L...3A}
---. 2020{\natexlab{b}}, \apjl, 892, L3, \dodoi{10.3847/2041-8213/ab75f5}

\bibitem[{{Abbott} {et~al.}(2021){Abbott}, {Abbott}, {Abraham}, {Acernese}, {Ackley}, {Adams}, {Adams}, {Adhikari}, {Adya}, {Affeldt}, {Agarwal}, {Agathos}, {Agatsuma}, {Aggarwal}, {Aguiar}, {Aiello}, {Ain}, {Ajith}, {Akutsu}, {Aleman}, {Allen}, {Allocca}, {Altin}, {Amato}, {Anand}, {Ananyeva}, {Anderson}, {Anderson}, {Ando}, {Angelova}, {Ansoldi}, {Antelis}, {Antier}, {Appert}, {Arai}, {Arai}, {Arai}, {Araki}, {Araya}, {Araya}, {Areeda}, {Ar{\`e}ne}, {Aritomi}, {Arnaud}, {Aronson}, {Arun}, {Asada}, {Asali}, {Ashton}, {Aso}, {Aston}, {Astone}, {Aubin}, {Aufmuth}, {Aultoneal}, {Austin}, {Babak}, {Badaracco}, {Bader}, {Bae}, {Bae}, {Baer}, {Bagnasco}, {Bai}, {Baiotti}, {Baird}, {Bajpai}, {Ball}, {Ballardin}, {Ballmer}, {Bals}, {Balsamo}, {Baltus}, {Banagiri}, {Bankar}, {Bankar}, {Barayoga}, {Barbieri}, {Barish}, {Barker}, {Barneo}, {Barone}, {Barr}, {Barsotti}, {Barsuglia}, {Barta}, {Bartlett}, {Barton}, {Bartos}, {Bassiri}, {Basti}, {Bawaj}, {Bayley}, {Baylor}, {Bazzan}, {B{\'e}csy}, {Bedakihale}, {Bejger},
  {Belahcene}, {Benedetto}, {Beniwal}, {Benjamin}, {Benkel}, {Bennett}, {Bentley}, {Benyaala}, {Bergamin}, {Berger}, {Bernuzzi}, {Berry}, {Bersanetti}, {Bertolini}, {Betzwieser}, {Bhandare}, {Bhandari}, {Bhattacharjee}, {Bhaumik}, {Bidler}, {Bilenko}, {Billingsley}, {Birney}, {Birnholtz}, {Biscans}, {Bischi}, {Biscoveanu}, {Bisht}, {Biswas}, {Bitossi}, {Bizouard}, {Blackburn}, {Blackman}, {Blair}, {Blair}, {Blair}, {Bobba}, {Bode}, {Boer}, {Bogaert}, {Boldrini}, {Bondu}, {Bonilla}, {Bonnand}, {Booker}, {Boom}, {Bork}, {Boschi}, {Bose}, {Bose}, {Bossilkov}, {Boudart}, {Bouffanais}, {Bozzi}, {Bradaschia}, {Brady}, {Bramley}, {Branch}, {Branchesi}, {Brau}, {Breschi}, {Briant}, {Briggs}, {Brillet}, {Brinkmann}, {Brockill}, {Brooks}, {Brooks}, {Brown}, {Brunett}, {Bruno}, {Bruntz}, {Bryant}, {Buikema}, {Bulik}, {Bulten}, {Buonanno}, {Buscicchio}, {Buskulic}, {Byer}, {Cadonati}, {Caesar}, {Cagnoli}, {Cahillane}, {Cain}, {Calder{\'o}n Bustillo}, {Callaghan}, {Callister}, {Calloni}, {Camp}, {Canepa},
  {Cannavacciuolo}, {Cannon}, {Cao}, {Cao}, {Cao}, {Capocasa}, {Capote}, {Carapella}, {Carbognani}, {Carlin}, {Carney}, {Carpinelli}, {Carullo}, {Carver}, {Casanueva Diaz}, {Casentini}, {Castaldi}, {Caudill}, {Cavagli{\`a}}, {Cavalier}, {Cavalieri}, {Cella}, {Cerd{\'a}-Dur{\'a}n}, {Cesarini}, {Chaibi}, {Chakravarti}, {Champion}, {Chan}, {Chan}, {Chan}, {Chan}, {Chandra}, {Chanial}, {Chao}, {Charlton}, {Chase}, {Chassande-Mottin}, {Chatterjee}, {Chaturvedi}, {Chatziioannou}, {Chen}, {Chen}, {Chen}, {Chen}, {Chen}, {Chen}, {Chen}, {Chen}, {Chen}, {Cheng}, {Cheong}, {Cheung}, {Chia}, {Chiadini}, {Chiang}, {Chierici}, {Chincarini}, {Chiofalo}, {Chiummo}, {Cho}, {Cho}, {Choate}, {Choudhary}, {Choudhary}, {Christensen}, {Chu}, {Chu}, {Chu}, {Chua}, {Chung}, {Ciani}, {Ciecielag}, {Cie{\'s}lar}, {Cifaldi}, {Ciobanu}, {Ciolfi}, {Cipriano}, {Cirone}, {Clara}, {Clark}, {Clark}, {Clarke}, {Clearwater}, {Clesse}, {Cleva}, {Coccia}, {Cohadon}, {Cohen}, {Cohen}, {Colleoni}, {Collette}, {Colpi}, {Compton}, {Constancio},
  {Conti}, {Cooper}, {Corban}, {Corbitt}, {Cordero-Carri{\'o}n}, {Corezzi}, {Corley}, {Cornish}, {Corre}, {Corsi}, {Cortese}, {Costa}, {Cotesta}, {Coughlin}, {Coughlin}, {Coulon}, {Countryman}, {Cousins}, {Couvares}, {Covas}, {Coward}, {Cowart}, {Coyne}, {Coyne}, {Creighton}, {Creighton}, {Criswell}, {Croquette}, {Crowder}, {Cudell}, {Cullen}, {Cumming}, {Cummings}, {Cuoco}, {Cury{\l}o}, {Dal Canton}, {D{\'a}lya}, {Dana}, {Daneshgaranbajastani}, {D'Angelo}, {Danilishin}, {D'Antonio}, {Danzmann}, {Darsow-Fromm}, {Dasgupta}, {Datrier}, {Dattilo}, {Dave}, {Davier}, {Davies}, {Davis}, {Daw}, {Dean}, {Debra}, {Deenadayalan}, {Degallaix}, {de Laurentis}, {Del{\'e}glise}, {Del Favero}, {de Lillo}, {de Lillo}, {Del Pozzo}, {Demarchi}, {de Matteis}, {D'Emilio}, {Demos}, {Dent}, {Depasse}, {de Pietri}, {De Rosa}, {de Rossi}, {Desalvo}, {de Simone}, {Dhurandhar}, {D{\'\i}az}, {Diaz-Ortiz}, {Didio}, {Dietrich}, {di Fiore}, {di Fronzo}, {di Giorgio}, {di Giovanni}, {di Girolamo}, {di Lieto}, {Ding}, {di Pace}, {di Palma},
  {di Renzo}, {Divakarla}, {Dmitriev}, {Doctor}, {D'Onofrio}, {Donovan}, {Dooley}, {Doravari}, {Dorrington}, {Drago}, {Driggers}, {Drori}, {Du}, {Ducoin}, {Dupej}, {Durante}, {D'Urso}, {Duverne}, {Dwyer}, {Easter}, {Ebersold}, {Eddolls}, {Edelman}, {Edo}, {Edy}, {Effler}, {Eguchi}, {Eichholz}, {Eikenberry}, {Eisenmann}, {Eisenstein}, {Ejlli}, {Enomoto}, {Errico}, {Essick}, {Estell{\'e}s}, {Estevez}, {Etienne}, {Etzel}, {Evans}, {Evans}, {Ewing}, {Fafone}, {Fair}, {Fairhurst}, {Fan}, {Farah}, {Farinon}, {Farr}, {Farr}, {Farrow}, {Fauchon-Jones}, {Favata}, {Fays}, {Fazio}, {Feicht}, {Fejer}, {Feng}, {Fenyvesi}, {Ferguson}, {Fernandez-Galiana}, {Ferrante}, {Ferreira}, {Fidecaro}, {Figura}, {Fiori}, {Fishbach}, {Fisher}, {Fittipaldi}, {Fiumara}, {Flaminio}, {Floden}, {Flynn}, {Fong}, {Font}, {Fornal}, {Forsyth}, {Franke}, {Frasca}, {Frasconi}, {Frederick}, {Frei}, {Freise}, {Frey}, {Fritschel}, {Frolov}, {Fronz{\'e}}, {Fujii}, {Fujikawa}, {Fukunaga}, {Fukushima}, {Fulda}, {Fyffe}, {Gabbard}, {Gadre}, {Gaebel},
  {Gair}, {Gais}, {Galaudage}, {Gamba}, {Ganapathy}, {Ganguly}, {Gao}, {Gaonkar}, {Garaventa}, {Garc{\'\i}a-N{\'u}{\~n}ez}, {Garc{\'\i}a-Quir{\'o}s}, {Garufi}, {Gateley}, {Gaudio}, {Gayathri}, {Ge}, {Gemme}, {Gennai}, {George}, {Gergely}, {Gewecke}, {Ghonge}, {Ghosh}, {Ghosh}, {Ghosh}, {Ghosh}, {Ghosh}, {Giacomazzo}, {Giacoppo}, {Giaime}, {Giardina}, {Gibson}, {Gier}, {Giesler}, {Giri}, {Gissi}, {Glanzer}, {Gleckl}, {Godwin}, {Goetz}, {Goetz}, {Gohlke}, {Goncharov}, {Gonz{\'a}lez}, {Gopakumar}, {Gosselin}, {Gouaty}, {Grace}, {Grado}, {Granata}, {Granata}, {Grant}, {Gras}, {Grassia}, {Gray}, {Gray}, {Greco}, {Green}, {Green}, {Gretarsson}, {Gretarsson}, {Griffith}, {Griffiths}, {Griggs}, {Grignani}, {Grimaldi}, {Grimes}, {Grimm}, {Grote}, {Grunewald}, {Gruning}, {Guerrero}, {Guidi}, {Guimaraes}, {Guix{\'e}}, {Gulati}, {Guo}, {Guo}, {Gupta}, {Gupta}, {Gupta}, {Gustafson}, {Gustafson}, {Guzman}, {Ha}, {Haegel}, {Hagiwara}, {Haino}, {Halim}, {Hall}, {Hamilton}, {Hammond}, {Han}, {Haney}, {Hanks}, {Hanna},
  {Hannam}, {Hannuksela}, {Hansen}, {Hansen}, {Hanson}, {Harder}, {Hardwick}, {Haris}, {Harms}, {Harry}, {Harry}, {Hartwig}, {Hasegawa}, {Haskell}, {Hasskew}, {Haster}, {Hattori}, {Haughian}, {Hayakawa}, {Hayama}, {Hayes}, {Healy}, {Heidmann}, {Heintze}, {Heinze}, {Heinzel}, {Heitmann}, {Hellman}, {Hello}, {Helmling-Cornell}, {Hemming}, {Hendry}, {Heng}, {Hennes}, {Hennig}, {Hennig}, {Hernandez Vivanco}, {Heurs}, {Hild}, {Hill}, {Himemoto}, {Hinderer}, {Hines}, {Hiranuma}, {Hirata}, {Hirose}, {Ho}, {Hochheim}, {Hofman}, {Hohmann}, {Holgado}, {Holland}, {Hollows}, {Holmes}, {Holt}, {Holz}, {Hong}, {Hopkins}, {Hough}, {Howell}, {Hoy}, {Hoyland}, {Hreibi}, {Hsieh}, {Hsu}, {Huang}, {Huang}, {Huang}, {Huang}, {Huang}, {Huang}, {H{\"u}bner}, {Huddart}, {Huerta}, {Hughey}, {Hui}, {Hui}, {Husa}, {Huttner}, {Huxford}, {Huynh-Dinh}, {Ide}, {Idzkowski}, {Iess}, {Ikenoue}, {Imam}, {Inayoshi}, {Inchauspe}, {Ingram}, {Inoue}, {Intini}, {Ioka}, {Isi}, {Isleif}, {Ito}, {Itoh}, {Iyer}, {Izumi}, {Jaberianhamedan}, {Jacqmin},
  {Jadhav}, {Jadhav}, {James}, {Jan}, {Jani}, {Janssens}, {Janthalur}, {Jaranowski}, {Jariwala}, {Jaume}, {Jenkins}, {Jeon}, {Jeunon}, {Jia}, {Jiang}, {Jin}, {Johns}, {Jones}, {Jones}, {Jones}, {Jones}, {Jones}, {Jonker}, {Ju}, {Jung}, {Jung}, {Junker}, {Kaihotsu}, {Kajita}, {Kakizaki}, {Kalaghatgi}, {Kalogera}, {Kamai}, {Kamiizumi}, {Kanda}, {Kandhasamy}, {Kang}, {Kanner}, {Kao}, {Kapadia}, {Kapasi}, {Karat}, {Karathanasis}, {Karki}, {Kashyap}, {Kasprzack}, {Kastaun}, {Katsanevas}, {Katsavounidis}, {Katzman}, {Kaur}, {Kawabe}, {Kawaguchi}, {Kawai}, {Kawasaki}, {K{\'e}f{\'e}lian}, {Keitel}, {Key}, {Khadka}, {Khalili}, {Khan}, {Khan}, {Khazanov}, {Khetan}, {Khursheed}, {Kijbunchoo}, {Kim}, {Kim}, {Kim}, {Kim}, {Kim}, {Kim}, {Kimball}, {Kimura}, {King}, {Kinley-Hanlon}, {Kirchhoff}, {Kissel}, {Kita}, {Kitazawa}, {Kleybolte}, {Klimenko}, {Knee}, {Knowles}, {Knyazev}, {Koch}, {Koekoek}, {Kojima}, {Kokeyama}, {Koley}, {Kolitsidou}, {Kolstein}, {Komori}, {Kondrashov}, {Kong}, {Kontos}, {Koper}, {Korobko}, {Kotake},
  {Kovalam}, {Kozak}, {Kozakai}, {Kozu}, {Kringel}, {Krishnendu}, {Kr{\'o}lak}, {Kuehn}, {Kuei}, {Kumar}, {Kumar}, {Kumar}, {Kumar}, {Kume}, {Kuns}, {Kuo}, {Kuo}, {Kuromiya}, {Kuroyanagi}, {Kusayanagi}, {Kwak}, {Kwang}, {Laghi}, {Lalande}, {Lam}, {Lamberts}, {Landry}, {Landry}, {Lane}, {Lang}, {Lange}, {Lantz}, {La Rosa}, {Lartaux-Vollard}, {Lasky}, {Laxen}, {Lazzarini}, {Lazzaro}, {Leaci}, {Leavey}, {Lecoeuche}, {Lee}, {Lee}, {Lee}, {Lee}, {Lee}, {Lee}, {Lehmann}, {Lema{\^\i}tre}, {Leon}, {Leonardi}, {Leroy}, {Letendre}, {Levin}, {Leviton}, {Li}, {Li}, {Li}, {Li}, {Li}, {Li}, {Lin}, {Lin}, {Lin}, {Lin}, {Lin}, {Linde}, {Linker}, {Linley}, {Littenberg}, {Liu}, {Liu}, {Liu}, {Liu}, {Llorens-Monteagudo}, {Lo}, {Lockwood}, {Lollie}, {London}, {Longo}, {Lopez}, {Lorenzini}, {Loriette}, {Lormand}, {Losurdo}, {Lough}, {Lousto}, {Lovelace}, {L{\"u}ck}, {Lumaca}, {Lundgren}, {Luo}, {Macas}, {Macinnis}, {MacLeod}, {MacMillan}, {Macquet}, {Maga{\~n}a Hernandez}, {Maga{\~n}a-Sandoval}, {Magazz{\`u}}, {Magee},
  {Maggiore}, {Majorana}, {Makarem}, {Maksimovic}, {Maliakal}, {Malik}, {Man}, {Mandic}, {Mangano}, {Mango}, {Mansell}, {Manske}, {Mantovani}, {Mapelli}, {Marchesoni}, {Marchio}, {Marion}, {Mark}, {M{\'a}rka}, {M{\'a}rka}, {Markakis}, {Markosyan}, {Markowitz}, {Maros}, {Marquina}, {Marsat}, {Martelli}, {Martin}, {Martin}, {Martinez}, {Martinez}, {Martinovic}, {Martynov}, {Marx}, {Masalehdan}, {Mason}, {Massera}, {Masserot}, {Massinger}, {Masso-Reid}, {Mastrogiovanni}, {Matas}, {Mateu-Lucena}, {Matichard}, {Matiushechkina}, {Mavalvala}, {McCann}, {McCarthy}, {McClelland}, {McClincy}, {McCormick}, {McCuller}, {McGhee}, {McGuire}, {McIsaac}, {McIver}, {McManus}, {McRae}, {McWilliams}, {Meacher}, {Mehmet}, {Mehta}, {Melatos}, {Melchor}, {Mendell}, {Menendez-Vazquez}, {Menoni}, {Mercer}, {Mereni}, {Merfeld}, {Merilh}, {Merritt}, {Merzougui}, {Meshkov}, {Messenger}, {Messick}, {Meyers}, {Meylahn}, {Mhaske}, {Miani}, {Miao}, {Michaloliakos}, {Michel}, {Michimura}, {Middleton}, {Milano}, {Miller}, {Millhouse},
  {Mills}, {Milotti}, {Milovich-Goff}, {Minazzoli}, {Minenkov}, {Mio}, {Mir}, {Mishkin}, {Mishra}, {Mishra}, {Mistry}, {Mitra}, {Mitrofanov}, {Mitselmakher}, {Mittleman}, {Miyakawa}, {Miyamoto}, {Miyazaki}, {Miyo}, {Miyoki}, {Mo}, {Mogushi}, {Mohapatra}, {Mohite}, {Molina}, {Molina-Ruiz}, {Mondin}, {Montani}, {Moore}, {Moraru}, {Morawski}, {More}, {Moreno}, {Moreno}, {Mori}, {Morisaki}, {Moriwaki}, {Mours}, {Mow-Lowry}, {Mozzon}, {Muciaccia}, {Mukherjee}, {Mukherjee}, {Mukherjee}, {Mukherjee}, {Mukund}, {Mullavey}, {Munch}, {Mu{\~n}iz}, {Murray}, {Musenich}, {Nadji}, {Nagano}, {Nagano}, {Nagar}, {Nakamura}, {Nakano}, {Nakano}, {Nakashima}, {Nakayama}, {Nardecchia}, {Narikawa}, {Naticchioni}, {Nayak}, {Nayak}, {Negishi}, {Neil}, {Neilson}, {Nelemans}, {Nelson}, {Nery}, {Neunzert}, {Ng}, {Ng}, {Nguyen}, {Nguyen}, {Nguyen}, {Nguyen Quynh}, {Ni}, {Nichols}, {Nishizawa}, {Nissanke}, {Nocera}, {Noh}, {Norman}, {North}, {Nozaki}, {Nuttall}, {Oberling}, {O'Brien}, {Obuchi}, {O'Dell}, {Ogaki}, {Oganesyan}, {Oh}, {Oh},
  {Oh}, {Ohashi}, {Ohishi}, {Ohkawa}, {Ohme}, {Ohta}, {Okada}, {Okutani}, {Okutomi}, {Olivetto}, {Oohara}, {Ooi}, {Oram}, {O'Reilly}, {Ormiston}, {Ormsby}, {Ortega}, {O'Shaughnessy}, {O'Shea}, {Oshino}, {Ossokine}, {Osthelder}, {Otabe}, {Ottaway}, {Overmier}, {Pace}, {Pagano}, {Page}, {Pagliaroli}, {Pai}, {Pai}, {Palamos}, {Palashov}, {Palomba}, {Pan}, {Panda}, {Pang}, {Pang}, {Pankow}, {Pannarale}, {Pant}, {Paoletti}, {Paoli}, {Paolone}, {Parisi}, {Park}, {Parker}, {Pascucci}, {Pasqualetti}, {Passaquieti}, {Passuello}, {Patel}, {Patricelli}, {Payne}, {Pechsiri}, {Pedraza}, {Pegoraro}, {Pele}, {Pe{\~n}a Arellano}, {Penn}, {Perego}, {Pereira}, {Pereira}, {Perez}, {P{\'e}rigois}, {Perreca}, {Perri{\`e}s}, {Petermann}, {Petterson}, {Pfeiffer}, {Pham}, {Phukon}, {Piccinni}, {Pichot}, {Piendibene}, {Piergiovanni}, {Pierini}, {Pierro}, {Pillant}, {Pilo}, {Pinard}, {Pinto}, {Piotrzkowski}, {Piotrzkowski}, {Pirello}, {Pitkin}, {Placidi}, {Plastino}, {Pluchar}, {Poggiani}, {Polini}, {Pong}, {Ponrathnam}, {Popolizio},
  {Porter}, {Powell}, {Pracchia}, {Pradier}, {Prajapati}, {Prasai}, {Prasanna}, {Pratten}, {Prestegard}, {Principe}, {Prodi}, {Prokhorov}, {Prosposito}, {Prudenzi}, {Puecher}, {Punturo}, {Puosi}, {Puppo}, {P{\"u}rrer}, {Qi}, {Quetschke}, {Quinonez}, {Quitzow-James}, {Raab}, {Raaijmakers}, {Radkins}, {Radulesco}, {Raffai}, {Rail}, {Raja}, {Rajan}, {Ramirez}, {Ramirez}, {Ramos-Buades}, {Rana}, {Rapagnani}, {Rapol}, {Ratto}, {Ray}, {Raymond}, {Raza}, {Razzano}, {Read}, {Rees}, {Regimbau}, {Rei}, {Reid}, {Reitze}, {Relton}, {Rettegno}, {Ricci}, {Richardson}, {Richardson}, {Richardson}, {Ricker}, {Riemenschneider}, {Riles}, {Rizzo}, {Robertson}, {Robie}, {Robinet}, {Rocchi}, {Rocha}, {Rodriguez}, {Rodriguez-Soto}, {Rolland}, {Rollins}, {Roma}, {Romanelli}, {Romano}, {Romel}, {Romero}, {Romero-Shaw}, {Romie}, {Rose}, {Rosi{\'n}ska}, {Rosofsky}, {Ross}, {Rowan}, {Rowlinson}, {Roy}, {Roy}, {Rozza}, {Ruggi}, {Ryan}, {Sachdev}, {Sadecki}, {Sadiq}, {Sago}, {Saito}, {Saito}, {Sakai}, {Sakai}, {Sakellariadou}, {Sakuno},
  {Salafia}, {Salconi}, {Saleem}, {Salemi}, {Samajdar}, {Sanchez}, {Sanchez}, {Sanchez}, {Sanchis-Gual}, {Sanders}, {Sanuy}, {Saravanan}, {Sarin}, {Sassolas}, {Satari}, {Sathyaprakash}, {Sato}, {Sato}, {Sauter}, {Savage}, {Savant}, {Sawada}, {Sawant}, {Sawant}, {Sayah}, {Schaetzl}, {Scheel}, {Scheuer}, {Schindler-Tyka}, {Schmidt}, {Schnabel}, {Schneewind}, {Schofield}, {Sch{\"o}nbeck}, {Schulte}, {Schutz}, {Schwartz}, {Scott}, {Scott}, {Seglar-Arroyo}, {Seidel}, {Sekiguchi}, {Sekiguchi}, {Sellers}, {Sengupta}, {Sennett}, {Sentenac}, {Seo}, {Sequino}, {Sergeev}, {Setyawati}, {Shaffer}, {Shahriar}, {Shams}, {Shao}, {Sharifi}, {Sharma}, {Sharma}, {Shawhan}, {Shcheblanov}, {Shen}, {Shibagaki}, {Shikauchi}, {Shimizu}, {Shimoda}, {Shimode}, {Shink}, {Shinkai}, {Shishido}, {Shoda}, {Shoemaker}, {Shoemaker}, {Shukla}, {Shyamsundar}, {Sieniawska}, {Sigg}, {Singer}, {Singh}, {Singh}, {Singha}, {Sintes}, {Sipala}, {Skliris}, {Slagmolen}, {Slaven-Blair}, {Smetana}, {Smith}, {Smith}, {Somala}, {Somiya}, {Son}, {Soni},
  {Soni}, {Sorazu}, {Sordini}, {Sorrentino}, {Sorrentino}, {Sotani}, {Soulard}, {Souradeep}, {Sowell}, {Spagnuolo}, {Spencer}, {Spera}, {Srivastava}, {Srivastava}, {Staats}, {Stachie}, {Steer}, {Steinlechner}, {Steinlechner}, {Stops}, {Stevenson}, {Stover}, {Strain}, {Strang}, {Stratta}, {Strunk}, {Sturani}, {Stuver}, {S{\"u}dbeck}, {Sudhagar}, {Sudhir}, {Sugimoto}, {Suh}, {Summerscales}, {Sun}, {Sun}, {Sunil}, {Sur}, {Suresh}, {Sutton}, {Suzuki}, {Suzuki}, {Swinkels}, {Szczepa{\'n}czyk}, {Szewczyk}, {Tacca}, {Tagoshi}, {Tait}, {Takahashi}, {Takahashi}, {Takamori}, {Takano}, {Takeda}, {Takeda}, {Talbot}, {Tanaka}, {Tanaka}, {Tanaka}, {Tanaka}, {Tanaka}, {Tanasijczuk}, {Tanioka}, {Tanner}, {Tao}, {Tapia}, {Tapia San Martin}, {Tasson}, {Telada}, {Tenorio}, {Terkowski}, {Test}, {Thirugnanasambandam}, {Thomas}, {Thomas}, {Thompson}, {Thondapu}, {Thorne}, {Thrane}, {Tiwari}, {Tiwari}, {Tiwari}, {Toland}, {Tolley}, {Tomaru}, {Tomigami}, {Tomura}, {Tonelli}, {Torres-Forn{\'e}}, {Torrie}, {Tosta E Melo},
  {T{\"o}yr{\"a}}, {Trapananti}, {Travasso}, {Traylor}, {Tringali}, {Tripathee}, {Troiano}, {Trovato}, {Trozzo}, {Trudeau}, {Tsai}, {Tsai}, {Tsang}, {Tsang}, {Tsao}, {Tse}, {Tso}, {Tsubono}, {Tsuchida}, {Tsukada}, {Tsuna}, {Tsutsui}, {Tsuzuki}, {Turconi}, {Tuyenbayev}, {Ubhi}, {Uchikata}, {Uchiyama}, {Udall}, {Ueda}, {Uehara}, {Ueno}, {Ueshima}, {Ugolini}, {Unnikrishnan}, {Uraguchi}, {Urban}, {Ushiba}, {Usman}, {Utina}, {Vahlbruch}, {Vajente}, {Vajpeyi}, {Valdes}, {Valentini}, {Valsan}, {van Bakel}, {van Beuzekom}, {van den Brand}, {van den Broeck}, {Vander-Hyde}, {van der Schaaf}, {van Heijningen}, {Vanosky}, {van Putten}, {Vardaro}, {Vargas}, {Varma}, {Vas{\'u}th}, {Vecchio}, {Vedovato}, {Veitch}, {Veitch}, {Venkateswara}, {Venneberg}, {Venugopalan}, {Verkindt}, {Verma}, {Veske}, {Vetrano}, {Vicer{\'e}}, {Viets}, {Villa-Ortega}, {Vinet}, {Vitale}, {Vo}, {Vocca}, {von Reis}, {von Wrangel}, {Vorvick}, {Vyatchanin}, {Wade}, {Wade}, {Wagner}, {Walet}, {Walker}, {Wallace}, {Wallace}, {Walsh}, {Wang}, {Wang},
  {Wang}, {Ward}, {Warner}, {Was}, {Washimi}, {Washington}, {Watchi}, {Weaver}, {Wei}, {Weinert}, {Weinstein}, {Weiss}, {Weller}, {Wellmann}, {Wen}, {We{\ss}els}, {Westhouse}, {Wette}, {Whelan}, {White}, {Whiting}, {Whittle}, {Wilken}, {Williams}, {Williams}, {Williamson}, {Willis}, {Willke}, {Wilson}, {Winkler}, {Wipf}, {Wlodarczyk}, {Woan}, {Woehler}, {Wofford}, {Wong}, {Wu}, {Wu}, {Wu}, {Wu}, {Wysocki}, {Xiao}, {Xu}, {Yamada}, {Yamamoto}, {Yamamoto}, {Yamamoto}, {Yamamoto}, {Yamashita}, {Yamazaki}, {Yang}, {Yang}, {Yang}, {Yang}, {Yang}, {Yap}, {Yeeles}, {Yelikar}, {Ying}, {Yokogawa}, {Yokoyama}, {Yokozawa}, {Yoon}, {Yoshioka}, {Yu}, {Yu}, {Yuzurihara}, {Zadro{\.z}ny}, {Zanolin}, {Zappa}, {Zeidler}, {Zelenova}, {Zendri}, {Zevin}, {Zhan}, {Zhang}, {Zhang}, {Zhang}, {Zhang}, {Zhang}, {Zhao}, {Zhao}, {Zhao}, {Zhao}, {Zhou}, {Zhu}, {Zhu}, {Zimmerman}, {Zlochower}, {Zucker}, {Zweizig}, {Ligo Scientific Collaboration}, {VIRGO Collaboration}, \& {KAGRA Collaboration}}]{2021ApJ...915L...5A}
{Abbott}, R., {Abbott}, T.~D., {Abraham}, S., {et~al.} 2021, \apjl, 915, L5, \dodoi{10.3847/2041-8213/ac082e}

\bibitem[{{Acernese} {et~al.}(2015){Acernese}, {Agathos}, {Agatsuma}, {Aisa}, {Allemandou}, {Allocca}, {Amarni}, {Astone}, {Balestri}, {Ballardin}, {Barone}, {Baronick}, {Barsuglia}, {Basti}, {Basti}, {Bauer}, {Bavigadda}, {Bejger}, {Beker}, {Belczynski}, {Bersanetti}, {Bertolini}, {Bitossi}, {Bizouard}, {Bloemen}, {Blom}, {Boer}, {Bogaert}, {Bondi}, {Bondu}, {Bonelli}, {Bonnand}, {Boschi}, {Bosi}, {Bouedo}, {Bradaschia}, {Branchesi}, {Briant}, {Brillet}, {Brisson}, {Bulik}, {Bulten}, {Buskulic}, {Buy}, {Cagnoli}, {Calloni}, {Campeggi}, {Canuel}, {Carbognani}, {Cavalier}, {Cavalieri}, {Cella}, {Cesarini}, {Mottin}, {Chincarini}, {Chiummo}, {Chua}, {Cleva}, {Coccia}, {Cohadon}, {Colla}, {Colombini}, {Conte}, {Coulon}, {Cuoco}, {Dalmaz}, {D'Antonio}, {Dattilo}, {Davier}, {Day}, {Debreczeni}, {Degallaix}, {Del{\'e}glise}, {Pozzo}, {Dereli}, {Rosa}, {Fiore}, {Lieto}, {Virgilio}, {Doets}, {Dolique}, {Drago}, {Ducrot}, {Endr{\H{o}}czi}, {Fafone}, {Farinon}, {Ferrante}, {Ferrini}, {Fidecaro}, {Fiori}, {Flaminio},
  {Fournier}, {Franco}, {Frasca}, {Frasconi}, {Gammaitoni}, {Garufi}, {Gaspard}, {Gatto}, {Gemme}, {Gendre}, {Genin}, {Gennai}, {Ghosh}, {Giacobone}, {Giazotto}, {Gouaty}, {Granata}, {Greco}, {Groot}, {Guidi}, {Harms}, {Heidmann}, {Heitmann}, {Hello}, {Hemming}, {Hennes}, {Hofman}, {Jaranowski}, {Jonker}, {Kasprzack}, {K{\'e}f{\'e}lian}, {Kowalska}, {Kraan}, {Kr{\'o}lak}, {Kutynia}, {Lazzaro}, {Leonardi}, {Leroy}, {Letendre}, {Li}, {Lieunard}, {Lorenzini}, {Loriette}, {Losurdo}, {Magazz{\`u}}, {Majorana}, {Maksimovic}, {Malvezzi}, {Man}, {Mangano}, {Mantovani}, {Marchesoni}, {Marion}, {Marque}, {Martelli}, {Martellini}, {Masserot}, {Meacher}, {Meidam}, {Mezzani}, {Michel}, {Milano}, {Minenkov}, {Moggi}, {Mohan}, {Montani}, {Morgado}, {Mours}, {Mul}, {Nagy}, {Nardecchia}, {Naticchioni}, {Nelemans}, {Neri}, {Neri}, {Nocera}, {Pacaud}, {Palomba}, {Paoletti}, {Paoli}, {Pasqualetti}, {Passaquieti}, {Passuello}, {Perciballi}, {Petit}, {Pichot}, {Piergiovanni}, {Pillant}, {Piluso}, {Pinard}, {Poggiani}, {Prijatelj},
  {Prodi}, {Punturo}, {Puppo}, {Rabeling}, {R{\'a}cz}, {Rapagnani}, {Razzano}, {Re}, {Regimbau}, {Ricci}, {Robinet}, {Rocchi}, {Rolland}, {Romano}, {Rosi{\'n}ska}, {Ruggi}, {Saracco}, {Sassolas}, {Schimmel}, {Sentenac}, {Sequino}, {Shah}, {Siellez}, {Straniero}, {Swinkels}, {Tacca}, {Tonelli}, {Travasso}, {Turconi}, {Vajente}, {van Bakel}, {van Beuzekom}, {van den Brand}, {Van Den Broeck}, {van der Sluys}, {van Heijningen}, {Vas{\'u}th}, {Vedovato}, {Veitch}, {Verkindt}, {Vetrano}, {Vicer{\'e}}, {Vinet}, {Visser}, {Vocca}, {Ward}, {Was}, {Wei}, {Yvert}, {{\.z}ny}, \& {Zendri}}]{2015CQGra..32b4001A}
{Acernese}, F., {Agathos}, M., {Agatsuma}, K., {et~al.} 2015, Classical and Quantum Gravity, 32, 024001, \dodoi{10.1088/0264-9381/32/2/024001}

\bibitem[{{Akutsu} {et~al.}(2021){Akutsu}, {Ando}, {Arai}, {Arai}, {Araki}, {Araya}, {Aritomi}, {Aso}, {Bae}, {Bae}, {Baiotti}, {Bajpai}, {Barton}, {Cannon}, {Capocasa}, {Chan}, {Chen}, {Chen}, {Chen}, {Chu}, {Chu}, {Eguchi}, {Enomoto}, {Flaminio}, {Fujii}, {Fukunaga}, {Fukushima}, {Ge}, {Hagiwara}, {Haino}, {Hasegawa}, {Hayakawa}, {Hayama}, {Himemoto}, {Hiranuma}, {Hirata}, {Hirose}, {Hong}, {Hsieh}, {Huang}, {Huang}, {Huang}, {Ikenoue}, {Imam}, {Inayoshi}, {Inoue}, {Ioka}, {Itoh}, {Izumi}, {Jung}, {Jung}, {Kajita}, {Kamiizumi}, {Kanda}, {Kang}, {Kawaguchi}, {Kawai}, {Kawasaki}, {Kim}, {Kim}, {Kim}, {Kim}, {Kimura}, {Kita}, {Kitazawa}, {Kojima}, {Kokeyama}, {Komori}, {Kong}, {Kotake}, {Kozakai}, {Kozu}, {Kumar}, {Kume}, {Kuo}, {Kuo}, {Kuroyanagi}, {Kusayanagi}, {Kwak}, {Lee}, {Lee}, {Lee}, {Leonardi}, {Lin}, {Lin}, {Lin}, {Liu}, {Luo}, {Marchio}, {Michimura}, {Mio}, {Miyakawa}, {Miyamoto}, {Miyazaki}, {Miyo}, {Miyoki}, {Morisaki}, {Moriwaki}, {Nagano}, {Nagano}, {Nakamura}, {Nakano}, {Nakano}, {Nakashima},
  {Narikawa}, {Negishi}, {Ni}, {Nishizawa}, {Obuchi}, {Ogaki}, {Oh}, {Oh}, {Ohashi}, {Ohishi}, {Ohkawa}, {Okutomi}, {Oohara}, {Ooi}, {Oshino}, {Pan}, {Pang}, {Park}, {Arellano}, {Pinto}, {Sago}, {Saito}, {Saito}, {Sakai}, {Sakai}, {Sakuno}, {Sato}, {Sato}, {Sawada}, {Sekiguchi}, {Sekiguchi}, {Shibagaki}, {Shimizu}, {Shimoda}, {Shimode}, {Shinkai}, {Shishido}, {Shoda}, {Somiya}, {Son}, {Sotani}, {Sugimoto}, {Suzuki}, {Suzuki}, {Tagoshi}, {Takahashi}, {Takahashi}, {Takamori}, {Takano}, {Takeda}, {Takeda}, {Tanaka}, {Tanaka}, {Tanaka}, {Tanaka}, {Tanaka}, {Tanioka}, {Tapia San Martin}, {Telada}, {Tomaru}, {Tomigami}, {Tomura}, {Travasso}, {Trozzo}, {Tsang}, {Tsubono}, {Tsuchida}, {Tsuzuki}, {Tuyenbayev}, {Uchikata}, {Uchiyama}, {Ueda}, {Uehara}, {Ueno}, {Ueshima}, {Uraguchi}, {Ushiba}, {van Putten}, {Vocca}, {Wang}, {Wu}, {Wu}, {Wu}, {Xu}, {Yamada}, {Yamamoto}, {Yamamoto}, {Yamamoto}, {Yokogawa}, {Yokoyama}, {Yokozawa}, {Yoshioka}, {Yuzurihara}, {Zeidler}, {Zhao}, \& {Zhu}}]{2021PTEP.2021eA101A}
{Akutsu}, T., {Ando}, M., {Arai}, K., {et~al.} 2021, Progress of Theoretical and Experimental Physics, 2021, 05A101, \dodoi{10.1093/ptep/ptaa125}

\bibitem[{{Arcavi} {et~al.}(2017{\natexlab{a}}){Arcavi}, {Hosseinzadeh}, {Howell}, {McCully}, {Poznanski}, {Kasen}, {Barnes}, {Zaltzman}, {Vasylyev}, {Maoz}, \& {Valenti}}]{2017Natur.551...64A}
{Arcavi}, I., {Hosseinzadeh}, G., {Howell}, D.~A., {et~al.} 2017{\natexlab{a}}, \nat, 551, 64, \dodoi{10.1038/nature24291}

\bibitem[{{Arcavi} {et~al.}(2017{\natexlab{b}}){Arcavi}, {McCully}, {Hosseinzadeh}, {Howell}, {Vasylyev}, {Poznanski}, {Zaltzman}, {Maoz}, {Singer}, {Valenti}, {Kasen}, {Barnes}, {Piran}, \& {Fong}}]{2017ApJ...848L..33A}
{Arcavi}, I., {McCully}, C., {Hosseinzadeh}, G., {et~al.} 2017{\natexlab{b}}, \apjl, 848, L33, \dodoi{10.3847/2041-8213/aa910f}

\bibitem[{{Artale} {et~al.}(2020{\natexlab{a}}){Artale}, {Bouffanais}, {Mapelli}, {Giacobbo}, {Sabha}, {Santoliquido}, {Pasquato}, \& {Spera}}]{2020MNRAS.495.1841A}
{Artale}, M.~C., {Bouffanais}, Y., {Mapelli}, M., {et~al.} 2020{\natexlab{a}}, \mnras, 495, 1841, \dodoi{10.1093/mnras/staa1252}

\bibitem[{{Artale} {et~al.}(2020{\natexlab{b}}){Artale}, {Mapelli}, {Bouffanais}, {Giacobbo}, {Pasquato}, \& {Spera}}]{2020MNRAS.491.3419A}
{Artale}, M.~C., {Mapelli}, M., {Bouffanais}, Y., {et~al.} 2020{\natexlab{b}}, \mnras, 491, 3419, \dodoi{10.1093/mnras/stz3190}

\bibitem[{{Artale} {et~al.}(2019){Artale}, {Mapelli}, {Giacobbo}, {Sabha}, {Spera}, {Santoliquido}, \& {Bressan}}]{2019MNRAS.487.1675A}
{Artale}, M.~C., {Mapelli}, M., {Giacobbo}, N., {et~al.} 2019, \mnras, 487, 1675, \dodoi{10.1093/mnras/stz1382}

\bibitem[{{Astropy Collaboration} {et~al.}(2013){Astropy Collaboration}, {Robitaille}, {Tollerud}, {Greenfield}, {Droettboom}, {Bray}, {Aldcroft}, {Davis}, {Ginsburg}, {Price-Whelan}, {Kerzendorf}, {Conley}, {Crighton}, {Barbary}, {Muna}, {Ferguson}, {Grollier}, {Parikh}, {Nair}, {Unther}, {Deil}, {Woillez}, {Conseil}, {Kramer}, {Turner}, {Singer}, {Fox}, {Weaver}, {Zabalza}, {Edwards}, {Azalee Bostroem}, {Burke}, {Casey}, {Crawford}, {Dencheva}, {Ely}, {Jenness}, {Labrie}, {Lim}, {Pierfederici}, {Pontzen}, {Ptak}, {Refsdal}, {Servillat}, \& {Streicher}}]{2013A&A...558A..33A}
{Astropy Collaboration}, {Robitaille}, T.~P., {Tollerud}, E.~J., {et~al.} 2013, \aap, 558, A33, \dodoi{10.1051/0004-6361/201322068}

\bibitem[{{Astropy Collaboration} {et~al.}(2018){Astropy Collaboration}, {Price-Whelan}, {Sip{\H{o}}cz}, {G{\"u}nther}, {Lim}, {Crawford}, {Conseil}, {Shupe}, {Craig}, {Dencheva}, {Ginsburg}, {VanderPlas}, {Bradley}, {P{\'e}rez-Su{\'a}rez}, {de Val-Borro}, {Aldcroft}, {Cruz}, {Robitaille}, {Tollerud}, {Ardelean}, {Babej}, {Bach}, {Bachetti}, {Bakanov}, {Bamford}, {Barentsen}, {Barmby}, {Baumbach}, {Berry}, {Biscani}, {Boquien}, {Bostroem}, {Bouma}, {Brammer}, {Bray}, {Breytenbach}, {Buddelmeijer}, {Burke}, {Calderone}, {Cano Rodr{\'\i}guez}, {Cara}, {Cardoso}, {Cheedella}, {Copin}, {Corrales}, {Crichton}, {D'Avella}, {Deil}, {Depagne}, {Dietrich}, {Donath}, {Droettboom}, {Earl}, {Erben}, {Fabbro}, {Ferreira}, {Finethy}, {Fox}, {Garrison}, {Gibbons}, {Goldstein}, {Gommers}, {Greco}, {Greenfield}, {Groener}, {Grollier}, {Hagen}, {Hirst}, {Homeier}, {Horton}, {Hosseinzadeh}, {Hu}, {Hunkeler}, {Ivezi{\'c}}, {Jain}, {Jenness}, {Kanarek}, {Kendrew}, {Kern}, {Kerzendorf}, {Khvalko}, {King}, {Kirkby}, {Kulkarni},
  {Kumar}, {Lee}, {Lenz}, {Littlefair}, {Ma}, {Macleod}, {Mastropietro}, {McCully}, {Montagnac}, {Morris}, {Mueller}, {Mumford}, {Muna}, {Murphy}, {Nelson}, {Nguyen}, {Ninan}, {N{\"o}the}, {Ogaz}, {Oh}, {Parejko}, {Parley}, {Pascual}, {Patil}, {Patil}, {Plunkett}, {Prochaska}, {Rastogi}, {Reddy Janga}, {Sabater}, {Sakurikar}, {Seifert}, {Sherbert}, {Sherwood-Taylor}, {Shih}, {Sick}, {Silbiger}, {Singanamalla}, {Singer}, {Sladen}, {Sooley}, {Sornarajah}, {Streicher}, {Teuben}, {Thomas}, {Tremblay}, {Turner}, {Terr{\'o}n}, {van Kerkwijk}, {de la Vega}, {Watkins}, {Weaver}, {Whitmore}, {Woillez}, {Zabalza}, \& {Astropy Contributors}}]{2018AJ....156..123A}
{Astropy Collaboration}, {Price-Whelan}, A.~M., {Sip{\H{o}}cz}, B.~M., {et~al.} 2018, \aj, 156, 123, \dodoi{10.3847/1538-3881/aabc4f}

\bibitem[{{Astropy Collaboration} {et~al.}(2022){Astropy Collaboration}, {Price-Whelan}, {Lim}, {Earl}, {Starkman}, {Bradley}, {Shupe}, {Patil}, {Corrales}, {Brasseur}, {N{\"o}the}, {Donath}, {Tollerud}, {Morris}, {Ginsburg}, {Vaher}, {Weaver}, {Tocknell}, {Jamieson}, {van Kerkwijk}, {Robitaille}, {Merry}, {Bachetti}, {G{\"u}nther}, {Aldcroft}, {Alvarado-Montes}, {Archibald}, {B{\'o}di}, {Bapat}, {Barentsen}, {Baz{\'a}n}, {Biswas}, {Boquien}, {Burke}, {Cara}, {Cara}, {Conroy}, {Conseil}, {Craig}, {Cross}, {Cruz}, {D'Eugenio}, {Dencheva}, {Devillepoix}, {Dietrich}, {Eigenbrot}, {Erben}, {Ferreira}, {Foreman-Mackey}, {Fox}, {Freij}, {Garg}, {Geda}, {Glattly}, {Gondhalekar}, {Gordon}, {Grant}, {Greenfield}, {Groener}, {Guest}, {Gurovich}, {Handberg}, {Hart}, {Hatfield-Dodds}, {Homeier}, {Hosseinzadeh}, {Jenness}, {Jones}, {Joseph}, {Kalmbach}, {Karamehmetoglu}, {Ka{\l}uszy{\'n}ski}, {Kelley}, {Kern}, {Kerzendorf}, {Koch}, {Kulumani}, {Lee}, {Ly}, {Ma}, {MacBride}, {Maljaars}, {Muna}, {Murphy}, {Norman},
  {O'Steen}, {Oman}, {Pacifici}, {Pascual}, {Pascual-Granado}, {Patil}, {Perren}, {Pickering}, {Rastogi}, {Roulston}, {Ryan}, {Rykoff}, {Sabater}, {Sakurikar}, {Salgado}, {Sanghi}, {Saunders}, {Savchenko}, {Schwardt}, {Seifert-Eckert}, {Shih}, {Jain}, {Shukla}, {Sick}, {Simpson}, {Singanamalla}, {Singer}, {Singhal}, {Sinha}, {Sip{\H{o}}cz}, {Spitler}, {Stansby}, {Streicher}, {{\v{S}}umak}, {Swinbank}, {Taranu}, {Tewary}, {Tremblay}, {de Val-Borro}, {Van Kooten}, {Vasovi{\'c}}, {Verma}, {de Miranda Cardoso}, {Williams}, {Wilson}, {Winkel}, {Wood-Vasey}, {Xue}, {Yoachim}, {Zhang}, {Zonca}, \& {Astropy Project Contributors}}]{2022ApJ...935..167A}
{Astropy Collaboration}, {Price-Whelan}, A.~M., {Lim}, P.~L., {et~al.} 2022, \apj, 935, 167, \dodoi{10.3847/1538-4357/ac7c74}

\bibitem[{{Barbieri} {et~al.}(2020){Barbieri}, {Salafia}, {Colpi}, {Ghirlanda}, \& {Perego}}]{2020arXiv200209395B}
{Barbieri}, C., {Salafia}, O.~S., {Colpi}, M., {Ghirlanda}, G., \& {Perego}, A. 2020, arXiv e-prints, arXiv:2002.09395, \dodoi{10.48550/arXiv.2002.09395}

\bibitem[{{Becker}(2015)}]{2015ascl.soft04004B}
{Becker}, A. 2015, {HOTPANTS: High Order Transform of PSF ANd Template Subtraction}, Astrophysics Source Code Library, record ascl:1504.004.
\newblock \doeprint{1504.004}

\bibitem[{{Belczynski} {et~al.}(2002){Belczynski}, {Kalogera}, \& {Bulik}}]{2002ApJ...572..407B}
{Belczynski}, K., {Kalogera}, V., \& {Bulik}, T. 2002, \apj, 572, 407, \dodoi{10.1086/340304}

\bibitem[{{Berthier} {et~al.}(2006){Berthier}, {Vachier}, {Thuillot}, {Fernique}, {Ochsenbein}, {Genova}, {Lainey}, \& {Arlot}}]{2006ASPC..351..367B}
{Berthier}, J., {Vachier}, F., {Thuillot}, W., {et~al.} 2006, in Astronomical Society of the Pacific Conference Series, Vol. 351, Astronomical Data Analysis Software and Systems XV, ed. C.~{Gabriel}, C.~{Arviset}, D.~{Ponz}, \& S.~{Enrique}, 367

\bibitem[{{Bertin}(2010)}]{2010ascl.soft10063B}
{Bertin}, E. 2010, {SCAMP: Automatic Astrometric and Photometric Calibration}, Astrophysics Source Code Library, record ascl:1010.063.
\newblock \doeprint{1010.063}

\bibitem[{{Bertin} \& {Arnouts}(1996)}]{1996A&AS..117..393B}
{Bertin}, E., \& {Arnouts}, S. 1996, \aaps, 117, 393, \dodoi{10.1051/aas:1996164}

\bibitem[{{Brammer} {et~al.}(2008){Brammer}, {van Dokkum}, \& {Coppi}}]{2008ApJ...686.1503B}
{Brammer}, G.~B., {van Dokkum}, P.~G., \& {Coppi}, P. 2008, \apj, 686, 1503, \dodoi{10.1086/591786}

\bibitem[{{Buckley} {et~al.}(2019){Buckley}, {Jha}, {Cooke}, \& {Mogotsi}}]{2019GCN.24205....1B}
{Buckley}, D.~A.~H., {Jha}, S.~W., {Cooke}, J., \& {Mogotsi}, M. 2019, GRB Coordinates Network, 24205, 1

\bibitem[{{Bulla}(2019)}]{2019MNRAS.489.5037B}
{Bulla}, M. 2019, \mnras, 489, 5037, \dodoi{10.1093/mnras/stz2495}

\bibitem[{{Chambers} {et~al.}(2016){Chambers}, {Magnier}, {Metcalfe}, {Flewelling}, {Huber}, {Waters}, {Denneau}, {Draper}, {Farrow}, {Finkbeiner}, {Holmberg}, {Koppenhoefer}, {Price}, {Rest}, {Saglia}, {Schlafly}, {Smartt}, {Sweeney}, {Wainscoat}, {Burgett}, {Chastel}, {Grav}, {Heasley}, {Hodapp}, {Jedicke}, {Kaiser}, {Kudritzki}, {Luppino}, {Lupton}, {Monet}, {Morgan}, {Onaka}, {Shiao}, {Stubbs}, {Tonry}, {White}, {Ba{\~n}ados}, {Bell}, {Bender}, {Bernard}, {Boegner}, {Boffi}, {Botticella}, {Calamida}, {Casertano}, {Chen}, {Chen}, {Cole}, {Deacon}, {Frenk}, {Fitzsimmons}, {Gezari}, {Gibbs}, {Goessl}, {Goggia}, {Gourgue}, {Goldman}, {Grant}, {Grebel}, {Hambly}, {Hasinger}, {Heavens}, {Heckman}, {Henderson}, {Henning}, {Holman}, {Hopp}, {Ip}, {Isani}, {Jackson}, {Keyes}, {Koekemoer}, {Kotak}, {Le}, {Liska}, {Long}, {Lucey}, {Liu}, {Martin}, {Masci}, {McLean}, {Mindel}, {Misra}, {Morganson}, {Murphy}, {Obaika}, {Narayan}, {Nieto-Santisteban}, {Norberg}, {Peacock}, {Pier}, {Postman}, {Primak}, {Rae}, {Rai},
  {Riess}, {Riffeser}, {Rix}, {R{\"o}ser}, {Russel}, {Rutz}, {Schilbach}, {Schultz}, {Scolnic}, {Strolger}, {Szalay}, {Seitz}, {Small}, {Smith}, {Soderblom}, {Taylor}, {Thomson}, {Taylor}, {Thakar}, {Thiel}, {Thilker}, {Unger}, {Urata}, {Valenti}, {Wagner}, {Walder}, {Walter}, {Watters}, {Werner}, {Wood-Vasey}, \& {Wyse}}]{2016arXiv161205560C}
{Chambers}, K.~C., {Magnier}, E.~A., {Metcalfe}, N., {et~al.} 2016, arXiv e-prints, arXiv:1612.05560, \dodoi{10.48550/arXiv.1612.05560}

\bibitem[{{Chang} {et~al.}(2019){Chang}, {Wolf}, {Onken}, {Scott}, {Lidman}, {Cooke}, {Webb}, {Gurri}, \& {Nordlander}}]{2019GCN.24260....1C}
{Chang}, S.~W., {Wolf}, C., {Onken}, C.~A., {et~al.} 2019, GRB Coordinates Network, 24260, 1

\bibitem[{{Chase} {et~al.}(2022){Chase}, {O'Connor}, {Fryer}, {Troja}, {Korobkin}, {Wollaeger}, {Ristic}, {Fontes}, {Hungerford}, \& {Herring}}]{2022ApJ...927..163C}
{Chase}, E.~A., {O'Connor}, B., {Fryer}, C.~L., {et~al.} 2022, \apj, 927, 163, \dodoi{10.3847/1538-4357/ac3d25}

\bibitem[{{Choi} \& {Im}(2017)}]{2017JKAS...50...71C}
{Choi}, C., \& {Im}, M. 2017, Journal of Korean Astronomical Society, 50, 71, \dodoi{10.5303/JKAS.2017.50.3.71}

\bibitem[{{Chornock} {et~al.}(2017){Chornock}, {Berger}, {Kasen}, {Cowperthwaite}, {Nicholl}, {Villar}, {Alexander}, {Blanchard}, {Eftekhari}, {Fong}, {Margutti}, {Williams}, {Annis}, {Brout}, {Brown}, {Chen}, {Drout}, {Farr}, {Foley}, {Frieman}, {Fryer}, {Herner}, {Holz}, {Kessler}, {Matheson}, {Metzger}, {Quataert}, {Rest}, {Sako}, {Scolnic}, {Smith}, \& {Soares-Santos}}]{2017ApJ...848L..19C}
{Chornock}, R., {Berger}, E., {Kasen}, D., {et~al.} 2017, \apjl, 848, L19, \dodoi{10.3847/2041-8213/aa905c}

\bibitem[{{Coughlin} {et~al.}(2019){Coughlin}, {Ahumada}, {Anand}, {De}, {Hankins}, {Kasliwal}, {Singer}, {Bellm}, {Andreoni}, {Cenko}, {Cooke}, {Copperwheat}, {Dugas}, {Jencson}, {Perley}, {Yu}, {Bhalerao}, {Kumar}, {Bloom}, {Anupama}, {Ashley}, {Bagdasaryan}, {Biswas}, {Buckley}, {Burdge}, {Cook}, {Cromer}, {Cunningham}, {D'A{\`\i}}, {Dekany}, {Delacroix}, {Dichiara}, {Duev}, {Dutta}, {Feeney}, {Frederick}, {Gatkine}, {Ghosh}, {Goldstein}, {Golkhou}, {Goobar}, {Graham}, {Hanayama}, {Horiuchi}, {Hung}, {Jha}, {Kong}, {Giomi}, {Kaplan}, {Karambelkar}, {Kowalski}, {Kulkarni}, {Kupfer}, {Masci}, {Mazzali}, {Moore}, {Mogotsi}, {Neill}, {Ngeow}, {Mart{\'\i}nez-Palomera}, {La Parola}, {Pavana}, {Ofek}, {Patil}, {Riddle}, {Rigault}, {Rusholme}, {Serabyn}, {Shupe}, {Sharma}, {Singh}, {Sollerman}, {Soon}, {Staats}, {Taggart}, {Tan}, {Travouillon}, {Troja}, {Waratkar}, \& {Yatsu}}]{2019ApJ...885L..19C}
{Coughlin}, M.~W., {Ahumada}, T., {Anand}, S., {et~al.} 2019, \apjl, 885, L19, \dodoi{10.3847/2041-8213/ab4ad8}

\bibitem[{{Coulter} {et~al.}(2017){Coulter}, {Foley}, {Kilpatrick}, {Drout}, {Piro}, {Shappee}, {Siebert}, {Simon}, {Ulloa}, {Kasen}, {Madore}, {Murguia-Berthier}, {Pan}, {Prochaska}, {Ramirez-Ruiz}, {Rest}, \& {Rojas-Bravo}}]{2017Sci...358.1556C}
{Coulter}, D.~A., {Foley}, R.~J., {Kilpatrick}, C.~D., {et~al.} 2017, Science, 358, 1556, \dodoi{10.1126/science.aap9811}

\bibitem[{{Cowperthwaite} {et~al.}(2017){Cowperthwaite}, {Berger}, {Villar}, {Metzger}, {Nicholl}, {Chornock}, {Blanchard}, {Fong}, {Margutti}, {Soares-Santos}, {Alexander}, {Allam}, {Annis}, {Brout}, {Brown}, {Butler}, {Chen}, {Diehl}, {Doctor}, {Drout}, {Eftekhari}, {Farr}, {Finley}, {Foley}, {Frieman}, {Fryer}, {Garc{\'\i}a-Bellido}, {Gill}, {Guillochon}, {Herner}, {Holz}, {Kasen}, {Kessler}, {Marriner}, {Matheson}, {Neilsen}, {Quataert}, {Palmese}, {Rest}, {Sako}, {Scolnic}, {Smith}, {Tucker}, {Williams}, {Balbinot}, {Carlin}, {Cook}, {Durret}, {Li}, {Lopes}, {Louren{\c{c}}o}, {Marshall}, {Medina}, {Muir}, {Mu{\~n}oz}, {Sauseda}, {Schlegel}, {Secco}, {Vivas}, {Wester}, {Zenteno}, {Zhang}, {Abbott}, {Banerji}, {Bechtol}, {Benoit-L{\'e}vy}, {Bertin}, {Buckley-Geer}, {Burke}, {Capozzi}, {Carnero Rosell}, {Carrasco Kind}, {Castander}, {Crocce}, {Cunha}, {D'Andrea}, {da Costa}, {Davis}, {DePoy}, {Desai}, {Dietrich}, {Drlica-Wagner}, {Eifler}, {Evrard}, {Fernandez}, {Flaugher}, {Fosalba}, {Gaztanaga}, {Gerdes},
  {Giannantonio}, {Goldstein}, {Gruen}, {Gruendl}, {Gutierrez}, {Honscheid}, {Jain}, {James}, {Jeltema}, {Johnson}, {Johnson}, {Kent}, {Krause}, {Kron}, {Kuehn}, {Nuropatkin}, {Lahav}, {Lima}, {Lin}, {Maia}, {March}, {Martini}, {McMahon}, {Menanteau}, {Miller}, {Miquel}, {Mohr}, {Neilsen}, {Nichol}, {Ogando}, {Plazas}, {Roe}, {Romer}, {Roodman}, {Rykoff}, {Sanchez}, {Scarpine}, {Schindler}, {Schubnell}, {Sevilla-Noarbe}, {Smith}, {Smith}, {Sobreira}, {Suchyta}, {Swanson}, {Tarle}, {Thomas}, {Thomas}, {Troxel}, {Vikram}, {Walker}, {Wechsler}, {Weller}, {Yanny}, \& {Zuntz}}]{2017ApJ...848L..17C}
{Cowperthwaite}, P.~S., {Berger}, E., {Villar}, V.~A., {et~al.} 2017, \apjl, 848, L17, \dodoi{10.3847/2041-8213/aa8fc7}

\bibitem[{{Craig} {et~al.}(2015){Craig}, {Crawford}, {Deil}, {Gomez}, {G{\"u}nther}, {Heidt}, {Horton}, {Karr}, {Nelson}, {Ninan}, {Pattnaik}, {Rol}, {Schoenell}, {Seifert}, {Singh}, {Sipocz}, {Stotts}, {Streicher}, {Tollerud}, {Walker}, \& {ccdproc contributors}}]{2015ascl.soft10007C}
{Craig}, M.~W., {Crawford}, S.~M., {Deil}, C., {et~al.} 2015, {ccdproc: CCD data reduction software}, Astrophysics Source Code Library, record ascl:1510.007.
\newblock \doeprint{1510.007}

\bibitem[{{Cutri} {et~al.}(2013){Cutri}, {Wright}, {Conrow}, {Fowler}, {Eisenhardt}, {Grillmair}, {Kirkpatrick}, {Masci}, {McCallon}, {Wheelock}, {Fajardo-Acosta}, {Yan}, {Benford}, {Harbut}, {Jarrett}, {Lake}, {Leisawitz}, {Ressler}, {Stanford}, {Tsai}, {Liu}, {Helou}, {Mainzer}, {Gettings}, {Gonzalez}, {Hoffman}, {Marsh}, {Padgett}, {Skrutskie}, {Beck}, {Papin}, \& {Wittman}}]{2013wise.rept....1C}
{Cutri}, R.~M., {Wright}, E.~L., {Conrow}, T., {et~al.} 2013, {Explanatory Supplement to the AllWISE Data Release Products}, Explanatory Supplement to the AllWISE Data Release Products, by R. M. Cutri et al.

\bibitem[{{D{\'a}lya} {et~al.}(2018){D{\'a}lya}, {Galg{\'o}czi}, {Dobos}, {Frei}, {Heng}, {Macas}, {Messenger}, {Raffai}, \& {de Souza}}]{2018MNRAS.479.2374D}
{D{\'a}lya}, G., {Galg{\'o}czi}, G., {Dobos}, L., {et~al.} 2018, \mnras, 479, 2374, \dodoi{10.1093/mnras/sty1703}

\bibitem[{{D{\'a}lya} {et~al.}(2022){D{\'a}lya}, {D{\'\i}az}, {Bouchet}, {Frei}, {Jasche}, {Lavaux}, {Macas}, {Mukherjee}, {P{\'a}lfi}, {de Souza}, {Wandelt}, {Bilicki}, \& {Raffai}}]{2022MNRAS.514.1403D}
{D{\'a}lya}, G., {D{\'\i}az}, R., {Bouchet}, F.~R., {et~al.} 2022, \mnras, 514, 1403, \dodoi{10.1093/mnras/stac1443}

\bibitem[{{de Jaeger} {et~al.}(2019){de Jaeger}, {Zheng}, {Stahl}, {Filippenko}, {Brink}, {Bigley}, {Blanchard}, {Blanchard}, {Bradley}, {Cargill}, {Casper}, {Cenko}, {Channa}, {Choi}, {Clubb}, {Cobb}, {Cohen}, {de Kouchkovsky}, {Ellison}, {Falcon}, {Fox}, {Fuller}, {Ganeshalingam}, {Gould}, {Graham}, {Halevi}, {Hayakawa}, {Hestenes}, {Hyland}, {Jeffers}, {Joubert}, {Kandrashoff}, {Kelly}, {Kim}, {Kim}, {Kumar}, {Leonard}, {Li}, {Lowe}, {Lu}, {Mason}, {McAllister}, {Mauerhan}, {Modjaz}, {Molloy}, {Perley}, {Pina}, {Poznanski}, {Ross}, {Shivvers}, {Silverman}, {Soler}, {Stegman}, {Taylor}, {Tang}, {Wilkins}, {Wang}, {Wang}, {Yuk}, {Yunus}, \& {Zhang}}]{2019MNRAS.490.2799D}
{de Jaeger}, T., {Zheng}, W., {Stahl}, B.~E., {et~al.} 2019, \mnras, 490, 2799, \dodoi{10.1093/mnras/stz2714}

\bibitem[{{Dichiara} {et~al.}(2019){Dichiara}, {Gatkine}, {Durbak}, {Troja}, {Cenko}, {Kutyrev}, \& {Veilleux}}]{2019GCN.24220....1D}
{Dichiara}, S., {Gatkine}, P., {Durbak}, J., {et~al.} 2019, GRB Coordinates Network, 24220, 1

\bibitem[{{Dimitriadis} {et~al.}(2019){Dimitriadis}, {Jones}, {Siebert}, {Brown}, {Arcavi}, {Bloom}, {Bostroem}, {Coulter}, {Drout}, {Ebeling}, {Filippenko}, {Foley}, {Howell}, {Hung}, {Jha}, {Kasen}, {Kilpatrick}, {Piro}, {Prochaska}, {Quataert}, {Ramirez-Ruiz}, {Riess}, {Rojas-Bravo}, {Sand}, {Scolnic}, {Siellez}, {Valenti}, \& {Zheng}}]{2019GCN.24358....1D}
{Dimitriadis}, G., {Jones}, D.~O., {Siebert}, M.~R., {et~al.} 2019, GRB Coordinates Network, 24358, 1

\bibitem[{{Drout} {et~al.}(2017){Drout}, {Piro}, {Shappee}, {Kilpatrick}, {Simon}, {Contreras}, {Coulter}, {Foley}, {Siebert}, {Morrell}, {Boutsia}, {Di Mille}, {Holoien}, {Kasen}, {Kollmeier}, {Madore}, {Monson}, {Murguia-Berthier}, {Pan}, {Prochaska}, {Ramirez-Ruiz}, {Rest}, {Adams}, {Alatalo}, {Ba{\~n}ados}, {Baughman}, {Beers}, {Bernstein}, {Bitsakis}, {Campillay}, {Hansen}, {Higgs}, {Ji}, {Maravelias}, {Marshall}, {Moni Bidin}, {Prieto}, {Rasmussen}, {Rojas-Bravo}, {Strom}, {Ulloa}, {Vargas-Gonz{\'a}lez}, {Wan}, \& {Whitten}}]{2017Sci...358.1570D}
{Drout}, M.~R., {Piro}, A.~L., {Shappee}, B.~J., {et~al.} 2017, Science, 358, 1570, \dodoi{10.1126/science.aaq0049}

\bibitem[{{Evans} {et~al.}(2017){Evans}, {Cenko}, {Kennea}, {Emery}, {Kuin}, {Korobkin}, {Wollaeger}, {Fryer}, {Madsen}, {Harrison}, {Xu}, {Nakar}, {Hotokezaka}, {Lien}, {Campana}, {Oates}, {Troja}, {Breeveld}, {Marshall}, {Barthelmy}, {Beardmore}, {Burrows}, {Cusumano}, {D'A{\`\i}}, {D'Avanzo}, {D'Elia}, {de Pasquale}, {Even}, {Fontes}, {Forster}, {Garcia}, {Giommi}, {Grefenstette}, {Gronwall}, {Hartmann}, {Heida}, {Hungerford}, {Kasliwal}, {Krimm}, {Levan}, {Malesani}, {Melandri}, {Miyasaka}, {Nousek}, {O'Brien}, {Osborne}, {Pagani}, {Page}, {Palmer}, {Perri}, {Pike}, {Racusin}, {Rosswog}, {Siegel}, {Sakamoto}, {Sbarufatti}, {Tagliaferri}, {Tanvir}, \& {Tohuvavohu}}]{2017Sci...358.1565E}
{Evans}, P.~A., {Cenko}, S.~B., {Kennea}, J.~A., {et~al.} 2017, Science, 358, 1565, \dodoi{10.1126/science.aap9580}

\bibitem[{{Flewelling} {et~al.}(2020){Flewelling}, {Magnier}, {Chambers}, {Heasley}, {Holmberg}, {Huber}, {Sweeney}, {Waters}, {Calamida}, {Casertano}, {Chen}, {Farrow}, {Hasinger}, {Henderson}, {Long}, {Metcalfe}, {Narayan}, {Nieto-Santisteban}, {Norberg}, {Rest}, {Saglia}, {Szalay}, {Thakar}, {Tonry}, {Valenti}, {Werner}, {White}, {Denneau}, {Draper}, {Hodapp}, {Jedicke}, {Kaiser}, {Kudritzki}, {Price}, {Wainscoat}, {Chastel}, {McLean}, {Postman}, \& {Shiao}}]{2020ApJS..251....7F}
{Flewelling}, H.~A., {Magnier}, E.~A., {Chambers}, K.~C., {et~al.} 2020, \apjs, 251, 7, \dodoi{10.3847/1538-4365/abb82d}

\bibitem[{{Fong} {et~al.}(2022){Fong}, {Nugent}, {Dong}, {Berger}, {Paterson}, {Chornock}, {Levan}, {Blanchard}, {Alexander}, {Andrews}, {Cobb}, {Cucchiara}, {Fox}, {Fryer}, {Gordon}, {Kilpatrick}, {Lunnan}, {Margutti}, {Miller}, {Milne}, {Nicholl}, {Perley}, {Rastinejad}, {Escorial}, {Schroeder}, {Smith}, {Tanvir}, \& {Terreran}}]{2022ApJ...940...56F}
{Fong}, W.-f., {Nugent}, A.~E., {Dong}, Y., {et~al.} 2022, \apj, 940, 56, \dodoi{10.3847/1538-4357/ac91d0}

\bibitem[{{Gal-Yam}(2021)}]{2021AAS...23742305G}
{Gal-Yam}, A. 2021, in American Astronomical Society Meeting Abstracts, Vol.~53, American Astronomical Society Meeting Abstracts, 423.05

\bibitem[{{Gehrels} {et~al.}(2016){Gehrels}, {Cannizzo}, {Kanner}, {Kasliwal}, {Nissanke}, \& {Singer}}]{2016ApJ...820..136G}
{Gehrels}, N., {Cannizzo}, J.~K., {Kanner}, J., {et~al.} 2016, \apj, 820, 136, \dodoi{10.3847/0004-637X/820/2/136}

\bibitem[{{Ginsburg} {et~al.}(2019){Ginsburg}, {Sip{\H{o}}cz}, {Brasseur}, {Cowperthwaite}, {Craig}, {Deil}, {Guillochon}, {Guzman}, {Liedtke}, {Lian Lim}, {Lockhart}, {Mommert}, {Morris}, {Norman}, {Parikh}, {Persson}, {Robitaille}, {Segovia}, {Singer}, {Tollerud}, {de Val-Borro}, {Valtchanov}, {Woillez}, {Astroquery Collaboration}, \& {a subset of astropy Collaboration}}]{2019AJ....157...98G}
{Ginsburg}, A., {Sip{\H{o}}cz}, B.~M., {Brasseur}, C.~E., {et~al.} 2019, \aj, 157, 98, \dodoi{10.3847/1538-3881/aafc33}

\bibitem[{{G{\'o}rski} {et~al.}(2005){G{\'o}rski}, {Hivon}, {Banday}, {Wandelt}, {Hansen}, {Reinecke}, \& {Bartelmann}}]{2005ApJ...622..759G}
{G{\'o}rski}, K.~M., {Hivon}, E., {Banday}, A.~J., {et~al.} 2005, \apj, 622, 759, \dodoi{10.1086/427976}

\bibitem[{{Henden} {et~al.}(2009){Henden}, {Welch}, {Terrell}, \& {Levine}}]{2009AAS...21440702H}
{Henden}, A.~A., {Welch}, D.~L., {Terrell}, D., \& {Levine}, S.~E. 2009, in American Astronomical Society Meeting Abstracts, Vol. 214, American Astronomical Society Meeting Abstracts \#214, 407.02

\bibitem[{{Holz} \& {Hughes}(2005)}]{2005ApJ...629...15H}
{Holz}, D.~E., \& {Hughes}, S.~A. 2005, \apj, 629, 15, \dodoi{10.1086/431341}

\bibitem[{{Hosseinzadeh} {et~al.}(2019){Hosseinzadeh}, {Cowperthwaite}, {Gomez}, {Villar}, {Nicholl}, {Margutti}, {Berger}, {Chornock}, {Paterson}, {Fong}, {Savchenko}, {Short}, {Alexander}, {Blanchard}, {Braga}, {Calkins}, {Cartier}, {Coppejans}, {Eftekhari}, {Laskar}, {Ly}, {Patton}, {Pelisoli}, {Reichart}, {Terreran}, \& {Williams}}]{2019ApJ...880L...4H}
{Hosseinzadeh}, G., {Cowperthwaite}, P.~S., {Gomez}, S., {et~al.} 2019, \apjl, 880, L4, \dodoi{10.3847/2041-8213/ab271c}

\bibitem[{{Hwang} {et~al.}(2021){Hwang}, {Im}, {Taak}, {Paek}, {Choi}, {Shin}, {Lee}, {Ji}, {Pak}, {Lee}, {Ahn}, {Han}, {Kim}, {Marshall}, {Johns-Krull}, {Gibson}, {Schmidt}, \& {Prochaska}}]{2021ApJ...908..113H}
{Hwang}, S., {Im}, M., {Taak}, Y.~C., {et~al.} 2021, \apj, 908, 113, \dodoi{10.3847/1538-4357/abcd9a}

\bibitem[{{Im} {et~al.}(2015){Im}, {Choi}, \& {Kim}}]{2015JKAS...48..207I}
{Im}, M., {Choi}, C., \& {Kim}, K. 2015, Journal of Korean Astronomical Society, 48, 207, \dodoi{10.5303/JKAS.2015.48.4.207}

\bibitem[{{Im} {et~al.}(2020){Im}, {Kim}, \& {Paek}}]{2020grbg.conf...25I}
{Im}, M., {Kim}, J., \& {Paek}, G. S.~H. 2020, in Gamma-ray Bursts in the Gravitational Wave Era 2019, 25--27

\bibitem[{{Im} {et~al.}(2019{\natexlab{a}}){Im}, {Kim}, {Paek}, {Lee}, {Kim}, \& {Mok}}]{2019GCN.24318....1I}
{Im}, M., {Kim}, J., {Paek}, G.~S.~H., {et~al.} 2019{\natexlab{a}}, GRB Coordinates Network, 24318, 1

\bibitem[{{Im} {et~al.}(2017){Im}, {Yoon}, {Lee}, {Lee}, {Kim}, {Lee}, {Kim}, {Troja}, {Choi}, {Lim}, {Ko}, \& {Shim}}]{2017ApJ...849L..16I}
{Im}, M., {Yoon}, Y., {Lee}, S.-K.~J., {et~al.} 2017, \apjl, 849, L16, \dodoi{10.3847/2041-8213/aa9367}

\bibitem[{{Im} {et~al.}(2019{\natexlab{b}}){Im}, {Paek}, {Lim}, {Paek}, {Shin}, {Kim}, {Pak}, {Hwang}, {Park}, {Kim}, {Choi}, {Lee}, {Kim}, {Mok}, \& {McDonald}}]{2019GCN.24183....1I}
{Im}, M., {Paek}, G.~S.~H., {Lim}, G., {et~al.} 2019{\natexlab{b}}, GRB Coordinates Network, 24183, 1

\bibitem[{{Im} {et~al.}(2019{\natexlab{c}}){Im}, {Choi}, {Hwang}, {Lim}, {Kim}, {Kim}, {Paek}, {Lee}, {Yoon}, {Jung}, {Sung}, {Jeon}, {Ehgamberdiev}, {Burhonov}, {Milzaqulov}, {Parmonov}, {Lee}, {Kang}, {Kim}, {Kwon}, {Pak}, {Ji}, {Lee}, {Park}, {Ahn}, {Byeon}, {Han}, {Gibson}, {Wheeler}, {Kuehne}, {Johns-Krull}, {Marshall}, {Hyun}, {Lee}, {Kim}, {Yoon}, {Paek}, {Shin}, {Taak}, {Kang}, {Choi}, {Jeong}, {Jung}, {Kim}, {Kim}, {Lee}, {Park}, {Park}, \& {O}}]{2019JKAS...52...11I}
{Im}, M., {Choi}, C., {Hwang}, S., {et~al.} 2019{\natexlab{c}}, Journal of Korean Astronomical Society, 52, 11, \dodoi{10.5303/JKAS.2019.52.1.11}

\bibitem[{{Im} {et~al.}(2021){Im}, {Kim}, {Lee}, {Lee}, {Pak}, {Shim}, {Sung}, {Kang}, {Kim}, {Heo}, {Hinse}, {Ishiguro}, {Lim}, {Ly}, {Paek}, {Seo}, {Yoon}, {Woo}, {Ahn}, {Cho}, {Choi}, {Han}, {Hwang}, {Ji}, {Lee}, {Lee}, {Lee}, {Kim}, {Kim}, {Kim}, {Kim}, {Jeong}, {Park}, {Paek}, {Kim}, \& {Park}}]{2021JKAS...54...89I}
{Im}, M., {Kim}, Y., {Lee}, C.-U., {et~al.} 2021, Journal of Korean Astronomical Society, 54, 89

\bibitem[{{Im} {et~al.}(2010){Im}, {Ko}, {Cho}, {Choi}, {Jeon}, {Lee}, \& {Ibrahimov}}]{2010JKAS...43...75I}
{Im}, M.-S., {Ko}, J.-W., {Cho}, Y.-S., {et~al.} 2010, Journal of Korean Astronomical Society, 43, 75, \dodoi{10.5303/JKAS.2010.43.3.075}

\bibitem[{{IPAC}(2020)}]{ipac_2mass_2020}
{IPAC}. 2020, 2MASS All-Sky Atlas Image Service,  IPAC, \dodoi{10.26131/IRSA121}

\bibitem[{{Jarrett} {et~al.}(2000){Jarrett}, {Chester}, {Cutri}, {Schneider}, {Skrutskie}, \& {Huchra}}]{2000AJ....119.2498J}
{Jarrett}, T.~H., {Chester}, T., {Cutri}, R., {et~al.} 2000, \aj, 119, 2498, \dodoi{10.1086/301330}

\bibitem[{{Jeon} {et~al.}(2016){Jeon}, {Im}, {Pak}, {Hyun}, {Kim}, {Kim}, {Lee}, \& {Park}}]{2016JKAS...49...25J}
{Jeon}, Y., {Im}, M., {Pak}, S., {et~al.} 2016, Journal of Korean Astronomical Society, 49, 25, \dodoi{10.5303/JKAS.2016.49.1.25}

\bibitem[{{Ji} {et~al.}(2018){Ji}, {Byeon}, {Lee}, {Park}, {Lee}, {Hwang}, {Choi}, {Gibson}, {Kuehne}, {Prochaska}, {Marshall}, {Im}, \& {Pak}}]{2018AAS...23144203J}
{Ji}, T.-G., {Byeon}, S., {Lee}, H.-I., {et~al.} 2018, in American Astronomical Society Meeting Abstracts, Vol. 231, American Astronomical Society Meeting Abstracts \#231, 442.03

\bibitem[{{Kanner} {et~al.}(2012){Kanner}, {Camp}, {Racusin}, {Gehrels}, \& {White}}]{2012ApJ...759...22K}
{Kanner}, J., {Camp}, J., {Racusin}, J., {Gehrels}, N., \& {White}, D. 2012, \apj, 759, 22, \dodoi{10.1088/0004-637X/759/1/22}

\bibitem[{{Kasen} {et~al.}(2017){Kasen}, {Metzger}, {Barnes}, {Quataert}, \& {Ramirez-Ruiz}}]{2017Natur.551...80K}
{Kasen}, D., {Metzger}, B., {Barnes}, J., {Quataert}, E., \& {Ramirez-Ruiz}, E. 2017, \nat, 551, 80, \dodoi{10.1038/nature24453}

\bibitem[{{Kasliwal} {et~al.}(2017){Kasliwal}, {Nakar}, {Singer}, {Kaplan}, {Cook}, {Van Sistine}, {Lau}, {Fremling}, {Gottlieb}, {Jencson}, {Adams}, {Feindt}, {Hotokezaka}, {Ghosh}, {Perley}, {Yu}, {Piran}, {Allison}, {Anupama}, {Balasubramanian}, {Bannister}, {Bally}, {Barnes}, {Barway}, {Bellm}, {Bhalerao}, {Bhattacharya}, {Blagorodnova}, {Bloom}, {Brady}, {Cannella}, {Chatterjee}, {Cenko}, {Cobb}, {Copperwheat}, {Corsi}, {De}, {Dobie}, {Emery}, {Evans}, {Fox}, {Frail}, {Frohmaier}, {Goobar}, {Hallinan}, {Harrison}, {Helou}, {Hinderer}, {Ho}, {Horesh}, {Ip}, {Itoh}, {Kasen}, {Kim}, {Kuin}, {Kupfer}, {Lynch}, {Madsen}, {Mazzali}, {Miller}, {Mooley}, {Murphy}, {Ngeow}, {Nichols}, {Nissanke}, {Nugent}, {Ofek}, {Qi}, {Quimby}, {Rosswog}, {Rusu}, {Sadler}, {Schmidt}, {Sollerman}, {Steele}, {Williamson}, {Xu}, {Yan}, {Yatsu}, {Zhang}, \& {Zhao}}]{2017Sci...358.1559K}
{Kasliwal}, M.~M., {Nakar}, E., {Singer}, L.~P., {et~al.} 2017, Science, 358, 1559, \dodoi{10.1126/science.aap9455}

\bibitem[{{Kasliwal} {et~al.}(2019){Kasliwal}, {Coughlin}, {Bellm}, {Singer}, {de}, {Andreoni}, {Duev}, {Anand}, {Ahumada}, {Cenko}, {Goldstein}, {Ho}, {Perley}, {Bhalerao}, {Kumar}, {Sharma}, {Goobar}, {Kaplan}, {Sollerman}, {Bloom}, {Bulla}, {Kawai}, {Yatsu}, {Murata}, {Hanayama}, {Horiuchi}, {Anupama}, {Rigault}, {Barbarino}, {Biswas}, {Cook}, \& {Helou}}]{2019GCN.24191....1K}
{Kasliwal}, M.~M., {Coughlin}, M.~W., {Bellm}, E.~C., {et~al.} 2019, GRB Coordinates Network, 24191, 1

\bibitem[{{Kilpatrick} {et~al.}(2017){Kilpatrick}, {Foley}, {Kasen}, {Murguia-Berthier}, {Ramirez-Ruiz}, {Coulter}, {Drout}, {Piro}, {Shappee}, {Boutsia}, {Contreras}, {Di Mille}, {Madore}, {Morrell}, {Pan}, {Prochaska}, {Rest}, {Rojas-Bravo}, {Siebert}, {Simon}, \& {Ulloa}}]{2017Sci...358.1583K}
{Kilpatrick}, C.~D., {Foley}, R.~J., {Kasen}, D., {et~al.} 2017, Science, 358, 1583, \dodoi{10.1126/science.aaq0073}

\bibitem[{{Kim} \& {Im}(2012)}]{2012AAS...21944127K}
{Kim}, D., \& {Im}, M. 2012, in American Astronomical Society Meeting Abstracts, Vol. 219, American Astronomical Society Meeting Abstracts \#219, 441.27

\bibitem[{{Kim} {et~al.}(2019){Kim}, {Im}, {Lee}, {Kim}, {Paek}, {Lim}, {Kim}, {Hwang}, {Choi}, {Paek}, {Shin}, {Park}, {Pak}, {Mok}, \& {Sung}}]{2019GCN.24216....1K}
{Kim}, J., {Im}, M., {Lee}, C.~U., {et~al.} 2019, GRB Coordinates Network, 24216, 1

\bibitem[{{Kim} {et~al.}(2021){Kim}, {Im}, {Paek}, {Lee}, {Kim}, {Chang}, {Choi}, {Hwang}, {Kang}, {Kim}, {Kim}, {Lee}, {Lim}, {Seo}, \& {Sung}}]{2021ApJ...916...47K}
{Kim}, J., {Im}, M., {Paek}, G. S.~H., {et~al.} 2021, \apj, 916, 47, \dodoi{10.3847/1538-4357/ac0446}

\bibitem[{{Kim} {et~al.}(2016{\natexlab{a}}){Kim}, {Jeon}, {Lee}, {Park}, {Ji}, {Hyun}, {Choi}, {Im}, \& {Pak}}]{2016PASP..128k5004K}
{Kim}, S., {Jeon}, Y., {Lee}, H.-I., {et~al.} 2016{\natexlab{a}}, \pasp, 128, 115004, \dodoi{10.1088/1538-3873/128/969/115004}

\bibitem[{{Kim} {et~al.}(2016{\natexlab{b}}){Kim}, {Lee}, {Park}, {Kim}, {Cha}, {Lee}, {Han}, {Chun}, \& {Yuk}}]{2016JKAS...49...37K}
{Kim}, S.-L., {Lee}, C.-U., {Park}, B.-G., {et~al.} 2016{\natexlab{b}}, Journal of Korean Astronomical Society, 49, 37, \dodoi{10.5303/JKAS.2016.49.1.37}

\bibitem[{{Klingler} {et~al.}(2019){Klingler}, {Kennea}, {Evans}, {Tohuvavohu}, {Cenko}, {Barthelmy}, {Beardmore}, {Breeveld}, {Brown}, {Burrows}, {Campana}, {Cusumano}, {D'A{\`\i}}, {D'Avanzo}, {D'Elia}, {de Pasquale}, {Emery}, {Garcia}, {Giommi}, {Gronwall}, {Hartmann}, {Krimm}, {Kuin}, {Lien}, {Malesani}, {Marshall}, {Melandri}, {Nousek}, {Oates}, {O'Brien}, {Osborne}, {Page}, {Palmer}, {Perri}, {Racusin}, {Siegel}, {Sakamoto}, {Sbarufatti}, {Tagliaferri}, \& {Troja}}]{2019ApJS..245...15K}
{Klingler}, N.~J., {Kennea}, J.~A., {Evans}, P.~A., {et~al.} 2019, \apjs, 245, 15, \dodoi{10.3847/1538-4365/ab4ea2}

\bibitem[{{Lang} {et~al.}(2010){Lang}, {Hogg}, {Mierle}, {Blanton}, \& {Roweis}}]{2010AJ....139.1782L}
{Lang}, D., {Hogg}, D.~W., {Mierle}, K., {Blanton}, M., \& {Roweis}, S. 2010, \aj, 139, 1782, \dodoi{10.1088/0004-6256/139/5/1782}

\bibitem[{{Lee} {et~al.}(2010){Lee}, {Im}, \& {Urata}}]{2010JKAS...43...95L}
{Lee}, I.-D., {Im}, M.-S., \& {Urata}, Y. 2010, Journal of Korean Astronomical Society, 43, 95, \dodoi{10.5303/JKAS.2010.43.3.095}

\bibitem[{{Levan} {et~al.}(2017){Levan}, {Lyman}, {Tanvir}, {Hjorth}, {Mandel}, {Stanway}, {Steeghs}, {Fruchter}, {Troja}, {Schr{\o}der}, {Wiersema}, {Bruun}, {Cano}, {Cenko}, {de Ugarte Postigo}, {Evans}, {Fairhurst}, {Fox}, {Fynbo}, {Gompertz}, {Greiner}, {Im}, {Izzo}, {Jakobsson}, {Kangas}, {Khandrika}, {Lien}, {Malesani}, {O'Brien}, {Osborne}, {Palazzi}, {Pian}, {Perley}, {Rosswog}, {Ryan}, {Schulze}, {Sutton}, {Th{\"o}ne}, {Watson}, \& {Wijers}}]{2017ApJ...848L..28L}
{Levan}, A.~J., {Lyman}, J.~D., {Tanvir}, N.~R., {et~al.} 2017, \apjl, 848, L28, \dodoi{10.3847/2041-8213/aa905f}

\bibitem[{{LIGO Scientific Collaboration} {et~al.}(2015){LIGO Scientific Collaboration}, {Aasi}, {Abbott}, {Abbott}, {Abbott}, {Abernathy}, {Ackley}, {Adams}, {Adams}, {Addesso}, {Adhikari}, {Adya}, {Affeldt}, {Aggarwal}, {Aguiar}, {Ain}, {Ajith}, {Alemic}, {Allen}, {Amariutei}, {Anderson}, {Anderson}, {Arai}, {Araya}, {Arceneaux}, {Areeda}, {Ashton}, {Ast}, {Aston}, {Aufmuth}, {Aulbert}, {Aylott}, {Babak}, {Baker}, {Ballmer}, {Barayoga}, {Barbet}, {Barclay}, {Barish}, {Barker}, {Barr}, {Barsotti}, {Bartlett}, {Barton}, {Bartos}, {Bassiri}, {Batch}, {Baune}, {Behnke}, {Bell}, {Bell}, {Benacquista}, {Bergman}, {Bergmann}, {Berry}, {Betzwieser}, {Bhagwat}, {Bhandare}, {Bilenko}, {Billingsley}, {Birch}, {Biscans}, {Biwer}, {Blackburn}, {Blackburn}, {Blair}, {Blair}, {Bock}, {Bodiya}, {Bojtos}, {Bond}, {Bork}, {Born}, {Bose}, {Brady}, {Braginsky}, {Brau}, {Bridges}, {Brinkmann}, {Brooks}, {Brown}, {Brown}, {Brown}, {Buchman}, {Buikema}, {Buonanno}, {Cadonati}, {Calder{\'o}n Bustillo}, {Camp}, {Cannon}, {Cao},
  {Capano}, {Caride}, {Caudill}, {Cavagli{\`a}}, {Cepeda}, {Chakraborty}, {Chalermsongsak}, {Chamberlin}, {Chao}, {Charlton}, {Chen}, {Cho}, {Cho}, {Chow}, {Christensen}, {Chu}, {Chung}, {Ciani}, {Clara}, {Clark}, {Collette}, {Cominsky}, {Constancio}, {Cook}, {Corbitt}, {Cornish}, {Corsi}, {Costa}, {Coughlin}, {Countryman}, {Couvares}, {Coward}, {Cowart}, {Coyne}, {Coyne}, {Craig}, {Creighton}, {Creighton}, {Cripe}, {Crowder}, {Cumming}, {Cunningham}, {Cutler}, {Dahl}, {Dal Canton}, {Damjanic}, {Danilishin}, {Danzmann}, {Dartez}, {Dave}, {Daveloza}, {Davies}, {Daw}, {DeBra}, {Del Pozzo}, {Denker}, {Dent}, {Dergachev}, {DeRosa}, {DeSalvo}, {Dhurandhar}, {D{\textasciiacute}{\i}az}, {Di Palma}, {Dojcinoski}, {Dominguez}, {Donovan}, {Dooley}, {Doravari}, {Douglas}, {Downes}, {Driggers}, {Du}, {Dwyer}, {Eberle}, {Edo}, {Edwards}, {Edwards}, {Effler}, {Eggenstein}, {Ehrens}, {Eichholz}, {Eikenberry}, {Essick}, {Etzel}, {Evans}, {Evans}, {Factourovich}, {Fairhurst}, {Fan}, {Fang}, {Farr}, {Farr}, {Favata}, {Fays},
  {Fehrmann}, {Fejer}, {Feldbaum}, {Ferreira}, {Fisher}, {Frei}, {Freise}, {Frey}, {Fricke}, {Fritschel}, {Frolov}, {Fuentes-Tapia}, {Fulda}, {Fyffe}, {Gair}, {Gaonkar}, {Gehrels}, {Gergely}, {Giaime}, {Giardina}, {Gleason}, {Goetz}, {Goetz}, {Gondan}, {Gonz{\'a}lez}, {Gordon}, {Gorodetsky}, {Gossan}, {Go{\ss}ler}, {Gr{\"a}f}, {Graff}, {Grant}, {Gras}, {Gray}, {Greenhalgh}, {Gretarsson}, {Grote}, {Grunewald}, {Guido}, {Guo}, {Gushwa}, {Gustafson}, {Gustafson}, {Hacker}, {Hall}, {Hammond}, {Hanke}, {Hanks}, {Hanna}, {Hannam}, {Hanson}, {Hardwick}, {Harry}, {Harry}, {Hart}, {Hartman}, {Haster}, {Haughian}, {Hee}, {Heintze}, {Heinzel}, {Hendry}, {Heng}, {Heptonstall}, {Heurs}, {Hewitson}, {Hild}, {Hoak}, {Hodge}, {Hollitt}, {Holt}, {Hopkins}, {Hosken}, {Hough}, {Houston}, {Howell}, {Hu}, {Huerta}, {Hughey}, {Husa}, {Huttner}, {Huynh}, {Huynh-Dinh}, {Idrisy}, {Indik}, {Ingram}, {Inta}, {Islas}, {Isler}, {Isogai}, {Iyer}, {Izumi}, {Jacobson}, {Jang}, {Jawahar}, {Ji}, {Jim{\'e}nez-Forteza}, {Johnson}, {Jones},
  {Jones}, {Ju}, {Haris}, {Kalogera}, {Kandhasamy}, {Kang}, {Kanner}, {Katsavounidis}, {Katzman}, {Kaufer}, {Kaufer}, {Kaur}, {Kawabe}, {Kawazoe}, {Keiser}, {Keitel}, {Kelley}, {Kells}, {Keppel}, {Key}, {Khalaidovski}, {Khalili}, {Khazanov}, {Kim}, {Kim}, {Kim}, {Kim}, {Kim}, {King}, {King}, {Kinzel}, {Kissel}, {Klimenko}, {Kline}, {Koehlenbeck}, {Kokeyama}, {Kondrashov}, {Korobko}, {Korth}, {Kozak}, {Kringel}, {Krishnan}, {Krueger}, {Kuehn}, {Kumar}, {Kumar}, {Kuo}, {Landry}, {Lantz}, {Larson}, {Lasky}, {Lazzarini}, {Lazzaro}, {Le}, {Leaci}, {Leavey}, {Lebigot}, {Lee}, {Lee}, {Lee}, {Leong}, {Levin}, {Levine}, {Lewis}, {Li}, {Libbrecht}, {Libson}, {Lin}, {Littenberg}, {Lockerbie}, {Lockett}, {Logue}, {Lombardi}, {Lormand}, {Lough}, {Lubinski}, {L{\"u}ck}, {Lundgren}, {Lynch}, {Ma}, {Macarthur}, {MacDonald}, {Machenschalk}, {MacInnis}, {Macleod}, {Maga{\~n}a-Sandoval}, {Magee}, {Mageswaran}, {Maglione}, {Mailand}, {Mandel}, {Mandic}, {Mangano}, {Mansell}, {M{\'a}rka}, {M{\'a}rka}, {Markosyan}, {Maros},
  {Martin}, {Martin}, {Martynov}, {Marx}, {Mason}, {Massinger}, {Matichard}, {Matone}, {Mavalvala}, {Mazumder}, {Mazzolo}, {McCarthy}, {McClelland}, {McCormick}, {McGuire}, {McIntyre}, {McIver}, {McLin}, {McWilliams}, {Meadors}, {Meinders}, {Melatos}, {Mendell}, {Mercer}, {Meshkov}, {Messenger}, {Meyers}, {Miao}, {Middleton}, {Mikhailov}, {Miller}, {Miller}, {Millhouse}, {Ming}, {Mirshekari}, {Mishra}, {Mitra}, {Mitrofanov}, {Mitselmakher}, {Mittleman}, {Moe}, {Mohanty}, {Mohapatra}, {Moore}, {Moraru}, {Moreno}, {Morriss}, {Mossavi}, {Mow-Lowry}, {Mueller}, {Mueller}, {Mukherjee}, {Mullavey}, {Munch}, {Murphy}, {Murray}, {Mytidis}, {Nash}, {Nayak}, {Necula}, {Nedkova}, {Newton}, {Nguyen}, {Nielsen}, {Nissanke}, {Nitz}, {Nolting}, {Normandin}, {Nuttall}, {Ochsner}, {O'Dell}, {Oelker}, {Ogin}, {Oh}, {Oh}, {Ohme}, {Oppermann}, {Oram}, {O'Reilly}, {Ortega}, {O'Shaughnessy}, {Osthelder}, {Ott}, {Ottaway}, {Ottens}, {Overmier}, {Owen}, {Padilla}, {Pai}, {Pai}, {Palashov}, {Pal-Singh}, {Pan}, {Pankow}, {Pannarale},
  {Pant}, {Papa}, {Paris}, {Patrick}, {Pedraza}, {Pekowsky}, {Pele}, {Penn}, {Perreca}, {Phelps}, {Pierro}, {Pinto}, {Pitkin}, {Poeld}, {Post}, {Poteomkin}, {Powell}, {Prasad}, {Predoi}, {Premachandra}, {Prestegard}, {Price}, {Principe}, {Privitera}, {Prix}, {Prokhorov}, {Puncken}, {P{\"u}rrer}, {Qin}, {Quetschke}, {Quintero}, {Quiroga}, {Quitzow-James}, {Raab}, {Rabeling}, {Radkins}, {Raffai}, {Raja}, {Rajalakshmi}, {Rakhmanov}, {Ramirez}, {Raymond}, {Reed}, {Reid}, {Reitze}, {Reula}, {Riles}, {Robertson}, {Robie}, {Rollins}, {Roma}, {Romano}, {Romanov}, {Romie}, {Rowan}, {R{\"u}diger}, {Ryan}, {Sachdev}, {Sadecki}, {Sadeghian}, {Saleem}, {Salemi}, {Sammut}, {Sandberg}, {Sanders}, {Sannibale}, {Santiago-Prieto}, {Sathyaprakash}, {Saulson}, {Savage}, {Sawadsky}, {Scheuer}, {Schilling}, {Schmidt}, {Schnabel}, {Schofield}, {Schreiber}, {Schuette}, {Schutz}, {Scott}, {Scott}, {Sellers}, {Sengupta}, {Sergeev}, {Serna}, {Sevigny}, {Shaddock}, {Shahriar}, {Shaltev}, {Shao}, {Shapiro}, {Shawhan}, {Shoemaker},
  {Sidery}, {Siemens}, {Sigg}, {Silva}, {Simakov}, {Singer}, {Singer}, {Singh}, {Sintes}, {Slagmolen}, {Smith}, {Smith}, {Smith}, {Smith-Lefebvre}, {Son}, {Sorazu}, {Souradeep}, {Staley}, {Stebbins}, {Steinke}, {Steinlechner}, {Steinlechner}, {Steinmeyer}, {Stephens}, {Steplewski}, {Stevenson}, {Stone}, {Strain}, {Strigin}, {Sturani}, {Stuver}, {Summerscales}, {Sutton}, {Szczepanczyk}, {Szeifert}, {Talukder}, {Tanner}, {T{\'a}pai}, {Tarabrin}, {Taracchini}, {Taylor}, {Tellez}, {Theeg}, {Thirugnanasambandam}, {Thomas}, {Thomas}, {Thorne}, {Thorne}, {Thrane}, {Tiwari}, {Tomlinson}, {Torres}, {Torrie}, {Traylor}, {Tse}, {Tshilumba}, {Ugolini}, {Unnikrishnan}, {Urban}, {Usman}, {Vahlbruch}, {Vajente}, {Valdes}, {Vallisneri}, {van Veggel}, {Vass}, {Vaulin}, {Vecchio}, {Veitch}, {Veitch}, {Venkateswara}, {Vincent-Finley}, {Vitale}, {Vo}, {Vorvick}, {Vousden}, {Vyatchanin}, {Wade}, {Wade}, {Wade}, {Walker}, {Wallace}, {Walsh}, {Wang}, {Wang}, {Wang}, {Ward}, {Warner}, {Was}, {Weaver}, {Weinert}, {Weinstein},
  {Weiss}, {Welborn}, {Wen}, {Wessels}, {Westphal}, {Wette}, {Whelan}, {Whitcomb}, {White}, {Whiting}, {Wilkinson}, {Williams}, {Williams}, {Williamson}, {Willis}, {Willke}, {Wimmer}, {Winkler}, {Wipf}, {Wittel}, {Woan}, {Worden}, {Xie}, {Yablon}, {Yakushin}, {Yam}, {Yamamoto}, {Yancey}, {Yang}, {Zanolin}, {Zhang}, {Zhang}, {Zhang}, {Zhang}, {Zhao}, {Zhou}, {Zhu}, {Zucker}, {Zuraw}, \& {Zweizig}}]{2015CQGra..32g4001L}
{LIGO Scientific Collaboration}, {Aasi}, J., {Abbott}, B.~P., {et~al.} 2015, Classical and Quantum Gravity, 32, 074001, \dodoi{10.1088/0264-9381/32/7/074001}

\bibitem[{{Lipunov} {et~al.}(2019){Lipunov}, {Gorbovskoy}, {Tyurina}, {Kornilov}, {Vlasenko}, {Vladimirov}, {Zimnukhov}, {Kuznetsov}, {Balanutsa}, {Chasovnikov}, {Topolev}, {Kuvshinov}, {Balakin}, {Petkov}, {Boliev}, {Kurenya}, {Rebolo}, {Serra}, {Lodieu}, {Israelian}, {Suarez-Andres}, {Buckley}, {Gress}, {Budnev}, {Ershova}, {Yurkov}, {Gabovich}, {Sergienko}, {Kobcev}, {Tlatov}, {Senik}, {Parhomenko}, {Dormidontov}, {Podesta}, {Lopez}, {Francile}, {Podesta}, \& {Levato}}]{2019GCN.24241....1L}
{Lipunov}, V., {Gorbovskoy}, E., {Tyurina}, N., {et~al.} 2019, GRB Coordinates Network, 24241, 1

\bibitem[{{Lipunov} {et~al.}(2017){Lipunov}, {Gorbovskoy}, {Kornilov}, {. Tyurina}, {Balanutsa}, {Kuznetsov}, {Vlasenko}, {Kuvshinov}, {Gorbunov}, {Buckley}, {Krylov}, {Podesta}, {Lopez}, {Podesta}, {Levato}, {Saffe}, {Mallamachi}, {Potter}, {Budnev}, {Gress}, {Ishmuhametova}, {Vladimirov}, {Zimnukhov}, {Yurkov}, {Sergienko}, {Gabovich}, {Rebolo}, {Serra-Ricart}, {Israelyan}, {Chazov}, {Wang}, {Tlatov}, \& {Panchenko}}]{2017ApJ...850L...1L}
{Lipunov}, V.~M., {Gorbovskoy}, E., {Kornilov}, V.~G., {et~al.} 2017, \apjl, 850, L1, \dodoi{10.3847/2041-8213/aa92c0}

\bibitem[{{Lundquist} {et~al.}(2019){Lundquist}, {Paterson}, {Fong}, {Sand}, {Andrews}, {Shivaei}, {Daly}, {Valenti}, {Yang}, {Christensen}, {Gibbs}, {Shelly}, {Wyatt}, {Eskandari}, {Kuhn}, {Amaro}, {Arcavi}, {Behroozi}, {Butler}, {Chomiuk}, {Corsi}, {Drout}, {Egami}, {Fan}, {Foley}, {Frye}, {Gabor}, {Green}, {Grier}, {Guzman}, {Hamden}, {Howell}, {Jannuzi}, {Kelly}, {Milne}, {Moe}, {Nugent}, {Olszewski}, {Palazzi}, {Paschalidis}, {Psaltis}, {Reichart}, {Rest}, {Rossi}, {Schroeder}, {Smith}, {Smith}, {Spekkens}, {Strader}, {Stark}, {Trilling}, {Veillet}, {Wagner}, {Weiner}, {Wheeler}, {Williams}, \& {Zabludoff}}]{2019ApJ...881L..26L}
{Lundquist}, M.~J., {Paterson}, K., {Fong}, W., {et~al.} 2019, \apjl, 881, L26, \dodoi{10.3847/2041-8213/ab32f2}

\bibitem[{{Mapelli} {et~al.}(2018){Mapelli}, {Giacobbo}, {Toffano}, {Ripamonti}, {Bressan}, {Spera}, \& {Branchesi}}]{2018MNRAS.481.5324M}
{Mapelli}, M., {Giacobbo}, N., {Toffano}, M., {et~al.} 2018, \mnras, 481, 5324, \dodoi{10.1093/mnras/sty2663}

\bibitem[{{McCully} {et~al.}(2017){McCully}, {Hiramatsu}, {Howell}, {Hosseinzadeh}, {Arcavi}, {Kasen}, {Barnes}, {Shara}, {Williams}, {V{\"a}is{\"a}nen}, {Potter}, {Romero-Colmenero}, {Crawford}, {Buckley}, {Cooke}, {Andreoni}, {Pritchard}, {Mao}, {Gromadzki}, \& {Burke}}]{2017ApJ...848L..32M}
{McCully}, C., {Hiramatsu}, D., {Howell}, D.~A., {et~al.} 2017, \apjl, 848, L32, \dodoi{10.3847/2041-8213/aa9111}

\bibitem[{{McCully} {et~al.}(2019){McCully}, {Hiramatsu}, {Hiramatsu}, {Howell}, {Arcavi}, {Drout}, {Burke}, {Peligrino}, {de Carvalho}, {Forster}, {Foley}, {Coulter}, {Kilpatrick}, {Sand}, {Valenti}, {Soares-Santos}, {Rembold}, {Resti}, {Kasen}, {Metzger}, {Piro}, {Quataert}, {Ramirez-Ruiz}, {Wheeler}, {Bauer}, {Brink}, {Cooke}, {Clocchiatti}, {Filippenko}, {Freedman}, {Garnavich}, {Horvath}, {Jha}, {Kirshner}, {Krisciunas}, {Lin}, {Madore}, {Makler}, {Prochaska}, {Riess}, {Sturani}, {Suntzeff}, {Tanaka}, {Tucker}, {Vinko}, {Wang}, {Brown}, {Contrerasi}, {D'Andrea}, {Dimitriadis}, {Jones}, {Lundquist}, {Narayan}, {Olivares}, {Palmese}, {Pan}, {Scolnic}, {Zheng}, {Bernardo}, {Bostroem}, {Berthier}, {Rodriguez}, {Rojas-Bravo}, {Siebert}, \& {Souza}}]{2019GCN.24295....1M}
{McCully}, C., {Hiramatsu}, D., {Hiramatsu}, D., {et~al.} 2019, GRB Coordinates Network, 24295, 1

\bibitem[{{McMahon} {et~al.}(2013){McMahon}, {Banerji}, {Gonzalez}, {Koposov}, {Bejar}, {Lodieu}, {Rebolo}, \& {VHS Collaboration}}]{2013Msngr.154...35M}
{McMahon}, R.~G., {Banerji}, M., {Gonzalez}, E., {et~al.} 2013, The Messenger, 154, 35

\bibitem[{{Metzger} \& {Berger}(2012)}]{2012ApJ...746...48M}
{Metzger}, B.~D., \& {Berger}, E. 2012, \apj, 746, 48, \dodoi{10.1088/0004-637X/746/1/48}

\bibitem[{{Metzger} {et~al.}(2018){Metzger}, {Thompson}, \& {Quataert}}]{2018ApJ...856..101M}
{Metzger}, B.~D., {Thompson}, T.~A., \& {Quataert}, E. 2018, \apj, 856, 101, \dodoi{10.3847/1538-4357/aab095}

\bibitem[{{Metzger} {et~al.}(2010){Metzger}, {Mart{\'\i}nez-Pinedo}, {Darbha}, {Quataert}, {Arcones}, {Kasen}, {Thomas}, {Nugent}, {Panov}, \& {Zinner}}]{2010MNRAS.406.2650M}
{Metzger}, B.~D., {Mart{\'\i}nez-Pinedo}, G., {Darbha}, S., {et~al.} 2010, \mnras, 406, 2650, \dodoi{10.1111/j.1365-2966.2010.16864.x}

\bibitem[{{Morokuma} {et~al.}(2019){Morokuma}, {Ohta}, {Yoshida}, {Aoki}, {Tanaka}, {Sasada}, {Nakaoka}, {Akitaya}, {Kawabata}, {Itoh}, \& {Utsumi}}]{2019GCN.24230....1M}
{Morokuma}, T., {Ohta}, K., {Yoshida}, M., {et~al.} 2019, GRB Coordinates Network, 24230, 1

\bibitem[{{Nugent} {et~al.}(2022){Nugent}, {Fong}, {Dong}, {Leja}, {Berger}, {Zevin}, {Chornock}, {Cobb}, {Kelley}, {Kilpatrick}, {Levan}, {Margutti}, {Paterson}, {Perley}, {Escorial}, {Smith}, \& {Tanvir}}]{2022ApJ...940...57N}
{Nugent}, A.~E., {Fong}, W.-F., {Dong}, Y., {et~al.} 2022, \apj, 940, 57, \dodoi{10.3847/1538-4357/ac91d1}

\bibitem[{{Ohgami} {et~al.}(2021){Ohgami}, {Tominaga}, {Utsumi}, {Niino}, {Tanaka}, {Banerjee}, {Hamasaki}, {Yoshida}, {Terai}, {Takagi}, {Morokuma}, {Sasada}, {Akitaya}, {Yasuda}, {Yanagisawa}, \& {Ohsawa}}]{2021PASJ...73..350O}
{Ohgami}, T., {Tominaga}, N., {Utsumi}, Y., {et~al.} 2021, \pasj, 73, 350, \dodoi{10.1093/pasj/psab002}

\bibitem[{{Ohgami} {et~al.}(2023){Ohgami}, {Becerra Gonz{\'a}lez}, {Tominaga}, {Morokuma}, {Utsumi}, {Niino}, {Tanaka}, {Banerjee}, {Poidevin}, {Acosta-Pulido}, {P{\'e}rez-Fournon}, {Mu{\~n}oz-Darias}, {Akitaya}, {Yanagisawa}, {Sasada}, {Yoshida}, {Simunovic}, {Ohsawa}, {Tanaka}, {Terai}, {Takagi}, \& {J-GEM Collaboration}}]{2023ApJ...947....9O}
{Ohgami}, T., {Becerra Gonz{\'a}lez}, J., {Tominaga}, N., {et~al.} 2023, \apj, 947, 9, \dodoi{10.3847/1538-4357/acbd42}

\bibitem[{Paek(2023{\natexlab{a}})}]{gppy}
Paek, G.~S. 2023{\natexlab{a}}, SilverRon/gppy: v0.1 Preliminary Release,  Zenodo, \dodoi{10.5281/ZENODO.8318777}

\bibitem[{Paek(2023{\natexlab{b}})}]{geckodigestor}
---. 2023{\natexlab{b}}, SilverRon/GeckoDigestor: v0.1 Preliminary Release,  Zenodo, \dodoi{10.5281/ZENODO.8321870}

\bibitem[{{Paek} {et~al.}(2019){Paek}, {Im}, {Lim}, {Paek}, {Shin}, {Kim}, {Pak}, {Hwang}, {Park}, {Kim}, {Choi}, {Lee}, {Kim}, \& {Mok}}]{2019GCN.24188....1P}
{Paek}, G.~S.~H., {Im}, M., {Lim}, G., {et~al.} 2019, GRB Coordinates Network, 24188, 1

\bibitem[{{Perley} {et~al.}(2019){Perley}, {Copperwheat}, \& {Taggart}}]{2019GCN.24204....1P}
{Perley}, D.~A., {Copperwheat}, C.~M., \& {Taggart}, K.~L. 2019, GRB Coordinates Network, 24204, 1

\bibitem[{{Petrov} {et~al.}(2022){Petrov}, {Singer}, {Coughlin}, {Kumar}, {Almualla}, {Anand}, {Bulla}, {Dietrich}, {Foucart}, \& {Guessoum}}]{2022ApJ...924...54P}
{Petrov}, P., {Singer}, L.~P., {Coughlin}, M.~W., {et~al.} 2022, \apj, 924, 54, \dodoi{10.3847/1538-4357/ac366d}

\bibitem[{{Phinney}(1991)}]{1991ApJ...380L..17P}
{Phinney}, E.~S. 1991, \apjl, 380, L17, \dodoi{10.1086/186163}

\bibitem[{{Pian} {et~al.}(2017){Pian}, {D'Avanzo}, {Benetti}, {Branchesi}, {Brocato}, {Campana}, {Cappellaro}, {Covino}, {D'Elia}, {Fynbo}, {Getman}, {Ghirlanda}, {Ghisellini}, {Grado}, {Greco}, {Hjorth}, {Kouveliotou}, {Levan}, {Limatola}, {Malesani}, {Mazzali}, {Melandri}, {M{\o}ller}, {Nicastro}, {Palazzi}, {Piranomonte}, {Rossi}, {Salafia}, {Selsing}, {Stratta}, {Tanaka}, {Tanvir}, {Tomasella}, {Watson}, {Yang}, {Amati}, {Antonelli}, {Ascenzi}, {Bernardini}, {Bo{\"e}r}, {Bufano}, {Bulgarelli}, {Capaccioli}, {Casella}, {Castro-Tirado}, {Chassande-Mottin}, {Ciolfi}, {Copperwheat}, {Dadina}, {De Cesare}, {di Paola}, {Fan}, {Gendre}, {Giuffrida}, {Giunta}, {Hunt}, {Israel}, {Jin}, {Kasliwal}, {Klose}, {Lisi}, {Longo}, {Maiorano}, {Mapelli}, {Masetti}, {Nava}, {Patricelli}, {Perley}, {Pescalli}, {Piran}, {Possenti}, {Pulone}, {Razzano}, {Salvaterra}, {Schipani}, {Spera}, {Stamerra}, {Stella}, {Tagliaferri}, {Testa}, {Troja}, {Turatto}, {Vergani}, \& {Vergani}}]{2017Natur.551...67P}
{Pian}, E., {D'Avanzo}, P., {Benetti}, S., {et~al.} 2017, \nat, 551, 67, \dodoi{10.1038/nature24298}

\bibitem[{{Piro} \& {Kollmeier}(2018)}]{2018ApJ...855..103P}
{Piro}, A.~L., \& {Kollmeier}, J.~A. 2018, \apj, 855, 103, \dodoi{10.3847/1538-4357/aaaab3}

\bibitem[{{Sasada} {et~al.}(2021){Sasada}, {Utsumi}, {Itoh}, {Tominaga}, {Tanaka}, {Morokuma}, {Yanagisawa}, {Kawabata}, {Ohgami}, {Yoshida}, {Abe}, {Adachi}, {Akitaya}, {Chong}, {Daikuhara}, {Hamasaki}, {Honda}, {Hosokawa}, {Iida}, {Imazato}, {Ishioka}, {Iwasaki}, {Jian}, {Kamei}, {Kanai}, {Kaneda}, {Kaneko}, {Katoh}, {Kawai}, {Kubota}, {Kubota}, {Mamiya}, {Matsubayashi}, {Morihana}, {Murata}, {Nagayama}, {Nakamura}, {Nakaoka}, {Niino}, {Nishinaka}, {Niwano}, {Nogami}, {Oasa}, {Oeda}, {Ogawa}, {Ohsawa}, {Ohta}, {Oide}, {Onozato}, {Sako}, {Saito}, {Sekiguchi}, {Shigeyama}, {Shigeyoshi}, {Shikauchi}, {Shiraishi}, {Suzuki}, {Takagi}, {Takahashi}, {Takarada}, {Takayama}, {Takeuchi}, {Tamura}, {Tanaka}, {Toma}, {Tozuka}, {Uchida}, {Uzawa}, {Yamanaka}, {Yasuda}, \& {Yatsu}}]{2021PTEP.2021eA104S}
{Sasada}, M., {Utsumi}, Y., {Itoh}, R., {et~al.} 2021, Progress of Theoretical and Experimental Physics, 2021, 05A104, \dodoi{10.1093/ptep/ptab007}

\bibitem[{{Schutz}(1986)}]{1986Natur.323..310S}
{Schutz}, B.~F. 1986, \nat, 323, 310, \dodoi{10.1038/323310a0}

\bibitem[{{Shappee} {et~al.}(2017){Shappee}, {Simon}, {Drout}, {Piro}, {Morrell}, {Prieto}, {Kasen}, {Holoien}, {Kollmeier}, {Kelson}, {Coulter}, {Foley}, {Kilpatrick}, {Siebert}, {Madore}, {Murguia-Berthier}, {Pan}, {Prochaska}, {Ramirez-Ruiz}, {Rest}, {Adams}, {Alatalo}, {Ba{\~n}ados}, {Baughman}, {Bernstein}, {Bitsakis}, {Boutsia}, {Bravo}, {Di Mille}, {Higgs}, {Ji}, {Maravelias}, {Marshall}, {Placco}, {Prieto}, \& {Wan}}]{2017Sci...358.1574S}
{Shappee}, B.~J., {Simon}, J.~D., {Drout}, M.~R., {et~al.} 2017, Science, 358, 1574, \dodoi{10.1126/science.aaq0186}

\bibitem[{{Skrutskie} {et~al.}(2006){Skrutskie}, {Cutri}, {Stiening}, {Weinberg}, {Schneider}, {Carpenter}, {Beichman}, {Capps}, {Chester}, {Elias}, {Huchra}, {Liebert}, {Lonsdale}, {Monet}, {Price}, {Seitzer}, {Jarrett}, {Kirkpatrick}, {Gizis}, {Howard}, {Evans}, {Fowler}, {Fullmer}, {Hurt}, {Light}, {Kopan}, {Marsh}, {McCallon}, {Tam}, {Van Dyk}, \& {Wheelock}}]{2006AJ....131.1163S}
{Skrutskie}, M.~F., {Cutri}, R.~M., {Stiening}, R., {et~al.} 2006, \aj, 131, 1163, \dodoi{10.1086/498708}

\bibitem[{{Smartt} {et~al.}(2017){Smartt}, {Chen}, {Jerkstrand}, {Coughlin}, {Kankare}, {Sim}, {Fraser}, {Inserra}, {Maguire}, {Chambers}, {Huber}, {Kr{\"u}hler}, {Leloudas}, {Magee}, {Shingles}, {Smith}, {Young}, {Tonry}, {Kotak}, {Gal-Yam}, {Lyman}, {Homan}, {Agliozzo}, {Anderson}, {Angus}, {Ashall}, {Barbarino}, {Bauer}, {Berton}, {Botticella}, {Bulla}, {Bulger}, {Cannizzaro}, {Cano}, {Cartier}, {Cikota}, {Clark}, {De Cia}, {Della Valle}, {Denneau}, {Dennefeld}, {Dessart}, {Dimitriadis}, {Elias-Rosa}, {Firth}, {Flewelling}, {Fl{\"o}rs}, {Franckowiak}, {Frohmaier}, {Galbany}, {Gonz{\'a}lez-Gait{\'a}n}, {Greiner}, {Gromadzki}, {Guelbenzu}, {Guti{\'e}rrez}, {Hamanowicz}, {Hanlon}, {Harmanen}, {Heintz}, {Heinze}, {Hernandez}, {Hodgkin}, {Hook}, {Izzo}, {James}, {Jonker}, {Kerzendorf}, {Klose}, {Kostrzewa-Rutkowska}, {Kowalski}, {Kromer}, {Kuncarayakti}, {Lawrence}, {Lowe}, {Magnier}, {Manulis}, {Martin-Carrillo}, {Mattila}, {McBrien}, {M{\"u}ller}, {Nordin}, {O'Neill}, {Onori}, {Palmerio}, {Pastorello},
  {Patat}, {Pignata}, {Podsiadlowski}, {Pumo}, {Prentice}, {Rau}, {Razza}, {Rest}, {Reynolds}, {Roy}, {Ruiter}, {Rybicki}, {Salmon}, {Schady}, {Schultz}, {Schweyer}, {Seitenzahl}, {Smith}, {Sollerman}, {Stalder}, {Stubbs}, {Sullivan}, {Szegedi}, {Taddia}, {Taubenberger}, {Terreran}, {van Soelen}, {Vos}, {Wainscoat}, {Walton}, {Waters}, {Weiland}, {Willman}, {Wiseman}, {Wright}, {Wyrzykowski}, \& {Yaron}}]{2017Natur.551...75S}
{Smartt}, S.~J., {Chen}, T.~W., {Jerkstrand}, A., {et~al.} 2017, \nat, 551, 75, \dodoi{10.1038/nature24303}

\bibitem[{{Smith} {et~al.}(2019){Smith}, {Young}, {McBrien}, {Gillanders}, {Srivastav}, {Smartt}, {O'Neil}, {Clark}, {Sim}, {Chambers}, {Huber}, {Magnier}, {Schultz}, {Denneau}, {Flewelling}, {Heinze}, {Tonry}, {Weiland}, {Rest}, {Stalder}, \& {Stubbs}}]{2019GCN.24210....1S}
{Smith}, K.~W., {Young}, D.~R., {McBrien}, O., {et~al.} 2019, GRB Coordinates Network, 24210, 1

\bibitem[{{Troja} {et~al.}(2017){Troja}, {Piro}, {van Eerten}, {Wollaeger}, {Im}, {Fox}, {Butler}, {Cenko}, {Sakamoto}, {Fryer}, {Ricci}, {Lien}, {Ryan}, {Korobkin}, {Lee}, {Burgess}, {Lee}, {Watson}, {Choi}, {Covino}, {D'Avanzo}, {Fontes}, {Gonz{\'a}lez}, {Khandrika}, {Kim}, {Kim}, {Lee}, {Lee}, {Kutyrev}, {Lim}, {S{\'a}nchez-Ram{\'\i}rez}, {Veilleux}, {Wieringa}, \& {Yoon}}]{2017Natur.551...71T}
{Troja}, E., {Piro}, L., {van Eerten}, H., {et~al.} 2017, \nat, 551, 71, \dodoi{10.1038/nature24290}

\bibitem[{{Utsumi} {et~al.}(2017){Utsumi}, {Tanaka}, {Tominaga}, {Yoshida}, {Barway}, {Nagayama}, {Zenko}, {Aoki}, {Fujiyoshi}, {Furusawa}, {Kawabata}, {Koshida}, {Lee}, {Morokuma}, {Motohara}, {Nakata}, {Ohsawa}, {Ohta}, {Okita}, {Tajitsu}, {Tanaka}, {Terai}, {Yasuda}, {Abe}, {Asakura}, {Bond}, {Miyazaki}, {Sumi}, {Tristram}, {Honda}, {Itoh}, {Itoh}, {Kawabata}, {Morihana}, {Nagashima}, {Nakaoka}, {Ohshima}, {Takahashi}, {Takayama}, {Aoki}, {Baar}, {Doi}, {Finet}, {Kanda}, {Kawai}, {Kim}, {Kuroda}, {Liu}, {Matsubayashi}, {Murata}, {Nagai}, {Saito}, {Saito}, {Sako}, {Sekiguchi}, {Tamura}, {Tanaka}, {Uemura}, \& {Yamaguchi}}]{2017PASJ...69..101U}
{Utsumi}, Y., {Tanaka}, M., {Tominaga}, N., {et~al.} 2017, \pasj, 69, 101, \dodoi{10.1093/pasj/psx118}

\bibitem[{{Verde} {et~al.}(2019){Verde}, {Treu}, \& {Riess}}]{2019NatAs...3..891V}
{Verde}, L., {Treu}, T., \& {Riess}, A.~G. 2019, Nature Astronomy, 3, 891, \dodoi{10.1038/s41550-019-0902-0}

\bibitem[{{Wang} \& {Chen}(2019)}]{2019ApJ...877..116W}
{Wang}, S., \& {Chen}, X. 2019, \apj, 877, 116, \dodoi{10.3847/1538-4357/ab1c61}

\bibitem[{{Wiersema} {et~al.}(2019){Wiersema}, {Levan}, {Fraser}, {Steeghs}, {Jonker}, {Malesani}, \& {Tanvir}}]{2019GCN.24209....1W}
{Wiersema}, K., {Levan}, A.~J., {Fraser}, M., {et~al.} 2019, GRB Coordinates Network, 24209, 1

\bibitem[{{Wollaeger} {et~al.}(2021){Wollaeger}, {Fryer}, {Chase}, {Fontes}, {Ristic}, {Hungerford}, {Korobkin}, {O'Shaughnessy}, \& {Herring}}]{2021ApJ...918...10W}
{Wollaeger}, R.~T., {Fryer}, C.~L., {Chase}, E.~A., {et~al.} 2021, \apj, 918, 10, \dodoi{10.3847/1538-4357/ac0d03}

\bibitem[{Wright {et~al.}(2019)Wright, Eisenhardt, Mainzer, Ressler, Cutri, Jarrett, Kirkpatrick, Padgett, McMillan, Skrutskie, Stanford, Cohen, Walker, Mather, Leisawitz, Gautier, McLean, Benford, Lonsdale, Blain, Mendez, Irace, Duval, Liu, Royer, Heinrichsen, Howard, Shannon, Kendall, Walsh, Larsen, Cardon, Schick, Schwalm, Abid, Fabinsky, Naes, \& Tsai}]{wright_allwise_2019}
Wright, E.~L., Eisenhardt, P. R.~M., Mainzer, A.~K., {et~al.} 2019, AllWISE Source Catalog,  IPAC, \dodoi{10.26131/IRSA1}

\bibitem[{{Wyatt} {et~al.}(2020){Wyatt}, {Tohuvavohu}, {Arcavi}, {Lundquist}, {Howell}, \& {Sand}}]{2020ApJ...894..127W}
{Wyatt}, S.~D., {Tohuvavohu}, A., {Arcavi}, I., {et~al.} 2020, \apj, 894, 127, \dodoi{10.3847/1538-4357/ab855e}

\bibitem[{{Yang} {et~al.}(2019){Yang}, {Sand}, {Valenti}, {Cappellaro}, {Tartaglia}, {Wyatt}, {Corsi}, {Reichart}, {Haislip}, {Kouprianov}, \& {(DLT40 Collaboration}}]{2019ApJ...875...59Y}
{Yang}, S., {Sand}, D.~J., {Valenti}, S., {et~al.} 2019, \apj, 875, 59, \dodoi{10.3847/1538-4357/ab0e06}

\bibitem[{{Yoshida} {et~al.}(2017){Yoshida}, {Utsumi}, {Tominaga}, {Morokuma}, {Tanaka}, {Asakura}, {Matsubayashi}, {Ohta}, {Abe}, {Chimasu}, {Furusawa}, {Itoh}, {Itoh}, {Kanda}, {Kawabata}, {Kawabata}, {Koshida}, {Koshimoto}, {Kuroda}, {Moritani}, {Motohara}, {Murata}, {Nagayama}, {Nakaoka}, {Nakata}, {Nishioka}, {Saito}, {Terai}, {Tristram}, {Yanagisawa}, {Yasuda}, {Doi}, {Fujisawa}, {Kawachi}, {Kawai}, {Tamura}, {Uemura}, \& {Yatsu}}]{2017PASJ...69....9Y}
{Yoshida}, M., {Utsumi}, Y., {Tominaga}, N., {et~al.} 2017, \pasj, 69, 9, \dodoi{10.1093/pasj/psw113}

\end{thebibliography}
\bibliographystyle{aasjournal}



\begin{figure}
\plotone{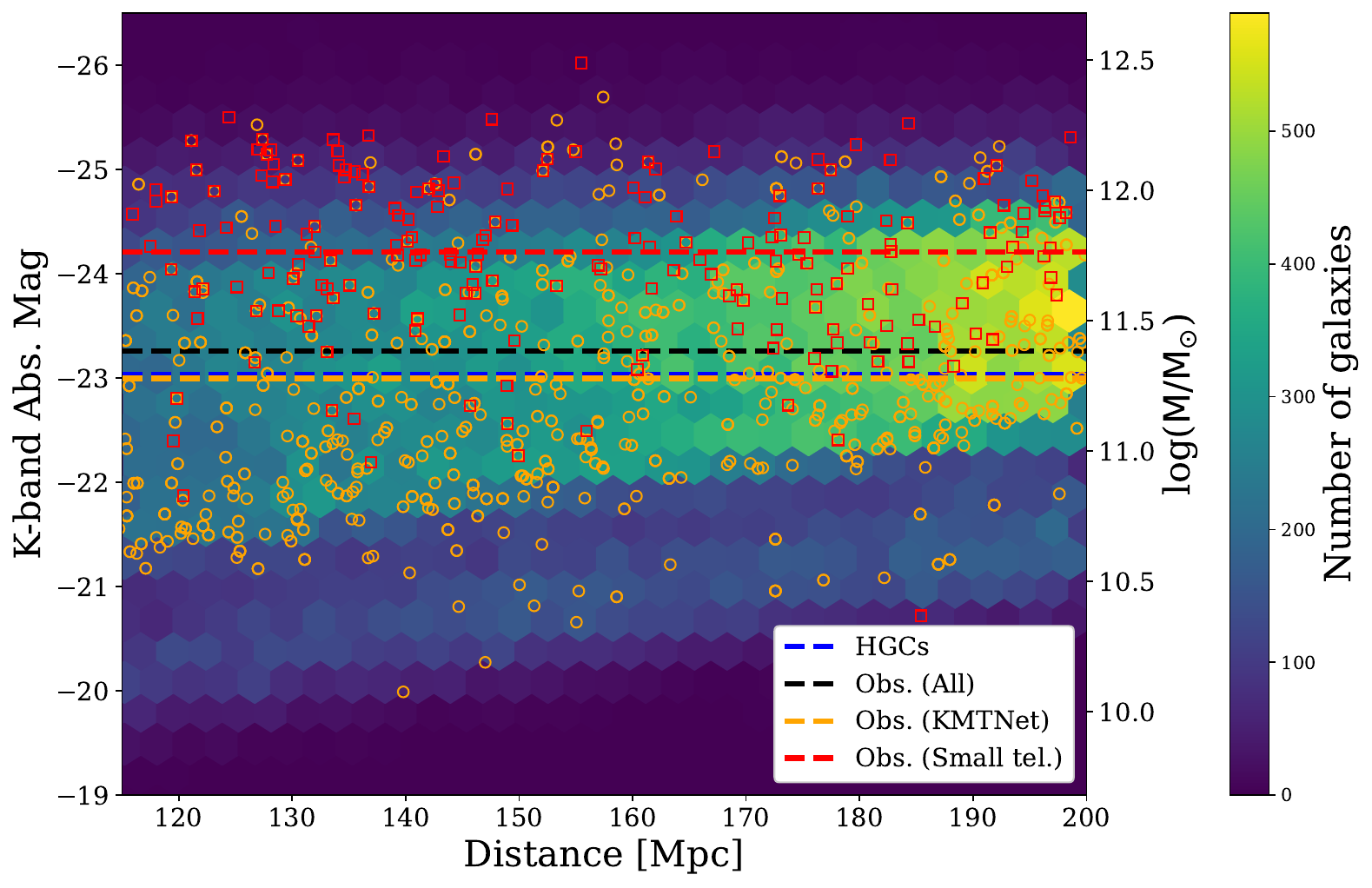}
\caption{The K-band luminosity and stellar mass distribution of observed host galaxy candidates. The hexagonal bin shows the number density of all galaxies within $\rm d=156\pm41\:Mpc$ in the modified GLADE catalog and the black horizontal dotted line shows the median value of their stellar mass. The blue horizontal dotted line shows the median value of the stellar mass of the host galaxy candidates of GW190425 selected by Section \ref{subsec:selection}. The red open rectangular point shows the host galaxy candidates observed by a small telescope (1-m class facilities) and the red horizontal dotted line shows the median value of their stellar mass. The yellow circular point shows the host galaxy candidates observed by the KMTNet and the yellow horizontal dotted line shows the median value of their stellar mass. The stellar mass is converted from the K-band luminosity, simply assuming the mass-to-light ratio as 1. 
\label{fig:hgcobs}}
\end{figure}



\begin{longrotatetable}
\begin{deluxetable}{ccccccc}
\tablecaption{Global network of telescopes in GECKO\label{tab:facility}}
\tablewidth{0.5\textwidth}
\tablehead{\colhead{Name} & \colhead{Location} & \colhead{Aperture Diameter (m)} & \colhead{FoV} & \colhead{Bands} & \colhead{Reference}}
\startdata
SAO\footnote{Seoul National University Astronomical Observatory}         & Korea        & 1m               & 21.2' $\times$ 21.2'                        & $BVRI$      & \citet{2021JKAS...54...89I} \\
SOAO\footnote{Sobaeksan Optical Astronomy Observatory}        & Korea        & 0.61m            & 17.6' $\times$ 17.6'                        & $UBVRI$     & - \\
DOAO\footnote{Deokheung Optical Astronomy Observatory}        & Korea        & 1m               & 13.2' $\times$ 17.6'                        & $BVRI$      & \citet{2021JKAS...54...89I} \\
MAAO\footnote{Miryang Arirang Astronomical Observatory}        & Korea        & 1m               & 28.2' $\times$ 28.2'                        & $UBVRI$      & Gu Lim et al., in preparation \\
DNSM\footnote{Daegu National Science Museum, newly added in O4}        & Korea        & 1m               & 15.0' $\times$ 15.0'                        & $BVRI$      & -\\
KHAO\footnote{Kyunghee Astronomical Observatory, newly added in O4}        & Korea        & 0.8m               & 23.7' $\times$ 23.7'                        & $BVRI$      & \citet{2021JKAS...54...89I}\\
        & Korea        & 0.4m               & 21.0' $\times$ 16'                        & $BVRI$      & \citet{2021JKAS...54...89I}\\
CBNUO\footnote{Chungbuk National University Observatory, newly added in O4}        & Korea        & 0.6m               & 1.19 $\times$ 1.19 sq. deg                        & $BVR$      & \citet{2021JKAS...54...89I}\\
LOAO\footnote{Lemmonsan Optical Astronomy Observatory}        & Arizona,USA  & 1m               & 28.1' $\times$ 28.1'                        & $UBVRI$     & \citet{2010JKAS...43...95L} \\
SQUEAN      & Texas,USA    & 2.1m             & 4.7' $\times$ 4.7'                          & $i$, medium & \citet{2016JKAS...49...25J} \\
McD30inch   & Texas,USA    & 0.8m             & 46.2' $\times$ 46.2'                        & $UBVRI$     & - \\
CCA250      & Texas,USA    & 0.25m            & 2.34 $\times$ 2.34 sq. deg              & $V$, medium & \citet{2021ApJ...908..113H} \\
LSGT\footnote{Lee Sang Gak Telescope}        & Australia    & 0.43m            & 15.7' $\times$ 15.7'                        & $ugriz$     & \citet{2015JKAS...48..207I, 2017JKAS...50...71C} \\
KMTNet-SAAO & South Africa & 1.6m             & 1 $\times$ 1 sq. deg ($\times$ 4 chips)       & $BVRI$      & \citet{2016JKAS...49...37K} \\
KMTNet-SSO  & Australia    & 1.6m             & 1 $\times$ 1 sq. deg ($\times$ 4 chips)       & $BVRI$      & - \\
KMTNet-CTIO & Chile        & 1.6m             & 1 $\times$ 1 sq. deg ($\times$ 4 chips)       & $BVRI$      & - \\
KCT\footnote{KIAS Chamnun Telescope, newly added in O4}         & Chile        & 0.36m            & 23.7' $\times$ 18.0'                        & $UBVRI$     & Gu Lim et al., in preparation \\
RASA36\footnote{newly added in O4}& Chile        & 0.36m            & 2.67 $\times$ 2.67 sq.deg              & $r$         & \citet{2021JKAS...54...89I} \\
MAO\footnote{Maidanak Astronomical Observatory}         & Uzbekistan   & 1.5m             & 18.1' $\times$ 18.1'                        & $BVRI$      & \citet{2010JKAS...43...75I} \\
7DT\footnote{7-Dimensional Telescope (in construction)}         & Chile        & 0.5m $\times$ 20 & 1.36 $\times$ 0.90 sq. deg            & medium ($\rm \Delta \lambda = 250 \AA$) & - \\
\enddata 
\end{deluxetable}
\end{longrotatetable}


\begin{deluxetable*}{cccccccc}
\tablecaption{Host galaxy candidates for GW170817\label{tab:gw170817}}
\tablewidth{0pt} 

\tablehead{\colhead{Rank} & \colhead{Name} & \colhead{R.A.} & \colhead{Dec.} & \colhead{K} & \colhead{Dist.} & \colhead{Score} & \colhead{Cumulative Score}\\ \colhead{ } & \colhead{ } & \colhead{J2000} & \colhead{J2000} & \colhead{ABmag} & \colhead{$\mathrm{Mpc}$} & \colhead{ } & \colhead{ }}
\small 
\startdata
1  & NGC4993 & 13:09:47.70 & $-$23:23:01.7 & 9.33 & 39.35 & 1.857$\times$$10^{-1}$ & 0.1857 \\
2  & NGC5061 & 13:18:05.05 & $-$26:50:13.9 & 7.37 & 20.32 & 1.014$\times$$10^{-1}$ & 0.2871 \\
3  & IC4197 & 13:08:04.32 & $-$23:47:48.6 & 9.35 & 41.06 & 9.189$\times$$10^{-2}$ & 0.3790 \\
4  & NGC5078 & 13:19:50.02 & $-$27:24:36.2 & 7.15 & 21.95 & 8.616$\times$$10^{-2}$ & 0.4651 \\
5  & NGC5124 & 13:24:50.27 & $-$30:18:27.9 & 9.00 & 59.08 & 7.741$\times$$10^{-2}$ & 0.5425 \\
6  & NGC4970 & 13:07:33.75 & $-$24:00:30.9 & 9.17 & 47.07 & 7.059$\times$$10^{-2}$ & 0.6131 \\
7  & NGC4756 & 12:52:52.62 & $-$15:24:47.8 & 9.20 & 57.42 & 5.419$\times$$10^{-2}$ & 0.6673 \\
8  & NGC4763 & 12:53:27.23 & $-$17:00:19.7 & 9.70 & 56.57 & 5.041$\times$$10^{-2}$ & 0.7177 \\
9  & NGC4968 & 13:07:05.97 & $-$23:40:37.3 & 9.80 & 40.49 & 4.362$\times$$10^{-2}$ & 0.7614 \\
10 & PGC046026 & 13:14:17.74 & $-$26:34:57.6 & 10.91 & 58.16 & 3.526$\times$$10^{-2}$ & 0.7966 \\
11 & ESO508-033 & 13:16:23.22 & $-$26:33:41.5 & 10.88 & 47.93 & 3.462$\times$$10^{-2}$ & 0.8312 \\
12 & IC4180 & 13:06:56.51 & $-$23:55:01.4 & 9.69 & 40.67 & 2.704$\times$$10^{-2}$ & 0.8583 \\
13 & ESO508-010 & 13:07:37.73 & $-$23:34:44.1 & 11.30 & 45.19 & 2.220$\times$$10^{-2}$ & 0.8805 \\
14 & 2MASS+13104593-2351566 & 13:10:45.93 & $-$23:51:56.6 & 11.62 & 38.73 & 1.840$\times$$10^{-2}$ & 0.8989 \\
15 & IC0874 & 13:19:00.52 & $-$27:37:42.5 & 9.77 & 24.79 & 1.496$\times$$10^{-2}$ & 0.9138 \\
16 & ESO575-053 & 13:05:04.93 & $-$22:23:02.2 & 11.26 & 33.71 & 1.415$\times$$10^{-2}$ & 0.9280 \\
17 & PGC046803 & 13:23:42.97 & $-$30:03:24.3 & 11.14 & 58.45 & 1.198$\times$$10^{-2}$ & 0.9400 \\
18 & PGC043908 & 12:54:28.86 & $-$16:21:02.9 & 11.54 & 57.36 & 1.003$\times$$10^{-2}$ & 0.9500 \\
19 & PGC044234 & 12:57:00.54 & $-$17:19:13.4 & 12.40 & 57.90 & 9.029$\times$$10^{-3}$ & 0.9590 \\
20 & 2MASXJ13242754-3025548 & 13:24:27.54 & $-$30:25:54.8 & 11.43 & 55.51 & 7.103$\times$$10^{-3}$ & 0.9661 \\
21 & PGC043823 & 12:53:42.31 & $-$15:16:55.7 & 11.55 & 56.95 & 6.523$\times$$10^{-3}$ & 0.9726 \\
22 & PGC183552 & 13:07:37.68 & $-$23:56:18.1 & 12.50 & 55.86 & 5.778$\times$$10^{-3}$ & 0.9784 \\
23 & ESO508-003 & 13:06:24.01 & $-$24:09:50.4 & 11.94 & 53.47 & 3.200$\times$$10^{-3}$ & 0.9816 \\
24 & ESO444-026 & 13:24:29.04 & $-$30:25:54.0 & 12.60 & 58.01 & 2.630$\times$$10^{-3}$ & 0.9843 \\
25 & ESO444-021 & 13:23:30.64 & $-$30:06:51.3 & 13.18 & 58.87 & 1.857$\times$$10^{-3}$ & 0.9861 \\
26 & IC0879 & 13:19:40.56 & $-$27:25:44.2 & 11.08 & 19.13 & 1.755$\times$$10^{-3}$ & 0.9879 \\
27 & PGC797164 & 13:08:42.48 & $-$23:46:32.6 & 13.83 & 36.55 & 1.647$\times$$10^{-3}$ & 0.9895 \\
28 & PGC043664 & 12:52:25.68 & $-$15:31:01.9 & 12.85 & 50.90 & 1.612$\times$$10^{-3}$ & 0.9911 \\
29 & ESO576-003 & 13:10:35.72 & $-$21:44:53.6 & 12.12 & 36.68 & 1.166$\times$$10^{-3}$ & 0.9923 \\
30 & PGC044500 & 12:58:45.67 & $-$17:32:34.3 & 14.28 & 53.76 & 1.144$\times$$10^{-3}$ & 0.9934 \\
31 & ESO508-019 & 13:09:51.84 & $-$24:14:21.7 & 14.30 & 39.47 & 1.065$\times$$10^{-3}$ & 0.9945 \\
32 & PGC83601 & 12:51:35.18 & $-$15:19:27.2 & 13.54 & 59.36 & 8.894$\times$$10^{-4}$ & 0.9954 \\
33 & PGC141607 & 13:25:37.68 & $-$30:09:39.1 & 13.76 & 56.35 & 8.087$\times$$10^{-4}$ & 0.9962 \\
34 & ESO508-024 & 13:10:45.84 & $-$23:51:56.2 & 14.40 & 29.06 & 8.015$\times$$10^{-4}$ & 0.9970 \\
35 & PGC044023 & 12:55:20.40 & $-$17:05:47.4 & 15.11 & 53.88 & 5.621$\times$$10^{-4}$ & 0.9976 \\
36 & PGC141609 & 13:25:41.28 & $-$30:16:37.3 & 14.30 & 56.51 & 4.971$\times$$10^{-4}$ & 0.9981 \\
37 & PGC141593 & 13:24:14.40 & $-$30:15:21.7 & 14.46 & 56.33 & 4.862$\times$$10^{-4}$ & 0.9985 \\
38 & PGC141595 & 13:24:17.04 & $-$30:14:50.1 & 14.65 & 59.36 & 4.541$\times$$10^{-4}$ & 0.9990 \\
39 & PGC910856 & 12:52:54.48 & $-$15:28:06.6 & 14.92 & 58.09 & 2.930$\times$$10^{-4}$ & 0.9993 \\
40 & PGC908166 & 12:51:48.96 & $-$15:39:34.8 & 15.02 & 58.44 & 2.809$\times$$10^{-4}$ & 0.9996 \\
41 & [BZZ2000]J132249.66-300651.8 & 13:22:49.68 & $-$30:06:51.5 & 15.56 & 59.04 & 2.161$\times$$10^{-4}$ & 0.9998 \\
42 & PGC043505 & 12:51:00.48 & $-$15:40:49.0 & 14.92 & 57.30 & 2.101$\times$$10^{-4}$ & 1.0000
\enddata
\end{deluxetable*}


\begin{longrotatetable}
\begin{deluxetable}{cccccccc}
\tablecaption{GECKO Observation of GW190425 \label{tab:mediandepth}}
\tablewidth{0pt}
\tablehead{\colhead{Observatory} & \colhead{Filter} & \colhead{$\Delta$t\footnote{Median value of the time difference between GW alert and observation time.}} & \colhead{Date-obs\footnote{Median value}} & \colhead{Depth\footnote{Median 5$\sigma$ upper limits of observed images}} & \colhead{Exposure Time} & \colhead{Number of Galaxies} & \colhead{Area}\\
           \colhead{ }           & \colhead{ }      & \colhead{days}             & \colhead{UTC}             & \colhead{ABmag}                  & \colhead{seconds}       & \colhead{}                   & \colhead{$\rm deg^{2}$}}

\startdata
SAO & $R$ & 1.261 & 2019-04-26 14:34:04 & 19.75 & 900 & 6 &  \\
LSGT & $R$ & 0.238 & 2019-04-25 14:00:57 & 19.46 & 180 & 0 &  \\
LSGT & $R$ & 1.355 & 2019-04-26 16:50:09 & 18.96 & 180 & 9 &  \\
LOAO & $R$ & 0.109 & 2019-04-25 10:55:12 & 19.83 & 60 & 16 &  \\
LOAO & $R$ & 1.058 & 2019-04-26 09:41:45 & 20.27 & 120 & 8 &  \\
LOAO & $R$ & 2.09 & 2019-04-27 10:27:50 & 20.22 & 120 & 0 &  \\
LOAO & $R$ & 3.109 & 2019-04-28 10:55:12 & 19.11 & 120 & 0 &  \\
KMTNet-SSO & $R$ & 0.221 & 2019-04-25 13:36:28 & 21.11 & 240 & 148 & 106.6 \\
KMTNet-SSO & $R$ & 1.238 & 2019-04-26 14:00:57 & 20.59 & 240 & 76 & 67.5 \\
KMTNet-SSO & $R$ & 2.032 & 2019-04-27 09:04:19 & 20.89 & 240 & 35 & 26.6 \\
KMTNet-CTIO & $R$ & 0.796 & 2019-04-26 03:24:28 & 21.39 & 240 & 178 & 128.1 \\
KMTNet-CTIO & $R$ & 2.631 & 2019-04-27 23:26:52 & 21.71 & 240 & 72 & 61.3 \\
KMTNet-SAAO & $R$ & 0.55 & 2019-04-25 21:30:14 & 21.68 & 240 & 175 & 126.4 \\
KMTNet-SAAO & $R$ & 1.539 & 2019-04-26 21:14:24 & 21.97 & 240 & 21 & 8.8 \\
KMTNet-SAAO & $R$ & 2.37 & 2019-04-27 17:11:02 & 21.76 & 240 & 65 & 43.3 \\
SQUEAN & $i$ & 0.101 & 2019-04-25 10:43:40 & 21.64 & 60-120 & 29 &  \\
SQUEAN & $i$ & 0.979 & 2019-04-26 07:48:00 & 22.32 & 60-180 & 65 &  \\
SQUEAN & $i$ & 1.934 & 2019-04-27 06:43:12 & 21.76 & 180 & 48 &  \\
SQUEAN & $i$ & 3.109 & 2019-04-28 10:55:12 & 22.0 & 180-300 & 11 &  \\
UKIRT & $J$ & 2.11 & 2019-04-27 10:56:38 & 21.1 & 60 & 22 & 4.2 \\
UKIRT & $J$ & 6.088 & 2019-05-01 10:24:57 & 21.31 & 60 & 13 & 1.9 \\
UKIRT & $J$ & 7.129 & 2019-05-02 11:24:00 & 21.23 & 60 & 1 & 0.1 \\
UKIRT & $K$ & 0.291 & 2019-04-25 15:18:00 & 20.62 & 60 & 5 & 1.0 \\
UKIRT & $K$ & 1.065 & 2019-04-26 09:51:50 & 20.86 & 60-100 & 15 & 2.4 \\
UKIRT & $K$ & 1.92 & 2019-04-27 06:23:02 & 20.76 & 60 & 2 & 0.2 \\
\enddata
\end{deluxetable}
\end{longrotatetable}
\begin{longrotatetable}
\begin{deluxetable}{ccccccccc}

\tablecaption{Photometry Table of Transients from GECKO\label{tab:transient}}
\tablewidth{700pt}

\tablehead{\colhead{Object} & \colhead{Observatory} & \colhead{Filter} & \colhead{Date-obs} & \colhead{Median $\Delta$t} & \colhead{R.A.} & \colhead{Dec.} & \colhead{Mag.} & \colhead{Mag. error}\\ \colhead{ } & \colhead{ } & \colhead{ } & \colhead{UTC} & \colhead{days} & \colhead{ } & \colhead{ } & \colhead{ABmag} & \colhead{ABmag}}

\startdata
AT 2019dta & KMTNet-CTIO & $R$ & 2019-04-26 03:11:02 & 0.787 & 16:42:54.80 & $-$19:41:22.3 & 19.73 & 0.06 \\
AT 2019dnv & KMTNet-SAAO & $R$ & 2019-04-25 21:38:33 & 0.556 & 16:56:08.22 & $-$08:11:54.6 & 19.32 & 0.05 \\
AT 2019dnv & KMTNet-CTIO & $R$ & 2019-04-26 03:31:59 & 0.801 & 16:56:08.23 & $-$08:11:54.7 & 19.18 & 0.04 \\
AT 2019dws & KMTNet-SSO  & $R$ & 2019-04-25 13:12:38 & 0.204 & 16:48:11.15 & $+$01:04:00.6 & 19.06 & 0.06 \\
AT 2019flz & KMTNet-SAAO & $R$ & 2019-04-27 17:26:17 & 2.380 & 05:45:20.45 & $-$26:50:51.0 & 18.53 & 0.02 \\
AT 2019hae & KMTNet-SAAO & $R$ & 2019-04-25 21:16:51 & 0.541 & 16:55:14.93 & $-$17:52:54.2 & 18.06 & 0.03 \\
AT 2019hae & KMTNet-CTIO & $R$ & 2019-04-26 03:16:57 & 0.791 & 16:55:14.94 & $-$17:52:54.2 & 18.09 & 0.03 \\
AT 2019ocg & KMTNet-CTIO & $R$ & 2019-04-26 03:16:57 & 0.791 & 16:52:45.01 & $-$19:05:38.8 & 20.11 & 0.05 \\
SN 2019dzk & SQUEAN      & $i$ & 2019-04-26 08:47:02 & 1.02 & 17:13:29.76  & $-$09:56:31.2 & 18.45 & 0.05 \\
SN 2019dzw & KMTNet-SAAO & $R$ & 2019-04-25 22:10:28 & 0.578 & 17:31:09.96 & $-$08:27:02.8 & 19.94 & 0.05 \\
SN 2019dzw & SQUEAN      & $i$ & 2019-04-26 08:38:24 & 1.014 & 17:31:09.60 & $-$08:27:54.0 & 19.57 & 0.04 \\
PS19qp     & UKIRT       & $K$ & 2019-04-26 15:09:15 & 1.285 & 17:01:18.33 & $-$07:00:10.1 & 20.26 & 0.13 \\
\textbf{GECKO190427a} & KMTNet-SAAO & $R$ & 2019-04-27 17:12:57 & 2.371 & 05:03:39.04 & $-$23:47:38.8 & 20.29 & 0.05 \\
\enddata

\end{deluxetable}
\end{longrotatetable}
\begin{deluxetable}{ccc}
\tablecaption{Photometric results of the host galaxy of GECKO190427a\label{tab:GECKO190427a}}
\tablewidth{0pt}

\tablehead{\colhead{Filter} & \colhead{Mag.} & \colhead{Mag. error}}
\startdata
Pan-STARRS $g$ & 18.62 & 0.03 \\
Pan-STARRS $r$ & 18.03 & 0.03 \\
Pan-STARRS $i$ & 17.71 & 0.02 \\
Pan-STARRS $z$ & 17.56 & 0.02 \\
Pan-STARRS $y$ & 17.36 & 0.03 \\
2MASS $J$ & 16.97 & 0.1 \\
2MASS $H$ & 16.96 & 0.1 \\
WISE $W1$ & 17.74 & 0.03 \\
WISE $W2$ & 18.02 & 0.05
\enddata
\tablecomments{Apparent magnitudes are corrected for the Galactic extinction.}
\end{deluxetable}


\end{document}